\documentclass{article}
\usepackage{my_style}
\newsavebox{\figboxA}\newsavebox{\figboxB}
\usetikzlibrary{fadings,shadings}
\begin{document}
    
\title{Topological phases and extendible states}
\author[1]{Isaac H.~Kim}
\author[2]{Alexei Kitaev}
\author[2]{Daniel Ranard}
\affil[1]{Department of Computer Science, University of California, Davis}
\affil[2]{California Institute of Technology, Pasadena, California}

\date{}
\maketitle
\begin{abstract}
    Gapped lattice Hamiltonians often exhibit topological properties at low energy, described by topological quantum field theory.  However, this connection is difficult to make precise.  Part of the difficulty is that not all gapped systems actually behave this way, with fractons providing a counterexample. We present locally testable conditions on the ground state or Hamiltonian that guarantee topological properties.  These conditions distinguish ordinary topological phases from systems like fractons, offering a candidate definition for a genuine topological phase. We find the following conditions are closely related, or in some cases equivalent: (1) the ability to extend local patches of the state using quantum channels, (2) the existence of the state on certain immersed manifolds, (3) the indistinguishability of zero-energy states on a ball, and (4) the mobility of localized excitations. We take the first, local extendibility, as our starting point. This condition is invariant under constant-depth circuits and provides a more flexible axiom for the recent entanglement bootstrap program. As one application, we build efficient quantum circuits that prepare the ground states of commuting Hamiltonians satisfying the third condition. We also initiate the use of our framework to extract topological invariants.  For instance, we propose a circuit-invariant definition of topological entanglement entropy. We leave the approximate setting to future work. 
\end{abstract}
\tableofcontents

\section{Introduction}
\label{sec:intro}
When a quantum system with local interactions has an energy gap, the low-energy physics is often ``topological'': various features depend only on topology.  In this case the ground state degeneracy depends only on the system's spatial topology, and isolated excitations obey braiding and fusion rules that depend only on the topology of their trajectories. Accordingly, a quantum phase of matter at zero temperature is often described by a topological quantum field theory~\cite{witten1988topological}.  

However, the connection between condensed matter systems and topological quantum field theory is difficult to make precise.  Part of the difficulty is that not all gapped lattice systems exhibit topological properties at low energy.  For instance, the system may be inhomogeneous, with a Hamiltonian that varies abruptly over space, so that it cannot be described by a single topological theory. One naive solution is to restrict attention to translation-invariant systems, but a few issues arise.  First, translation symmetry is awkward to formulate without a regular lattice, or on manifolds of non-trivial topology that may not admit a translation group.  More importantly, translation invariance does \textit{not} guarantee purely topological properties at low energy, with counterexamples found in three or more spatial dimensions. The known counterexamples are called ``fracton'' models \cite{pretko2020fracton}, characterized by excitations that are immobile or may only move in certain directions. They exhibit some limited topological features \cite{shirley2018fracton}, but they are not described by a topological field theory; for instance, their ground space degeneracy is sensitive to geometry.

The purpose of this paper is to offer conditions on a lattice Hamiltonian or ground state that indicate a genuine topological phase.  These conditions will rule out both fractons and systems with inhomogeneities like domain walls.  We hope the resulting class of systems captures all phases of matter that one might call ``topological.'' Once properly delineated, these phases may ultimately prove amenable to classification by topological quantum field theory.  

Recent work made progress toward these goals, using a set of techniques called the entanglement bootstrap \cite{shi2020fusion, shi2021entanglement,huang2023knots}. That work considers quantum states satisfying certain strict conditions on the entanglement entropy. However, their assumptions on entanglement entropies are overly restrictive in two ways.  First, they demand exact equalities that only hold precisely for toy models. The distinction between approximate and exact conditions can be physically meaningful, especially when discussing chiral topological phases, where certain exact conditions are necessarily violated for all systems \cite{lemm2019spectral, kim2024strict}.  Second, the assumptions of Ref.~\cite{shi2020fusion} may also fail more dramatically: they are violated with large error even for some states in the trivial phase. More generally, the conditions of Ref.~\cite{shi2020fusion} may always be broken by constant-depth unitary circuits, even though the topological phase is unchanged.  Thus they do not hold for all realizations of a topological phase.  These violations are related to the phenomenon of ``spurious'' topological entanglement entropy \cite{Zou2016,Williamson2019,Kato2020,kim2023universal}.  

In the present work, we remedy the above situation by initiating a framework that is invariant under constant-depth circuits.  Our discussion applies to lattice systems of any dimension, though we mainly illustrate results in two spatial dimensions.  We will not discuss the small errors that arise when our assumptions are not exactly satisfied.  We expect to address the approximate case in future work. Here we focus on the physical ideas and some rigorous results in the simplified setting with zero error.

In particular, we study quantum states that we call locally purifiable and locally extendible (Section~\ref{sec:formal-intro}), a variant of the conditions considered in Ref.~\cite{shi2020fusion}. We begin with a set of locally consistent density matrices that satisfy local extendibility, adding local purifiability where it is needed, and derive their consequences. We prove that the existence of a globally consistent state follows from this local data.  The global state satisfies many of the properties discussed in Ref.~\cite{shi2020fusion}, but in a form that is invariant under constant-depth circuits. In particular, we show the isomorphism theorem of information convex sets~\cite{shi2020fusion}, which implies anyon mobility (Section~\ref{sec:mobility}) and suggests the homogeneity of the underlying topological phase. We also develop a method for gluing density matrices defined over different regions, which serves the purpose of the merging theorem of Ref.~\cite{shi2020fusion} in a form that is robust to constant-depth circuits. We show that for the special case of commuting local Hamiltonians, a variant of local topological quantum order condition~\cite{michalakis2013stability} implies the ground state satisfies our conditions (Section~\ref{sec:extendibility_ltqo}). Moreover, such ground states can be prepared by efficient quantum circuits (Subsection~\ref{subsec:sequential_circuit}). (In a separate work, some of us have presented an application to the quantum circuit learning problem~\cite{kim2024learning}.)  Finally, we define a quantity invariant under constant-depth quantum circuits, which reveals the same information as intended by the topological entanglement entropy~\cite{Kitaev2006,Levin2006}, and therefore may be viewed as a proposed circuit-invariant definition of the topological entanglement entropy (Section \ref{sec:tee_invariant}). In future work, we will pursue questions about the classification of topological phases within this framework.

\section{Conceptual overview} \label{sec:overview}
We begin with some heuristic discussion on what makes a gapped phase of matter genuinely ``topological.''
Recall that two gapped Hamiltonians are said to be in the same phase if there exists a continuous path of gapped Hamiltonians from one to the other~\cite{bachmann2012automorphic}. In a roughly equivalent formulation \cite{hastings2005quasiadiabatic, chen2010local}, two gapped ground states are said to be in the same phase if there exists a constant-depth, ``quasi-local'' circuit that approximately connects them.  For simplicity, we will not be concerned with quasi-locality, referring mainly to constant-depth circuits with strict locality.  
Let us examine the notion of homogeneity without translation invariance. Here we use ``homogeneity'' to mean roughly ``all spatial regions are in the same phase,'' contrasted with an inhomogeneous patchwork of distinct phases.

First consider a translation-invariant system in a non-trivial topological phase, such as the 2D toric code on the plane.  A small extensive perturbation of the Hamiltonian will not close the gap \cite{bravyi2011short}, including when the perturbation is disordered or spatially varying.  Then the system becomes non-translation-invariant, but we might say it remains ``homogeneous'': the entire plane remains in the same topological phase.  Relatedly, the excitations are mobile; there exist string operators to move the anyons throughout the plane. Rather than a small perturbation, we can also imagine applying an arbitrary, non-translation-invariant constant-depth circuit to the ground state. Then while the state looks quite different in each region, the topological phase again remains homogeneous.

On the other hand, if the Hamiltonian is sufficiently perturbed on a single half-plane, the system may remain gapped but exhibit distinct topological phases on each half, separated by a gapped domain wall that restricts the movement of some anyons. 
More generally, the system may be littered with domain walls, with different patches in various phases. Then we say the system is inhomogeneous. 

\begin{figure}[t]
    \centering
    \tikzset{every picture/.style={line width=0.75pt}} 

\begin{tikzpicture}[x=0.75pt,y=0.75pt,yscale=-1,xscale=1]

\draw  [fill=tikzblue  ,fill opacity=1 ] (44.14,80.83) -- (105.94,80.83) .. controls (109.75,80.83) and (112.83,83.92) .. (112.83,87.73) -- (112.83,101.27) .. controls (110.82,102.7) and (109.5,105.05) .. (109.5,107.71) .. controls (109.5,110.37) and (110.82,112.72) .. (112.83,114.14) -- (112.83,114.33) -- (128.67,114.33) .. controls (132.58,114.33) and (135.75,117.5) .. (135.75,121.41) -- (135.75,128.92) .. controls (135.75,132.83) and (132.58,136) .. (128.67,136) -- (112.83,136) -- (112.83,136.16) .. controls (111.1,137.6) and (110,139.78) .. (110,142.21) .. controls (110,144.64) and (111.1,146.81) .. (112.83,148.26) -- (112.83,160.94) .. controls (112.83,164.75) and (109.75,167.83) .. (105.94,167.83) -- (44.14,167.83) .. controls (40.34,167.83) and (37.25,164.75) .. (37.25,160.94) -- (37.25,87.73) .. controls (37.25,83.92) and (40.34,80.83) .. (44.14,80.83) -- cycle ;
\draw  [fill=tikzblue  ,fill opacity=0 ][dash pattern={on 0.75pt off 0.75pt}] (108.08,100.83) -- (112.83,100.83) -- (112.83,101.08) .. controls (110.82,102.51) and (109.5,104.86) .. (109.5,107.52) .. controls (109.5,110.18) and (110.82,112.53) .. (112.83,113.95) -- (112.83,114.14) -- (128.67,114.14) .. controls (132.58,114.14) and (135.75,117.31) .. (135.75,121.22) -- (135.75,128.73) .. controls (135.75,132.64) and (132.58,135.81) .. (128.67,135.81) -- (112.83,135.81) -- (112.83,135.97) .. controls (111.1,137.41) and (110,139.59) .. (110,142.02) .. controls (110,144.45) and (111.1,146.62) .. (112.83,148.07) -- (112.83,148.33) -- (108.08,148.33) .. controls (105.69,148.33) and (103.75,146.39) .. (103.75,144) -- (103.75,105.17) .. controls (103.75,102.77) and (105.69,100.83) .. (108.08,100.83) -- cycle ;
\draw    (107.75,125.33) -- (129.75,125.33) ;
\draw [shift={(131.75,125.33)}, rotate = 180] [color={rgb, 255:red, 0; green, 0; blue, 0 }  ][line width=0.75]    (10.93,-3.29) .. controls (6.95,-1.4) and (3.31,-0.3) .. (0,0) .. controls (3.31,0.3) and (6.95,1.4) .. (10.93,3.29)   ;
\draw  [fill=tikzblue  ,fill opacity=0.94 ] (255.61,95.11) .. controls (269.35,108.71) and (269.48,130.87) .. (255.89,144.61) .. controls (252.18,148.35) and (247.85,151.09) .. (243.22,152.81) -- (243.22,152.83) -- (227.53,159.56) .. controls (222.33,161.8) and (216.26,159.46) .. (213.98,154.35) .. controls (211.7,149.23) and (214.08,143.27) .. (219.29,141.04) -- (236.08,133.83) .. controls (238.05,133.11) and (239.9,131.95) .. (241.47,130.36) .. controls (247.19,124.57) and (247.14,115.25) .. (241.36,109.53) .. controls (238.36,106.57) and (234.42,105.15) .. (230.52,105.28) -- (229.86,85.02) .. controls (239.13,84.72) and (248.5,88.08) .. (255.61,95.11) -- cycle ;
\draw  [fill=tikzblue  ,fill opacity=0.94 ] (207.58,95.13) .. controls (194.8,108.72) and (194.68,130.88) .. (207.33,144.63) .. controls (210.77,148.37) and (214.81,151.11) .. (219.12,152.83) -- (219.11,152.84) -- (233.71,159.58) .. controls (238.56,161.81) and (244.2,159.48) .. (246.32,154.37) .. controls (248.44,149.25) and (246.23,143.29) .. (241.38,141.06) -- (225.76,133.85) .. controls (223.92,133.12) and (222.2,131.97) .. (220.74,130.38) .. controls (215.42,124.59) and (215.46,115.27) .. (220.85,109.55) .. controls (223.63,106.58) and (227.3,105.17) .. (230.93,105.3) -- (231.55,85.04) .. controls (222.92,84.74) and (214.2,88.1) .. (207.58,95.13) -- cycle ;
\draw  [draw opacity=0][fill=tikzblue  ,fill opacity=1 ] (226.31,86.08) .. controls (229.56,85.58) and (232.84,85.56) .. (236.04,86) -- (232.56,105.01) .. controls (231.23,104.91) and (229.87,105) .. (228.52,105.28) -- cycle ;
\draw  [draw opacity=0][fill=tikzblue  ,fill opacity=1 ] (334.5,90.72) .. controls (334.5,87.01) and (337.51,84) .. (341.22,84) -- (419.7,84) .. controls (423.41,84) and (426.42,87.01) .. (426.42,90.72) -- (426.42,159.45) .. controls (426.42,163.16) and (423.41,166.17) .. (419.7,166.17) -- (341.22,166.17) .. controls (337.51,166.17) and (334.5,163.16) .. (334.5,159.45) -- cycle ;
\draw  [color={rgb, 255:red, 0; green, 0; blue, 0 }  ,draw opacity=1 ][fill={rgb, 255:red, 0; green, 0; blue, 0 }  ,fill opacity=0.13 ][line width=0.75]  (352.71,105.77) .. controls (355.53,103.25) and (371.35,93.05) .. (367.76,103.34) .. controls (364.17,113.63) and (366.4,115.38) .. (371.76,123.21) .. controls (377.11,131.04) and (361.06,132.56) .. (354.28,123.21) .. controls (347.5,113.86) and (349.9,108.29) .. (352.71,105.77) -- cycle ;
\draw  [color={rgb, 255:red, 0; green, 0; blue, 0 }  ,draw opacity=1 ][line width=0.75]  (350.28,103.77) .. controls (354.29,100.69) and (376.81,88.23) .. (371.7,100.8) .. controls (366.59,113.38) and (369.76,115.51) .. (377.39,125.08) .. controls (385.01,134.64) and (362.16,136.5) .. (352.52,125.08) .. controls (342.87,113.65) and (346.27,106.85) .. (350.28,103.77) -- cycle ;
\draw  [draw opacity=0][fill=tikzblue  ,fill opacity=1 ] (506.5,90.72) .. controls (506.5,87.01) and (509.51,84) .. (513.22,84) -- (591.7,84) .. controls (595.41,84) and (598.42,87.01) .. (598.42,90.72) -- (598.42,159.45) .. controls (598.42,163.16) and (595.41,166.17) .. (591.7,166.17) -- (513.22,166.17) .. controls (509.51,166.17) and (506.5,163.16) .. (506.5,159.45) -- cycle ;
\draw [color={rgb, 255:red, 220; green, 116; blue, 116 }  ,draw opacity=0.41 ][line width=3]    (520.32,138.44) -- (530.13,147.94) ;
\draw [color={rgb, 255:red, 220; green, 116; blue, 116 }  ,draw opacity=0.41 ][line width=3]    (520.47,147.78) -- (529.66,138.44) ;

\draw [color={rgb, 255:red, 221; green, 38; blue, 38 }  ,draw opacity=0.8 ][line width=3]    (575.32,94.44) -- (585.13,103.94) ;
\draw [color={rgb, 255:red, 221; green, 38; blue, 38 }  ,draw opacity=0.8 ][line width=3]    (575.47,103.78) -- (584.66,94.44) ;

\draw  [draw opacity=0] (524.93,135.49) .. controls (526.17,130.28) and (528.33,125.19) .. (531.45,120.46) .. controls (540.3,107.06) and (554.82,99.65) .. (569.75,99.3) -- (570.8,146.45) -- cycle ; \draw   (524.93,135.49) .. controls (526.17,130.28) and (528.33,125.19) .. (531.45,120.46) .. controls (540.3,107.06) and (554.82,99.65) .. (569.75,99.3) ;  
\draw    (568.75,99.3) -- (572.27,98.79) ;
\draw [shift={(574.25,98.5)}, rotate = 171.7] [color={rgb, 255:red, 0; green, 0; blue, 0 }  ][line width=0.75]    (10.93,-3.29) .. controls (6.95,-1.4) and (3.31,-0.3) .. (0,0) .. controls (3.31,0.3) and (6.95,1.4) .. (10.93,3.29)   ;

\draw (23.5,29.5+13) node [anchor=north west][inner sep=0.75pt]    {Local extendibility};
\draw (191,28+13) node [anchor=north west][inner sep=0.75pt]    {$ \begin{array}{l}
\textrm{Existence of}\\
\textrm{immersions}
\end{array}$};
\draw (507.5,27+13) node [anchor=north west][inner sep=0.75pt]    {$ \begin{array}{l}
\textrm{Mobility of} \\
\textrm{excitations}
\end{array}$};
\draw (320,27.5+13) node [anchor=north west][inner sep=0.75pt]    {$ \begin{array}{l}
\textrm{Unique ground state} \\
\textrm{on balls (LTQO)}
\end{array}$};
\draw (61.5,172) node [anchor=north west][inner sep=0.75pt]    {(a)};
\draw (220,171.5) node [anchor=north west][inner sep=0.75pt]    {(b)};
\draw (367.5,170.5) node [anchor=north west][inner sep=0.75pt]    {(c)};
\draw (542.5,171.5) node [anchor=north west][inner sep=0.75pt]    {(d)};

\end{tikzpicture}
    \caption{These conditions are loosely equivalent (subject to further assumptions) and each guarantees a homogeneous topological phase. \textbf{(a)} Local extendibility requires the existence of local channels that can extend a patch of the state in any direction. \textbf{(b)} The existence of immersions refers to the existence of a new state on an immersed region that locally matches the original state. \textbf{(c)} Our formulation of LTQO requires that for every topological ball $A$, every ground state of the Hamiltonian $H_A$ must have identical reduced state on the interior of $A$.  
    LTQO guarantees homogeneity for systems on a sphere, otherwise it must be supplemented with a complementary condition (Definition \ref{definition:spherical_ltqo}).  \textbf{(d)} The mobility of excitations means that localized excitations can be moved in any direction with local operations.}
    \label{fig:homogeneity-conditions}
\end{figure}
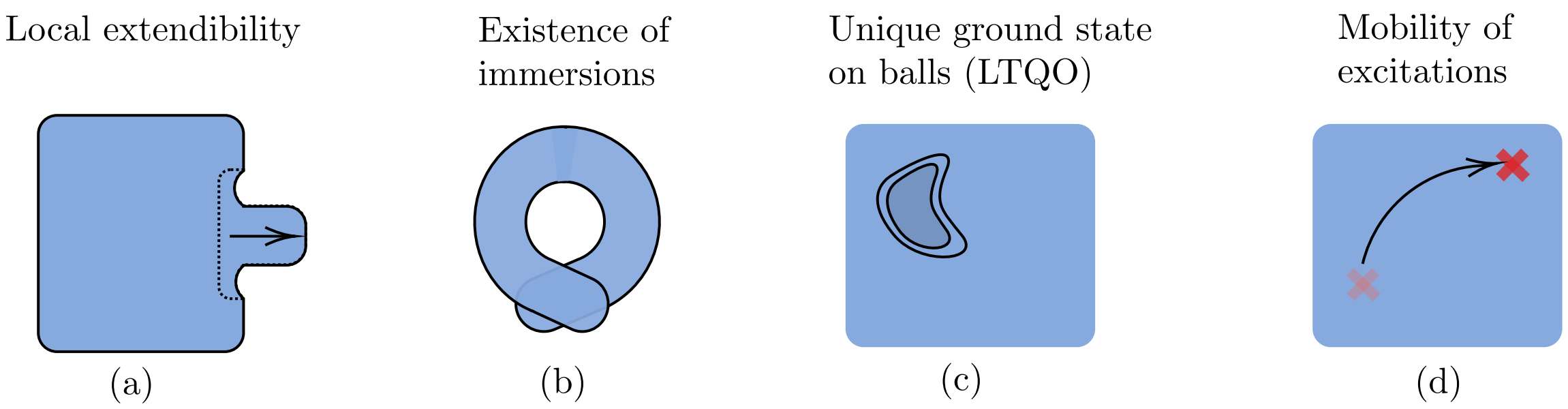

Given a gapped ground state without translation invariance, how can we distinguish the inhomogeneous case from the homogeneous case? This paper explores several related conditions for homogeneous topological phases.  These conditions are loosely equivalent, though some of the implications are subject to further assumptions or caveats.
The conditions are (a) local extendibility, (b) existence of immersions, (c) unique ground state on balls, or local topological quantum order (LTQO), and (d) mobility of excitations.  
A cartoon and brief description of each appears in Fig.~\ref{fig:homogeneity-conditions}.  

Ultimately, local extendibility will constitute our starting point in Section~\ref{sec:formal-intro}, but for the present conceptual overview, we treat the notions in Fig.~\ref{fig:homogeneity-conditions} on equal footing.  
These conditions refer to the state or Hamiltonian, though they may be re-formulated to refer only to the state.  They are satisfied by the product state, or any state obtained from a product state by a constant-depth circuit: they are not meant to imply \textit{non-trivial} topological order, but merely a homogeneous topological phase, including the trivial phase.  

We begin by discussing the property of local topological quantum order (LTQO), because it is often most straightforward to diagnose inhomogeneity by observing a violation of LTQO. The formulation of LTQO refers to a frustration-free Hamiltonian with zero-energy ground state.  We then require that for all topological balls $A$, 
every ground state of Hamiltonian $H_A$ is indistinguishable on the interior of $A$. In that restricted sense, the ground space on the ball is nondegenerate.  The LTQO condition was first introduced in Ref.~\cite{michalakis2013stability} to prove the stability of the gap, where it was required to hold for geometric balls; crucially, here we demand it to \emph{any} topological ball.

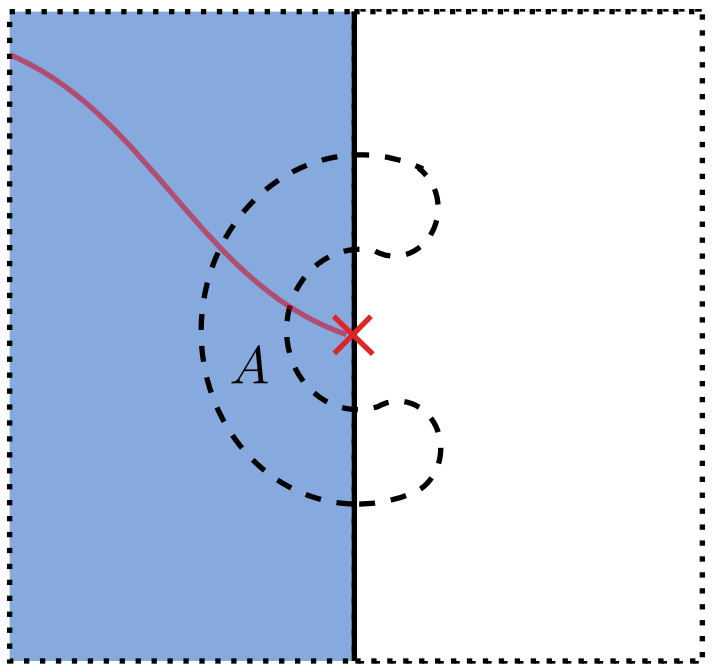
\begin{figure}[htbp]
    \centering
    \tikzset{every picture/.style={line width=1pt}} 

\begin{tikzpicture}[x=0.75pt,y=0.75pt,yscale=-1,xscale=1, line width=1pt]

\draw  [dotted][fill=tikzblue ] (70.5,15) -- (243.5,15) -- (243.5,177.94) -- (70.5,177.94) -- cycle ;
\draw  [dotted][fill=white] (157,15) -- (244.32,15) -- (244.32,177.94) -- (157,177.94) -- cycle ;
\draw    (157,15) -- (157,177.94) ;
\draw  [dashed] (156.5,138.54) .. controls (134.64,137.78) and (117.67,117.56) .. (118.6,93.37) .. controls (119.53,69.19) and (138,50.2) .. (159.86,50.95) .. controls (164.75,51.12) and (169.39,52.27) .. (173.65,54.21) -- (173.5,54.59) .. controls (175.6,56.24) and (177.17,58.72) .. (177.76,61.73) .. controls (179.02,68.18) and (175.31,74.59) .. (169.46,76.06) .. controls (166.91,76.7) and (164.36,76.29) .. (162.21,75.08) .. controls (161.16,74.83) and (160.07,74.69) .. (158.95,74.65) .. controls (148.92,74.3) and (140.44,83.02) .. (140.01,94.12) .. controls (139.59,105.22) and (147.37,114.5) .. (157.41,114.85) .. controls (159.32,114.92) and (161.17,114.66) .. (162.93,114.11) .. controls (163.74,113.64) and (164.63,113.28) .. (165.57,113.04) .. controls (171.41,111.57) and (177.17,115.61) .. (178.43,122.05) .. controls (179.7,128.5) and (175.98,134.92) .. (170.14,136.38) -- (170.14,136.38) -- (170.14,136.38) .. controls (165.85,137.94) and (161.25,138.71) .. (156.5,138.54) -- cycle ;
\draw [color={rgb, 255:red, 221; green, 38; blue, 38 }  ,draw opacity=1 ]   (151.82,91.44) -- (161.63,100.94) ;
\draw [color={rgb, 255:red, 221; green, 38; blue, 38 }  ,draw opacity=1 ]   (151.97,100.78) -- (161.16,91.44) ;

\draw [color={rgb, 255:red, 208; green, 2; blue, 27 }  ,draw opacity=0.55 ]   (70.82,25.94) .. controls (105.82,40.94) and (118.32,83.44) .. (155,96) ;

\draw (124,97.5) node [anchor=north west][inner sep=0.75pt]    {$A$};

\end{tikzpicture}
    \caption{An example of an inhomogeneous system. The left side is the toric code and the right side is a product state. The LTQO condition is violated on a topological ball $A$, because this region can detect a presence of boundary excitation (red cross).}
    \label{fig:boundary-excitation}
\end{figure}

We illustrate how violations of LTQO diagnose the presence of a domain wall.  Imagine the toric code with gapped boundary on the left half-plane, with a product state on the right half-plane, as in Fig.~\ref{fig:boundary-excitation}. 
  This system can be arranged with commuting parent Hamiltonian and unique gapped ground state.
   Consider the region $A$ of Fig.~\ref{fig:boundary-excitation}.  This region is a topological ball, but it violates LTQO.  That is, the Hamiltonian on $A$ has the property that not all states of zero energy on $A$ look identical on the interior of $A$.  For instance, different zero-energy states on $A$ can be obtained by placing a boundary excitation outside of $A$ (shown in red).

While we only impose LTQO for balls, we can gain intuition by considering the analogous property for an annular subregion $A$ of the 2D plane.  When the system is topologically ordered, there exist multiple zero-energy states on annulus $A$ that are distinguishable on the interior of $A$, thus violating LTQO for $A$.  In particular, consider the distinct zero-energy states on $A$ obtained by choosing an anyon to place outside the annulus at its center.  These states are distinguishable on the interior of $A$ because the anyon type is measurable there.

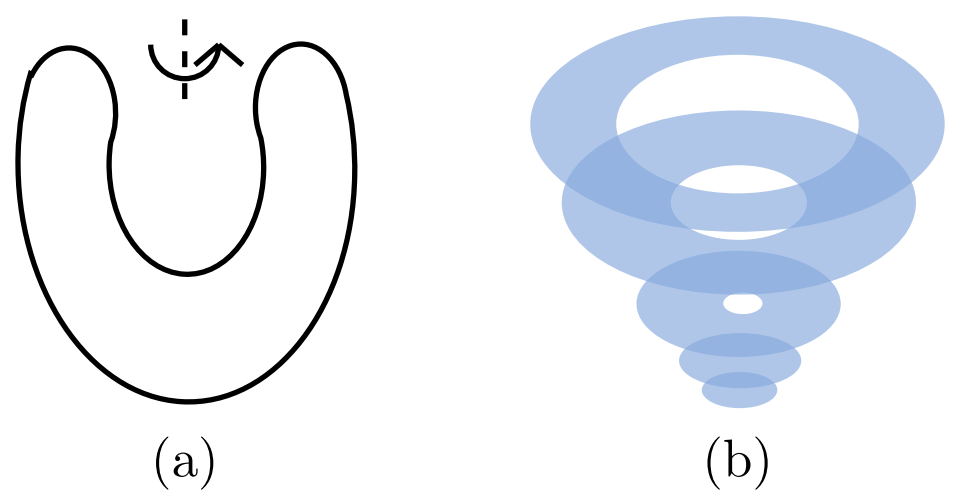
\begin{figure}[t]
    \centering
  \centering
    \tikzset{every picture/.style={line width=1pt}} 

\begin{tikzpicture}[x=0.75pt,y=0.75pt,yscale=-1,xscale=1]

\draw   (116.65,62.99) .. controls (115.96,95.65) and (96.49,120.99) .. (73.16,119.59) .. controls (49.83,118.19) and (31.48,90.58) .. (32.18,57.93) .. controls (32.33,50.62) and (33.43,43.69) .. (35.29,37.33) -- (35.66,37.55) .. controls (37.24,34.42) and (39.64,32.08) .. (42.53,31.2) .. controls (48.75,29.31) and (54.94,34.86) .. (56.37,43.59) .. controls (56.99,47.41) and (56.6,51.21) .. (55.44,54.43) .. controls (55.2,56) and (55.06,57.63) .. (55.03,59.29) .. controls (54.71,74.29) and (63.13,86.96) .. (73.84,87.6) .. controls (84.55,88.24) and (93.48,76.61) .. (93.8,61.62) .. controls (93.86,58.76) and (93.61,55.99) .. (93.08,53.38) .. controls (92.62,52.15) and (92.27,50.83) .. (92.04,49.43) .. controls (90.61,40.7) and (94.49,32.09) .. (100.71,30.21) .. controls (106.92,28.33) and (113.12,33.87) .. (114.54,42.6) -- (114.54,42.6) -- (114.54,42.6) .. controls (116.05,49.01) and (116.8,55.88) .. (116.65,62.99) -- cycle ;
\draw  [draw opacity=0][fill=tikzblue  ,fill opacity=0.67 ] (198.04,109.3) .. controls (198.04,105.48) and (204.9,102.39) .. (213.37,102.39) .. controls (221.84,102.39) and (228.7,105.48) .. (228.7,109.3) .. controls (228.7,113.12) and (221.84,116.21) .. (213.37,116.21) .. controls (204.9,116.21) and (198.04,113.12) .. (198.04,109.3) -- cycle ;
\draw  [draw opacity=0][fill=tikzblue  ,fill opacity=0.67 ] (212.98,81.73) .. controls (227.14,81.73) and (238.61,87.69) .. (238.61,95.05) .. controls (238.61,102.41) and (227.14,108.38) .. (212.98,108.38) .. controls (198.83,108.38) and (187.36,102.41) .. (187.36,95.05) .. controls (187.36,87.69) and (198.83,81.73) .. (212.98,81.73) -- cycle (209.13,94.92) .. controls (209.13,96.41) and (211.33,97.62) .. (214.06,97.62) .. controls (216.78,97.62) and (218.98,96.41) .. (218.98,94.92) .. controls (218.98,93.43) and (216.78,92.22) .. (214.06,92.22) .. controls (211.33,92.22) and (209.13,93.43) .. (209.13,94.92) -- cycle ;
\draw  [draw opacity=0][fill=tikzblue  ,fill opacity=0.67 ] (213.06,46.52) .. controls (237.6,46.52) and (257.5,56.86) .. (257.5,69.63) .. controls (257.5,82.39) and (237.6,92.73) .. (213.06,92.73) .. controls (188.51,92.73) and (168.61,82.39) .. (168.61,69.63) .. controls (168.61,56.86) and (188.51,46.52) .. (213.06,46.52) -- cycle (195.97,69.63) .. controls (195.97,74.8) and (203.62,78.99) .. (213.06,78.99) .. controls (222.49,78.99) and (230.14,74.8) .. (230.14,69.63) .. controls (230.14,64.45) and (222.49,60.26) .. (213.06,60.26) .. controls (203.62,60.26) and (195.97,64.45) .. (195.97,69.63) -- cycle ;
\draw  [draw opacity=0][fill=tikzblue  ,fill opacity=0.67 ] (212.7,22.89) .. controls (241.42,22.89) and (264.7,35) .. (264.7,49.93) .. controls (264.7,64.86) and (241.42,76.97) .. (212.7,76.97) .. controls (183.98,76.97) and (160.7,64.86) .. (160.7,49.93) .. controls (160.7,35) and (183.98,22.89) .. (212.7,22.89) -- cycle (182.28,49.93) .. controls (182.28,59.53) and (195.9,67.32) .. (212.7,67.32) .. controls (229.5,67.32) and (243.12,59.53) .. (243.12,49.93) .. controls (243.12,40.33) and (229.5,32.55) .. (212.7,32.55) .. controls (195.9,32.55) and (182.28,40.33) .. (182.28,49.93) -- cycle ;
\draw  [draw opacity=0][fill=tikzblue  ,fill opacity=0.67 ] (203.75,116.67) .. controls (203.75,114.16) and (207.99,112.13) .. (213.23,112.13) .. controls (218.46,112.13) and (222.7,114.16) .. (222.7,116.67) .. controls (222.7,119.18) and (218.46,121.21) .. (213.23,121.21) .. controls (207.99,121.21) and (203.75,119.18) .. (203.75,116.67) -- cycle ;

\begin{scope}[shift={(74,30)}, xscale=0.15, yscale=0.3]
  \draw[dashed] (0,1.2 cm) -- (0,-0.8 cm);

  \begin{scope}[scale=1.5]
    \draw (-1cm,0) arc (180:0:1cm and 0.5cm);
    \draw (1cm,0) -- (1.7cm, 0.3cm);
    \draw (1cm,0) -- (0.3cm, 0.3cm);
  \end{scope}

  \node[] () at (0, 350) {(a)};
  \node[] () at (925, 350) {(b)};
\end{scope}

\end{tikzpicture}
    \caption{\textbf{(a)} We illustrate a non-convex 3D ball given by revolving the horseshoe region around the vertical axis.  \textbf{(b)} For a 3D system composed of stacked horizontal 2D planes, the ball intersects the plane in both disks (bottom) and annuli (top).}
    \label{fig:bad_ball}
\end{figure}

We recall that even translation invariance does not guarantee topological properties.
Perhaps the simplest example is obtained in three spatial dimensions by stacking copies of a 2D topologically ordered system. For instance, consider a 3D cubic lattice, where each horizontal plane is a copy of the 2D toric code on the plane. 
To emphasize this system lacks topological properties, note local excitations have restricted mobility (they can only move along planes), and the system cannot be adiabatically continued to a rotated version of itself.  It turns out this sort of inhomogeneity is also diagnosed by a violation of LTQO on a topological ball. Consider the non-convex 3D ball indicated in Fig.~\ref{fig:bad_ball}(a) by revolving the horseshoe around the vertical axis.   Fig.~\ref{fig:bad_ball}(b) shows the ball composed of cross-sections with the 2D planes.  The ball violates LTQO because each 2D annulus cross-section individually violates LTQO, due to the possibility of an anyon present at its center. 

\begin{figure}[t]
    \centering
    \input{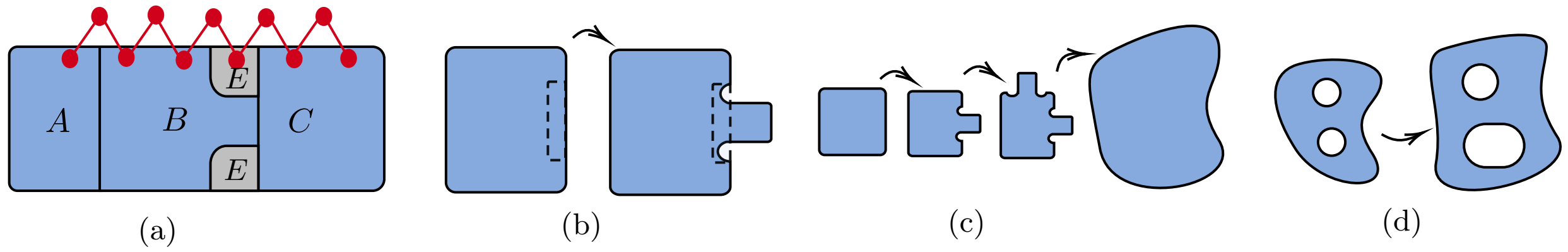}
    \caption{\textbf{(a)} Consider a 2D product state, augmented by a 1D cluster state with qubits on the red vertices.  This state can be prepared by a constant-depth circuit, but it has nonzero conditional mutual information $I(A:C|BE)$ even for large regions \cite{Zou2016,Williamson2019}.
    Thus there is no recovery map $R_{BE \to BEC}$ such that $R_{BE \to BEC}[\rho_{ABE}] = \rho_{ABEC}$.  However, there exists a channel $R_{BE \to BC}$ such that $R_{BE \to BC}[\rho_{ABE}] = \rho_{ABC}$. This may be viewed as a partial recovery map that recovers $C$ while discarding $E$. \textbf{(b)} A partial recovery map acts on the dashed region to extend the state to a larger region, while discarding a small subsystem.  \textbf{(c)} A sequence of partial recovery maps can extend a state onto a larger patch.  We can achieve net growth by choosing the discarded subsystems to be smaller than the added subsystems. \textbf{(d)} The space of zero-energy states on a topologically non-trivial region is mapped isomorphically to the corresponding space on a deformed region (Section \ref{sec:isomorphism}), which allows one to extract topological invariants \cite{shi2020fusion}.}
    \label{fig:isomorphism-growth-intro}
\end{figure}

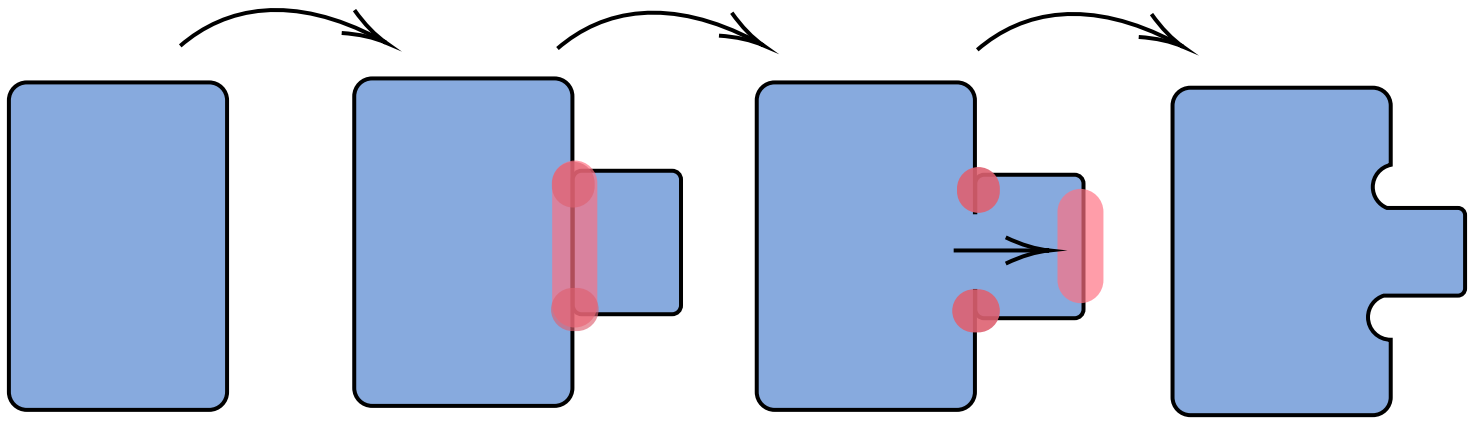
\begin{figure}[t]
    \centering
    \tikzset{every picture/.style={line width=0.75pt}} 

\begin{tikzpicture}[x=0.75pt,y=0.75pt,yscale=-1,xscale=1]

\draw  [color={rgb, 255:red, 0; green, 0; blue, 0 }  ,draw opacity=1 ][fill=tikzblue  ,fill opacity=1 ][line width=0.75]  (110.33,48.98) .. controls (110.33,46.51) and (112.34,44.5) .. (114.81,44.5) -- (160.6,44.5) .. controls (163.08,44.5) and (165.08,46.51) .. (165.08,48.98) -- (165.08,122.19) .. controls (165.08,124.66) and (163.08,126.67) .. (160.6,126.67) -- (114.81,126.67) .. controls (112.34,126.67) and (110.33,124.66) .. (110.33,122.19) -- cycle ;
\draw  [color={rgb, 255:red, 0; green, 0; blue, 0 }  ,draw opacity=1 ][fill=tikzblue  ,fill opacity=1 ][line width=0.75]  (165.17,69.89) .. controls (165.17,68.66) and (166.16,67.67) .. (167.39,67.67) -- (190.11,67.67) .. controls (191.34,67.67) and (192.33,68.66) .. (192.33,69.89) -- (192.33,101.44) .. controls (192.33,102.67) and (191.34,103.67) .. (190.11,103.67) -- (167.39,103.67) .. controls (166.16,103.67) and (165.17,102.67) .. (165.17,101.44) -- cycle ;
\draw  [draw opacity=0][fill={rgb, 255:red, 255; green, 113; blue, 130 }  ,fill opacity=0.69 ][line width=0.75]  (160,70.78) .. controls (160,67.65) and (162.54,65.11) .. (165.67,65.11) -- (165.67,65.11) .. controls (168.8,65.11) and (171.33,67.65) .. (171.33,70.78) -- (171.33,101.33) .. controls (171.33,104.46) and (168.8,107) .. (165.67,107) -- (165.67,107) .. controls (162.54,107) and (160,104.46) .. (160,101.33) -- cycle ;
\draw  [color={rgb, 255:red, 0; green, 0; blue, 0 }  ,draw opacity=1 ][fill=tikzblue  ,fill opacity=1 ][line width=0.75]  (211.33,49.98) .. controls (211.33,47.51) and (213.34,45.5) .. (215.81,45.5) -- (261.6,45.5) .. controls (264.08,45.5) and (266.08,47.51) .. (266.08,49.98) -- (266.08,123.19) .. controls (266.08,125.66) and (264.08,127.67) .. (261.6,127.67) -- (215.81,127.67) .. controls (213.34,127.67) and (211.33,125.66) .. (211.33,123.19) -- cycle ;
\draw  [color={rgb, 255:red, 0; green, 0; blue, 0 }  ,draw opacity=1 ][fill=tikzblue  ,fill opacity=1 ][line width=0.75]  (266.17,70.89) .. controls (266.17,69.66) and (267.16,68.67) .. (268.39,68.67) -- (291.11,68.67) .. controls (292.34,68.67) and (293.33,69.66) .. (293.33,70.89) -- (293.33,102.44) .. controls (293.33,103.67) and (292.34,104.67) .. (291.11,104.67) -- (268.39,104.67) .. controls (267.16,104.67) and (266.17,103.67) .. (266.17,102.44) -- cycle ;
\draw  [draw opacity=0][fill={rgb, 255:red, 222; green, 97; blue, 113 }  ,fill opacity=0.9 ][line width=0.75]  (266.95,66.74) .. controls (269.92,66.74) and (272.33,69.14) .. (272.33,72.1) -- (272.33,72.85) .. controls (272.33,75.82) and (269.93,78.23) .. (266.96,78.23) -- (266.96,78.23) .. controls (264,78.23) and (261.59,75.83) .. (261.59,72.87) -- (261.59,72.12) .. controls (261.59,69.15) and (263.99,66.74) .. (266.95,66.74) -- cycle ;
\draw  [draw opacity=0][fill={rgb, 255:red, 222; green, 97; blue, 113 }  ,fill opacity=0.9 ][line width=0.75]  (266.91,97.41) .. controls (269.9,97.4) and (272.33,99.82) .. (272.33,102.81) -- (272.33,102.81) .. controls (272.33,105.8) and (269.91,108.23) .. (266.92,108.23) -- (265.81,108.23) .. controls (262.82,108.23) and (260.39,105.81) .. (260.39,102.82) -- (260.39,102.82) .. controls (260.39,99.84) and (262.81,97.41) .. (265.8,97.41) -- cycle ;
\draw  [draw opacity=0][fill={rgb, 255:red, 255; green, 113; blue, 130 }  ,fill opacity=0.69 ][line width=0.75]  (286.8,77.99) .. controls (286.8,74.8) and (289.38,72.22) .. (292.57,72.22) -- (292.57,72.22) .. controls (295.75,72.22) and (298.33,74.8) .. (298.33,77.99) -- (298.33,95.12) .. controls (298.33,98.31) and (295.75,100.89) .. (292.57,100.89) -- (292.57,100.89) .. controls (289.38,100.89) and (286.8,98.31) .. (286.8,95.12) -- cycle ;
\draw  [draw opacity=0][fill=tikzblue  ,fill opacity=1 ][line width=0.75]  (250.17,80.13) .. controls (250.17,79.28) and (250.85,78.59) .. (251.7,78.59) -- (275.8,78.59) .. controls (276.65,78.59) and (277.33,79.28) .. (277.33,80.13) -- (277.33,95.8) .. controls (277.33,96.65) and (276.65,97.33) .. (275.8,97.33) -- (251.7,97.33) .. controls (250.85,97.33) and (250.17,96.65) .. (250.17,95.8) -- cycle ;
\draw    (260.75,87.67) -- (282.75,87.67) ;
\draw [shift={(284.75,87.67)}, rotate = 180] [color={rgb, 255:red, 0; green, 0; blue, 0 }  ][line width=0.75]    (10.93,-3.29) .. controls (6.95,-1.4) and (3.31,-0.3) .. (0,0) .. controls (3.31,0.3) and (6.95,1.4) .. (10.93,3.29)   ;
\draw    (161.33,37) .. controls (172.71,27.14) and (189.47,23.73) .. (211.31,35.3) ;
\draw [shift={(213,36.22)}, rotate = 209.2] [color={rgb, 255:red, 0; green, 0; blue, 0 }  ][line width=0.75]    (10.93,-3.29) .. controls (6.95,-1.4) and (3.31,-0.3) .. (0,0) .. controls (3.31,0.3) and (6.95,1.4) .. (10.93,3.29)   ;
\draw  [color={rgb, 255:red, 0; green, 0; blue, 0 }  ,draw opacity=1 ][fill=tikzblue  ,fill opacity=1 ][line width=0.75]  (23.67,49.98) .. controls (23.67,47.51) and (25.67,45.5) .. (28.15,45.5) -- (73.94,45.5) .. controls (76.41,45.5) and (78.42,47.51) .. (78.42,49.98) -- (78.42,123.19) .. controls (78.42,125.66) and (76.41,127.67) .. (73.94,127.67) -- (28.15,127.67) .. controls (25.67,127.67) and (23.67,125.66) .. (23.67,123.19) -- cycle ;
\draw    (66.67,36.33) .. controls (78.04,26.47) and (94.8,23.06) .. (116.64,34.64) ;
\draw [shift={(118.33,35.56)}, rotate = 209.2] [color={rgb, 255:red, 0; green, 0; blue, 0 }  ][line width=0.75]    (10.93,-3.29) .. controls (6.95,-1.4) and (3.31,-0.3) .. (0,0) .. controls (3.31,0.3) and (6.95,1.4) .. (10.93,3.29)   ;
\draw    (266.67,37.33) .. controls (278.04,27.47) and (294.8,24.06) .. (316.64,35.64) ;
\draw [shift={(318.33,36.56)}, rotate = 209.2] [color={rgb, 255:red, 0; green, 0; blue, 0 }  ][line width=0.75]    (10.93,-3.29) .. controls (6.95,-1.4) and (3.31,-0.3) .. (0,0) .. controls (3.31,0.3) and (6.95,1.4) .. (10.93,3.29)   ;
\draw  [color={rgb, 255:red, 0; green, 0; blue, 0 }  ,draw opacity=1 ][fill=tikzblue  ,fill opacity=1 ][line width=0.75]  (320.15,46.83) -- (365.94,46.83) .. controls (368.41,46.83) and (370.42,48.84) .. (370.42,51.31) -- (370.42,66.19) .. controls (367.85,66.74) and (365.92,69.02) .. (365.92,71.75) .. controls (365.93,74.13) and (367.4,76.17) .. (369.48,77) -- (387.28,77) .. controls (388.28,77) and (389.08,77.81) .. (389.08,78.8) -- (389.08,97.2) .. controls (389.08,98.19) and (388.28,99) .. (387.28,99) -- (368.69,99) .. controls (366.39,99.72) and (364.72,101.87) .. (364.72,104.41) .. controls (364.72,107.54) and (367.26,110.07) .. (370.39,110.07) -- (370.42,110.07) -- (370.42,124.52) .. controls (370.42,126.99) and (368.41,129) .. (365.94,129) -- (320.15,129) .. controls (317.67,129) and (315.67,126.99) .. (315.67,124.52) -- (315.67,51.31) .. controls (315.67,48.84) and (317.67,46.83) .. (320.15,46.83) -- cycle ;
\draw  [draw opacity=0][fill={rgb, 255:red, 222; green, 97; blue, 113 }  ,fill opacity=0.65 ][line width=0.75]  (165.29,65.41) .. controls (168.25,65.4) and (170.66,67.8) .. (170.66,70.77) -- (170.66,71.52) .. controls (170.66,74.49) and (168.26,76.89) .. (165.3,76.9) -- (165.3,76.9) .. controls (162.33,76.9) and (159.93,74.5) .. (159.92,71.53) -- (159.92,70.78) .. controls (159.92,67.82) and (162.32,65.41) .. (165.29,65.41) -- cycle ;
\draw  [draw opacity=0][fill={rgb, 255:red, 222; green, 97; blue, 113 }  ,fill opacity=0.65 ][line width=0.75]  (166.25,97.07) .. controls (169.23,97.07) and (171.66,99.49) .. (171.66,102.48) -- (171.66,102.48) .. controls (171.66,105.47) and (169.24,107.89) .. (166.25,107.9) -- (165.14,107.9) .. controls (162.15,107.9) and (159.72,105.48) .. (159.72,102.49) -- (159.72,102.49) .. controls (159.72,99.5) and (162.14,97.08) .. (165.13,97.07) -- cycle ;

\end{tikzpicture}
    \caption{We illustrate a heuristic argument that mobility of excitations implies local extendibility.  Given a zero-energy state on a patch, we extend the state to a larger patch with zero energy, using a local channel composed of three steps.  First, we attach a small new patch with a zero-energy state, unentangled with the original patch.  This may result in nonzero energy density (red) at the interface, interpreted as excitations. Second, we use the mobility of excitations to apply a channel that pushes the energy to the right. Energy near the two darker red points may not be movable, due to entanglement with the complement.
    Third, we erase a region including the red region, to leave an enlarged patch with zero energy.}
    \label{fig:energy-pushing}
\end{figure}

Similarly, for fracton models such as Ref.~\cite{Vijay2016}, there is a nonlocal stabilizer on a plane that cannot be generated by stabilizer generators strictly supported in the ball. Thus while fracton models may satisfy LTQO for convex balls, it appears they always violate LTQO for some topological ball.  In fact, for any commuting Hamiltonian satisfying LTQO on a sphere, or satisfying both LTQO and the related condition \ref{definition:spherical_ltqo}, all excitations are fully mobile (Proposition~\ref{prop:mobility} together with Theorem~\ref{thm:LTQO_local_extendibility}), forbidding immobile excitations like fractons.

Among all the conditions listed in Figure \ref{fig:homogeneity-conditions}, local extendibility is most useful for proving non-trivial consequences. This condition is a weaker version of the axiom \textbf{A1} of Ref.~\cite{shi2020fusion}.
A precise definition of local extendibility is given in Definition \ref{def:loc_extend}, but the functional consequence is that a patch of the ground state can be extended in any direction to a slightly larger patch, by acting locally with a quantum channel. 
Crucially, the quantum channel is allowed to discard a subsystem, so that the patch partly grows and partly shrinks (Fig.~\ref{fig:isomorphism-growth-intro}(b)).  Local extendibility is weaker (easier to satisfy) than the axiom \textbf{A1} of Ref.~\cite{shi2020fusion} precisely because the former allows this partial shrinking.  The need to allow shrinking is demonstrated by the example in Fig.~\ref{fig:isomorphism-growth-intro}(a), which exhibits a state in the trivial phase that violates \textbf{A1} but satisfies local extendibility.
By acting with a sequence of extension maps, despite the partial shrinking at each step, ultimately the state can be extended to a strictly larger region (Fig.~\ref{fig:isomorphism-growth-intro}(c)). 

An important concept from Refs.~\cite{shi2019characterizing, shi2020fusion} is the ``information convex set'' associated to a region $A$.  This is essentially the set of density matrices obtained by taking zero-energy states on $A$ and tracing out the boundary region.  From this perspective, our formulation of LTQO states that for each topological ball, the information convex set contains a single point. For topologically non-trivial regions, the information convex may be non-trivial, but the isomorphism theorem of Ref.~\cite{shi2020fusion} uses their axiom \textbf{A1} to show it remains invariant when smoothly deforming the region (Fig.~\ref{fig:isomorphism-growth-intro}(d)). We show that the isomorphism theorem also follows from the weaker assumption of local extendibility (Section~\ref{sec:isomorphism}).  It follows that localized excitations are mobile in all directions (Section \ref{sec:mobility}). This supports the idea that local extendibility guarantees a homogeneous phase. In particular, local extendibility rules out fracton phases with immobile excitations.

For a converse statement, we give a physical argument that local extendibility follows from the mobility of excitations, elaborated in Fig.~\ref{fig:energy-pushing}, though it is only a heuristic one. That picture also suggests an intuition for why the extension map must allow partial shrinking.

We have emphasized a desire not to assume translation invariance as a proxy for homogeneous topological phases: it is ill-defined on arbitrary lattice systems, and it is both too strong (ruling out homogeneous phases with disorder) and too weak (allowing fractons in higher dimensions).  However, it is interesting to ask whether translation invariance guarantees a homogeneous topological phase in two spatial dimensions. We leave this as an open question.

\begin{figure}\centering
\begin{tabular}{c@{\qquad}c}
\includegraphics{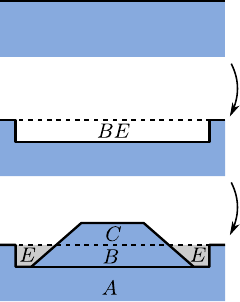} & \includegraphics{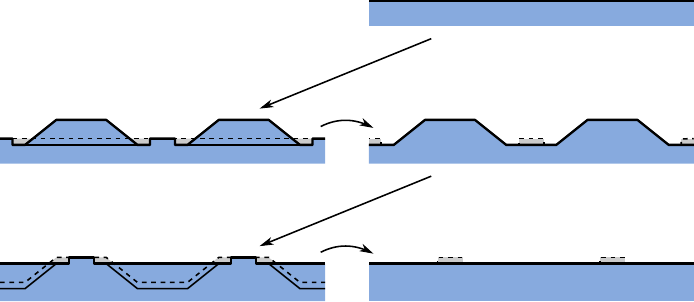}
\\[2pt]
(a) & (b)
\\[10pt]
\multicolumn{2}{c}{\raisebox{10pt}{\hbox{(c)}}\qquad\includegraphics{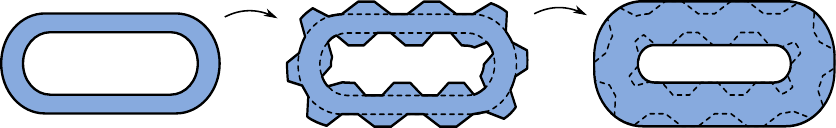}}
\end{tabular}
\caption{\textbf{(a)} To extend a region along part of its boundary, we imagine making a dent $BE$ and depositing a bump $BC$ on its bottom. \textbf{(b)} Multiple extension steps are applied in parallel at different places of the boundary, followed by the removal of protrusions. Then the procedure is repeated along the new (uneven) boundary. \textbf{(c)} The same process on a larger scale, omitting some details.}
\label{fig:growth0}
\end{figure}

Before developing formalism, we preview an essential method: extending the quantum state to a larger region, by sequentially applying the local extendibility axiom. 
The extension procedure may be compared to the technological process of growing a layer of material atop an existing piece. First, the surface is cleaned by removing a thin layer over the growth area, with some margin, and then a thicker layer is deposited (Fig.~\ref{fig:growth0}(a)). In our case, no explicit cleanup is performed, but the quantum channel $R_{BE\to BC}$ that implements the growth has the flexibility to first discard any portion of near-boundary state, at the cost of losing region $E$. The state can be extended along the whole boundary by applying multiple instances of the channel in parallel, removing some protrusions, and repeating the procedure (Fig.~\ref{fig:growth0}(b)). We formulate this precisely in any number of spatial dimensions.

The paper is organized as follows. First, we consider some choices in the design of a mathematical theory and formulate our core assumptions in Section~\ref{sec:formal-intro}. We focus in particular on the local extendibility assumption, whose general properties are discussed in Section~\ref{subsec:circuit_invariance}. In Section~\ref{sec:isomorphism}, we obtain some corollaries and prove the isomorphism theorem. The sketch of this theorem is explained in Section~\ref{subsec:iso_sketch} and a more rigorous formulation hinges on the geometric notions developed in Section~\ref{sec:geom}. In Section~\ref{sec:extension}, we show that the local data equipped with the local extendibility axiom determines a unique global state, which is extendible at all larger scales. Sections~\ref{sec:cmi} and~\ref{sec:rev_extendibility_merging} develop a machinery for gluing states defined on subregions along their boundaries. Section~\ref{sec:extendibility_ltqo} shows that, for commuting Hamiltonians, the local topological quantum order assumption~\cite{michalakis2013stability} implies local extendibility. Section~\ref{sec:mobility} shows the mobility of excitations is implied by the extendibility axiom. In Section~\ref{sec:tee_invariant}, we propose a circuit-invariant definition of topological entanglement entropy~\cite{Kitaev2006,Levin2006}.

\section{Further discussion and formal setting} \label{sec:formal-intro}

Let us fix basic notation and terminology. We consider spin systems in $\RR^d$, possibly infinite at the outset but restricted to bounded regions.%
\footnote{A \emph{region} is an arbitrary subset of some fixed geometric space, usually $\RR^d$.} Formally, a \emph{spin}, or qudit, is a point $x$ together with a finite-dimensional Hilbert space $\calH_x$. A \emph{spin system} on a metric space $X$ is a family $(\calH_x:\,x\in J)$, where $J\subseteq X$ has finite intersection with any ball. Thus, each bounded region $Y$ contains finitely many spins. The corresponding Hilbert space and the algebra of operators on it are $\calH_Y=\bigotimes_{x\in J\cap Y}\calH_x$ and $\Alg_Y=\LL(\calH_Y)$, respectively.%
\footnote{$\LL(\calH)$ stands for the $C^*$-algebra of linear operators on a finite-dimensional Hilbert space $\calH$. For $Y \cap J = \emptyset$, we take $\calH_Y = \CC.$ } We also assume that $\dim\calH_Y$ is bounded uniformly over all unit balls $Y$. A \emph{quantum state} is a positive linear functional $\rho$ on the operator algebra such that $\rho(\one)=1$. The set of states is denoted by $\St_Y=\St(\Alg_Y)$. Although we work in the finite setting, where states are represented by density matrices, it is still useful to view them as functionals. Given a subregion $Z\subseteq Y$, we have $\Alg_Z\subseteq\Alg_Y$ by implicitly tensoring operators on $\calH_Z$ with $\one_{Y\setminus Z}$. The restriction of a $\rho\in\St_Y$ to $\Alg_Z$ is denoted by $\rho|_{Z}=\Tr_{Y\setminus Z}\rho$. States $\rho_1,\rho_2$ on regions $Y_1,Y_2$ are called \emph{consistent} with each other if $\rho_1|_{Y_1\cap Y_2}=\rho_2|_{Y_1\cap Y_2}$.

As stated in the introduction, our main goal is to define a class of local Hamiltonians or their ground states that may be called ``topological''. The definition constructed in this paper captures only a subset of such states, namely, those for which the local conditions hold exactly, with no correlation tails beyond a certain length $r_0$. While this restriction does not preclude nontrivial examples such as Levin-Wen models, typical topological states (and in particular, all chiral ones) have exponentially decaying correlations. In future work, we will address this issue by replacing exact conditions with approximate ones, such that small correlation tails beyond $r_0$ can be neglected. The density of spins is completely arbitrary. (If all spins are farther apart than $r_0$, then only product pure states will meet the definition.) For technical reasons, we use three different lengths, $r_0\ll r_1\ll r_2$, where $x\ll y$ means that $x\le cy$ for a sufficiently small positive $c$ that depends only on the number of spatial dimensions $d$. Thus,
\begin{equation} \label{eq:scales}
r_0\le c_1r_1,\qquad r_1\le c_2r_2,
\end{equation}
where the constants $c_1$ and $c_2$ can be adjusted later to make the proofs work.

As a fundamental requirement, the definition should be in the form of local data and conditions. Although the collection of Hamiltonian terms is a kind of local data, reasoning about ground states requires nonlocal calculations. Even if a good candidate state is known, it may be difficult to check that there is no state with lower energy. Therefore, we will use a different type of data, for which local checks are possible. Specifically, a hypothetical state $\rho$ on an arbitrarily large ball is described by a family of consistent density matrices $\sigma_K$ on some small balls $K$. These data amount to a linear functional on a subspace of the operator algebra, namely, the linear span of the operators fitting in individual balls.

Note that the mutual consistency between the states $\sigma_K$ implies neither the existence nor the uniqueness of a consistent global state. To make this inference and to establish nice topological properties, we will impose further conditions on $\sigma_K$. Each of them posits the existence of a quantum channel acting in a subset of $K$ and transforming $\sigma_K$ in a certain way. The subsets $L$ involved are smaller than $r_1$ but their geometric features are larger than $r_0$. For example, region $BE$ in Fig.~\ref{fig:growth0}(a) should have thickness at least $r_0$ but the whole structure should be at most $r_1$ in size. Technically, states on various regions are required to be consistent with the local data on balls of radius $r_1$, but the data themselves are defined on balls of radius $r_2$ (which means stronger mutual consistency).

The exact geometric details are somewhat arbitrary, but our definition of a topological state has two important properties. First, it is geometrically robust, meaning that all subsets of $\RR^d$ involved may be deformed by $\eps$ in the Hausdorff metric as long as $\eps\ll r_0$. For this reason, we need not distinguish between open and closed regions. As another corollary, the set of conditions can be replaced with a locally finite one, such that there is a constant number of conditions per $r_0^d$ volume. The second property is that the conditions reproduce themselves on a bigger scale. We will elaborate on both points and other geometric questions later.

Now, we describe the general form of local conditions, which are associated with small regions $L\subseteq\RR^d$ partitioned in certain ways. The following terminology will be used. A \emph{purification} of a state $\rho\in\St_L$ is a pure state $\overline{\rho}=\ket{\psi}\bra{\psi}$ for some $\ket{\psi}\in\calH_L\otimes\calH_A$ such that $\overline{\rho}|_L=\rho$. Here $A$ is an \emph{abstract system}, i.e., just a label for the newly introduced Hilbert space $\calH_A$.%
\footnote{When referring to Hilbert spaces in this way, canonical isomorphisms, e.g., between $\calH_{AB}=\calH_A\otimes\calH_B$ and $\calH_{BA}=\calH_B\otimes\calH_A$ are implicit. Thus, one may write $\ket{\xi}\otimes\ket{\eta}\in\calH_{ABCD}$ for $\ket{\xi}\in\calH_{AC}$ and $\ket{\eta}\in\calH_{BD}$. But to use the same space twice, one needs two copies with different labels.} A purification is called \emph{minimal} if $\supp(\overline{\rho}|_A)=\calH_A$. A \emph{quantum channel} is a completely positive, trace-preserving linear map. For convenience, its domain and codomain may be indicated with a subscript: $R_{L\to M}\colon\LL(\calH_L)\to\LL(\calH_M)$.

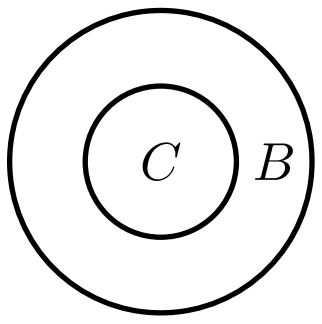
\begin{figure}
    \centering
    \begin{tikzpicture}[line width=1pt]
    \draw[] (0,0) circle (1cm);
    \draw[] (0,0) circle (0.5cm);
    \node[] () at (0,0) {$C$};
    \node[] () at (0.75, 0) {$B$};
    \end{tikzpicture}
    \caption{The regions used in the local purifiability condition. In higher dimensions, $C$ and $BC$ are concentric balls.}
    \label{fig:A0}
\end{figure}

Let us begin with a simple condition type, \emph{local purifiability}, which is equivalent to axiom \textbf{A0} of Ref.~\cite{shi2020fusion}. It is supplementary to local extendibility and used to translate between local consistency conditions of the form $\rho|_L=\sigma_L$, where $\sigma_L\in\St_L$ is a reference state, and subspace constraints like ``$\rho|_L$ is supported by $\calL$'' for some $\calL\subseteq\calH_L$. The latter in turn can be converted to a frustration-free Hamiltonian that stabilizes the states in question. The local purifiability is expected to hold (approximately, up to exponentially small correlation tails) for any gapped system with a unique ground state, whether or not it is topological. We will eventually require this condition for regions $L=BC$ as in Fig.~\ref{fig:A0},%
\footnote{The shorthand $BC=B\cup C$ is used when $B$ and $C$ are disjoint.} where $B$ is a thickened spherical shell surrounding a ball $C$, but the present discussion is more general. 

\begin{Definition}[Local purifiability]\label{def:loc_puri}
Let $B$ and $C$ be disjoint regions, $\sigma$ a state on some region containing $BC$, and $\overline{\sigma} \in \St_{ABC}$ a purification of $\sigma|_{BC}$ using some abstract system $A$. The state $\sigma$ is said to be \emph{$(B,C)$-purifiable} if any of the following equivalent conditions holds:
\begin{itemize}
\item[(1)]\, $\overline{\sigma}|_{AC} =\overline{\sigma}|_A\otimes\overline{\sigma}|_C$.
\item[(2)] The Hilbert space $\calH_B$ admits a unitary isomorphism to $(\calH_{B_1}\otimes\calH_{B_2})\oplus\calH_{B^\perp}$ which takes $\sigma|_{BC}$ to $\rho\otimes\overline{\gamma}$, where $\rho$ is some state on $B_1$ and $\overline{\gamma}$ is a pure state on $B_2C$.
\item[(3)] There exists a quantum channel $R_{B\to BC}$ such that $(\id_A\otimes R_{B\to BC})[\overline{\sigma}|_{AB}]=\overline{\sigma}$.
\end{itemize}
\end{Definition}
The term ``local purifiability'' arises from point~(2): the state $\sigma|_C$ is purified by its surroundings $B$. Point~(3) means that a channel can locally repair a ``hole'' in $\overline{\sigma}|_{AB}$ created by erasing $C$, and this formulation mirrors the definition of local extendibility below.

\begin{proof}[Proof of the equivalence between conditions~(1)--(3).]

(1) $\Rightarrow$ (2). Let $\overline{\rho}\in\St_{AB_1}$ and $\overline{\gamma}\in\St_{B_2C}$ be minimal purifications of $\overline{\sigma}|_A$ and $\overline{\sigma}|_C$ with abstract systems $B_1$ and $B_2$, respectively. Then $\overline{\rho}\otimes\overline{\gamma}$ is a minimal purification of $\overline{\sigma}|_{AC} =\overline{\sigma}|_A\otimes\overline{\sigma}|_C$ with $B_1B_2$. The state $\overline{\sigma}$ is a purification of $\overline{\sigma}|_{AC}$ with $B$, and hence, there is a linear isometry $V\colon\calH_{B_1}\otimes\calH_{B_2}\to\calH_B$ that takes $\overline{\rho}\otimes\overline{\gamma}$ to $\overline{\sigma}$. It remains to set $\rho=\Tr_A\overline{\rho}=\overline{\rho}|_{B_1}$.

(2) $\Rightarrow$ (3). Without loss of generality, we may take the isomorphism to be the identity so that $\calH_B=(\calH_{B_1}\otimes\calH_{B_2})\oplus\calH_{B^\perp}$ and $\sigma|_{BC}=\rho\otimes\overline{\gamma}$. Then $\overline{\sigma}=\overline{\rho}\otimes\overline{\gamma}$, where $\overline{\rho}$ is a purification of $\rho$. It is sufficient to define the channel on density matrices supported by $\calH_{B_1B_2}\supseteq\supp(\sigma|_{B})$ and extend it to $\LL(\calH_B)$ in an arbitrary way. For any $X\in\LL(\calH_{B_1B_2})$, we set $R_{B\to BC}[X]=(\Tr_{B_2}X)\otimes\overline{\gamma}$.

(3) $\Rightarrow$ (1). Let $\overline{\sigma}_{AB'C'}$ be a copy of $\overline{\sigma}=\overline{\sigma}_{ABC}$, so that $(\id_A\otimes R_{B\to B'C'})[\overline{\sigma}|_{AB}]=\overline{\sigma}_{AB'C'}$. The channel can be represented by a linear isometry $V_{B\to B'C'F'}\colon\calH_B\to\calH_{B'C'F'}$ for some auxiliary system $F'$, i.e.\ $R_{B\to B'C'}[X]=\Tr_{F'}(V_{B\to B'C'F'}XV_{B\to B'C'F'}^\dag)$. Thus, condition~(3) becomes
\[
\Tr_{CF'}\underbrace{\bigl((1_{AC}\otimes V_{B\to B'C'F'})
\overline{\sigma}_{ABC}(1_{AC}\otimes V_{B\to B'C'F'})^\dag\bigr)}
_{\mu_{ACB'C'F'}}
=\overline{\sigma}_{AB'C'}.
\]
A partial trace of a state is pure only if that state is a tensor product; therefore, $\mu_{ACB'C'F'}=\nu_{CF'}\otimes\overline{\sigma}_{AB'C'}$ for some $\nu_{CF'}\in\St_{CF'}$. Tracing out $B'C'F'$, we get
\[
\Tr_{B'C'F'}\mu_{ACB'C'F'}
=(\Tr_{F'}\nu_{CF'})\otimes(\Tr_{B'C'}\overline{\sigma}_{AB'C'}).
\]
The left-hand side is equal to $\Tr_B\overline{\sigma}_{ABC}=\overline{\sigma}|_{AC}$, and the right-hand side is a product of two states, which implies condition~(1).
\end{proof}

Other equivalent characterizations of local purifiability follow from this straightforward fact.

\begin{Lemma}\label{lem:supp_puri}
Let $\overline{\rho}$ be a pure state of system $AL$, and let $R',R''$ be quantum channels from $L$ to $M$ (or even arbitrary linear maps $\LL(\calH_L)\to\LL(\calH_M)$). Then $(\id_A\otimes R')[\overline{\rho}] =(\id_A\otimes R'')[\overline{\rho}]$ if and only if $R'$ and $R''$ agree on density matrices supported by $\calL=\supp(\overline{\rho}|_L)$.
\end{Lemma}

\begin{Proposition}\label{prop:supp}
Let $\sigma$ be a state on some region containing $BC$, and let $\calL=\supp(\sigma|_{BC})$. The $(B,C)$-purifiability of $\sigma$ (see conditions~(1)--(3) in Definition~\ref{def:loc_puri}) is equivalent to any of these conditions:
\begin{itemize}
\item[(4)] There exists a channel $R_{B\to BC}$ such that $R_{B\to BC}[\tau|_B]=\tau$ for all states $\tau\in\St_{BC}$ supported by $\calL$.
\item[(5)] All states $\tau\in\St_{BC}$ supported by $\calL$ satisfy $\tau|_C=\sigma|_C$.
\end{itemize}
\end{Proposition}
\begin{proof}

(3) $\Leftrightarrow$ (4). Use Lemma~\ref{lem:supp_puri} with $R'\colon X\mapsto R_{B\to BC}[\Tr_CX]$ and $R''=\id_{BC}$.

(1) $\Leftrightarrow$ (5). Use Lemma~\ref{lem:supp_puri} with $R'\colon X\mapsto \Tr_BX$ and $R''\colon X\mapsto(\Tr X)\,\sigma|_C$.
\end{proof}

Point~(5) is useful when matching an arbitrary state $\rho$ against a reference state $\sigma$.

\begin{Corollary}\label{cor:supp_puri}
Let $\sigma$ be a $(B,C)$-purifiable state, $\calL_{BC}=\supp(\sigma|_{BC})$, and let $\rho$ be an arbitrary state on a region containing $BC$. Then the following implications are true:
\begin{equation}
\rho|_{BC}=\sigma|_{BC} \qquad\Rightarrow\qquad
\rho|_{BC} \text{ is supported by } \calL_{BC} \qquad\Rightarrow\qquad
\rho|_{C}=\sigma|_{C}.
\medskip
\end{equation}
\end{Corollary}

\begin{figure}\centering
\begin{tabular}{c@{\qquad}c@{\qquad}c}
\raisebox{10pt}{\includegraphics{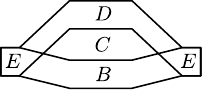}} &
\includegraphics{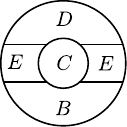} &
\raisebox{-5pt}{\includegraphics{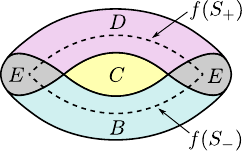}}
\\[2pt]
(a) & (b) & (c)
\end{tabular}
\caption{\textbf{(a)} A typical $(B,C,D,E)$ configuration used to extend states to bigger regions as shown in Fig.~\ref{fig:growth0}. \textbf{(b)} Another possible geometry. \textbf{(c)} General description of such configurations: regions $BE$ and $DE$ are neighborhoods of two arcs or deformed disks, $f(S_-)$ and $f(S_+)$.}
\label{fig:extend}
\end{figure}

The main defining condition for topological states, \emph{local extendibility}, is a weaker version of axiom \textbf{A1} of Ref.~\cite{shi2020fusion}. (Sometimes we refer to the present condition as  \textbf{A1}$^\prime$.)  It is expected to hold for disjoint regions $B$, $C$, $D$, $E$ arranged as in Fig.~\ref{fig:extend}.  There $C$ is a ball surrounded by thickened spherical shell $BDE$, where $B,D$ are thickened hemispheres, separated by a thickened equator $E$.

To describe the topology more generally, consider a continuous embedding $f\colon S^{d-1}\to\RR^d$. 
The sphere $S^{d-1}$ is defined by the equation $x_1^2+\dots+x_d^2=1$ and viewed as the union of the ``southern hemisphere'' $S_-=\{x\in S^{d-1}:\,x_d\le 0\}$ and the  ``northern hemisphere'' $S_+=\{x\in S^{d-1}:\,x_d\ge 0\}$. The regions $BE$ and $DE$ are neighborhoods of $f(S_-)$ and $f(S_+)$, respectively. This data determine all the regions: $E=BE\cap DE$,\, $B=BE\setminus E$,\, $D=DE\setminus E$, and $C$ is the part of $\RR^{d}\setminus(BE\cup DE)$ that lies inside $f(S^{d-1})$. We will later be more specific about the region choice, though Proposition~\ref{prop:g_mono} below allows for much flexibility. The following definition is completely general.

\begin{Definition}[Local extendibility]\label{def:loc_extend}
Let $B$, $C$, $D$, $E$ be disjoint regions, $\sigma$ a state on some region containing $BCDE$, and $\overline{\sigma} \in \St_{ABCDE}$ a purification of $\sigma|_{BCDE}$ using some abstract system $A$. The state $\sigma$ is said to be \emph{$(B,C,D,E)$-extendible} if there exists a quantum channel $R_{BE\to BC}$ such that
\begin{equation}\label{eq:loc_extend}
(\id_A\otimes R_{BE\to BC})[\overline{\sigma}|_{ABE}]=\overline{\sigma}|_{ABC}.
\end{equation}
\end{Definition}
\noindent Note that if Eq.~\eqref{eq:loc_extend} holds for one purification $\overline{\sigma}$, it holds using the same channel $R_{BE\to BC}$ for any purification (because all purifications are related by isometries), and it also holds for non-pure extensions of $\sigma|_{BCDE}$ (by taking a further purification). We refer to the channel $R_{BE \to BC}$ as an extension map or recovery map.

The following property may aid intuition.
\begin{Proposition}\label{prop:extendible-match} Let $\rho, \tau \in \St_{ABCDE}$ such that $\tau$ is $(B,C,D,E)$-extendible, $\rho|_{ABE}=\tau|_{ABE}$, and $\rho|_{BCDE}=\tau|_{BCDE}$. Then $\rho|_{ABC}=\tau|_{ABC}$.  
\end{Proposition}
\noindent That is, when extendible states match on overlapping regions $ABE$ and $BCDE$, they also match on a subset $ABC$ of the union. To see this, let $R_{BE \to BC}$ witness the $(B,C,D,E)$-extendibility of $\tau$.  Then 
\begin{align}
\tau|_{ABC}=(\id_A \otimes R_{BE \to BC})[\tau|_{ABE}] = (\id_A \otimes R_{BE \to BC})[\rho|_{ABE}] = \rho|_{ABC}
\end{align}
with the final equality because $\rho$ is also an extension of $\tau|_{BCDE}$, using the observation that Eq.~\eqref{eq:loc_extend} holds for any extension.

We name a few simple examples. The state $\frac{1}{2}( |0\rangle\langle0|^{\otimes 4} + |1\rangle\langle1|^{\otimes 4} )$ on qubits $BCDE$ is $(B,C,D,E)$-extendible but not  $(B,C)$-purifiable. Both $(B,C)$-purifiability and $(B,C,D,E)$-extendibility are invariant under unitaries on individual subsystems.  
If $\rho|_{BC} = \rho|_B \otimes \rho|_C$, then $\rho$ is $(B,C)$-purifiable iff $\rho|_C$ is pure. Likewise, if $\rho|_{BCDE} = \rho|_{BDE} \otimes \rho|_C$,  then $\rho$ is $(B,C,D,E)$-extendible iff $\rho|_C$ is pure.  

As with purifiability, extendibility may be seen as a property of the support subspace $\calL=\supp(\sigma|_{BCDE})$, rather than a property of the specific state.  Analogous to Proposition~\ref{prop:supp}(5), here we have:
\begin{Proposition} \label{prop:extendibility-mixture} 
The $(B,C,D,E)$-extendibility property for $\sigma|_{BCDE}$ is equivalent to extendibility for all $\tau\in\St_{BCDE}$ supported by $\calL=\supp(\sigma|_{BCDE})$.  
\end{Proposition}
\noindent This follows from  Lemma~\ref{lem:supp_puri} applied to the two channels $R_{BE \to BC} \circ \Tr_{CD}$ and $\Tr_{DE}$ from $BCDE$ to $BC$.  As one consequence, if a positive mixture of states $\sigma = \sum_i p_i \sigma_i$ is extendible, then so are the components $\sigma_i$.

\begin{figure}\centering
\begin{tabular}{c@{\qquad}c}
\raisebox{10pt}{\includegraphics{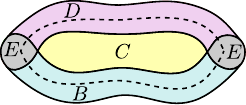}} & \includegraphics{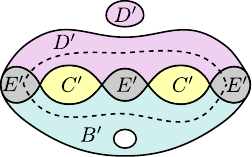}
\\[2pt]
(a) & (b)
\end{tabular}
\caption{An example of configurations satisfying conditions~\eqref{g_mono}.}
\label{fig:extend1}
\end{figure}

The following is an important property for our geometric manipulations.
\begin{Proposition}[Geometric monotonicity]\label{prop:g_mono}
Suppose that
\begin{equation}\label{g_mono}
B'E'\supseteq BE,\qquad D'E'\supseteq DE,\qquad
C'\subseteq C,\qquad B'C'D'E'\supseteq BCDE.
\end{equation}
Then any $(B,C,D,E)$-extendible state $\sigma$ on a region containing $B'C'D'E'$ is also $(B',C',D',E')$-extendible.
\end{Proposition}
We denote condition~\eqref{g_mono} as $(B,C,D,E)\cge(B',C',D',E')$, which indicates the strength of the condition rather than the size of the subset. Assuming the first two inclusions in \eqref{g_mono}, the other two are equivalent to the equation $C'=C\setminus B'D'E'$, which is automatically satisfied if both $(B,C,D,E)$ and $(B',C',D',E')$ are constructed using the previous topological recipe with the same embedding of $S^{d-1}$. An example is shown in Fig.~\ref{fig:extend1}. Thus, it is sufficient to require local extendibility for simpler shapes as in Fig.~\ref{fig:extend1}(a), and it will also hold for more complex shapes as in Fig.~\ref{fig:extend1}(b).

\begin{proof}
 Let $F=B'\setminus BC$. First, we observe that $F$ is contained in $G=B'C'D'E'\setminus BCDE$. This can be seen as follows: $F=B'\setminus BCD'E'\subseteq G$ since $B'\cap D'E'=\emptyset$ and $D'E'\supseteq DE$. Next,
\begin{equation}\label{g_mono_1}
F\cap BE=\emptyset,\qquad FBE\subseteq B'E'.
\end{equation}
Indeed, $F\cap BE=(F\cap B)\cup(F\cap E)\subseteq \emptyset\cup(F\cap E')=\emptyset$, where we have used the fact that $E=BE\cap DE\subseteq B'E'\cap D'E'=E'$ and that $B'$ and $E'$ are disjoint. The second part of \eqref{g_mono_1} holds because both $F$ and $BE$ are contained in $B'E'$. Finally,
\begin{equation}\label{g_mono_2}
F\cap BC=\emptyset,\qquad FBC\supseteq B'C',
\end{equation}
where the first part is obvious and the second is true because $FBC=B'\cup BC$ clearly includes $B'$ and also includes $C'$ due to the condition $C'\subseteq C$.

Let $\overline{\sigma}'$ be a purification of $\sigma|_{B'C'D'E'}$ with an abstract system $A'$. The last condition in \eqref{g_mono} implies that $B'C'D'E'=GBCDE$. Therefore, $\overline{\sigma}'$ is also a purification of $\sigma|_{BCDE}$ with system $A=A'G$, and so the condition that $\sigma|_{BCDE}$ is $(B,C,D,E)$-extendible is given by the equation $(\id_{A'G}\otimes R_{BE\to BC})[\overline{\sigma}'|_{A'GBE}]
=\overline{\sigma}'|_{A'GBC}$. Replacing $G$ with $F\subseteq G$, we get \begin{equation}\label{g_mono_3}
(\id_{A'F}\otimes R_{BE\to BC})[\overline{\sigma}'|_{A'FBE}]
=\overline{\sigma}'|_{A'FBC}.
\end{equation}

Eqs.~\eqref{g_mono_1} and \eqref{g_mono_2} imply that this quantum channel is well-defined:
\begin{equation}
R'_{B'E'\to B'C'}[X]
=\bigl((\id_{F}\otimes R_{BE\to BC})[X|_{FBE}]
\bigr)\bigr|_{B'C'}\qquad (X\in\LL(\calH_{B'E'})).
\end{equation}
Using Eq.~\eqref{g_mono_3}, we conclude that
\[
(\id_{A'}\otimes R'_{B'E'\to B'C'})[\overline{\sigma}'|_{A'B'E'}]
=\bigl((\id_{A'F}\otimes R_{BE\to BC})[\overline{\sigma}'|_{A'FBE}]
\bigr)\bigr|_{A'B'C'}
=\overline{\sigma}'|_{A'B'C'},
\]
and hence, $\sigma$ is $(B',C',D',E')$-extendible.
\medskip
\end{proof}

\begin{figure}
\pgfdeclarepatternformonly{mydots}{\pgfqpoint{-1pt}{-1pt}}{\pgfqpoint{1pt}{1pt}}{\pgfqpoint{4pt}{4pt}}
{%
  \pgfpathcircle{\pgfqpoint{0pt}{0pt}}{.5pt}%
  \pgfusepath{fill}%
}%
\pgfdeclarepatternformonly{mysparsedots}{\pgfqpoint{-1pt}{-1pt}}{\pgfqpoint{1pt}{1pt}}{\pgfqpoint{7pt}{7pt}}
{%
  \pgfpathcircle{\pgfqpoint{0pt}{0pt}}{.5pt}%
  \pgfusepath{fill}%
}%

\tikzset{
    partial ellipse/.style args={#1:#2:#3}{
        insert path={+ (#1:#3) arc (#1:#2:#3)}
    }
}
\begin{tikzpicture}[scale=1.125]
\draw[pattern=mydots, draw=none] (-4,-2) rectangle ++(6,4);
\draw[] (-2, 0) circle (1cm);
\node[below, fill=white, inner sep=1pt] () at (-2, -1.125) {$K\in \Gamma_2$};
\draw[Stealth-Stealth] (-3.25, -1) --++ (0, 1);
\node[fill=white, inner sep=1pt] () at (-3.5, -0.5) {$r_2$};

\begin{scope}[xshift=-0.5cm]
\draw[Stealth-Stealth] (0.5, -0.375) --++ (0, 0.375);
\node[fill=white, inner sep=1pt] () at (0.25, -0.1875) {$r_1$};
\draw[blue] (1, 0) circle (0.375cm);
\node[below, fill=white, inner sep=1pt] () at (1,-0.5) {$L\in \Gamma_1$};
\begin{scope}[xshift=1cm, yshift=0.05cm,scale=0.625]
\begin{scope}[rotate=-30]
\draw[line width=0.5pt, red] (0,0) ellipse [x radius= 0.45cm, y radius=0.225cm];
\draw[line width=0.5pt, red] (0,0) ellipse [x radius= 0.3cm, y radius=0.15cm];
\draw[line width=0.5pt, red] (-0.375cm, 0) circle (0.075cm);
\draw[line width=0.5pt, red] (0.375cm, 0) circle (0.075cm);
\end{scope}
\end{scope}
\draw[] (1.375, 0) -- ++ (1.5,0);
\begin{scope}[xshift=0.25cm]
\draw[pattern=mysparsedots, draw=blue] (4.625,0) circle (2cm); 
\begin{scope}[xshift=4.7cm, yshift=0.4cm, scale=3.33333]
\begin{scope}[rotate=-30]
\draw[line width=0.5pt, red] (0,0) ellipse [x radius= 0.45cm, y radius=0.225cm];
\draw[line width=0.5pt, red] (0,0) ellipse [x radius= 0.3cm, y radius=0.15cm];
\draw[line width=0.5pt, red] (-0.375cm, 0) circle (0.075cm);
\draw[line width=0.5pt, red] (0.375cm, 0) circle (0.075cm);

\begin{scope}[shift={(-0.48cm,-0.22cm)}] 
  \coordinate (pIn)  at (0.30cm, 0);
  \coordinate (pOut) at (0.33cm,0.10cm);

  \draw[Stealth-Stealth, line width=0.5pt]
    (pIn) -- (pOut)
    node[midway, xshift=-10pt, yshift = 5pt, xshift = 6pt, 
inner sep=1pt] {\footnotesize $r_0$};
\end{scope}

\end{scope}
\end{scope}
\node[fill=white, inner sep=1pt] () at (4.7, 0.4) {\footnotesize $C$};
\node[fill=white, inner sep=1pt] () at (5.3, -0.41) {\footnotesize $B$};
\node[fill=white, inner sep=1pt] () at (5.775, -0.225) {\footnotesize $E$};
\node[fill=white, inner sep=1pt] () at (4.1, 1.21) {\footnotesize $D$};
\node[fill=white, inner sep=1pt] () at (3.625, 1.025) {\footnotesize $E$};
\node[fill=white, inner sep=1pt] () at (4.7, -1) { $(B,C,D,E)\in \Lambda$};
\end{scope}
\end{scope}
\node[] (0,0) at (1, -2.5cm) {(a)};

\begin{scope}[xshift=9cm]

\draw[line width=1pt, fill=green!30!white] (0,0) circle (1.5cm);
\draw[line width=1pt, fill=white] (0,0) circle (0.75cm);

\draw[pattern=mydots, draw=none] (-2,-2) rectangle ++(4,4);

\foreach \t in {0, ..., 9}
{
\draw[blue] (\t * 40:0.55cm) circle (0.25cm);
}
\foreach \t in {0, ..., 12}
{
\draw[blue] (\t * 30:0.835cm) circle (0.25cm);
}
\foreach \t in {0, ..., 18}
{
\draw[blue] (\t * 20:1.125cm) circle (0.25cm);
}
\foreach \t in {0, ..., 24}
{
\draw[blue] (\t * 15:1.4125cm) circle (0.25cm);
}
\foreach \t in {0, ..., 30}
{
\draw[blue] (\t * 12:1.7cm) circle (0.25cm);
}
\node[] (0,0) at (0, -2.5cm) {(b)};
\end{scope}
\end{tikzpicture}
\caption{\textbf{(a)} Organization of local data. \textbf{(b)} Regions used in the construction of $\IC(Y)$, where $Y$ is the green annulus. $\Gamma_1(Y)$ consists of $L\in \Gamma_1$ that have nontrivial intersections with $Y$. }
\label{fig:data_org}
\end{figure}
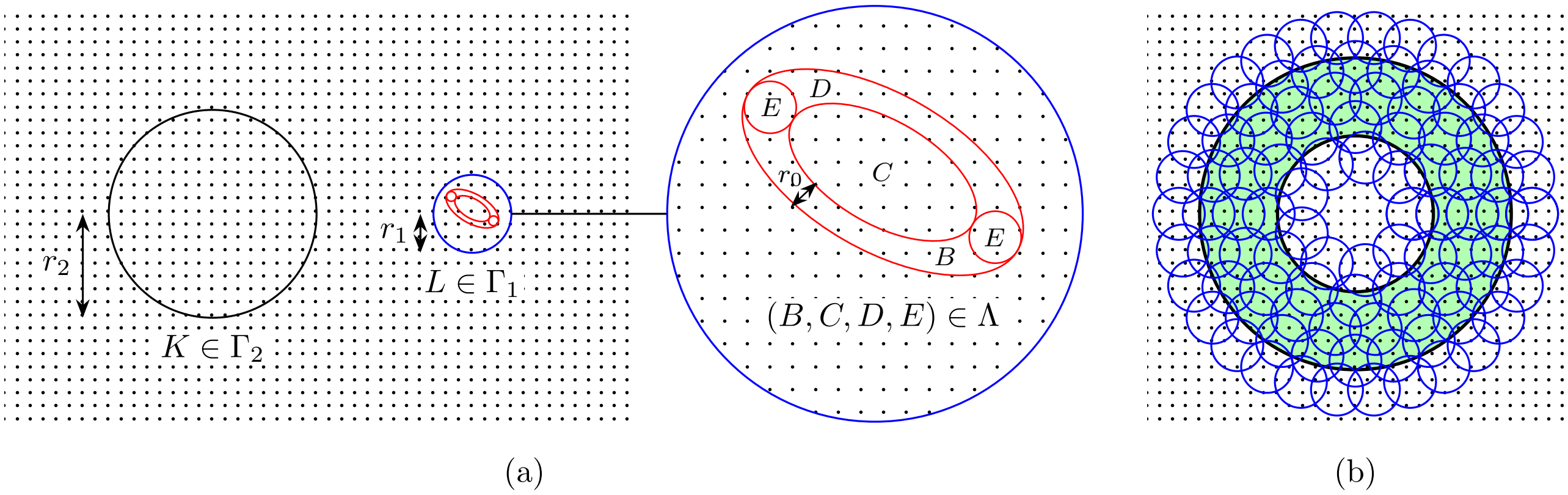

Let us now specify the exact organization of local data and conditions (see Fig.~\ref{fig:data_org}(a) for an illustration). The standard definition depends only on the three lengths $r_0\ll r_1\ll r_2$ explained earlier. However, we will allow more options such that, for example, the data and conditions could be required only in some region and multiple slightly different $(B,C,D,E)$ configurations could be replaced with a single one. The local data is a family $(\sigma_K:\, K\in\Gamma_2)$ of pairwise consistent states on some set $\Gamma_2$ of regions, which are called \emph{data patches}. The standard for $\Gamma_2$ is the set of all balls of radius $r_2$. We also introduce a set  $\Gamma_1$ of \emph{reference patches} (for matching global states against the data), the standard being all balls of radius $r_1$. 
In general, each reference patch $L$ is contained in some data patch $K$, and the corresponding state is $\sigma_L=(\sigma_K)|_L$. Finally, let $\Lambda$ be a set of configurations such that for each $(B,C,D,E)\in\Lambda$, the region $BCDE$ is contained in some reference patch $L$ and $\sigma_L$ is $(B,C,D,E)$-extendible. The standard choice is described in section~\ref{sec:micro}, where we supplement the topological recipe with some geometric constraints. This whole structure is called \emph{$\Lambda$-extendible $(\Gamma_1,\Gamma_2)$-data}, or just \emph{extendible local data} in the standard case.  While we imagine local data for balls covering $\RR^d$, our results concern local constructions on finite regions; it would suffice to specify local data covering a large bounded domain.

We summarize this discussion for easy reference, specializing to the standard case.
\begin{Definition}[Extendible local data]  
\label{def:extendible-data}
Given $r_2 \gg r_1 \gg r_0$, extendible local data refers to a family of pairwise consistent states $\{\sigma_K\}$ on all $r_2$-balls $K \subset \RR^d$ (``data patches''), such that each $\sigma_K$ is $(B,C,D,E)$-extendible for each configuration $(B,C,D,E) \in \Lambda$ with $BCDE \subset K$.  The configurations $\Lambda$ are specified in Definition~\ref{def:Lambda-micro} and depend on $r_0, r_1$: each $BCDE$ is contained in an $r_1$-ball (``reference patch''), and $r_0$ determines the thickness of $B,D,E$. 
\end{Definition}

Local purifiability is a supplementary condition and is not assumed by default. When it is needed, a set $\Lambda'$ of $(B,C)$ pairs is chosen such that for each $(B,C)\in\Lambda'$, the region $BC$ is contained in some data patch $K$ and $\sigma_K$ is $(B,C)$-purifiable. The standard is to have one such pair for each reference patch $L$, specifically, $C=L$ and $B=L'\setminus L$, where $L'$ is the $r_0$-neighborhood of $L$.

A key concept is that of the \emph{information convex set} from Refs.~\cite{shi2019characterizing,shi2020fusion}, which we state in a more general form. For a given region $Y$, let $Y^+\supseteq Y$ be a slightly bigger region and $\Gamma_1(Y)$ some set of reference patches contained in $Y^+$. The standard choices are (see Fig.~\ref{fig:data_org})
\begin{equation}\label{Gamma1_Yplus}
\Gamma_1(Y)=\{L\in\Gamma_1:\, L\cap Y\not=\emptyset\},\qquad
Y^+=Y\cup\bigcup_{L\in\Gamma_1(Y)}L.
\end{equation}
with $\Gamma_1$ consisting of $r_1$-balls; then equivalently, $Y^+ = \Ba_{2r_1}(Y)$, the $2r_1$-neighborhood of $Y$.

\begin{Definition}\label{def:ICS}
The \emph{information convex set} of region $Y$ is the set $\IC(Y)$ of states $\rho\in\St_Y$ that admit an extension $\rho_+\in\St_{Y^+}$ satisfying $(\rho_+)|_L=\sigma_L$ for all $L \in\Gamma_1(Y)$. 
\end{Definition}
By construction, $\IC(Y)$ is a convex subset of $\St_Y$.
The need to consider $\rho_+$ arises because the mere consistency of $\rho$ with all $\sigma_L$ does not prevent distant parts of the boundary of $Y$ from being entangled with each other. 

We will establish certain properties of the information convex sets of ``tame'' regions, namely, embedded $d$-dimensional manifolds $Y$ with boundary $\partial Y$ of sufficient regularity. The definition will also be extended to immersed manifolds with boundary. Generalizing the isomorphism theorem of Ref.~\cite{shi2020fusion} to our setting, we will prove that the information convex set is invariant under small deformations of the boundary, hence also invariant under regular homotopies of the region. A related property is that $\IC(Y)$ is stable to parameter changes such as a small decrease of $r_1$, and consequently, of $Y^+$.

The structure of the information convex set was described in Ref.~\cite{shi2020fusion}. In future work, we will demonstrate that a similar description holds when assuming local purifiability in addition to local extendibility. (Our working argument begins with applying Corollary~\ref{cor:supp_puri} to reformulate the definition of $\IC(Y)$ in terms of subspace constraints of the form ``$\rho|_{L}$ is supported by subspace $\calL_{L}$,'' rather than using constraints $\rho|_{L} = \sigma_L$ specified by reference states $\sigma_L$.  
Ultimately, we intend to revisit this topic using a set of such constraints, or even a frustration-free Hamiltonian, as the defining local data.) 
The claim, not proven here, is that $\IC(Y)$ can be identified with the set of states on a finite-dimensional $C^*$-algebra of the form
\begin{equation} \label{eq:IC-structure}
\calA=\bigoplus_{s\in S}\LL(V_s).
\end{equation}
Here $S$ is some set of ``topological sectors'' associated with $\partial Y$. For example, let $Y\subseteq\RR^2$ be a disk with two holes. Then $\partial Y$ is the union of three circles labeled by some ``anyonic charges'' $a$, $b$, $c$. In this case, $s=(a,b,c)$, and $V_s$ is the corresponding fusion space. A state on the algebra $\calA$ is given by assigning some probabilities $p_s$ and conditional states $\tau_s\in\St(\LL(V_s))$ to each topological sector $s$.

If we just consider extendible local data without local purifiability, this allows probabilistic mixtures of states corresponding to vacua in distinct topological phases.
The description in Eq.~\eqref{eq:IC-structure} would still hold if all such mixtures were allowed. However, in the setting based on conditions of the form $\rho|_L=\sigma_L$, the probabilities of the vacua are fixed. The alternative setting mentioned above, based on subspace constraints, does allow arbitrary mixtures. 

\subsection{Sketch of the isomorphism theorem}
\label{subsec:iso_sketch}

The full statement and proof of the isomorphism theorem in Section~\ref{sec:isomorphism} requires Section~\ref{sec:geom} to formalize the geometries used for the small growth steps seen in Fig.~\ref{fig:growth0}.
Because the isomorphism theorem is one of our core technical results, in this section we sketch a somewhat complete but less formal argument, leaving some geometric details to the imagination.

We assume extendible local data (Definition~\ref{def:extendible-data}) and consider the information convex set of ``tame'' regions (Definition~\ref{def:macro-region}). These are embedded $d$-dimensional manifolds $X$ with a sufficiently smooth boundary $\partial X$ (with radius of curvature at least $r_2$) and tubular neighborhood (of thickness at least $r_2$). Tameness ensures that patches of $X$ at scale $r_1 \ll r_2$ that include a piece of the boundary look approximately like a Euclidean half-space. This allows pictures like Fig.~\ref{fig:growth0}, where the boundary initially appears flat.
While the tameness requirement could be weakened, some regularity condition is necessary; the theorem would break for certain jagged or thin regions.
\begin{Theorem}[Isomorphism theorem, simplified] \label{thm:isomorphism-informal}
When tame regions $X$ and $Y$ are isotopic,
the information convex sets $\IC(X)$ and $\IC(Y)$ are isomorphic: there exist quantum channels $\Phi_{X \to Y} : \IC(X) \to \IC(Y)$ and $\Phi_{Y \to X} : \IC(Y) \to \IC(X)$ that compose both ways to the identity. 
\end{Theorem}
\noindent Here isotopic means connected by a homotopy of tame regions. The more general and precise statement is
Theorem~\ref{thm:isomorphism}, which allows immersed regions and elaborates on
compositional properties. 
A channel \(\Phi_{X \to Y} : \IC(X) \to \IC(Y)\) denotes the restriction to
\(\IC(X)\) of a quantum channel \(\St_X \to \St_Y\), with image contained in
\(\IC(Y)\).
 Naively, $\IC(X)$ may be empty, because we do not assume the existence of a global state consistent with the local data. Later, we find a global state on $\RR^d$ indeed exists---say, in the sense that $\IC(X)$ has a unique state on arbitrarily large geometric balls.  

Because any isotopy can be decomposed into a sequence of small deformations, to prove the isomorphism theorem it suffices to consider $X$ and $Z$ related by a small deformation.  Let $Y=X^{+r}$ be obtained from $X$ by extending $\partial X$ outward a constant distance $r$, so that $X,Z \subset Y$.  Thus it suffices to show $\IC(X) \cong \IC(Y)$ whenever $X \subset Y$, because then we can apply this to both $X \subset Y$ and $Z \subset Y$ to yield $\IC(X) \cong \IC(Z)$. For the remainder of our sketch, we therefore consider tame isotopic regions
\begin{equation}
X \subset Y = X^{+r_1}.
\end{equation}
 Below we define  $\Phi_{Y \to X} : \IC(Y) \to \IC(X)$ via restriction and define  $\Phi_{X \to Y} : \IC(X) \to \St_Y$ via growth maps.  Then we show the following, in order:
\begin{align}
\label{eq:XtoYtoX}   \Phi_{X\to Y}\Phi_{Y\to X} & = \mathrm{id}_{\IC(Y)}, \\
\label{eq:SigmaXtoSigmaY}  \Phi_{X\to Y}(\IC(X)) & \subseteq \IC(Y), \\
\label{eq:YtoXtoY}  \Phi_{Y\to X}\Phi_{X\to Y} & = \mathrm{id}_{\IC(X)}.
\end{align}

By Definition~\ref{def:ICS}, each state in $\IC(Y)$ has an extension $\rho \in \St_{Y^+}$ that is consistent with reference patches, so we can define the restriction
\begin{equation}
\Phi_{Y \to X}[\rho|_Y]=\rho|_X, \qquad \Phi_{Y \to X} : \IC(Y) \to \IC(X),  \qquad X \subset Y
\end{equation}
and verify that indeed $\rho|_X \in \IC(X)$.  To define $\Phi_{X \to Y}$ requires more care. 

\begin{figure}[t]
    \centering
   \input{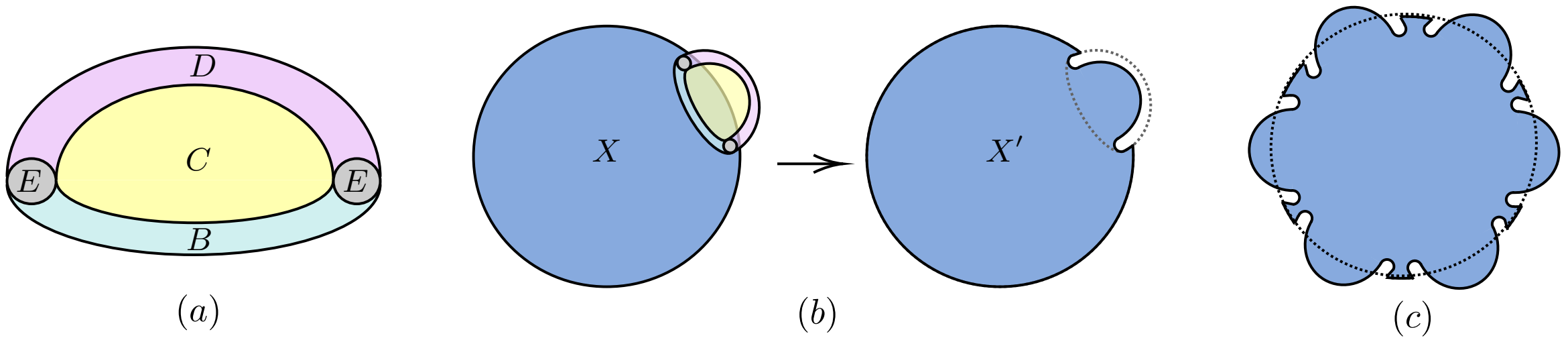} 
    \caption{\textbf{(a)} Regions for $(B,C,D,E)$-extendibility. \textbf{(b)} Let $\rho$ be any $(B,C,D,E)$-extendible state defined on $X \cup BCDE$ with  $BE \subset X$. Then $\rho|_{X'} =(\id_{X \setminus BCDE} \otimes R_{BE \to BC})[\rho|_{X \setminus CD}]$, where $X'=(X \setminus DE) \cup C$.  This follows from the extendibility property~\eqref{eq:loc_extend} with $A=X \backslash BCDE$. \textbf{(c)} Growing with one layer of a circuit.}
    \label{fig:growth-step}
\end{figure}

First we describe an elementary growth step $X_i \to X_{i+1}$, where $X_{i+1}$ is a slight extension of $X_i$.
 Recall that for each configuration $(B,C,D,E) \in \Lambda$, $BCDE$ is contained in a reference patch $L$ and local data $\sigma_L$ is $(B,C,D,E)$-extendible.  Choose a configuration $(B_i,C_i,D_i,E_i) \in \Lambda$ with $B_i E_i \subset X_i$, and let 
\begin{equation} X_{i+1} = (X_i \setminus D_iE_i) \cup C_i.
\end{equation}
Then, suppressing identities, we have a channel
\begin{equation}
R^i = R_{B_iE_i \to B_iC_i} \circ \Tr_{X_i \cap (C_iD_i)} : \St_{X_i} \to \St_{X_{i+1}}.
\end{equation}
(Sometimes we choose to arrange $X_i \cap (C_iD_i) = \emptyset$, in which case $ \Tr_{X_i \cap (C_iD_i)}$ above is trivial.)
Fig.~\ref{fig:growth-step}(b) depicts such a growth step $X \to X'$ for the case of a ball $X$.  For any state $\rho$ defined on $X_i \cup B_iC_iD_iE_i$ that is $(B_i,C_i,D_i,E_i)$-extendible,\begin{equation} 
\rho|_{X_{i+1}} = R^i [\rho|_{X_i}]
\end{equation}
This follows directly from the extendibility property~\eqref{eq:loc_extend} using $A=X_i \backslash B_iC_iD_iE_i$. Assume we have regions $X_0,\ldots,X_n$ and configurations $(B_i,C_i,D_i,E_i)$ for $i=0,\ldots,n-1$, satisfying the above relations. Then for any locally extendible state $\rho$ defined on all these regions, 
\begin{equation} \label{eq:X0Xn}
\rho|_{X_n} = R^{n-1} \circ \ldots \circ R^0 [\rho|_{X_0}].
\end{equation}

The above is completely general. Returning to the discussion of isotopic tame regions $X \subset Y$, we seek such a sequence $(X_i)_i$ ending with $X_n$ satisfying $Y \subset X_n \subset Y^+$.
Then we can define
\begin{equation} \label{eq:XtoY-R}
\Phi_{X \to Y} = \Tr_{X_n \backslash Y} \circ R^{n-1} \circ \ldots \circ R^0, \qquad \Phi_{X \to Y} :  \IC(X) \to \St_Y.
\end{equation}
(Note that the definition of $\Phi_{X\to Y}$ naively depends on choices in the circuit construction; after proving the theorem, we can conclude that $\Phi_{X\to Y}$ is independent of these choices on $\IC(X)$, because it is an inverse to $\Phi_{Y \to X}$.)
From Eq.~\eqref{eq:X0Xn}, if each $B_iC_iD_iE_i \subset L_i$ for some reference patch $L_i$ overlapping $Y$,
 it follows that for all $\rho \in \St_{Y^+}$ consistent with reference patches,
\begin{equation} \label{eq:XtoY}
\Phi_{X \to Y}[\rho|_X] = \rho|_Y.
\end{equation}

Intuitively, by choosing $C$ sufficiently larger than $E$, we can achieve the desired cumulative growth $X_0 \to \cdots \to X_n$.
  We must postpone a careful proof, especially since we have not yet precisely defined the allowed configurations $(B,C,D,E) \in \Lambda$ for which we impose extendibility.  However, the basic idea is straightforward.
We collect individual growth steps (or ``gates'') for disjoint regions $\{B_iC_i D_i E_i\}_i$ into a single ``layer,'' where the associated channels are applied in parallel, as in Fig.\ref{fig:growth-step}(c). Multiple layers constitute a ``geometric circuit,'' analogous to traditional quantum circuits.  We have flexibility in the exact choice of $(B,C,D,E)$ shapes.  Fig.~\ref{fig:growth0}(b) shows more carefully how to obtain net growth, in that case using two layers.   We will define configurations $(B,C,D,E) \in \Lambda$ at scale $\lesssim r_1$ so that $X$ need only be $r_1$-tame; then $\partial X$ is approximately locally flat at this scale, and a circuit of constant depth can extend $\partial X$ a distance $r_1$ to obtain $Y=X^{+r_1}$.

 Eq.~\eqref{eq:XtoY} says that the channel \(\Phi_{X\to Y}\) recovers an arbitrary state \(\rho \in \IC(Y)\) after it has been restricted to region \(X\), i.e.,  Eq.~\eqref{eq:XtoYtoX}.
However, we have not yet shown that every \(\rho_0 \in \IC(X)\) remains consistent through the growth process, i.e., Eq.~\eqref{eq:SigmaXtoSigmaY}.

First we argue the following proposition:
\begin{Proposition}[Informal] \label{prop:ball-unique-informal}
If region $Y$ is tame at scale $r_1$ and isotopic to a geometric ball,
then $\IC(Y)$ has at most one state.
\end{Proposition}
\noindent 
To see this, we take $X \subset Y$ to be an $r_1$-ball isotopic to $Y$ via an isotopy contained in $Y$, and define $\Phi_{X \to Y}$ similarly to Eq.~\eqref{eq:XtoY-R}.  Recall Eq.~\eqref{eq:XtoY-R} was described for the case $Y=X^{+r_1}$ with $X$ $r_1$-tame.  However, here we have a more general isotopy $X \to Y$.  We therefore define $\Phi_{X \to Y}$ by a sequence of maps $\Phi_{X' \to X'^{+r_1}}$ interspersed with additional erasures, trimming $X'^{+r_1}$ to smooth and reshape the boundary, thereby maintaining tameness at scale $r_1$ before each new growth step $\Phi_{X' \to X'^{+r_1}}$.
Then for all $\rho \in \St_{Y^+}$ consistent with reference patches, Eq.~\eqref{eq:XtoY} yields $\rho|_Y = \Phi_{X \to Y}[\rho|_X] = \Phi_{X \to Y}[\sigma_X] $ for reference patch $\sigma_X$.  This yields an expression for $\rho|_Y$ that is independent of $\rho$; it follows $\IC(Y)$ has at most one state.
This concludes the argument for Prop.~\ref{prop:ball-unique-informal}.

\begin{figure}[t]
    \centering
   \input{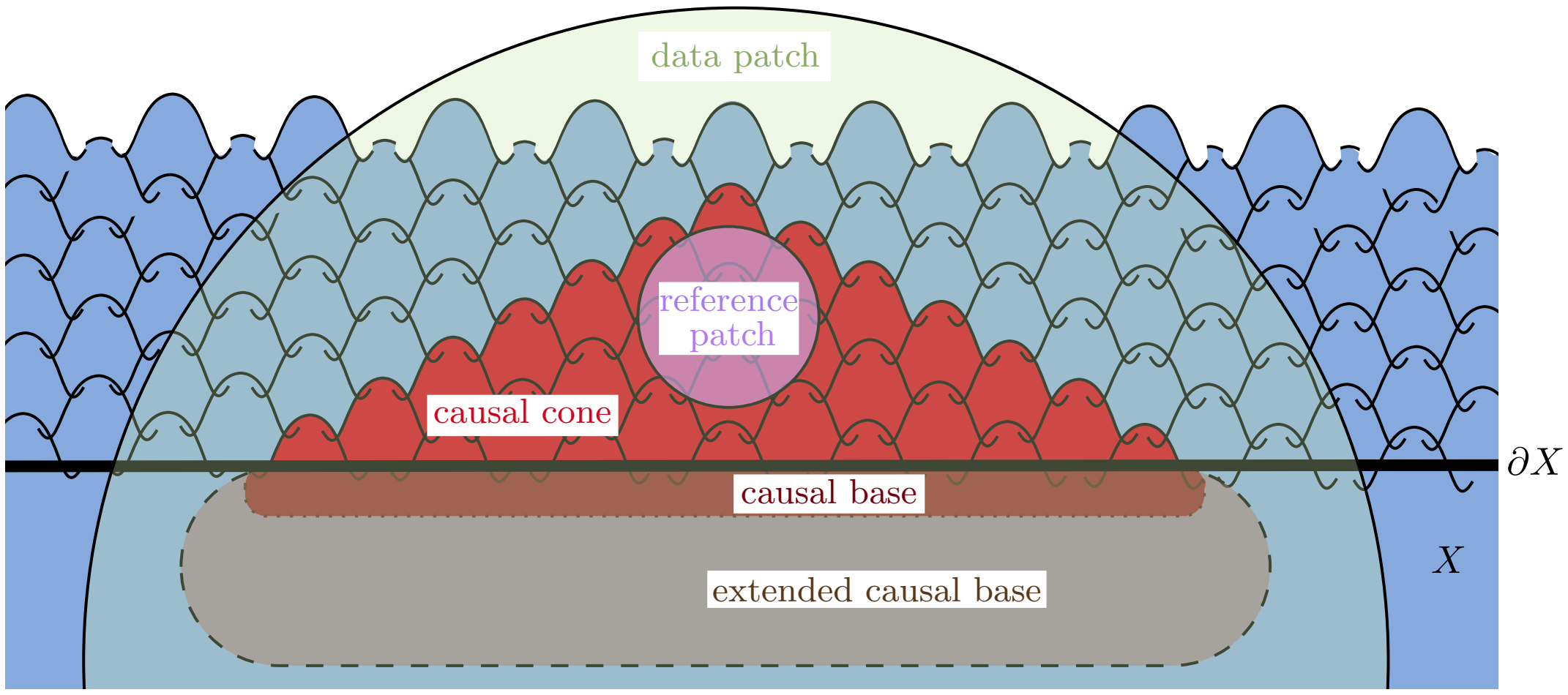} 
    \caption{We illustrate a step in the proof of the isomorphism theorem.  The region $X$ below the thick horizontal line $\partial X$ is extended upward with a circuit of channels. To verify the extended state is consistent with a reference patch, we consider the causal cone of the reference patch. The state on the reference patch can be obtained by acting on the causal base, a subset of $X$, using only the channels in the causal cone.}
    \label{fig:extend-with-cone}
\end{figure}

Next we show Eq.~\eqref{eq:SigmaXtoSigmaY}.  Given $\rho_X \in \IC(X)$, we must show $\Phi_{X \to Y}[\rho_X] \in \IC(Y)$, i.e., the grown state is consistent with reference patches. This is somewhat delicate. After applying $R_{BE \to BC}$, it is straightforward to show that consistency is maintained on reference patches either containing $BCDE$ or disjoint from $BCDE$, but we must worry about patches with partial overlap. 
 The resolution is to upgrade consistency with reference patches to consistency with larger patches, using Proposition~\ref{prop:ball-unique-informal} above, combined with a lightcone argument illustrated in Fig.~\ref{fig:extend-with-cone}. 

Recall $X \subset Y=X^{+r_1}$.  Now it becomes important that $X$ is $r_2$-tame with $r_2 \gg r_1$.
Define $\Phi_{X \to X^{+3r_1}} : \IC(X) \to \St_{X^{+3r_1}}$ by a circuit of channels that grows slightly past $X^{+3r_1}$ and then trims.  Given $\rho_X \in \IC(X)$, we want to show that the state
\begin{equation}
\rho'  = \Phi_{X \to X^{+3r_1}}[\rho_X] \in \St_{X^{+3r_1}}
\end{equation}
is consistent with reference patches overlapping $X^{+r_1}$, to conclude $\rho'|_{X^{+r_1}} \in \IC(X^{+r_1})$.  

Let $L$ be an $r_1$-ball overlapping $X^{+r_1}$.  We must show $\rho'|_L = \sigma_L$ for reference patch $\sigma_L$.   The ``causal cone'' (or past light cone) of $L$ with respect to the circuit refers to the union $\cup_i B_iC_iD_iE_i$ of all the gates that influence $L$, as in Fig.~\ref{fig:extend-with-cone}.   The ``causal base'' $P$ is the intersection of the causal cone with $X$; then $\rho'|_L$ is determined entirely by $\rho_X|_P$. 
Equivalently, in the Heisenberg picture, the adjoint channel of the circuit maps $\Alg_L$ into $\Alg_P$, so expectation values of observables on $L$ only depend on the input's reduced state on $P$.

 Let $\wt P$ be the ``extended causal base,'' a smooth ball with $ P \subset \wt P \subset  X$,  like the smooth dashed region surrounding the causal base in Fig.~\ref{fig:extend-with-cone}.
We take $\wt P$ tame at scale $r_1$---that is, tame enough to apply Proposition~\ref{prop:ball-unique-informal}, but still small enough to fit easily in an $r_2$-ball.  

Let $K$ be such an $r_2$-ball that contains the reference patch $L$, the causal cone, and the neighborhood $\wt P^+$ of the extended causal base. 
Note that $\wt P \subset X$ and $\rho_X \in \IC(X)$ imply $\rho_X|_{\wt P} \in \IC(\wt P)$. Since $K$ contains $\wt P^+$, the data patch $\sigma_K$ also witnesses that $\sigma_K|_{\wt P} \in \IC(\wt P)$. Proposition~\ref{prop:ball-unique-informal} therefore implies
\begin{equation}
\rho_X|_{\wt P} = \sigma_K|_{\wt P}.
\end{equation}
 In particular, $\rho_X$ matches the data patch on the causal base, $\rho_X|_P = \sigma_K|_P$.   Recall $\rho'|_L$ is determined entirely by the action of the gates in the causal cone on $\rho_X|_P$. Substituting $\rho_X|_P = \sigma_K|_P$, we find  $\rho'|_L = \sigma_K|_L = \sigma_L$.  We conclude
\begin{equation}
\rho'|_{X^{+r_1}} \in \IC(X^{+r_1}),
\end{equation}
concluding the argument for Eq.~\eqref{eq:SigmaXtoSigmaY}.

Finally we show  Eq.~\eqref{eq:YtoXtoY}.
Choose $W \subset X$ isotopic to $X$, so that $W \subset X \subset Y$.  Allow a slight buffer, $W^{+r_1} \subset X$, so that $\Phi_{X \to Y}$ is supported on $X \setminus W$.
Then
\begin{align}
\Phi_{X\to W} = \Phi_{X\to W}\Phi_{Y\to X}\Phi_{X\to Y} 
\end{align}
because $\Phi_{X\to Y}$  acts as the identity on $W$, while $\Phi_{Y \to X}$ and $\Phi_{X \to W}$ are just restrictions.
Composing with $\Phi_{W \to X}$ on the left and using the fact that $\Phi_{W\to X}\Phi_{X\to W} = \id_{\IC(X)}$, we conclude that  $\Phi_{Y\to X}\Phi_{X\to Y} = \id_{\IC(X)}.$

\section{General properties of extendible states}
\label{subsec:circuit_invariance}

This section addresses circuit-invariance of local extendibility and purifiability, as well as an equivalent reformulation of extendibility via immersed regions. 

First we show that when a state is locally extendible, it remains locally extendible after action by a constant-depth circuit, using regions that are modified by a constant distance. Similar circuit invariance holds for local purifiability. Both proofs rely on the geometric monotonicity property of Proposition~\ref{prop:g_mono}, along with invariance under unitaries supported within individual subsystems.  We begin with the simpler case of local purifiability.   In this section, $X^{+r}$ for region $X$ denotes the closed $r$-neighborhood of $X$. 

\begin{Theorem}[Circuit-invariance of local purifiability]\label{thm:circuit-invariance-A0}
    Consider disjoint regions $(B,C)$ and, separately, disjoint regions $(B',C')$.  Consider a unitary circuit $U$ composed of $\ell$ layers of gates, commuting within each layer, each gate with support of diameter at most $s$.  Define light-cone radius $r := \ell s$.  Assume  $B^{+r}  \subseteq B'$, $C' \subseteq C$, $(BC)^{+r} \subseteq B'C'$, and $C^{+s} \subseteq BC$. Let $\sigma$ be a state on $(B'C')^{+r}$ that is $(B,C)$-purifiable, and let $\sigma' = U \sigma U^\dagger$.  Then $\sigma'$ is $(B',C')$-purifiable.
\end{Theorem}
\noindent We are especially interested in the case $C$ and $BC$ are both topological balls, with $B$ a thickened spherical shell. In general, we write $(B,C)\cge(B',C')$ when $B \subseteq B'$, $C' \subseteq C$, and $BC\subseteq B'C'$, in which case $(B,C)$-purifiability implies $(B',C')$-purifiability.  Because $(B,C)$-purifiability may be seen as a special case of $(B,C,D,E)$-extendibility with $D,E$ empty, this monotonicity property may be seen as a special case of Proposition~\ref{prop:g_mono}. 
\begin{proof}
Without loss of generality, set $B' = B^{+r}$ and $C' = C \setminus B'$; the desired result follows by monotonicity. We specialize to the case of $\ell=1$, i.e., a single layer $U = \prod_i U_i$ of commuting unitaries $U_i$ of support diameter $s$. (The $U_i$ will not require disjoint support.)
 The case $\ell>1$ then follows directly by $\ell$ repeated applications of the case $\ell=1$.

Let $\sigma_0 = U_0 \sigma U_0^\dagger$ where $U_0$ consists of gates supported outside $BC$.  Thus $\sigma_0$ remains $(B,C)$-purifiable.  Let $U_1$ consist of gates supported within $C$.  Then $\sigma_1 = U_1 \sigma_0 U_1^\dagger$ remains $(B,C)$-purifiable, and $\sigma_1$ is also $(B',C')$-purifiable, because $(B,C) \cge (B',C')$.  Let $\sigma_2 = U_2 \sigma_1 U_2^\dagger$, where $U_2$ consists of the remaining gates supported within $B'$.  Then $\sigma_2$ is $(B',C')$-purifiable.   

What remains to show is that actually $\sigma_2 = \sigma'$, i.e., we have applied all the gates of $U$.  (Their order does not matter because they commute.)  It suffices to show that for every subset $S$ of diameter at most $s$, then either $S \subseteq (BC)^c$, $S \subseteq C$, or $S \subseteq B'$, because these cases are handled by $U_0, U_1, U_2$, respectively.  Consider such a subset $S$. If $S \subseteq (BC)^c$ or $S \subseteq C$, we are done.  Otherwise, if $S \not \subset C$ but $S$ overlaps $C$, then $S $ overlaps $C^{+s} \setminus C$ because $\mathrm{diam}(S) \leq s$.  Note $C^{+s} \setminus C \subseteq B$ using $C^{+s} \subseteq BC$, so $S$ overlaps  $B$.  Using $\mathrm{diam}(S)\leq s$ and $B^{+s} \subseteq B'$, that implies $S \subseteq B'$.
\end{proof}

\noindent The analogous proof for extendible states is similar but more elaborate. 

\begin{Theorem}[Circuit-invariance of extendibility]\label{thm:circuit-invariance}
    Consider disjoint regions $(B,C,D,E)$ and, separately, disjoint regions $(B',C',D',E')$.  Consider a unitary circuit $U$ composed of $\ell$ layers of gates, commuting within each layer, each gate with support of diameter at most $s$.  Define light-cone radius $r := \ell s$.  Assume $(BE)^{+2r}  \subseteq B'E'$, $(DE)^{+2r} \subseteq D'E'$, $C' \subseteq C$, $(BCDE)^{+2r} \subseteq B'C'D'E'$, and $C^{+s} \subseteq BCDE$. Let $\sigma$ be a state on $(B'C'D'E')^{+r}$ that is $(B,C,D,E)$-extendible, and let $\sigma' = U \sigma U^\dagger$.  Then $\sigma'$ is $(B',C',D',E')$-extendible.
    \end{Theorem}

\noindent We are especially interested in the case that $(B,C,D,E)$ and $(B',C',D',E')$ are defined using neighborhoods of hemispheres, as introduced above Definition~\ref{def:loc_extend} and elaborated in Section~\ref{sec:micro}.  If we define $(B,C,D,E)$ and $(B',C',D',E')$ by $t$-neighborhoods and $t'$-neighborhoods, respectively, of the same hemispheres, then the various hypotheses above hold whenever $t' > t + 2r$ and $t>s$.   To obtain a recovery map $R_{B'E' \to B'C'}$ for $\sigma'$, given the recovery map $R_{BE \to BC}$ for $\sigma$, one may attempt a direct construction as follows: $R_{B'E' \to B'C'}$ acts by (1) removing circuit gates from a neighborhood of $BE$, (2) applying $R_{BE \to BC}$, (3) re-applying some missing gates, and then (4) restricting to $B'C'$.  However, we opt for a less direct proof which is easier to make precise.   Alternatively, the curious reader may note an independent  proof by using the criterion for extendibility in Theorem~\ref{thm:immersion1}.

The hypotheses $(BE)^{+2r}  \subseteq B'E'$, $(DE)^{+2r} \subseteq D'E'$ could be replaced with $(BE)^{+r}  \subseteq B'E'$, $(DE)^{+r} \subseteq D'E'$ along with $\wt{E}^{+r} \subseteq E'$ where $\wt{E} = (BE)^{+r} \cap (DE)^{+r}$.

Recall that in general, we write $(B,C,D,E)\cge(B',C',D',E')$ if the inclusions~\eqref{g_mono} hold.  Then by Proposition~\ref{prop:g_mono}, $(B,C,D,E)$-extendibility implies $(B',C',D',E')$-extendibility. 
\begin{proof}[Proof of Theorem~\ref{thm:circuit-invariance}]
Without loss of generality, modify $B',C',D',E'$ by setting composite regions $B'E' = (BE)^{+2r}$, $D'E' = (DE)^{+2r}$, then setting $E' = (B'E') \cap (D'E')$ and $C' = C \setminus (B'D'E')$; the desired result follows by monotonicity.
We specialize to the case of $\ell=1$, i.e., a single layer $U = \prod_i U_i$ of commuting unitaries $U_i$ of support diameter $s$. (The $U_i$ will not require disjoint support.)
The case $\ell>1$ then follows directly by $\ell$ repeated applications of the case $\ell=1$.

We will define regions $(B_1, C_1, D_1, E_1)$ via an enlargement of $BE$ and $DE$, such that
\[
(B,C,D,E) \cge (B_1,C_1,D_1,E_1) \cge (B',C',D',E').
\]
Extendibility on $(B,C,D,E)$ therefore implies extendibility on $(B_1,C_1,D_1,E_1)$, and hence on $(B',C',D',E')$.
In particular, define composite regions $B_1E_1 = (BE)^{+s}$, $D_1 E_1 = (DE)^{+s}$, and then define disjoint regions $B_1=(B_1E_1)\setminus E_1$, $D_1=(D_1E_1)\setminus E_1$, $E_1 = (B_1 E_1) \cap (D_1 E_1)$, $C_1 = C \setminus (B_1 D_1 E_1)$.  

Note $(B,C,D,E)$-extendibility is invariant under unitaries supported within $B$, $C$, $D$, or $E$, or outside of $BCDE$. By assumption $\sigma$ is $(B,C,D,E)$-extendible.    Let $\sigma_0 = U_0 \sigma U_0^\dagger$ where $U_0$ consists of gates supported outside $BCDE$. Thus $\sigma_0$ remains $(B,C,D,E)$-extendible. Let $\sigma_1 = U_1 \sigma_0 U_1^\dagger$ where $U_1$ consists of gates supported within $C$.  Then $\sigma_1$ is still $(B,C,D,E)$-extendible, hence also $(B_1,C_1,D_1,E_1)$-extendible.  Let $U_2$ consist of the remaining gates of $U$ that are fully supported within one of $B_1,C_1,D_1,$ or $E_1$.  Then $\sigma_2 = U_2 \sigma_1 U_2^\dagger$ is  $(B_1,C_1,D_1,E_1)$-extendible, hence also $(B',C',D',E')$-extendible.  Let $U_3$ consist of the remaining gates of $U$ that are fully supported within $E'$.  Then $\sigma_3 = U_3 \sigma_2 U_3^\dagger$ is  $(B',C',D',E')$-extendible.

What remains to show is that in fact $\sigma_3 = \sigma'$, i.e., we have applied all the gates in $U$.  (Their order does not matter because they commute.) It suffices to show that for every subset $S$ with $\mathrm{diam}(S) \leq s$, it holds $S \subseteq A$ for at least one $A  \in \{(BCDE)^c\} \cup \{C\} \cup \{B_1, D_1,E_1\} \cup \{E'\}$, which are handled by $U_0, U_1, U_2, U_3,$ respectively. If $S \subseteq (BCDE)^c$, we are done. 
Otherwise, $S$ overlaps at least one of $B,C, D,E$.   We treat each case below.  We repeatedly use the fact that for any region $Z$, $Z \cap S \neq \emptyset$ and $\mathrm{diam}(S) \leq s$ together imply $S \subseteq Z^{+s}$.
 If $S$ overlaps $E$, then $S \subseteq E^{+s} \subseteq (BE)^{+s} \cap (DE)^{+s} = E_1$.  If $S$ overlaps  $B$, then $S \subseteq B^{+s} \subseteq B_1 E_1$, and either $S\subseteq B_1$ or $S \cap E_1 \neq \emptyset$. In the latter case, $S \subseteq E_1^{+s} \subseteq E'$.  The case $S$ overlaps $D$ follows symmetrically. Finally, if $S$ overlaps $C$, then using $C^{+s} \subseteq BCDE$ and $\mathrm{diam}(S) \leq s$, either $S \subseteq C$ or $S$ overlaps $C^{+s} \backslash C$.  In the latter case, $C^{+s} \setminus C \subseteq BDE$ implies $S$ overlaps $BDE$, treated previously.
\end{proof}

\begin{remark}
Theorem \ref{thm:circuit-invariance} can be generalized to the case that $\sigma'$ is related to $\sigma$ by a locality-preserving automorphism, also known as a quantum cellular automaton (QCA).
 We sketch a proof for readers familiar with the subject.  For any QCA $\alpha$ with light-cone $r$ on lattice Hilbert space $\mathcal{H}$, we consider a doubled space $\mathcal{H}_1 \otimes \mathcal{H}_2$ with two copies of $\mathcal{H}$ seen as two stacked layers.  Then the QCA $\alpha \otimes \alpha^{-1}$ can be implemented by a two-layer circuit whose gates have diameter at most $2r$ and commute within each layer \cite{gross2012index}, for any spatial dimension.  We use the straightforward facts that (1) local extendibility is preserved under stacking, and (2) when a state of the form $\rho_1 \otimes \rho_2$ is locally extendible, then so are $\rho_1$ and $\rho_2$.
By (1), the state $\sigma_1 \otimes \sigma_2$ is $(B_1 B_2,C_1 C_2,D_1 D_2,E_1 E_2)$-extendible, so we can apply Theorem~\ref{thm:circuit-invariance} to show that $\alpha(\sigma_1) \otimes \alpha^{-1}(\sigma_2)$ is $(B'_1B'_2,C'_1C'_2,D'_1D'_2,E'_1E'_2)$-extendible. Then by (2), $\alpha(\sigma_1)$ is $(B',C',D',E')$-extendible, as desired. Analogous remarks apply to Theorem~\ref{thm:circuit-invariance-A0}.
\end{remark}

We discuss some additional miscellaneous properties of local extendibility. It behaves nicely with tensor products, which corresponds to stacking sheets of material.
\begin{Proposition}\sloppy Let $\rho$ and $\rho'$ be states on distinct systems $BCDE$ and $B'C'D'E'$.  Then $\rho \otimes \rho'$ is $(BB', CC', DD', EE')$-extendible if and only if both $\rho$ is $(B,C,D,E)$-extendible and $\rho'$ is $(B',C',D',E')$-extendible.
\end{Proposition}
That is, the extendibility of the stacked systems is equivalent to the extendibility of the individual systems.    The proof is straightforward.  To show the individual systems are extendible given the extendibility of their product, one constructs a recovery map for an individual system that first tensors the state with the other system and then applies the joint-system recovery map.

The mixture of extendible states may not be extendible: for instance, the maximally mixed state on $BCDE$ is not extendible.  However, the converse holds; see Proposition \ref{prop:extendibility-mixture}.

The  extendibility condition on $BCDE$ can be checked with a semidefinite program, elaborated in Ref.~\cite{kim2023universal}.  Therefore, given a classical numerical representation of the state on $BCDE$, the condition can be checked in a time that scales polynomially in the Hilbert space dimension.

We show that the notion of local extendibility is equivalent to the existence of some state on an immersed manifold, which can serve as a physical motivation for this condition. Let us begin with the following abstract statement.

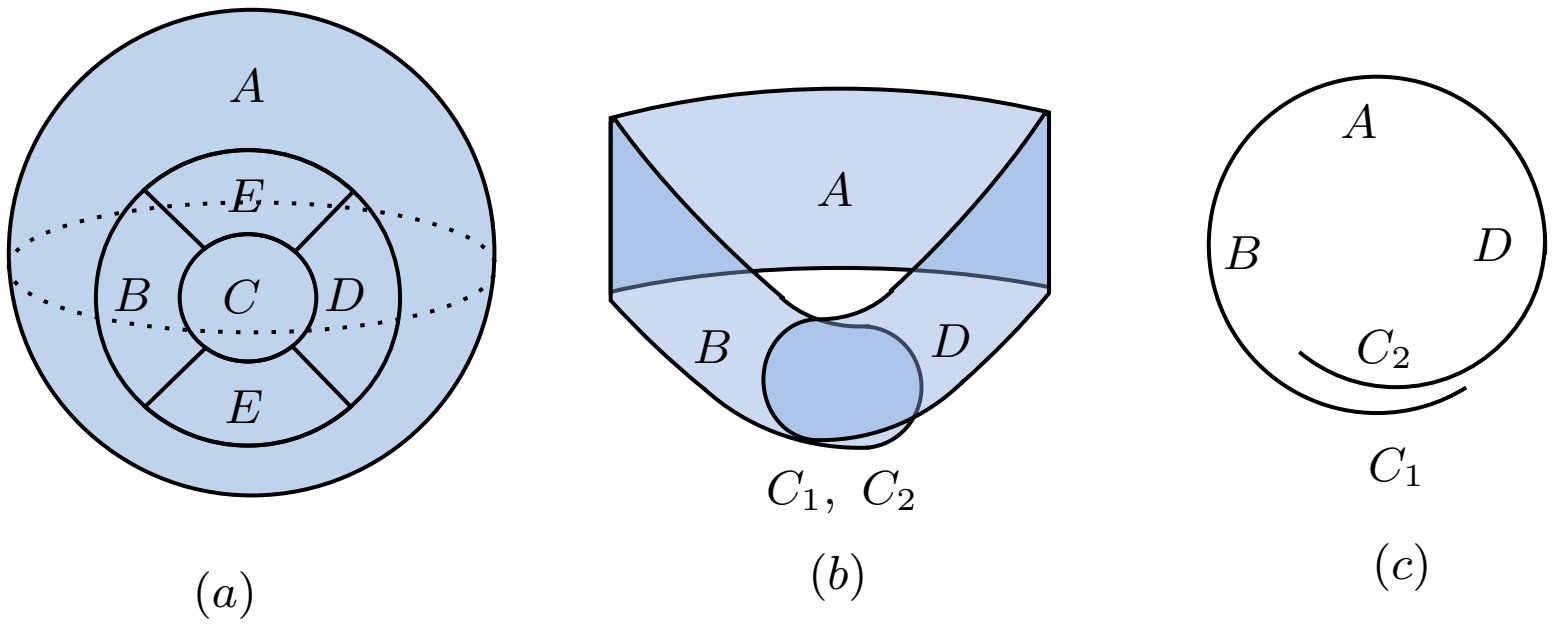
\begin{figure}[t]
    \centering
    \tikzset{every picture/.style={line width=0.75pt}} 

\begin{tikzpicture}[x=0.75pt,y=0.75pt,yscale=-1,xscale=1]

\draw  [fill=tikzblue  ,fill opacity=0.43 ] (175.5,37.39) -- (175.5,81.01) .. controls (191.57,77.23) and (211.57,74.99) .. (233.25,74.99) .. controls (252.52,74.99) and (270.46,76.76) .. (285.5,79.81) -- (285.5,35.98) .. controls (269.46,32.13) and (251.8,30) .. (233.25,30) .. controls (212.56,30) and (192.97,32.65) .. (175.5,37.39) -- cycle ;
\draw  [fill=tikzblue  ,fill opacity=0.43 ] (237.99,120.13) .. controls (222.84,120.13) and (209.11,114.53) .. (199.13,105.46) -- (199.12,105.47) .. controls (194.98,102.11) and (190.86,98.5) .. (186.81,94.67) .. controls (182.83,90.9) and (179.06,87.04) .. (175.5,83.14) -- (175.5,38.59) -- (176.46,37.49) .. controls (183.98,48.12) and (194.43,60.1) .. (206.84,71.85) .. controls (210.77,75.58) and (214.71,79.09) .. (218.6,82.38) -- (218.44,82.57) .. controls (223.16,86.92) and (230.17,89.68) .. (237.99,89.68) .. controls (238.64,89.68) and (239.29,89.66) .. (239.93,89.62) -- (239.93,89.73) .. controls (247.53,90.35) and (253.5,96.91) .. (253.5,104.9) .. controls (253.5,112.9) and (247.53,119.46) .. (239.93,120.08) -- (239.93,120.1) .. controls (239.77,120.11) and (239.61,120.11) .. (239.45,120.11) .. controls (239.21,120.13) and (238.98,120.13) .. (238.74,120.13) .. controls (238.66,120.13) and (238.58,120.13) .. (238.49,120.13) .. controls (238.33,120.13) and (238.16,120.13) .. (237.99,120.13) -- cycle ;
\draw  [fill=tikzblue  ,fill opacity=0.43 ] (228.08,118.21) .. controls (242,118.21) and (254.61,112.63) .. (263.79,103.59) -- (263.8,103.61) .. controls (267.61,100.25) and (271.38,96.66) .. (275.1,92.84) .. controls (278.76,89.09) and (282.23,85.25) .. (285.5,81.36) -- (285.5,36.98) -- (284.62,35.89) .. controls (277.71,46.48) and (268.1,58.41) .. (256.7,70.12) .. controls (253.09,73.83) and (249.47,77.33) .. (245.9,80.6) -- (246.05,80.79) .. controls (241.71,85.13) and (235.27,87.87) .. (228.08,87.87) .. controls (227.48,87.87) and (226.89,87.85) .. (226.3,87.82) -- (226.3,87.92) .. controls (219.32,88.54) and (213.83,95.08) .. (213.83,103.04) .. controls (213.83,111.01) and (219.32,117.54) .. (226.3,118.16) -- (226.3,118.18) .. controls (226.45,118.19) and (226.6,118.19) .. (226.74,118.2) .. controls (226.96,118.21) and (227.18,118.21) .. (227.4,118.21) .. controls (227.47,118.21) and (227.55,118.21) .. (227.62,118.21) .. controls (227.77,118.21) and (227.93,118.21) .. (228.08,118.21) -- cycle ;
\draw  [fill=tikzblue  ,fill opacity=0.52 ] (24.47,71.15) .. controls (24.47,37.49) and (51.75,10.21) .. (85.4,10.21) .. controls (119.05,10.21) and (146.33,37.49) .. (146.33,71.15) .. controls (146.33,104.8) and (119.05,132.08) .. (85.4,132.08) .. controls (51.75,132.08) and (24.47,104.8) .. (24.47,71.15) -- cycle ;
\draw  [dash pattern={on 0.84pt off 2.51pt}] (24.47,74.85) .. controls (24.47,65.84) and (51.82,58.55) .. (85.57,58.55) .. controls (119.31,58.55) and (146.67,65.84) .. (146.67,74.85) .. controls (146.67,83.85) and (119.31,91.15) .. (85.57,91.15) .. controls (51.82,91.15) and (24.47,83.85) .. (24.47,74.85) -- cycle ;
\draw   (84.49,45.42) .. controls (105.57,45.42) and (122.65,62.02) .. (122.65,82.49) .. controls (122.65,102.97) and (105.57,119.57) .. (84.49,119.57) .. controls (63.42,119.57) and (46.33,102.97) .. (46.33,82.49) .. controls (46.33,62.02) and (63.42,45.42) .. (84.49,45.42) -- cycle (67.36,82.49) .. controls (67.36,91.32) and (75.03,98.47) .. (84.49,98.47) .. controls (93.96,98.47) and (101.63,91.32) .. (101.63,82.49) .. controls (101.63,73.67) and (93.96,66.52) .. (84.49,66.52) .. controls (75.03,66.52) and (67.36,73.67) .. (67.36,82.49) -- cycle ;
\draw   (84.49,45.42) .. controls (94.77,45.42) and (104.09,49.36) .. (110.95,55.78) -- (96.29,70.91) .. controls (93.22,68.19) and (89.06,66.52) .. (84.49,66.52) .. controls (80.37,66.52) and (76.59,67.88) .. (73.64,70.13) -- (58.44,55.4) .. controls (65.26,49.21) and (74.42,45.42) .. (84.49,45.42) -- cycle ;
\draw   (84.49,119.57) .. controls (74.54,119.57) and (65.48,115.87) .. (58.68,109.81) -- (73.89,95.04) .. controls (76.8,97.19) and (80.49,98.47) .. (84.49,98.47) .. controls (88.73,98.47) and (92.61,97.03) .. (95.6,94.66) -- (110.3,109.8) .. controls (103.51,115.87) and (94.45,119.57) .. (84.49,119.57) -- cycle ;

\draw  [draw opacity=0] (390.19,104.98) .. controls (383.7,109.06) and (376.02,111.41) .. (367.79,111.41) .. controls (344.48,111.41) and (325.59,92.52) .. (325.59,69.21) .. controls (325.59,45.9) and (344.48,27) .. (367.79,27) .. controls (391.1,27) and (410,45.9) .. (410,69.21) .. controls (410,70.03) and (409.98,70.86) .. (409.93,71.67) -- (367.79,69.21) -- cycle ; \draw   (390.19,104.98) .. controls (383.7,109.06) and (376.02,111.41) .. (367.79,111.41) .. controls (344.48,111.41) and (325.59,92.52) .. (325.59,69.21) .. controls (325.59,45.9) and (344.48,27) .. (367.79,27) .. controls (391.1,27) and (410,45.9) .. (410,69.21) .. controls (410,70.03) and (409.98,70.86) .. (409.93,71.67) ;  
\draw  [draw opacity=0] (409.93,66.61) .. controls (409.95,67.05) and (409.95,67.49) .. (409.95,67.93) .. controls (409.95,88.36) and (393.21,104.91) .. (372.55,104.91) .. controls (363.3,104.91) and (354.83,101.59) .. (348.3,96.08) -- (372.55,67.93) -- cycle ; \draw   (409.93,66.61) .. controls (409.95,67.05) and (409.95,67.49) .. (409.95,67.93) .. controls (409.95,88.36) and (393.21,104.91) .. (372.55,104.91) .. controls (363.3,104.91) and (354.83,101.59) .. (348.3,96.08) ;  

\draw (225.61,44.64+5) node [anchor=north west][inner sep=0.75pt]  [font=\small]  {$A$};
\draw (194.61,84.14+5) node [anchor=north west][inner sep=0.75pt]  [font=\small]  {$B$};
\draw (254.11,82.64+5) node [anchor=north west][inner sep=0.75pt]  [font=\small]  {$D$};
\draw (210.+3,119.64+5) node [anchor=north west][inner sep=0.75pt]  [font=\small]  {$C_{1} ,\ C_{2}$};
\draw (76.61,71.64+5) node [anchor=north west][inner sep=0.75pt]  [font=\small]  {$C$};
\draw (49.11,71.14+5) node [anchor=north west][inner sep=0.75pt]  [font=\small]  {$B$};
\draw (102.11,71.14+5) node [anchor=north west][inner sep=0.75pt]  [font=\small]  {$D$};
\draw (78.11,18.64+5) node [anchor=north west][inner sep=0.75pt]  [font=\small]  {$A$};
\draw (77.11,99.14+5) node [anchor=north west][inner sep=0.75pt]  [font=\small]  {$E$};
\draw (77.61,46.14+5) node [anchor=north west][inner sep=0.75pt]  [font=\small]  {$E$};
\draw (357.11,27.64+5) node [anchor=north west][inner sep=0.75pt]  [font=\small]  {$A$};
\draw (327.61,60.64+5) node [anchor=north west][inner sep=0.75pt]  [font=\small]  {$B$};
\draw (390.11,58.64+5) node [anchor=north west][inner sep=0.75pt]  [font=\small]  {$D$};
\draw (356.11+5,84.14+5) node [anchor=north west][inner sep=0.75pt]  [font=\small]  {$C_{2}$};
\draw (359.11+5,113.64+5) node [anchor=north west][inner sep=0.75pt]  [font=\small]  {$C_{1}$};
\draw (69.11,149.5) node [anchor=north west][inner sep=0.75pt]  [font=\small]  {$( a)$};
\draw (223.61,145) node [anchor=north west][inner sep=0.75pt]  [font=\small]  {$( b)$};
\draw (365.11,142.5) node [anchor=north west][inner sep=0.75pt]  [font=\small]  {$( c)$};

\end{tikzpicture}
    \caption{We illustrate how subsystems $ABC_1 C_2 D$ of Theorem~\ref{thm:immersion1} may be viewed as a certain region immersed on a sphere. \textbf{(a)} Consider ball $BCDE \subset S^d$ together with complement $A$ to form  sphere. \textbf{(b)} Deforming the regions for visual aid, we draw $ABC_1 C_2 D$ as a ball-shaped region that wraps the sphere and partially overlaps itself.  The overlapping parts are $C_1, C_2$.  \textbf{(c)} Schematic of regions $ABC_1C_2D$ as viewed from above.}
    \label{fig:immersion-3D}
\end{figure}

\begin{Theorem}[Recovery maps and immersed states]
\label{thm:immersion1}
    Let $\bar{\sigma} \in \St_{ABCDE}$ be a pure state. There exists a channel $\Phi_{BE\to BC}$ such that
    \begin{equation}
        \bar{\sigma}|_{ABC} = \id_A\otimes \Phi_{BE\to BC}(\bar{\sigma}|_{ABE}) 
        \label{eq:extendibility_2}
    \end{equation}
    if and only if there exists a state $\rho \in \St_{ABC_1C_2D}$ such that $\rho|_{ABC_1} = \bar{\sigma}|_{ABC}$ and $\rho|_{AC_2D} = \bar{\sigma}|_{ACD}$, where $C_1$ and $C_2$ are copies of $C$.
\end{Theorem}
Before the proof, let us see why we call $ABC_1 C_2D$ an immersed region.  Imagine $A$ as the complement of the disk $BCDE$, so that $ABCDE$ forms a sphere (Fig.~\ref{fig:immersion-3D}(a)).  Then $ABC_1 C_2D$ is an immersed region (Fig.~\ref{fig:immersion-3D}(b,c)), and the corresponding state  $\rho|_{ABC_1C_2D}$ is locally consistent with the original state. We say immersed because the mapping from ball $ABC_1C_2D$ to sphere $ABCDE$ is an immersion of manifolds but not an embedding, i.e., locally injective but not globally injective: regions $C_1, C_2$ are both mapped to the region $C$.  Immersed states in the context of the entanglement bootstrap were studied in~\cite{Shi2020a}; also see \cite{shi2023remote}.

\begin{proof}
First we assume the existence of channel  $\Phi_{BE\to BC}$ satisfying Eq.~\eqref{eq:extendibility_2} and define $\rho \in \St_{ABC_1C_2D}$,
\begin{align} \label{eq:rho_ABCCD}
    \rho = \id_{AC_2D} \otimes \Phi_{BE\to BC_1}(\bar{\sigma}|_{AC_2 D BE})
\end{align} 
Tracing out $BC_1$ on both sides, we obtain $\rho|_{AC_2D} = \bar{\sigma}|_{ACD}$. Alternatively, tracing out $C_2 D$ on both sides of Eq.~\eqref{eq:rho_ABCCD}, and applying Eq.~\eqref{eq:extendibility_2}, we obtain $\rho|_{ABC_1} = \bar{\sigma}|_{ABC}$.  This completes one direction of the proof.

Next we assume the existence of a state $\rho \in \St_{ABC_1C_2D}$ such that $\rho|_{ABC_1} = \bar{\sigma}|_{ABC}$ and $\rho|_{AC_2D} = \bar{\sigma}|_{ACD}$, and we seek to define a channel $\Phi_{BE \to BC}$ satisfying Eq.~\eqref{eq:extendibility_2}.  Let $\bar{\rho} \in \St_{ABC_1C_2DF}$ be a purification of $\rho \in \St_{ABC_1C_2D}$, and choose the purifying system $F$ large enough such that subsystem $FC_1$ is at least as large as $E$. Note $\bar{\rho}$ and $\bar{\sigma}$ are both purifications of $\rho|_{AC_2D} = \bar{\sigma}|_{ACD}$. Using the fact that any two purifications are related by an isometry between the purifying systems, there exists an isometry  $V : BE \to BC_1F$ such that $ V \bar{\sigma}|_{ABC_2 DE}  V^\dagger= \bar{\rho}|_{ABC_1 C_2DF}$.   Define the quantum channel $\Phi_{BE\to BC_1} = \Tr_F( V(\cdot)V^\dagger)$.  Then $\Phi_{BE\to BC_1}[\bar{\sigma}|_{ABC_2 DE}] = \rho|_{ABC_1 C_2D}$.  Tracing out $C_2 D$ from both sides, and using the assumption $\rho|_{ABC_1} = \bar{\sigma}|_{ABC}$, we obtain $\Phi_{BE\to BC_1}[\bar{\sigma}|_{ABE}] = \bar{\sigma}|_{ABC}$, as desired.
\end{proof}

 Note the local extendibility condition is symmetric under the exchange of $B$ and $D$: the existence of a recovery map $\Phi_{BE\to BC}$ implies the existence of a recovery map $\Phi_{DE \to CD}$ and vice versa. This is a corollary of Theorem~\ref{thm:immersion1}, because the existence of a suitable immersed state $\rho \in \St_{ABC_1C_2D}$ is manifestly invariant under exchanging $B \leftrightarrow D$.

The existence of the immersed state $ABC_1 C_2 D$ in Theorem~\ref{thm:immersion1} is something we anticipate to hold on physical grounds, for systems with homogeneous topological properties. More concretely, given local Hamiltonian terms on $BCDE$, we can envision extending the Hamiltonian to the sphere $ABCDE$, and moreover extending it to the immersed region $X = ABC_1C_2D$ of Fig.~\ref{fig:immersion-3D}(b,c). Further extend the Hamiltonian to a slight enlargement of the region, $X^+ \supseteq X$. Consider the Hamiltonian $H_{X^+}$ with some ground state $\rho \in \St_{X^+}$, and finally erase the boundary region to obtain $\rho|_X$. Heuristically, we expect this state to be consistent with both $\bar{\sigma}|_{ABC}$ and $\bar{\sigma}|_{ACD}$, because the additional terms in the Hamiltonian far away from a given region are expected to make little difference in the reduced density matrices of that region. 

\section{Geometric definitions and constructions} \label{sec:geom}

In this section, we define the geometries involved in the state extension process shown in Fig.~\ref{fig:growth0}. We have found it convenient to distinguish two types of shapes. The initial and the final states are defined on embedded or immersed manifolds with sufficiently regular boundaries such as the first and the last annuli in Fig.~\ref{fig:growth0}(c). On the other hand, regions $B$, $C$, $D$, $E$ in the definition of local extendibility may be less regular; for example, they may have corners. We will refer to the former and the latter as macro- and micro-geometries, respectively.

It may be possible and in some respects advantageous to work with just one type of shapes in different sizes. But even with our definitions, the macro-geometries can approximate the micro-geometries scaled up by a sufficiently large factor $s$. Together with the macro version of local conditions (see Section~\ref{sec:extension}), this implies that suitably enlarged local data are $(s B,s C,s D,s E)$-extendible for all micro-configurations $(B,C,D,E)$. Thus, local extendibility is reproduced on bigger scales. We will sketch the approximation procedure, as well as constructing a locally finite version of the configuration set.

In this section, we do distinguish between open and closed sets. (However, when passing from geometry to quantum systems, only the spins in the interior of a region need to be included.) This is some common notation, applicable to $\RR^d$ with the Euclidean metric $\dist(x,y)=|x-y|$ and other metric spaces. The open and closed balls of radius $r$ centered at $x$ are denoted by $\Ba_r(x)$ and $\bar{\Ba}_r(x)$, respectively. If $X$ is a subset of the metric space, then $\Ba_r(X)$ stands for the open $r$-neighborhood of $X$, whereas $\bar{\Ba}_r(X)=\bigcup_{x\in X}\bar{\Ba}_r(x)$. The closed unit ball $\bar{\Ba}_1(0)$ in $\RR^d$ is also denoted by $D^d$. 

\subsection{Micro-geometries}\label{sec:micro}

In the previous discussion of $\Lambda$-extendible local data, we did not specify the set $\Lambda$ of configurations $(B,C,D,E) \in \Lambda$ on which we impose the $(B,C,D,E)$-extendibility condition. 
First we recall the general idea for specifying these regions, elaborating on the discussion above Definition~\ref{def:loc_extend}.  
These configurations depend on a thickness parameter $a$ (the standard choice being $a=r_0$) and some set of embeddings $f\colon S^{d-1}\to\RR^d$ specified below. Given an embedding $f$, the disjoint regions $B$, $C$, $D$, $E$ are defined by 
\begin{equation}\label{eq:microconfig_BEDE}
BE=\Ba_a(f(S_-)),\qquad DE=\Ba_a(f(S_+)),
\end{equation}
for hemispheres $S_-$ and $S_+$, which implicitly define $E=BE\cap DE$,\, $B=BE\setminus E$,\,and $D=DE\setminus E$, while $C$ is the part of $\RR^{d}\setminus(BE\cup DE)$ that lies inside $f(S^{d-1})$.   The condition ``$x$ lies inside $f(S^{d-1})$'' is understood as $\lk(f,x)\not=0$, where the \emph{linking number} $\lk(f,x)$ is the degree of the map $y\mapsto\frac{f(y)-x}{|f(y)-x|}\, \colon S^{d-1}\to S^{d-1}$. It is well-defined whenever $f\colon S^{d-1}\to\RR^d$ is a continuous map (not necessarily an embedding) and $x\notin f(S^{d-1})$.  We denote these regions $\lambda(a,f)=(B,C,D,E)$. Writing $\mathrm{Int}(S^{d-1},f) := \{x \in \RR^d \setminus f(S^{d-1}) \colon \lk(f,x) \not= 0\}$, we summarize this discussion:
\begin{Definition}[Regions specified by an embedding]
\label{def:BCDE-region-recipe}
Given a continuous map $f\colon S^{d-1}\to\RR^d$ and thickness parameter $a >0$, we associate disjoint regions $(B,C,D,E)$, denoted $\lambda(a,f)=(B,C,D,E)$, defined by
\[
E=\Ba_a(f(S_-)) \cap \Ba_a(f(S_+)),\qquad B=\Ba_a(f(S_-)) \setminus E,\qquad D=\Ba_a(f(S_+)) \setminus E,\qquad C=\mathrm{Int}(S^{d-1},f) \setminus \Ba_a(f(S^{d-1})).
\]
\end{Definition}

To proceed, let us identify the hemispheres $S_-$ and $S_+$ with two copies of the closed unit disk $D^{d-1} \subset \RR^{d-1}$ and denote the restrictions of $f$ to those disks by $f_\mp$. Thus $BE=\Ba_a(f_-(D^{d-1}))$ and $DE=\Ba_a(f_+(D^{d-1}))$
where $f_-$ and $f_+$ make up a map of a sphere to $\RR^d$, i.e.,
\begin{equation}
f_\mp\colon D^{d-1}\to\RR^d,\qquad
f_-(x)=f_+(x)\,\text{ for all }\, x\in\partial D^{d-1},
\end{equation}
chosen to be an embedding.

We define a standard set $\Delta(r_1)$ of such maps that is further constrained as follows. We assume that there is a linear isometry $v\colon\RR^{d-1}\to \RR^d$ such that%
\footnote{The constants $1/2$ and $1/6$ in Eq.~\eqref{fmp_Lip} are by themselves arbitrary, but their choice is coordinated with other choices in the paper.}
\begin{equation}\label{fmp_Lip}
\KLip\bigl(r^{-1}f_\mp - v\bigr)\le \frac{1}{2},\qquad r=\frac{r_1}{6},
\end{equation}
where $\KLip(g) = \sup_{x\not=y}\frac{|g(x)-g(y)|}{|x-y|}$ is the Lipschitz constant of function $g$. It follows that both sets $f_-(D^{d-1})$ and $f_+(D^{d-1})$ fit in the ball of radius $(3/2)\cdot 2r=r_1/2$ centered at $f_\mp(x_0)$, where $x_0\in\partial D^{d-1}$ is arbitrary.
Then when $a \le r_1/2$, region $BCDE$ fits in a reference patch of radius $r_1$.

Now we can define the standard configurations $(B,C,D,E)$ on which we impose extendibility.
\begin{Definition}[Standard configurations]\label{def:Lambda-micro}
For length scales $r_0 \ll r_1$, the standard set of configurations $(B,C,D,E)$ is
\begin{equation}\label{std_Lambda}
\Lambda(r_0,r_1)=\{\lambda(r_0,f):\, f\in\Delta(r_1)\}
\end{equation}
where $\Delta(r_1)$ is the set of sphere embeddings characterized by the Lipschitz condition~\eqref{fmp_Lip} and $\lambda(r_0,f)=(B,C,D,E)$ is described by Definition~\ref{def:BCDE-region-recipe}.
\end{Definition}
\noindent Each configuration $(B,C,D,E)$ above is contained in some $r_1$-ball.  For intuition, note the Lipschitz condition~\eqref{fmp_Lip} allows the embedded hemispheres to be nonsmooth and slightly jagged or crumpled, but it also requires them to remain approximately aligned with a common flat hyperplane, so the embedded sphere fits within a pancake-like shape.

\begin{figure}\centering
\begin{tabular}{c@{\qquad\qquad}c}
\raisebox{10pt}{\includegraphics{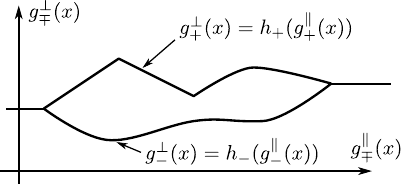}} & \includegraphics{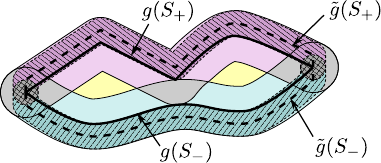}
\\[2pt]
(a) & (b)
\end{tabular}
\caption{\textbf{(a)} Representing embedded disks by the graphs of functions $h_-$, $h_+$. \textbf{(b)} The configuration $\lambda(r_0/r,g)$ may have a disconnected region $C$ (yellow). However, the modified embedded disks $\wt{g}(S_-)$, $\wt{g}(S_+)$  with thinner neighborhoods $\wt{B}\wt{E}$, $\wt{D}\wt{E}$ (shown by hatch patterns) are such that the enclosed region $\wt{C}$ is homeomorphic to a ball.}
\label{fig:gmp}
\end{figure}

For the subsequent discussion, it is convenient to represent $f_-(D^{d-1})$ and $f_+(D^{d-1})$ as graphs of some functions. To obtain such a representation, let us first extend $f_\mp$ from $D^{d-1}$ to $\RR^{d-1}$ by setting
\begin{equation}
f_-(x) = f_+(x) = f_{\mp}(x/|x|) - rv(x/|x|) + rv(x)\qquad \text{if }\, |x|\ge 1.
\end{equation}
The Lipschitz condition~\eqref{fmp_Lip} is still satisfied. Then we apply a linear coordinate change. Let $\wt{v}$ be an isometry of $\RR^d$ such that $\wt{v}(x_1,\dots,x_{d-1},0)=v(x_1,\dots,x_{d-1})$, and let
\begin{equation}\label{gmp_def}
g_\mp(x)=\wt{v}^{-1}(r^{-1}f_\mp(x))
=\bigl(g_\mp^\parallel(x),\ts g_\mp^\perp(x)\bigr)\in\RR^{d-1}\oplus\RR.
\end{equation}
In this notation, the Lipschitz bound applies to the map $x\mapsto g_\mp(x)-(x,0)$, and hence,
\begin{equation}
\KLip\bigl(g_\mp^\parallel-\id\bigr)\le\frac{1}{2},\qquad
\KLip\bigl(g_\mp^\perp\bigr)\le\frac{1}{2}.
\end{equation}
Due to Lemma~\ref{lem:Lipschitz} below, the function $g_\mp^\parallel\colon\RR^{d-1}\to\RR^{d-1}$ has an inverse with the Lipschitz constant at most $2$. Furthermore, it is isotopic to the identity (by linear interpolation), and therefore, the linking number between $(g_\mp^\parallel)|_{\partial D^{d-1}}$ and $g_\mp^\parallel(x)$ is equal to $\lk(\id|_{\partial D^{d-1}},x)$. It follows that both $g_-^\parallel$ and $g_+^\parallel$ map the interior of $D^{d-1}$ (i.e., the open ball $\Ba_1(0)$) onto the set $Z$ of points that lie inside $g_\mp^\parallel(\partial D^{d-1})$. Finally, we express $g_\mp^\perp(x)$ in terms of $g_\mp^\parallel(x)$ using the functions
\begin{equation}\label{hmp_def}
h_\mp(z)=g_\mp^\perp\bigl((g_\mp^\parallel)^{-1}(z)\bigr)\qquad
\bigl(h_\mp\colon\RR^{d-1}\to\RR,\qquad
\KLip(h_\mp)\le 1\bigr)
\end{equation}
(see Fig.~\ref{fig:gmp}(a)). Since $f_-$ and $f_+$ make up a sphere embedding, $h_-(z)\not=h_+(z)$ for all $z\in Z$. Without loss of generality, we may assume that $h_+(z)-h_-(z)$ is positive. Thus,
\begin{equation}\label{hmp_prop}
h_+(z)-h_-(z) \begin{dcases}
{}>0 & \text{if }\, z\in Z=g_\mp^\parallel(\Ba_1(0)),\\
{}=0 & \text{otherwise}.
\end{dcases}
\end{equation}

\begin{Lemma}\label{lem:Lipschitz}
Let $g$ be a map of a Euclidean space to itself such that $\KLip(g-\id)\le c<1$, where $\id$ is the identity map. Then $g$ is bijective, and $\KLip(g^{-1})\le(1-c)^{-1}$.
\end{Lemma}
\begin{proof}
For each given $y$, we solve the equation $g(x)=y$ by iteration. Let $x_0=y$ and $x_n=x_{n-1}+y-g(x_{n-1})$ for $n=1,2,\ldots$. If $n\ge 2$, then $|x_n-x_{n-1}|=|(x_{n-1}-g(x_{n-1}))-(x_{n-2}-g(x_{n-2}))|\le c|x_{n-1}-x_{n-2}|$. Therefore, the sequence $(x_n)$ converges to some $x$, and clearly, $g(x)=y$. The uniqueness of the solution and the bound $\KLip(g^{-1})\le(1-c)^{-1}$ are straightforward.
\end{proof}

With these tools, we can develop some variations of the standard configuration set, summarized later by Proposition~\ref{prop:nicer-extendibility-assumption}.
Slightly different sets of configurations are useful for three purposes:
\begin{enumerate}
\item We will show that using smooth maps $f_\mp$ does not change the strength of the local extendibility conditions.
\item The standard definition is highly redundant and includes infinitely many configurations in a finite region. Although this configuration set is reduced to a finite number of conditions because the region contains finitely many spins, it is preferable to achieve such reduction by geometric means, such that the number of conditions does not depend on the density of spins.
\item  The standard definition allows for a disconnected region $C$ as in Fig.~\ref{fig:extend1}(b). Such shapes are never used, and we could explicitly prohibit them. Instead, we will show how to replace them with nicer-looking shapes. This is largely an aesthetic issue.
\end{enumerate}

All these problems are solved using geometric approximations and monotonicity.  We write $(B,C,D,E)\cge(B',C',D',E')$ if the inclusions~\eqref{g_mono} hold, and $\cle$ for the reverse relation. By Proposition~\ref{prop:g_mono}, this implies that the extendibility condition for the first configuration is at least as strong as that for the second configuration. For configuration sets, $\wt{\Lambda}\cge\Lambda$ denotes that for each $\lambda\in\Lambda$, there is some $\wt{\lambda}\in\wt{\Lambda}$ such that $\wt{\lambda}\cge\lambda$. In each of the previously mentioned cases, our goal is to find some $\wt{\Lambda}$ that would meet certain criteria while being sufficiently strong, $\wt{\Lambda}\cge\Lambda(r_0,r_1)$. In cases~1 and~2, we will also have $\wt{\Lambda}\cle\Lambda(r_0-\eps,r_1)$, where $\eps$ can be made arbitrarily small.

A solution to problem~3 is shown in Fig.~\ref{fig:gmp}(b). More formally, we consider the function $g\colon S^{d-1}\to\RR^d$ whose restrictions $g_-$, $g_+$ to the two hemispheres are given by  Eq.~\eqref{gmp_def}. This $g$ is just $f$ in different Euclidean coordinates at scale factor $r^{-1}$. Let $b=2r_0/(3r)$, and let us define new functions $\wt{g}\colon S^{d-1}\to\RR^d$ and $u\colon S^{d-1}\to S^{d-1}$ such that for each $x$, the distance between $\wt{g}(x)$ and $g(u(x))$ is at most $b$:
\begin{equation}
\wt{g}_\mp(x)=\begin{dcases}
g_\mp(2x)\mp (0,b) & \text{if } |x|\le 1/2,\\
g_\mp(x/|x|)\mp (0,2(1-|x|)b) & \text{if } |x|\ge 1/2,
\end{dcases}\qquad\qquad
u_\mp(x)=\begin{dcases}
2x & \text{if } |x|\le 1/2,\\
x/|x| & \text{if } |x|\ge 1/2.
\end{dcases}
\end{equation}
It follows from Lemma~\ref{lem:ggtilde} below that $\lambda(a,\wt{g})\cge\lambda(a+b,g)$ for all $a>0$. We take $a=b/2$: the modified hemispheres are separated by a perpendicular distance of at least $2b/\sqrt{1+\KLip(h_\mp)^2}\ge\sqrt2\,b>2a$, so $\wt{C}$ is connected. Passing to the original coordinates, we get $\lambda(r_0/3,\wt{f})\cge\lambda(r_0,f)$.

\begin{Lemma}\label{lem:ggtilde}
Let $g,\wt{g}\colon S^{d-1}\to\RR^d$ and $u\colon S^{d-1}\to S^{d-1}$ be continuous maps. Suppose that $u$ is homotopic to the identity and that
\begin{equation}\label{ggtilde_cond}
u(S_-)\subseteq S_-,\qquad u(S_+)\subseteq S_+,\qquad
|\wt{g}(x)-g(u(x))|\le b\qquad \text{for all }\, x.
\end{equation}
Then $\lambda(a,\wt{g})\cge\lambda(a+b,g)$ for any positive $a$.
\end{Lemma}
\begin{proof}
Let $(\wt{B},\wt{C},\wt{D},\wt{E})=\lambda(a,\wt{g})$ and $(B,C,D,E)=\lambda(a+b,g)$. Conditions~\eqref{ggtilde_cond} imply that any point in $\wt{g}(S_\mp)$ is within distance $b$ from some point in $g(S_\mp)$; hence, $BE=\Ba_{a+b}(g(S_-))$ includes $\wt{B}\wt{E}=\Ba_a(\wt{g}(S_-))$, and similarly $DE\supseteq\wt{D}\wt{E}$. To show that $C\subseteq\wt{C}$ and $BCDE\supseteq\wt{B}\wt{C}\wt{D}\wt{E}$, we argue that $\lk(\wt{g},x)=\lk(g,x)$ for all $x\notin BDE=\Ba_{a+b}(g(S^{d-1}))$. Indeed, consider the homotopy of $\wt{g}$ to $g\circ u$ by linear interpolation, and further to $g$ using the homotopy between $u$ and the identity. Both homotopies keep the image of the sphere within distance $b$ from $g(S^{d-1})$, and thus, it never passes through $x$.
\end{proof}

Problem~1 is solved by the approximation of the functions $h_\mp$ defined by Eq.~\eqref{hmp_def} with smooth functions using the following relatively straightforward lemma.

\begin{Lemma}\label{lem:smoothing}
Let $\ph\colon\RR^{d-1}\to\RR$ be a smoothing kernel, namely, a smooth, nonnegative function supported by $D^{d-1}$ and such that $\int\ph(x)\,dx=1$. For any $\delta>0$, we set $\ph_\delta(x)=\delta^{1-d}\ph(x/\delta)$. Then the convolution $h_\delta=\ph_\delta*h$ of $\ph_\delta$ with an arbitrary Lipschitz-continuous function $h$ is smooth and satisfies the bounds
\begin{equation}\label{smoothing}
|h_\delta(x)-h(x)|\le\delta\,\KLip(h),\qquad
|h'_\delta(x)|\le \KLip(h),\qquad
\|h''_\delta(x)\|\le A\delta^{-1}\KLip(h),
\end{equation}
where $A=\int\|\ph''(x)\|\,dx$.
\end{Lemma}

Due to Eq.~\eqref{hmp_prop}, $(h_+)_\delta(x)-(h_-)_\delta(x)$ is nonnegative for all $x$ and vanishes if $x\notin\Ba_\delta(g_\mp^\parallel(\Ba_1(0)))\subseteq g_\mp^\parallel(\Ba_{1+2\delta}(0))$. Now, we apply Lemma~\ref{lem:ggtilde} to $u=\id$ and the map $\wt{g}$ composed of these functions:
\begin{equation}
\wt{g}_\mp^\parallel(x)= g_\mp^\parallel((1+2\delta)x) , \qquad
\wt{g}_\mp^\perp(x)=(h_\mp)_\delta\bigl(g_\mp^\parallel((1+2\delta)x)\bigr).
\end{equation}
(Although $\wt{g}$ is not an embedding because $(h_+)_\delta(x)-(h_-)_\delta(x)$ vanishes for some $x\in g_\mp^\parallel(\Ba_{1+2\delta}(0))$, this can be fixed by an arbitrarily small change to $(h_\mp)_\delta$.) It follows from the first inequality in~\eqref{smoothing} and the bounds $\KLip(h_\mp)\le 1$ and $\KLip(g_\mp^\perp)\le 1/2$ that $|\wt{g}_\mp^\perp(x)-g_\mp^\perp(x)|\le 2\delta$. The parallel components differ by $|g_\mp^\parallel((1+2\delta)x)-g_\mp^\parallel(x)|\le 3\delta$, so $|\wt{g}_\mp(x)-g_\mp(x)|\le 4\delta$ and Lemma~\ref{lem:ggtilde} gives $\lambda(a,\wt{g})\cge\lambda(a+4\delta,g)$, and a similar relation holds when $\wt{g}$ and $g$ are swapped. Using $\delta=\eps/(8r)$, we get
\begin{equation}
\lambda(r_0-\eps,f)\cge\lambda(r_0-\eps/2,\wt{f})\cge\lambda(r_0,f).
\end{equation}
Thus, the configuration set $\wt{\Lambda}=\{\lambda(r_0-\eps/2,\wt{f}):\,f\in\Delta(r_1)\}$ is almost equal in strength to the standard set $\Lambda(r_0,r_1)$, namely $\Lambda(r_0-\eps,r_1)\cge\wt{\Lambda}\cge\Lambda(r_0,r_1)$. This solves problem~1: the hemispheres $\wt{f}(S_\mp)$ are graphs of smooth functions. The maps $\wt{f}_\mp$ themselves are not smooth, but $\lambda(a,\wt{f})$ depends on $\wt{f}$ only through the sets $\wt{f}(S_\mp)$ and the linking number.

Finally, let us consider problem~2. A configuration set $\Lambda$ is called \emph{uniformly locally finite} if there exist some $a>0$ and $N\in\NN$ such that each open ball $Q$ of radius $a$ overlaps with at most $N$ configurations $(B,C,D,E)\in\Lambda$, meaning $Q\cap BCDE\not=\emptyset$. (The seemingly stronger property $\forall a'\,\exists N\ldots$ follows because every $a'$-ball can be covered by a certain number of $a$-balls.) Roughly speaking, such a set $\Lambda$ has a bounded number of configurations per $r_1^d$ volume. We will approximate the standard configuration set with a uniformly locally finite one.

Let us fix some $\eps>0$ and a maximal set $L\subseteq\RR^d$ of points such that the distance between any two distinct points is ${}\ge\eps/2$. Due to the maximality, for each $y\in\RR^d$ there is some $x\in L$ such that $|x-y|<\eps/2$. Associated with each region $Y$ and each configuration $(B,C,D,E)$ is a ``signature'',
\begin{equation}\label{signature}
\sig(Y)=\Ba_{\eps/2}(Y)\cap L,\qquad
\sig(B,C,D,E)=\bigl(\sig(BE),\, \sig(DE),\,
L\setminus\sig(\RR^d\setminus C),\, \sig(BCDE)\bigr).
\end{equation}
Now, let  $\Lambda=\Lambda(r_0-\eps,r_1)$, and let us choose a subset $\wt{\Lambda}\subseteq\Lambda$ that has exactly one configuration with each given signature in the set $\{\sig(\lambda):\,\lambda\in\Lambda\}$. We argue that $\wt{\Lambda}$ is uniformly locally finite. Indeed, consider all configurations in $\Lambda$ that overlap with a ball $Q$ of some fixed radius, say, $a=r_1$. These configurations and their $\eps/2$-neighborhoods lie in a ball of radius $O(r_1)$, which contains $O((r_1/\eps)^d)$ points of $L$. Thus, there are $N=\exp\bigl(O((r_1/\eps)^d)\bigr)$ possible signatures for each choice of $Q$. It remains to show that $\wt{\Lambda}$ approximates the standard configuration set. The relation $\Lambda(r_0-\eps,r_1)\cge\wt{\Lambda}$ is obvious, and we also claim that $\wt{\Lambda}\cge\Lambda(r_0,r_1)$.

\begin{proof}
By construction, for each configuration $(B,C,D,E)=\lambda(r_0-\eps,f)\in\Lambda$, there is some $(\wt{B},\wt{C},\wt{D},\wt{E})\in\wt{\Lambda}$ with the same signature. First, we observe that if $\sig(Y)=\sig(Z)$, then $Y\subseteq\Ba_\eps(Z)$. (Indeed, for each $y\in Y$ there is some $x\in\sig(Y)$ such that $|x-y|<\eps/2$. This $x$ also belongs to $\sig(Z)$, and hence, $|x-z|<\eps/2$ for some $z\in Z$.) Therefore, the condition $\sig(\wt{B},\wt{C},\wt{D},\wt{E})=\sig(B,C,D,E)$ implies that
\begin{equation}
\wt{B}\wt{E}\subseteq\Ba_\eps(BE),\qquad \wt{D}\wt{E}\subseteq\Ba_\eps(DE),\qquad
\RR^d\setminus\wt{C}\subseteq\Ba_\eps(\RR^d\setminus C),\qquad
\wt{B}\wt{C}\wt{D}\wt{E}\subseteq\Ba_\eps(BCDE).
\end{equation}
On the other hand, the configuration $(B',C',D',E')=\lambda(r_0,f)$ has the following properties. The regions $B'E'$ and $D'E'$ clearly include $\Ba_\eps(BE)$ and $\Ba_\eps(DE)$, respectively. In addition, $\RR^d\setminus C'\supseteq\Ba_\eps(\RR^d\setminus C)$ and $B'C'D'E'\supseteq\Ba_\eps(BCDE)$ due to Lemma~\ref{lem:thickening} below. Thus, $(\wt{B},\wt{C},\wt{D},\wt{E})\cge(B',C',D',E')$.
\end{proof}

Above we used the following lemma about how to satisfy the inclusions \eqref{g_mono} with quantitative margins. 
\begin{Lemma}[Thickening]\label{lem:thickening}
Let $B$, $C$, $D$, $E$ be disjoint regions such that $C$ and $A=\RR^d\setminus BCDE$ do not touch each other, i.e., their closures do not overlap. Consider another configuration $(B',C',D',E')$ such that $B'E'$ and $D'E'$ include the (open) $\eps$-neighborhoods of $BE$ and $DE$, respectively, whereas $C'=C\setminus B'D'E'$. Then $\RR^d\setminus C'$ includes the $\eps$-neighborhood of $\RR^d\setminus C$, and $B'C'D'E'$ includes the $\eps$-neighborhood of $BCDE$.
\end{Lemma}
\begin{proof}
The problem is symmetric under the swap of regions $C$ and $A$, so it is sufficient to prove the first statement: $\RR^d\setminus C'$ includes the open $\eps$-neighborhood of $\RR^d\setminus C$. In other words, we need to show that if $x\in C'$ and $y\in\RR^d\setminus C=BDEA$, then $|x-y|\ge\eps$.

If $y\in BDE$, then $|x-y|\ge\eps$ because the open $\eps$-neighborhood of $BDE$ is contained in $B'D'E'$ but $x\notin B'D'E'$. So let us consider the alternative, $y\in A$. Since $x\in C'\subseteq C$ while $C$ and $A$ do not touch each other, the line interval between $x$ and $y$ contains some point $z\in BDE$. The distance between $x$ and $z$ is at least $\eps$ because $x$ does not belong to the $\eps$-neighborhood of $BDE$. It follows that the distance between $x$ and $y$ is greater than $\eps$.
\end{proof}

Combining the three constructions above---(i) smoothing the hemispheres (Lemma~\ref{lem:smoothing}), (ii) guaranteeing $C$ is connected via the modification in Fig.~\ref{fig:gmp}(b) (Lemma~\ref{lem:ggtilde}), and (iii) reducing to a uniformly locally finite subfamily by the signature argument---we can choose $(B,C,D,E)$ regions that satisfy all three desiderata simultaneously. Thus we can summarize our discussion with the following result. Recall we use $ x \ll y$ to mean that $x\le cy$ for a sufficiently small universal constant $c>0$ that depends only on the number of spatial dimensions $d$.
\begin{Proposition}[Extendibility using modified regions] \label{prop:nicer-extendibility-assumption}
Assume $r_0' \ll r_0 \ll r_1$ and consider the standard $(B,C,D,E)$-configurations $\Lambda(r_0,r_1)$ (Definition~\ref{def:Lambda-micro}).  Then there exists a set of configurations $\wt{\Lambda}$ such that extendibility for $\wt{\Lambda}$ implies extendibility for  $\Lambda(r_0,r_1)$, and such that: (1) each $(B,C,D,E)\in \wt{\Lambda}$ takes the form $\lambda(r_0',f)$ for $f_{\mp}$ whose hemispheres $f(S_\mp)$ are graphs of smooth functions, with $BCDE$ contained in some $r_1$-ball,
(2) the set  $\wt{\Lambda}$ is uniformly locally finite, i.e., the number of configurations intersecting any ball of fixed radius is bounded by a constant independent of the position of the ball, (3) each $(B,C,D,E) \in \wt{\Lambda}$ has $C$ connected, and (4) after translating each configuration in $\wt{\Lambda}$ so that one containing $r_1$-ball is centered at the origin, only finitely many normalized configurations arise.
\end{Proposition}
Thus when discussing extendible local data (Definition~\ref{def:extendible-data}), it is sufficient to assume extendibility only for regions satisfying the above properties.

\subsection{Macro-geometries}\label{sec:macro}

We now consider smooth $d$-dimensional manifolds $M$ with boundary that are embedded as closed subsets or properly immersed%
\footnote{A map between topological spaces is called \emph{proper} if the inverse image of every compact set is compact. An \emph{immersion} of smooth manifolds is a smooth map $f$ whose derivative $f'(x)$ (which is a linear map between the tangent spaces) is injective at every point $x$. For example, the inclusion of an open ball in $\RR^d$ and the universal cover of a closed annulus are immersions but not proper.} in $\RR^d$. Let us illustrate some aspects of the latter setting. If $f\colon M\to\RR^d$ is an immersion, a spin located at point $x$ is replaced with multiple copies, one for each $\wt{x}\in f^{-1}(x)$. The set $f^{-1}(x)$ is discrete because $f$ is injective in some neighborhood of each of its elements $\wt{x}$. The condition that $f$ is proper implies that $f^{-1}(x)$ is finite. The pullback of the Euclidean metric on $\RR^d$ by $f$ gives a Riemannian metric on $M$, and we will use the corresponding path distance $\dist(\cdot,\cdot)$. Let us discuss some additional constraints on $M$, and in particular, on its boundary $N=\partial M$.

\begin{figure}\centering
\begin{tabular}{c@{\qquad\qquad}c}
\includegraphics{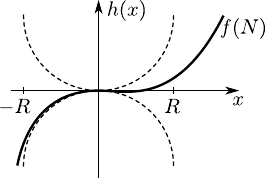} & \includegraphics{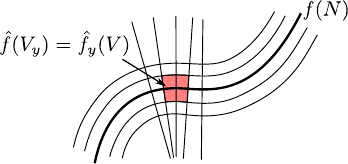}
\\[2pt]
(a) & (b)
\end{tabular}
\caption{\textbf{(a)} The immersed manifold $f(N)$ is represented as a graph of function $h$, which is well-defined in the domain $|x|<R$. \textbf{(b)} Illustration of the map $\hat{f}\colon (x,t)\mapsto f(x)+t\ts n(x)$: the grid lines are the images of $t=\const$ and $x=\const$ lines. The image of the low-distortion domain $V_y$ is also shown.}
\label{fig:fy}
\end{figure}

Let $N$ be a $(d-1)$-manifold without boundary, $f\colon N\to\RR^d$ a proper immersion, and $(n(x)\colon x\in N)$ a unit normal vector (pointing outward relative to $M$). We require that the minimum curvature radius of $f(N)$ is at least $R$, where $R$ is fixed. To describe the geometry around a given point $y\in N$, let us change the coordinates in $\RR^d$ by a translation and a rotation that move the origin to $f(y)$ and align the last axis with $n(y)$. The immersion $f$ followed by the projection onto the first $d-1$ coordinates is a map $\phi_y\colon N\to\RR^{d-1}$ which is invertible in some neighborhood of $y$, and thus, can serve as a local coordinate system. The expression of $f$ in these coordinates, given by the function $f_y=f\circ\phi_y^{-1}$, has the form
\begin{equation}\label{local_bmap}
f_y(x)=\begin{pmatrix}x\\ h(x)\end{pmatrix}\in\RR^{d-1}\oplus\RR,\qquad
h(0)=h'(0)=0.
\end{equation}
(Writing elements of $\RR^d$ as column vectors is consistent with the subsequent use of the Einstein notation.) It is evident from Fig.~\ref{fig:fy}(a) and confirmed by the calculation below that $h(x)$ is well-defined in the domain $|x|<R$, i.e., the open ball $U=\Ba_R(0)\subseteq\RR^{d-1}$. The domain $U$ represents some neighborhood $U_y$ of $y$ with $0$ representing $y$, meaning that $U=\phi_y(U_y)$ and $\phi_y(y)=0$.

In more detail, let $(x^\alpha)$ be an arbitrary local coordinate system on $N$, and let us use the Kronecker delta as the metric tensor on $\RR^d$. Then the Riemannian metric on $N$ and the extrinsic curvature are as follows:
\begin{equation}
g_{\alpha\beta}(x)
=\delta_{ab}\,(\partial_{\alpha}f^a(x))\,(\partial_{\beta}f^b(x)),\qquad
K_{\alpha\beta}(x)
=-(\partial_{\alpha}\partial_{\beta}f^a(x))\,n_a(x).
\end{equation}
The exact constraint on the curvature is an inequality between symmetric tensors:
\begin{equation}\label{ex_curv}
-R^{-1}g(x)\le K(x)\le R^{-1}g(x)\qquad\, \text{for all }\, x\in N.
\end{equation}
It can also be stated as the operator norm of $\kap=g^{-1/2}Kg^{-1/2}$ being less than or equal to $R^{-1}$. Accordingly, for a $(d-1)$-manifold $N$, possibly with boundary, we say $f : N \to \RR^d$ is an \emph{$r$-tame hypersurface} when it is smooth%
\footnote{Smoothness could be replaced with $C^{1,1}$ regularity, say, because a compact $C^{1,1}$ manifold with bounded curvature can be arbitrarily well approximated by a smooth manifold with the same curvature bound, but we work with smoothness for convenience.}
with curvature $K$ satisfying $-r^{-1} g\le K \le r^{-1} g$; for an \emph{$r$-tame embedded hypersurface} we require in addition that $\hat f$ be injective on $N\times(-r,r)$, as discussed below. In the special coordinates \eqref{local_bmap}, the relevant tensors are
\begin{equation}\label{ex_curv1}
g=\one+\ts(h')^\trans h',\qquad n^\trans=q^{-1/2}(-h',1),\qquad
K=-q^{-1/2}\ts h'',\qquad
\text{where}\qquad q=1+h'\ts(h')^\trans.
\end{equation}
These equations and the constraint on the curvature imply the inequality
\begin{equation}\label{ex_curv2}
\|h''\|\le R^{-1}\bigl(1+|h'|^2\bigr)^{3/2}.
\end{equation}
The inverse of the map $\phi_y$, or equivalently, the function $h$ clearly exists in $\Ba_r(0)$ for a sufficiently small $r$. It is easy to show that $r$ can be increased as long as $|h'|$ is bounded in the corresponding ball. (Indeed, $f_y(\Ba_r(0))$ fits in some closed ball $Z$, and $f^{-1}(Z)$ is compact because $f$ is proper. Then one argues that $\phi_y'(x)$ is nondegenerate for all $x$ on the boundary of $\phi_y^{-1}(\Ba_r(0))\subseteq f^{-1}(Z)$.) Integrating the inequality\eqref{ex_curv2}, we obtain the bound
\begin{equation}\label{hprime_bound}
|h'(x)|\le\frac{|x|}{\sqrt{R^2-|x|^2}},
\end{equation}
which shows that $r$ can be as large as $R$.\medskip

Now, let us consider the following extension of the immersion $f$ (see Fig.~\ref{fig:fy}(b)):
\begin{equation}\label{fhat}
\hat{f}\colon N\times(-R,R)\to\RR^{d},\qquad
\hat{f}(x,t)=f(x)+t\ts n(x).
\end{equation}
It is also an immersion, i.e., $\partial\hat{f}(x,t)/\partial (x,t)$ is nondegenerate. Indeed, the vectors $\partial\hat{f}(x,t)/\partial x^\alpha=\partial_\alpha f(x)+t\ts\partial_\alpha n(x)$ are orthogonal to $\partial\hat{f}(x,t)/\partial t=n(x)$ for all $\alpha$, so it is sufficient to check that $\partial\hat{f}(x,t)/\partial x$ is nondegenerate. Using the formula $\partial_\beta n^b=(\partial_\gamma f^b)\ts g^{\gamma\alpha} K_{\alpha\beta}$ and the expression $K=g^{1/2}\kap g^{1/2}$, we get
\begin{equation}\label{fhat_x}
\frac{\partial\hat{f}(x,t)}{\partial x}=f'+t\ts f'g^{-1}K
=f'g^{-1/2}(\one+t\kap)\ts g^{1/2},
\end{equation}
which is nondegenerate because $|t|<R$ and $\|\kap\|\le R^{-1}$. Let us also discuss pushing the immersed manifold by distance $t$ in the normal direction, i.e., replacing $f$ with $\wt{f}\colon x\mapsto\hat{f}(x,t)$. The corresponding metric and extrinsic curvature tensors are $\wt{g}=g^{1/2}(\one+t\kap)^2g^{1/2}$ and $\wt{K}=g^{1/2}(\one+t\kap)\kap g^{1/2}$, respectively. This implies a geometrically intuitive result:
\begin{equation}
-(R-t)^{-1}\wt{g}(x)\le\wt{K}(x)\le(R+t)^{-1}\wt{g}(x).
\end{equation}

In local coordinates, $\hat{f}$ is represented by the map $\hat{f}_y=\hat{f}\circ(\phi_y^{-1}\times\id)$, which is close to the identity map in a small neighborhood of the origin. The difference between those maps is quantified by the norm of its derivative. We estimate it using Eqs.\ \eqref{fhat_x}, \eqref{ex_curv1}, the fact that $|h'g^{-1/2}|=|h'|q^{-1/2}$ and $\|f'g^{-1/2}\|=1$, and finally, the bound \eqref{hprime_bound}:
\begin{equation}
\begin{aligned}
\biggl\|\frac{\partial\hat{f}_y(x,t)}{\partial(x,t)}-\one\biggr\|^2
&\le\biggl\|\frac{\partial\hat{f}_y(x,t)}{\partial x}
-\begin{pmatrix}\one\\ 0\end{pmatrix}\biggr\|^2
+\biggl\|\frac{\partial\hat{f}_y(x,t)}{\partial t}
-\begin{pmatrix}0\\ 1\end{pmatrix}\biggr\|^2
=\biggl\|\begin{pmatrix}0\\ h'\end{pmatrix}+f'g^{-1/2}t\kap g^{1/2}\biggr\|^2
+\biggl\|n-\begin{pmatrix}0\\ 1\end{pmatrix}\biggr\|^2
\\[2pt]
&\hspace{-20pt}\le |h'|^2+2|h'|^2|t/R|+|t/R|^2q +2\bigl(1-q^{-1/2}\bigr)
\le\frac{|x|^2(1+2|t/R|)+|t|^2}{R^2-|x|^2}
+2\biggl(1-\frac{\sqrt{R^2-|x|^2}}{R}\biggr)
\end{aligned}
\end{equation}
In particular, if $|x|\le R/6$ and $|t|\le R/6$, then $\|\partial\hat{f}_y(x,t)/\partial(x,t)-\one\|\le 0.31$, which is small enough for our purposes. Thus,
\begin{equation}\label{distortion_bound}
\KLip\bigl((\hat{f}_y-\id)|_V\bigr)
=\sup\left\{
\biggl\|\frac{\partial\hat{f}_y(x,t)}{\partial(x,t)}-\one\biggr\|:\:
(x,t)\in V\right\}\le 0.31 <\frac{1}{2},\qquad
V=\bar{\Ba}_{R/6}(0)\times[-R/6,R/6].
\end{equation}
We call $V$ or the corresponding neighborhood $V_y=(\phi_y^{-1}\times\id)(V)$ of $(y,0)\in N\times(-R,R)$ the \emph{low-distortion domain}.\medskip

\begin{figure}\centering
\begin{tabular}{c@{\qquad}c@{\qquad}c}
\includegraphics{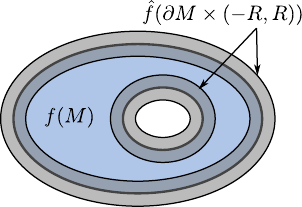} & 
\includegraphics{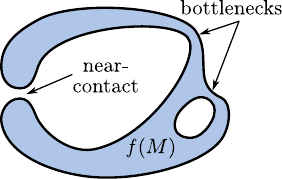} &
\includegraphics{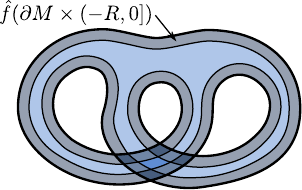}
\\[2pt]
(a) & (b) & (c)
\end{tabular}
\caption{\textbf{(a)} The tubular neighborhood $\hat{f}(\partial M\times(-R,R))$ of the boundary of an embedded manifold $f(M)$. \textbf{(b)}~Unwanted situations prevented by the tubular neighborhood, namely, narrow places in $f(M)$ (bottlenecks) or its complement (near-contacts). \textbf{(c)}~The inner collar of an immersed manifold.}
\label{fig:embimm}
\end{figure}

In the intended applications, the manifold $N$ is the boundary of a compact $d$-dimensional manifold $M$, and the map $f$ extends to an embedding or an immersion $f\colon M\to\RR^d$. For embeddings, we require that the map $\hat{f}\colon N\times(-R,R)\to\RR^d$ is injective. This prevents bottlenecks and near-contacts, i.e., situations where pushing the boundary in or out by a small amount could change the topology (see Fig.~\ref{fig:embimm}(a,\,b)). In the immersion setting, we only require the existence of an inner collar as defined below and illustrated by Fig.~\ref{fig:embimm}(c). Near-contacts are not an issue because, with suitable curvature bounds on $N$, $M$ can always be enlarged by attaching the outer collar $N\times[0,R)$ to it and extending the immersion by setting $f(x,t)=\hat{f}(x,t)$ for $(x,t)$ in the collar.

\begin{Definition}[Inner collar] \label{def:inner-collar}
Let $f\colon M\to\RR^d$ be a proper immersion of a $d$-dimensional manifold $M$ with the boundary $N=\partial M$ and the outward unit normal vector $(n(x):\,x\in N)$. An \emph{inner $R$-collar} is the image of a diffeomorphism $u\colon N\times(-R,0]\to M$ such that $u(x,0)=x$ and $f(u(x,t))=\hat{f}(x,t)=f(x)+t\ts n(x)$ for all $x\in N$ and $t\in(-R,0]$.
\end{Definition}

Note that such a diffeomorphism $u$ is unique, and its existence implies the inequality $K(x)\le R^{-1}g(x)$ because otherwise $\partial\hat{f}(x,t)/\partial x$ (see Eq.~\eqref{fhat_x}) would be degenerate for some $t\in(-R,0]$. We also remark about the path distance $\dist(\cdot,\cdot)$ on $M$ induced by the Riemannian metric. Specifically, the distance between $u(x,-l)$ and $N$ is equal to $l$. Indeed, a path of length $l$ is given by $t\mapsto u(x,t)$ for $t\in[-l,0]$. On the other hand, the shortest path is mapped by $f$ to some line interval starting at $\hat{f}(x,-l)$ and being perpendicular to the boundary; therefore, the path itself has the form $t\mapsto u(x',t)$ for $t\in[-l',0]$. But $u$ is bijective, so $(x',-l')=(x,-l)$.\medskip

Finally, let us discuss the relation of $M$ to reference patches. To use the definition of the information convex set with the standard $\Gamma_1(M)$ and $M^+$ (see Eq.~\eqref{Gamma1_Yplus}), we must assume that $R\ge 2r_1$. If $M$ is embedded in $\RR^d$, the relevant reference patches are contained in $M^{+R}=\Ba_R(M)$. In the case of immersion, each patch is replaced with zero or more ``admissible copies'', which are contained in a suitable manifold $M^{+R}\supseteq M$. The exact construction of $M^{+R}$ is as follows.

\begin{Definition}[Outer collar] \label{def:outer-collar}
Let $f\colon M\to\RR^d$ be a proper immersion of a $d$-dimensional manifold $M$ with the boundary $N=\partial M$ and outward unit normal vector $(n(x):\,x\in N)$, and suppose that $-R^{-1}g(x)\le K(x)$ for all $x\in N$. Glue $M$ with the \emph{outer $R$-collar} $N\times[0,R)$ by identifying each point $x\in N$ in the former with $(x,0)$ in the latter. Denote the resulting set by $M^{+R}$, and let $f_{+R}\colon M^{+R}\to\RR^d$ be the extension of $f$ with $\hat{f}\colon(x,t)\mapsto f(x)+t\ts n(x)$ for $(x,t)$ in the collar. The smooth structure and the Riemannian metric on $M^{+R}$ are induced by those on $\RR^d$.
\end{Definition}
\noindent Note that $M^{+R}$ as a smooth manifold does not depend on $f$, but the induced metric does.  Now we can discuss admissible copies, where $f$ acts injectively.
\begin{Definition} \label{def:admissible}
For immersion $f_{+R} \colon M^{+R} \to\RR^d$, an \emph{admissible copy} of a connected set $L\subseteq\RR^d$ is a subset $\wt{L}\subseteq M^{+R}$ that is mapped by $f_{+R}$ homeomorphically onto $L$.
\end{Definition}
\noindent We find that when metric balls in $M^{+R}$ stay away from the boundary, they are admissible copies of balls in $\RR^d$.
\begin{Proposition}[Admissible balls]\label{prop:admissible-ball}
Let $x \in M^{+R}$ and let $r \in (0,R]$. Then $\Ba_r(x)$ is an admissible copy of $\Ba_r(f_{+R}(x))$ iff $\dist(x,M) \leq R - r$. 
\end{Proposition}
\noindent Here the metric ball $\Ba_r(x)$ is defined using the metric on $M^{+R}$ induced by $f_{+R}$. The proof follows from a slightly more general fact which may be more transparent.
\begin{Lemma}\label{lemma:admissible-ball-general}
Let $X$ be a $d$-dimensional Riemannian manifold with boundary, and let $f : X \to \RR^d$ be a proper immersion, and a local isometry on the interior.  Then $f|_{\Ba_r(x)}$ is a homeomorphism onto $\Ba_r(f(x))$ iff $r \le \mathrm{dist}(x, \partial X)$.
\end{Lemma}
\noindent Then Proposition~\ref{prop:admissible-ball} follows by taking $X = M^{+R}$ with closed collar. (If only the open collar exists, one may consider $X = M^{+(R-\eps)}$ with closed collar and take $\eps \downarrow 0$.) Here we use the fact that  $\dist(x,M) + \dist(x,\partial(M^{+R})) = R$ for all $x \in M^{+R}\setminus M$; see the analogous discussion below Definition~\ref{def:inner-collar}.
\begin{proof}[Proof of Lemma~\ref{lemma:admissible-ball-general}] 
First we take $x \in X$ with $\dist(x,\partial X) \ge r$ and show $f|_{\Ba_r(x)}$ is a homeomorphism onto $\Ba_r(f(x))$.  
Note $\dist(f(x),f(x')) \le \dist(x,x')$ for all $x,x' \in X$ because $f$ maps paths between $x,x'$ to paths of equal length between $f(x),f(x')$. Then $f(\Ba_r(x)) \subset \Ba_r(f(x))$.  So $\Ba_r(x) \subset f^{-1}(\Ba_r(f(x))) \subset f^{-1}(\bar{\Ba}_r(f(x))) =: K$.  Note $K$ is compact because $f$ is proper, hence $\bar{\Ba}_r(x)$ is compact. Any geodesic starting at $x$ can be continued to length at least $r$.  Assume to the contrary that the geodesic cannot be continued past distance $r'$ for some $r' < r$.  Then it lies inside $\bar{\Ba}_r(x)$, so the geodesic approaches a limit point $z \in \bar{\Ba}_r(x)$, and $z \in \partial X$ (otherwise the geodesic could be continued), contradicting $\dist(x,\partial X) \ge r$. 

To show $f|_{\Ba_r(x)}$ is surjective onto $\Ba_r(f(x))$, consider any $y \in \Ba_r(f(x))$ and the straight geodesic segment $\alpha$ from $f(x) \to y$.  Let $\gamma$ be the geodesic in $X$ starting at $x$, with initial velocity lifted from $\alpha$, i.e., $\gamma'(0) = (df_x)^{-1}(\alpha'(0))$.  Since $f$ is a local isometry in $\Int(X)$, $f \circ \gamma$ is a geodesic in $\RR^d$, with the same initial point and initial velocity as $\alpha$; thus $f \circ \gamma = \alpha$. By the above discussion, $\gamma$ is defined up to time at least
$\dist(f(x),y) < r$, so $\gamma(\dist(f(x),y)) \in \Ba_r(x)$ and maps to $y$.  Hence $f|_{\Ba_r(x)}$ is surjective and $f(\Ba_r(x)) = \Ba_r(f(x))$.

Below we use the general fact that if $\bar{\Ba}_r(x)$ compact, then $x$ is connected to $x'$ by a geodesic for all $x' \in \Ba_r(x)$. To see this, take a sequence of unit-speed paths $x \to x'$ within $\Ba_r(x)$ with length approaching $\dist(x,x') < r$.  The Arzel$\grave{\textrm{a}}$–Ascoli theorem says that a family of uniformly bounded, uniformly equicontinuous functions on a compact space has a uniformly convergent subsequence.  Thus there exists a path $x \to x'$ of length $\dist(x,x')$, and this minimal path must be a geodesic.

To show  $f|_{\Ba_r(x)}$ is injective, consider $x_1, x_2$ such that $f(x_1)=f(x_2) \in \Ba_r(f(x))$.  Consider unit-speed geodesics $\gamma_1, \gamma_2$ from $x \to x_1$ and $x \to x_2$, respectively. (These exist by the fact above.) Because $f$ is a local isometry on $\Int(X)$, $f \circ \gamma_1$ and $f \circ \gamma_2$ are both unit-speed geodesics from $f(x)$ to $f(x_1)=f(x_2)$, hence the same geodesic, $f \circ \gamma_1= f \circ \gamma_2$.  Because $f$ is a local diffeomorphism around $x$, $f \circ \gamma_1= f \circ \gamma_2$ implies $\gamma_1$ and $\gamma_2$ have the same initial velocity, hence $\gamma_1 = \gamma_2.$  In particular, $x_1=x_2$, and $f|_{\Ba_r(x)}$ is a bijection.  Because it is a local diffeomorphism, and therefore an open map, it is also a diffeomorphism.

In the other direction, if $r > \mathrm{dist}(x, \partial X)$, then $\Ba_r(x)$ includes part of $\partial X$, so it is not homeomorphic to $\Ba_r(f(x))$.
\end{proof} 

Later we will need the fact that two nearby immersions $f, f' : M \to \RR^d$ are essentially related by a diffeomorphism of $M$. 
\begin{Proposition}\label{prop:small-diffeomorphism}
For $d$-dimensional manifold $M$ with boundary, consider proper immersions $f, f' : M \to \RR^d$ with outer $r$-collars and induced metrics $g_f,g_{f'}$ on $M^{+r}$, satisfying
\begin{align} \label{eq:ff-nearby}
\|f_{+r}-f'_{+r}\|_{C^0} < \delta, \qquad g_f  c^{-2} \le g_{f'} \le g_f c^2,
\end{align}
for some constants $c \ge 1$ and $\delta \le r/(3c)$.
Then there exists submanifold $\wt{M} \subset M^{+\delta}$ and diffeomorphism $\phi : M \to \wt{M}$ such that for all $x \in M$,  $\dist_f(\phi(x),x) < \delta$ and
\begin{align} \label{eq:ffphi}
f' = f_{+r}|_{\phi(M)} \circ \phi.
\end{align}
Moreover, if $\psi : M \to \hat{M} \subset M^{+r}$ is any diffeomorphism such that
$f' = f_{+r}|_{\psi(M)} \circ \psi$ and $\dist_f(\psi(x),x) < r/3$ for all $x\in M$,
then $\psi = \phi$.
\end{Proposition}
\noindent In Eq.~\eqref{eq:ff-nearby}, $C^0$ is the sup norm and the second expression is an inequality of bilinear forms; $\dist_f$ denotes the distance with respect to induced metric $g_f$.
\begin{proof}
First we construct $\phi$.  For each $x \in M$, consider the ball $B_x = \Ba^f_{\delta}(x) \subset M^{+r}$, where superscript $f$ denotes using the metric induced by $f$. Then $f_{+r}|_{B_x}$ acts as a homeomorphism onto $\Ba_{\delta}(f(x))$.  Note $f'(x) \in \Ba_{\delta}(f(x))$ because $\| f'(x) - f(x) \| < \delta$. Then define $\phi(x) = (f_{+r}|_{B_x})^{-1}(f'(x))$.  
By construction, $\dist_f(\phi(x),x) < \delta$, using metric induced by $f$. We set $\wt{M} = \phi(M)$, so that $f' = f_{+r}|_{\wt{M}} \circ \phi$ and $\wt{M} \subset M^{+\delta}$.

Note $f_{+r}, f'_{+r}$ are local diffeomorphisms.  Locally, $\phi$ can be obtained by composing $f'$ with a local inverse of $f_{+r}$, so $\phi$ is also a local diffeomorphism.  

To show $\phi$ is injective, consider $x_1, x_2 \in M$ such that $\phi(x_1)=\phi(x_2)=z$.  Then by $f' = f_{+r}|_{\wt{M}} \circ \phi$ we have $f'(x_1)=f'(x_2)$.  
Note $x_1, x_2 \in \Ba^f_{\delta}(z)$, because $\dist_f(\phi(x),x) < \delta$ for all $x \in M$.  By the metric comparison in Eq.~\eqref{eq:ff-nearby}, $\Ba^f_{\delta}(z) \subset \Ba^{f'}_{c \delta}(z)$, where the latter ball uses metric induced by $f'$.  Then $x_1, x_2 \in \Ba^{f'}_{c\delta }(z)$.  Note $\dist_{f'}(z,M) \le \dist_{f'}(z,x_1) \le c\, \dist_{f}(z,x_1) \le c \delta \le r - c\delta $, using the assumption  $2 c \delta \le r$.  Then by Proposition~\ref{prop:admissible-ball},  $\dist_{f'}(z,M) \le r - c\delta$ implies $\Ba^{f'}_{c \delta}(z)$ is an admissible ball for $f'$.  Then $f'$ is injective on $\Ba^{f'}_{c \delta}(z)$, so $f'(x_1)=f'(x_2)$ implies $x_1=x_2$.  The local diffeomorphism $\phi$ is then injective, hence a diffeomorphism onto its image, as desired.

Finally, let $\psi : M \to \hat{M} \subset M^{+r}$ be a diffeomorphism such that $f' = f_{+r}|_{\psi(M)} \circ \psi$ and $\dist_f(\psi(x),x) < r/3$ for all $x\in M$.  Then $f'(x) = f_{+r}(\phi(x)) = f_{+r}(\psi(x))$.  Because $\Ba^f_{r-\delta}(\phi(x))$ is admissible with respect to $f$,  either $\phi(x) = \psi(x)$ or $\dist_f(\phi(x),\psi(x)) \ge r - \delta$. But $\dist_f(\phi(x),\psi(x)) < \dist_f(\phi(x),x) +  \dist_f(x,\psi(x)) < \delta + r/3 \le r-\delta$ (using $\delta \le r/3$), so $\phi(x) = \psi(x)$.  
\end{proof}

We can use the uniqueness property of $\phi$ in Proposition~\ref{prop:small-diffeomorphism} to see that these diffeomorphisms compose nicely. Given  $f, f' : M \to \RR^d$ as in the proposition, let us denote the diffeomorphism $\phi$ as $\phi_{f' \to f_{+r}}$. 
 Consider three nearby immersions $f, f' , f'': M \to \RR^d$ with outer $2r$-collars, satisfying the analog of Eq.~\eqref{eq:ff-nearby} pairwise with the same constants $c, \delta$, and with $2 c\delta \le r/3$.  Then we find
\begin{equation} \label{eq:phi-compose}
\phi_{f'' \to f_{+2r}} = \phi_{f'_{+r} \to f_{+2r}} \circ \phi_{f'' \to f'_{+r}},
\end{equation}
by calculating $\dist_f(\phi_{f'_{+r} \to f_{+2r}} \circ \phi_{f'' \to f'_{+r}}(x) ,x) < \delta + c\delta \le r/3$ and applying the uniqueness property of $\phi$.

\subsection{Geometric circuits}

In this section, we formalize the extension process shown in Fig.~\ref{fig:growth0}, focusing on its geometric aspects and postponing the discussion of the corresponding quantum states. The process is described by a sequence of regions $Y_0,\dots,Y_n$, such that $Y_k$ differs from $Y_{k-1}$ by a set of local changes in disjoint reference patches and the removal of some parts. The individual changes are called ``geometric gates'', a set of parallel changes a ``layer'', and the whole sequence of layers a ``geometric circuit''. These concepts are very general and do not depend on any assumptions about the ambient space.

\begin{Definition}\label{def:geom_gate}
A \emph{geometric gate} is a triple $(G,F,F')$ of regions such that $F,F'\subseteq G$. It transforms an arbitrary region $Y\supseteq F$ into $(Y\setminus G)\cup F'$ or any smaller region. Thus, the gate \emph{transforms region $Y$ into region $Y'$} if
\begin{equation}\label{geom_gate}
Y\supseteq F,\qquad Y'\cap G\subseteq F',\qquad
Y'\setminus G \subseteq Y\setminus G.
\end{equation}
A \emph{layer} is a collection of geometric gates $(G_j,F_j,F'_j)$ such that the regions $G_j$ are disjoint. It can be treated as a single gate with $G=\bigcup_j G_j$,\, $F=\bigcup_j F_j$, and $F'=\bigcup_j F'_j$. A \emph{geometric circuit} transforming $Y_0$ into $Y_{n}$ is a sequence of $n$ layers, with the $k$-th layer transforming $Y_{k-1}$ into $Y_k$. The number $n$ is called the circuit \emph{depth}.
\end{Definition}

\begin{figure}\centering
\includegraphics{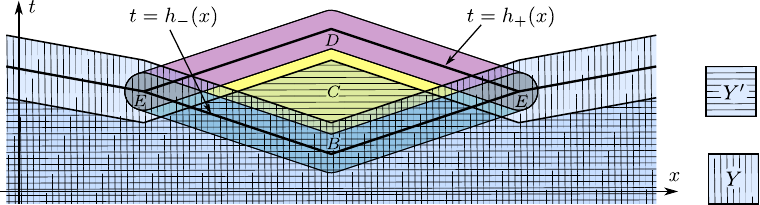}
\caption{The geometric gate $(BCDE,BE,BC)$ transforms region $Y$ into region $Y'$ according to Definition~\ref{def:geom_gate}, which combines the gate action with arbitrary shrinking. The new region has grown in the middle but retreated in other places.}
\label{fig:geom_gate}
\end{figure}

For example, consider the situation shown in Fig.~\ref{fig:gmp}(a), and with some additional features, in Fig.~\ref{fig:geom_gate}. Both figures include the graphs of functions $h_-,h_+\colon\RR^{d-1}\to\RR$, which are constructed as follows. Let $g_-,g_+\colon\RR^{d-1}\to\RR^{d-1}\times\RR$ be some functions that are equal outside the unit ball $D^{d-1}$ and on its boundary, and let us use the notation $g_\mp(x)=(g_\mp^\parallel(x),g_\mp^\perp(x))$. It is assumed that the Lipschitz constant of the map $x\mapsto g_\mp(x)-(x,0)$ is at most $1/2$, which implies that $g_\mp^\perp(x)=h_\mp(g_\mp^\parallel(x))$ for some $h_\mp$. We also assume that $h_+(x)\ge h_-(x)$ for all $x$. In this setting, let $g$ be the continuous map of $S^{d-1}$ to $\RR^d$ made of two hemisphere maps, namely, $g_\mp$ restricted to $D^{d-1}$. Recall that the configuration $(B,C,D,E)=\lambda(a,g)$ is given by $BE=\Ba_a(g_-(D^{d-1}))$ and $DE=\Ba_a(g_+(D^{d-1}))$. Suppose that
\begin{equation} \label{eq:geom-gate-1}
Y\supseteq\Ba_{a}(Y_-),\qquad \Ba_{a}(Y')\subseteq Y_+,\qquad
\text{where}\qquad
Y_\mp=\{(x,t)\in\RR^{d-1}\times\RR:\, t\le h_\mp(x)\},
\end{equation}
see Fig.~\ref{fig:geom_gate}.
Then the geometric gate $(BCDE,BE,BC)$ transforms $Y$ into $Y'$. Indeed, $Y\supseteq\Ba_a(Y_-)\supseteq BE$ and $\RR^d\setminus Y'\supseteq\Ba_a(\RR^d\setminus Y_+)\supseteq DE$, which implies the first two conditions in Eq.~\eqref{geom_gate}. The third condition is satisfied because $Y'\setminus Y\subseteq Y_+\setminus Y_-\subseteq BCDE$.\medskip

Now, let $f\colon M\to\RR^d$ be a proper immersion of a $d$-dimensional manifold $M$ with boundary $N=\partial M$, and $(n(x):\,x\in N)$ the unit normal vector pointing outward at $f(x)$. We assume the boundary curvature bound $-(2r_1)^{-1}g\le K\le r_1^{-1}g$ and the existence of an inner $r_1$-collar. The tubular neighborhood $T$ (i.e., the union of the inner and outer collars) is identified with $N\times(-r_1,2r_1)$ so that the extended immersion $f_{+2r_1}\colon M^{+2r_1}\to\RR^d$ restricted to $T$ agrees with $\hat{f}\colon (x,t)\mapsto f(x)+t\ts n(x)$. All regions to be discussed are subsets of $M^{+2r_1}$, and distances are understood as path distances in that ambient space. The regions $Y_k$ associated with a geometric circuit will have the form%
\footnote{Alternatively, one can consider the region $\Int(Y_k)$ defined by the strict inequality $t<w_k(x)$. The gates we will use work for both open and closed regions.} 
\begin{equation}\label{Yk_vs_wk}
Y_k=(M\setminus T)\cup\{(x,t)\in T:\, t\le w_k(x)\},\qquad w_k\colon N\to(-r_1,r_1).
\end{equation} 

\begin{Theorem}[Constructing geometric circuits] \label{thm:geom_circuit}
The immersed manifold $M$ can be transformed into $\bar{\Ba}_{r_1}(M)\subseteq M^{+2r_1}$ by a constant depth geometric circuit whose gates lie in admissible copies of open balls of radius $r_1$ and are themselves copies of $(BCDE,BE,BC)$ for some standard configurations $(B,C,D,E)\in\Lambda(r_0,r_1)$ defined in Section~\ref{sec:micro}.  The gates are supported in $f(T)$ for tubular neighborhood $T$ given by the union of the inner $r_1$-collar and outer $2r_1$-collar. (It is assumed that $r_0\le c_1r_1$ for a sufficiently small positive constant $c_1$; ``constants'' may depend on the number of spatial dimensions $d$.) 
\end{Theorem}

Let us first sketch the proof. In terms of the functions $w_k$ in Eq.~\eqref{Yk_vs_wk}, each circuit layer effects a transition from $w_{k-1}(x)=w_-(x)+2r_0$ to $w_{k}(x)=w_+(x)-2r_0$, where $w_+(x)-w_-(x)$ is supported by a union of disjoint balls. In each of those ball, the picture is similar to Fig.~\ref{fig:geom_gate}. Thus, $Y_k$ differs from $Y_{k-1}$ by a set of bumps and a general shrinkage by a small amount, namely, $4r_0$. Note that the $\pm2r_0$ terms in the previous formulas include some safety margins, which would not be needed if $\partial Y_{k-1}$ and $\partial Y_{k}$ were almost flat. In that case, $\pm r_0$ would suffice, and the shrinkage would be only by $2r_0$. The proof details are concerned with making sure that the boundary never becomes too rough. In the end, it should advance uniformly by $r_1$, and splitting this height between individual bumps amounts to a partition of unity on $N$. To organize the gates into layers, the following lemma is used.

\begin{Lemma}\label{lem:gcirc1}
Let $M$ be as described above, and let $0<a\le b\le 2r_1$. Consider a set of points $S\subseteq N=\partial M$ such that the distance between any two distinct points is ${}\ge a$. Then the number of points in $S\cap\Ba_b(\wt{x})$ is bounded by $O((b/a)^d)$ for all $\wt{x}\in\partial M$. Furthermore, $S$ can be partitioned into $O((b/a)^d)$ subsets such that the distance between any two distinct points in each subset is ${}\ge b$.
\end{Lemma}
\begin{proof}
Since $b\le 2r_1$, the ball $\Ba_b(\wt{x})\subseteq M^{+2r_1}$ is an admissible copy of $\Ba_b(x)\subseteq\RR^d$, where $x=f(\wt{x})$. By replacing each $\wt{y}\in S\cap\Ba_b(\wt{x})$ with $y=f(\wt{y})$, the first statement in question is reduced to an elementary fact: any subset of $\Ba_b(x)\subseteq\RR^d$ whose elements are at least distance $a$ apart has at most $n=O((b/a)^d)$ elements.

Now, let $E$ be the set of pairs of distinct points $\wt{y}_1,\wt{y}_2\in S$ such that $\dist(\wt{y}_1,\wt{y}_2)<b$. Then the graph $(S,E)$ has degree ${}\le n-1$, so the second required statement follows from the next lemma.
\end{proof}

\begin{Lemma}\label{lem:gcirc2}
The vertex set of an undirected graph of degree ${}\le n-1$ can be partitioned into $n$ independent sets.
\end{Lemma}
\begin{proof}
Denote the graph by $(V_n,E_n)$ and let $S_n\subseteq V_n$ be a maximal independent set. Due to the maximality, any vertex $x\notin S_n$ has at least one adjacent vertex in $S_n$; hence, deleting $S_n$ results in a subgraph $(V_{n-1},E_{n-1})$ of degree ${}\le n-2$. This procedure can be repeated $n$ times, after which the vertices are exhausted.
\end{proof}

\begin{proof}[Proof of Theorem~\ref{thm:geom_circuit}.]
We will construct a geometric circuit of constant depth $n$ that transforms $M$ into the region $\bar{\Ba}_{cr_1}(M)$ for some constant $c$ such that $0<c \le 1/8$. The gates of that circuit will lie in open balls of radius $r_1/2$ with centers on $N=\partial M$. By repeating the procedure $\lceil c^{-1}\rceil$ times, the region can be extended to $\bar{\Ba}_{r_1}(M)$.

As already mentioned, we generally use the path distance $\dist(\cdot,\cdot)$ on $M^{+2r_1}$. However, the Lipschitz constants of functions on $N=\partial M$ are more naturally defined in terms of the corresponding path distance $\dist_N$ so that for sufficiently regular functions, $\KLip(v)=\sup_{z\in N}|v'(z)|$. The two distances are almost equal locally, in the vicinity of a given point $y\in N$. For an accurate comparison, let us use the coordinate chart $(U_y,\phi_y)$ defined in the paragraph surrounding Eq.~\eqref{local_bmap}. That equation expresses $f(z)$ for $z\in U_y$ as $f_y(x)$, where $x=\phi_y(z)$. If $x_1=\phi_y(z_1)$ and $x_2=\phi_y(z_2)$, then
\begin{equation}\label{dist_vs_lc}
\begin{aligned}
|x_1-x_2| &\le |f(z_1)-f(z_2)| \le \dist(z_1,z_2) \le \dist_N(z_1,z_2)
\\
&\le \left(\sup_{|x|\le r}\sqrt{1+|h'(x)|^2}\right)|x_1-x_2|
\le \frac{|x_1-x_2|}{\sqrt{1-r^2/r_1^2}}\qquad
\text{if }\, x_1,x_2\in\bar{\Ba}_r(0),
\end{aligned}
\end{equation}
where the last inequality follows from Eq.~\eqref{hprime_bound} with $R=r_1$. In what follows, we consider functions on $\bar{\Ba}_r(0)$ for $r=r_1/6$.

Let $S\subseteq N$ be a maximal set of points such that the distance between any two distinct points is at least $r_1/8$. For each $y\in S$, let
\begin{equation}\label{eq:qy_def}
q_y\colon N\to\RR,\qquad
q_y(z)=\begin{cases}
r_1/6-|\phi_y(z)| & \text{if } z\in U_y \text{ and } |\phi_y(z)|\le r_1/6,\\
0 & \text{otherwise}.
\end{cases}
\end{equation}
Because of the distance bound~\eqref{dist_vs_lc}, the condition $|\phi_y(z)|\le r_1/6$ implies that $\dist(z,y)\le r_1/5$. Thus, $\supp(q_y)\subseteq\bar{\Ba}_{r_1/5}(y)$. The following functions have the same support and form a partition of unity on $N$:
\begin{equation}\label{eq:vy_def}
v_y(z)=\frac{q_y(z)}{q(z)},\qquad
\text{where}\qquad q(z)=\sum_{y\in S}q_y(z).
\end{equation}
By Lemma~\ref{lem:gcirc1}, the number of points $y\in S$ such that $\dist(z,y)\le r_1/5$ is bounded by some constant $m$; hence, the sum in the expression of $q(z)$ has at most $m$ terms. On the other hand, due to the maximality of $S$, for any $z\in N$ there is some $y\in S$ such that $\dist(z,y)<r_1/8$. But $|\phi_y(z)|\le\dist(z,y)$; therefore, $q(z)>r_1/24$. Using this inequality and the bound $|q'_y(z)|\le 1$, it is easy to prove the following:
\begin{equation}
\text{given any }\, W\subseteq S,\qquad
\KLip\biggl(\,\sum_{y\in W}v_y\biggr)\le \frac{A}{r_1},
\end{equation}
where $A=24m$ is a constant.

By the last statement in Lemma~\ref{lem:gcirc1}, $S$ can be partitioned into a constant number of subset $S_1,\dots,S_n$ such that the distance between any two distinct points in $S_k$ is at least $r_1$. We define regions $Y_0,\dots,Y_n$ by Eq.~\eqref{Yk_vs_wk} with
\begin{equation}\label{wk_r0}
w_k(z)=\sum_{j=1}^{k}
\biggl(\frac{4}{3}cr_1\sum_{y\in S_j}v_y(z)-4r_0\biggr),\qquad
r_0\le\frac{c}{12n}r_1.
\end{equation}
Note that $w_n(z)\ge cr_1$ for all $z\in N$ because the functions $v_y$ form a partition of unity. We assume that $c\le 1/8$ so that $|w_k(z)|\le r_1/6$. Another constraint on $c$ will be spelled out later.

We now construct a layer of geometric gates associated with the set $S_k$. Let
\begin{equation}
w_-(z)=w_{k-1}(z)-2r_0,\qquad w_+(z)=w_{k}(z)+2r_0,
\end{equation}
so that $w_+(z)-w_-(z)=(4/3)cr_1\sum_{y\in S_k}v_y(z)$ is supported by certain neighborhoods of the points $y\in S_k$. Specifically, representing the functions $w_\mp$ in local coordinates as $w_{\mp,y}(x)=w_\mp(\phi_y^{-1}(x))$, we get
\begin{equation}
w_{+,y}(x)-w_{-,y}(x) \begin{cases}
>0 &\text{if }|x|<r_1/6,\\
=0 & \text{otherwise}.
\end{cases}
\end{equation}
From now on, we restrict the domain of $w_{\mp,y}$ and similar functions to $\bar{\Ba}_r(0)$ for $r=r_1/6$. Thus, $\KLip(w_{\mp,y})\le \wt{A}c$, where $\wt{A}=(4/3)A/\sqrt{1-r^2/r_1^2}=\const$. The following functions interpolate between $w_\mp$ on the one hand and $w_{k-1}$, $w_k$ on the other: \begin{equation}
\hat{w}_\mp(z,s)=(z,\,w_\mp(z)+s),\qquad
\hat{w}_{\mp,y}(x,s)=(x,\,w_{\mp,y}(x)+s)\qquad
\bigl(x=\phi_y(z)\in\bar{\Ba}_r(0),\qquad s\in[-2r_0,2r_0]\bigr).
\end{equation}
Note that $|w_\mp(z)+s|\le r$ for the same reason as $|w_k(z)|\le r$ (which we have seen earlier), and therefore, the point $(x,w_{\mp,y}(x)+s)$ belongs to the low-distortion domain $V$ of the map $\hat{f}_y=\hat{f}\circ(\phi_y^{-1}\times\id)$, where $\hat{f}\colon (z,t)\to f(z)+t\ts n(z)$, see Eq.~\eqref{distortion_bound}. It follows that the map $\hat{f}_y\circ\hat{w}_{\mp,y}$ is close to identity, namely,
\begin{equation}
\KLip\bigl(\hat{f}_y\circ\hat{w}_{\mp,y}-\id\bigr)
\le \bigl(1+\KLip(\hat{f}_y-\id)\bigr)\bigl(1+\KLip(w_{\mp,y})\bigr)-1
\le 0.31+1.31\wt{A}c.
\end{equation}
Let us choose $c$ such that the last number does not exceed $1/2$. Then the Lipschitz constant of the difference between $g_\mp\colon x\mapsto r^{-1}\hat{f}_y(\hat{w}_{\mp,y}(rx,0))$ and $x\mapsto(x,0)$ is also bounded by $1/2$, and hence, $(B,C,D,E)=\lambda(r_0,rg)$ belongs to the standard configuration set $\Lambda(r_0,r_1)$. Furthermore, $BCDE\subseteq\Ba_{r_0}(\hat{f}_y(V)) \subseteq\Ba_{r_0+1.31\sqrt{2}\ts r}(f(y)) \subseteq\Ba_{r_1/2}(f(y))$, so the gates corresponding to different elements $y\in S_k$ do not overlap.

The bound $\KLip(\hat{f}_y\circ\hat{w}_{\mp,y}-\id)\le 1/2$ (which holds for all $y\in N$) also implies that the map $\hat{f}_y\circ\hat{w}_{\mp,y} =\hat{f}\circ\hat{w}_\mp\circ(\phi_y^{-1}\times\id)$ cannot shrink distances more than by factor of $2$. But $\hat{f}\colon T\to\RR^d$ (where $T$ is the tubular neighborhood) is a local isometry because it agrees $f_{+2r_1}$, which is a local isometry by definition. Therefore, the images of the sets $\bar{\Ba}_{r}(0)\times\{0\}$ and  $\bar{\Ba}_{r}(0)\times\{\pm 2r_0\}$ under the map $\hat{w}_\mp\circ(\phi_y^{-1}\times\id)$ are separated by distance $\ge r_0$. It follows that for all $y,z\in N$, the distance between the points $(y,w_\mp(y))$ and $(z,w_\mp(z)\pm 2r_0)$ is at least $r_0$. Indeed, if the distance were less than $r_0$, then both $\phi_y(y)=0$ and $\phi_y(z)$ would be in $\Ba_{r_0}(0)\subseteq\bar{\Ba}_r(0)$, and hence, the points in question would belong to the respective images considered previously, which is a contradiction. Thus,
\begin{equation}
Y_{k-1}\supseteq\Ba_{r_0}(Y_-),\qquad \Ba_{r_0}(Y_{k})\subseteq Y_+,\qquad
\text{where}\qquad Y_\mp=(M\setminus T)\cup\{(z,t)\in T:\, t\le w_\mp(z)\}.
\end{equation}
We conclude that the layer of gates transforms $Y_{k-1}$ into $Y_k$.
\end{proof}

\begin{remark}[Working directly with codimension-1 manifolds]\label{rmk:geom_circuit_collar}
Theorem~\ref{thm:geom_circuit} may instead be formulated in terms of a proper immersion $f \colon N \to  \RR^d$, for a $(d-1)$-dimensional manifold $N$ equipped with a smooth choice of unit normal along $f$.  We call this an \emph{immersed oriented hypersurface}. In our usual application, $N = \partial M$ is the boundary of the $d$-dimensional immersed manifold $M$.  However, we will encounter situations where, for instance, $\partial M$ has multiple connected components and we want Theorem~\ref{thm:geom_circuit} applied to a single component.  Formally, this requires some generalization. We use the notation \(N^{(-a,b)} \cong N \times (-a,b)\) to denote a collar that extends \(N\) on both sides.
We assume $f$ satisfies curvature bound $-(2r_1)^{-1}g\le K\le r_1^{-1}g$. Then in collar coordinates
\((x,t) \in N \times (-a,b)\), the immersion $f$ extends to an immersion \( f_{(-a,b)} \colon N^{(-a,b)} \to \RR^d \) by
\((x,t) \mapsto f(x) + t\,n(x)\). Then one can reformulate Theorem~\ref{thm:geom_circuit} for $f : N \to \RR^d$ as above, working
with \(N^{(-r_1,2r_1)}\) to produce a geometric circuit that transforms \(N^{(-r_1,0]}\) to
\(N^{(-r_1,r_1]}\).
\end{remark}

The geometric circuit of Theorem~\ref{thm:geom_circuit} has constant depth, independent of the size of the immersed region $M$. Any constant-size subset $X$ of the extended collar $\partial M^{[0,r_1]}$ will therefore be ``influenced'' only by preceding gates supported on some constant-size region $X^{C}$, the ``causal cone'' or ``past lightcone'' of $X$.  (This property will be important later when we use geometric circuits to define circuits of quantum channels, where the use of the word ``influence'' is more clear.)
More concretely, consider a geometric circuit transforming
\(
Y_0 \;\to\; Y_1 \;\to\; \cdots \;\to\; Y_n,
\)
with $n$ layers as in Definition~\ref{def:geom_gate} and Theorem~\ref{thm:geom_circuit}. In layer $i$, the gates are indexed by $j$ and have the form
\[
(G_{ij},F_{ij},F'_{ij}) \;=\; (B_{ij}C_{ij}D_{ij}E_{ij},\,B_{ij}E_{ij},\,B_{ij}C_{ij}),
\]
with $G_{ij}=B_{ij}C_{ij}D_{ij}E_{ij}$ contained in an admissible ball of radius $r_1$ and with the $G_{ij}$ disjoint for fixed $i$.

Given a region $X \subseteq Y_n$ of radius at most $r_1$, we define its \emph{backwards causal cone} as follows. Let $\mathcal{G}_1$ be the set of gates $j$ in the final layer $n$ whose supports $G_{nj}$ intersect $X$. Inductively, for $k=2,\ldots,n$, let $\mathcal{G}_k$ be the set of gates in layer $n+1-k$ whose supports intersect either $X$ or the supports of some gate in $\mathcal{G}_1 \cup \cdots \cup \mathcal{G}_{k-1}$. Set
\[
\mathcal{G} \;=\; \bigcup_{k=1}^n \mathcal{G}_k.
\]
Thus $\mathcal{G}$ consists of precisely those gates that can influence the region $X$. Let
\[
X^{C} \;=\; X \cup \bigcup_{(G_{ij},F_{ij},F'_{ij})\in \mathcal{G}} G_{ij}
\]
be the union of their supports with $X$, and define the \emph{causal base} of $X$ as
\[
X^{B} \;=\; Y_0 \cap X^{C}.
\]
Then the base $X^{B} \subset Y_0$ serves as the portion of $Y_0$ that influences $X$.  
Specifically, the gates of $\mathcal{G}$ form a circuit transforming $\wt{Y}_0 \to \ldots \to \wt{Y}_n$ with $\wt{Y}_0=X^B \subset Y_0$ and $\wt{Y}_n = X \subset Y_n$.
It is straightforward to bound the size of these regions.
\begin{Remark}[Bounded causal cone] \label{rmk:causal_cone}
For a geometric circuit of depth $n$ as above, there exists a constant $c_{\mathrm{cone}}>0$ depending only on $n$ (namely, $c_{\mathrm{cone}}=2(2n+1)$ works) such that, for any region $X \subseteq Y_n$ of radius at most $r_1$, the associated causal cone $X^{C}$ satisfies
\[ \mathrm{diam}(X^{C}) \le c_{\mathrm{cone}}\, r_1. \] Then the causal base $X^{B} \subset X^{C}$ satisfies the same bound. More generally, for any region $X\subseteq Y_n$, the causal cone satisfies $X^{C}\subseteq\Ba_{2nr_1}(X)$.
\end{Remark}
\noindent
Indeed, the region $X$ lies in a ball of radius $r_1$, and each of the $n$ layers enlarges the causal cone by at most the diameter $2r_1$ of a gate on either side, all distances being path distances in $M^{+2r_1}$.

We will also need a version of Theorem~\ref{thm:geom_circuit} that grows a region along a subset of its boundary only. This discussion is similar to the setup surrounding Theorem~\ref{thm:geom_circuit}, with the main difference being that $N\subset \partial M$ is a $(d-1)$-dimensional hypersurface with its own boundary, $\partial N \neq \varnothing$. In this context, we modify the notion of inner and outer $R$-collar by restricting the domain of $\hat{f}: (x,t) \mapsto f(x)+ t\ts n(x)$ to $N\times (-R, 0]$ and $N\times [0, R)$, respectively. We refer to these as \emph{inner/outer $R$-collars along $N$}. Note the notion of extended immersion also generalizes straightforwardly (Definition~\ref{def:outer-collar}). We refer to this as an \emph{extended immersion along $N$}. The notion of tameness is modified accordingly.
\begin{Definition}[Locally tame regions]
    \label{def:locally_tame}
    For $d$-manifold $M$ with boundary, we say $f:M\to \mathbb{R}^d$ is $r$-tame along $N\subseteq \partial M$ if $f|_N$ is an $r$-tame hypersurface and $M$ admits an inner $r$-collar along $N$.
\end{Definition}

In what follows, we consider a proper immersion $f:M\to \mathbb{R}^d$ which is $R$-tame along $N\subseteq \partial M$. We assume that $N$ is closed in $\partial M$, so that $f|_N$ is proper. Here $M$ is allowed to be a manifold with corners, with $N$ contained in the smooth part of $\partial M$. We denote by $M^{+R}_N$ the manifold with corners obtained by attaching the outer $R$-collar along $N$, as in Definition~\ref{def:outer-collar}. Here the gluing is restricted to $N$ and the extended immersion is described by $f_{+R}: M^{+R}_N\to\RR^d$. Admissible copies (Definition~\ref{def:admissible}) should be understood with respect to this version of $f_{+R}$. The union $T$ of the inner and outer $R$-collars along $N$ is identified with $N\times(-R,R)$ so that $f_{+R}$ restricted to $T$ agrees with $\hat{f}$. All regions below are subsets of $M^{+R}_N$; $\dist$ denotes the path distance in $M^{+R}_N$ and $\dist_N$ the path distance on $N$.

The following is a generalization of Proposition~\ref{prop:admissible-ball} to the current setup.
\begin{Lemma}\label{lem:collar_local}Let $x\in M^{+R}_N$ and $r\in(0,R]$. Then $\Ba_r(x)$ is an admissible copy of $\Ba_r(f_{+R}(x))$ iff $\dist(x,M)\le R-r$ and $\dist(x,\partial(M^{+R}_N))\ge r$. In particular, these conditions hold for $x\in N$ satisfying $\dist_N(x,\partial N)>2r$ for $r\le R/2$.
\end{Lemma}

\begin{proof}
The first statement follows from Lemma~\ref{lemma:admissible-ball-general} as in the proof of Proposition~\ref{prop:admissible-ball}. Specifically, one can truncate the outer collar at height $t=R'=R-\eps$, making it closed. Then the restriction of $f_{+R}$ is proper because $f$ and $f|_N$ are, and it remains a local isometry on the interior. The criterion follows by taking $\eps\downarrow 0$. While the truncated space has corners, they lie on the boundary, so the proof of Lemma~\ref{lemma:admissible-ball-general} is unaffected.

For the last statement, the first condition holds because $\dist(x, M)=0$. For the second condition, we consider a path $\gamma$ of length $\ell\le r$ starting at $x$, and let $s^*$ be the supremum over $s$ satisfying $\gamma([0,s]) \subseteq T$. Since $\partial\hat{f}/\partial t=n$ is a unit vector normal to the hypersurfaces $N\times\{t\}$, the height along $\gamma$ varies by at most $\ell$, so $|t|\le r\le R/2$ on $[0,s^*)$. The metric induced on each such hypersurface is $g^{1/2}(\one+t\kap)^2g^{1/2}\ge(1-|t|/R)^2\ts g\ge g/4$, where $g$ is the metric on $N$ and $\|\kap\|\le R^{-1}$; see the discussion following Eq.~\eqref{fhat_x}. Hence the projection of $\gamma([0,s^*))$ to $N$ has length at most $2\ell\le 2r<\dist_N(x,\partial N)$.

Since $N$ is closed in $\partial M$ and $f$ is proper, $N$ is complete, so the projected path converges in $N$. By the bounds above, its limit lies in $N\setminus\partial N$, and the height converges to a value in $[-R/2,R/2]$. Hence $\gamma(s^*)$ lies in the set $\{(y,t): y\in N\setminus\partial N,\ |t|<R\}$, which is open in $M^{+R}_N$. In particular, $s^*=\ell$. Thus no path of length $\le r$ from $x$ reaches $\partial(M^{+R}_N)$, and therefore $\dist(x,\partial(M^{+R}_N))\ge r$.

\end{proof}

\begin{Theorem}
    \label{thm:geom_circuit_local}
Let $f:M\to \mathbb{R}^d$ and $N\subseteq \partial M$ be as described above, with $R\ge 2r_1$. There is a constant $c>0$ depending only on $d$ such that the manifold can be transformed into $f':M'\to \mathbb{R}^d$ by a constant-depth geometric circuit whose gates lie in admissible copies of open balls of radius $r_1$ centered at points of $f(N)$ and are themselves copies of $(BCDE,BE,BC)$ for some standard configurations $(B,C,D,E)\in\Lambda(r_0,r_1)$. For every open $\widetilde{N}\subseteq N$ with $\dist_N(\widetilde{N}, \partial N)> 3r_1$, the following holds:
\begin{enumerate}
\item $\widetilde{N} \times (0, cr_1] \subseteq M'\setminus M \subseteq N \times (0, cr_1]$,
\item $M\setminus M'\subseteq (N\setminus \widetilde{N}) \times (-\frac{cr_1}{3}, 0]$. 
\end{enumerate}
In particular, $f': M'\to \mathbb{R}^d$ is $(R-r_1)$-tame along $\widetilde{N}\times\{cr_1\}\subseteq \partial M'$.
\end{Theorem}
\begin{proof}

We mostly follow the proof of Theorem~\ref{thm:geom_circuit}. As in that proof, the geometric circuit will sequentially transform a region from $Y_0$ to $Y_n$; see Eq.~\eqref{Yk_vs_wk}. In this discussion, the tubular neighborhood $T$ appearing in Eq.~\eqref{Yk_vs_wk} should be identified with $N\times (-R, R)$, where $N \subseteq \partial M$ is as described above. In Eq.~\eqref{Yk_vs_wk} we also replace $M\setminus T$ by its closure in $M$, so that each $Y_k$ is closed in $M^{+R}_N$. This adds only points of $\partial N\times(-R,0]$, which lie outside all gates. We set $M'=Y_n$ and $f'=f_{+R}|_{M'}$.

The gates of the circuit are also defined similarly. They are associated with a set of points $S'$ as follows. Let $S\subseteq \{z\in N:\,\dist_N(z,\partial N)\ge r_1/2\}$ be a maximal set of points with pairwise distances at least $r_1/8$. We define $S'=\{y\in S:\,\dist_N(y,\partial N)\ge 5r_1/2\} \subseteq S$. Let $q_y$ and $q$ be as in Eqs.~\eqref{eq:qy_def} and~\eqref{eq:vy_def}, where the chart $(U_y,\phi_y)$ of Eq.~\eqref{local_bmap} is now defined in $N\setminus\partial N$. We set $v_y=q_y/q$ where $q>0$ and $v_y=0$ elsewhere, so that the functions $v_y$ with $y\in S'$ are supported in $\{\dist_N(\cdot\,,\partial N)\ge 23r_1/10\}\subseteq\{\dist_N(\cdot\,,\partial N)\ge r_1\}$.

On $\{\dist_N(\cdot\,,\partial N)\ge r_1\}$, the two bounds used in the proof of Theorem~\ref{thm:geom_circuit} hold with the same constants: (i) $q(z)>r_1/24$ and (ii) the number of $y\in S$ with $q_y(z)\neq 0$ is bounded by a constant, the balls used in the proofs being admissible copies by Lemma~\ref{lem:collar_local}. Consequently, $\KLip\bigl(\sum_{y\in W}v_y\bigr)\le A/r_1$ for any $W\subseteq S'$, with the same constant $A$. Likewise, $\Ba_{r_1}(y)$ is an admissible copy for every $y\in S'$ by the same lemma. Therefore, as in Lemmas~\ref{lem:gcirc1} and~\ref{lem:gcirc2}, $S'$ can be partitioned into a constant number $n$ of subsets $S_1,\dots,S_n$ with pairwise distances at least $r_1$ within each.

The circuit can now be specified as in the proof of Theorem~\ref{thm:geom_circuit}. The regions in Eq.~\eqref{Yk_vs_wk} are chosen as follows:
\begin{equation}\label{eq:wk_local}
w_k(z)=\sum_{j=1}^{k}\biggl((cr_1+4nr_0)\sum_{y\in S_j}v_y(z)-4r_0\biggr),\qquad
r_0\le\frac{c}{12n}r_1.
\end{equation}
The gates of the $k$-th layer, for each $y\in S_k$, are built from
\begin{equation}
w_-(z)=w_{k-1}(z)-2r_0,\qquad w_+(z)=w_{k}(z)+2r_0,
\end{equation}
whose difference in the local coordinates at $y$ satisfies
\begin{equation}
w_{+,y}(x)-w_{-,y}(x)\begin{cases}
>0 &\text{if }|x|<r_1/6,\\
=0 &\text{otherwise}.
\end{cases}
\end{equation}
Each gate is a copy of $(BCDE,BE,BC)$ for a standard configuration in $\Lambda(r_0,r_1)$, with $BCDE\subseteq\Ba_{r_1/2}(f(y))$, taken inside the admissible copy of $\Ba_{r_1}(f(y))$ provided by Lemma~\ref{lem:collar_local}. Since, within an admissible ball, $\dist$ coincides with the Euclidean distance to the center, the copy lies in $\Ba_{r_1/2}(y)$, and the gates of one layer are disjoint.

As in the proof of Theorem~\ref{thm:geom_circuit}, each layer of this circuit transforms $Y_{k-1}$ into $Y_k$. This can be verified locally: the containments $Y_{k-1}\supseteq\Ba_{r_0}(Y_-)$ and $\Ba_{r_0}(Y_k)\subseteq Y_+$ of that proof are only needed inside the gates, which lie in the admissible balls $\Ba_{r_1}(y)$ with $y\in S_k\subseteq S'$, while the third condition of Eq.~\eqref{geom_gate} follows from $Y_k\setminus Y_{k-1}\subseteq Y_+\setminus Y_-\subseteq BCDE$. Furthermore, the two claims about $M'$ follow from the properties of $w_n$. By Eq.~\eqref{eq:wk_local}, $w_n=(cr_1+4nr_0)\chi-4nr_0$, where $\chi=\sum_{y\in S'}v_y\in[0,1]$. In particular, $-cr_1/3\le w_n\le cr_1$. If $\dist_N(z,\partial N)\ge 27r_1/10$, then every $y\in S$ with $q_y(z)\neq 0$ satisfies $\dist_N(y,\partial N)\ge 5r_1/2$, so $y\in S'$ and $\chi(z)=1$; hence $w_n=cr_1$ on $\widetilde{N}$, which proves the first part. For the second part, note $w_n(z)<0$ requires $\chi(z)<1/4$, hence $\dist_N(z,\partial N)<27r_1/10<3r_1$ and $z\notin\widetilde{N}$; together with $w_n\ge -cr_1/3$, the claim follows.

Finally, $f'$ is $(R-r_1)$-tame along $\widetilde{N}\times\{cr_1\}$. Indeed, over $\{\dist_N(\cdot\,,\partial N)\ge 27r_1/10\}$, the boundary of $M'$ is obtained by pushing $N$ outward by $cr_1$; by the curvature bound following Eq.~\eqref{fhat_x}, the hypersurface $\widetilde{N}\times\{cr_1\}$ is $(R-cr_1)$-tame, and $\widetilde{N}\times(cr_1-(R-r_1),\,cr_1]\subseteq M'$ provides the inner $(R-r_1)$-collar.
\end{proof}

\begin{figure}[t]
    \centering
    \includegraphics{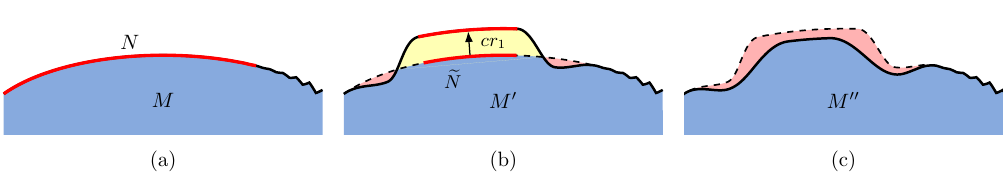}%
    \caption{\textbf{(a)} A boundary of $M$ may be partitioned into a tame part ($N$) and the part that may be jagged. \textbf{(b)} After applying a constant-depth geometric circuit, $\widetilde{N} \subset N$ advances uniformly by $cr_1$. Outside $\widetilde{N}$, the advance tapers off and eventually dips below the original baseline. \textbf{(c)} Applying a partial trace, the region near $\partial N$ can be smoothed.\label{fig:geom_circuit_local}}
\end{figure}

The circuits in Theorem~\ref{thm:geom_circuit_local} can break the tameness of the region in the vicinity of $\partial N$. Nonetheless, one can apply a partial trace to recover the tameness, setting up for the transformation at the next step; see Fig.~\ref{fig:geom_circuit_local}(c).

\section{Isomorphism theorem for deforming regions}\label{sec:isomorphism}

Throughout this section, we consider extendible local data, introduced in Section~\ref{sec:formal-intro}.  Recall this data is organized by length scales $r_0 \ll r_1 \ll r_2$.  We have ``data patches'' given by states $\sigma_K \in \St_K$ for all $r_2$-balls $K \subset \RR^d$, and we have ``reference patches'' given by states $\sigma_L \in \St_L$ for all $r_1$-balls $L \subset \RR^d$; these states are all pairwise consistent on overlaps.   For each reference patch $L$, for all standard configurations $(B,C,D,E)\in \Lambda(r_0,r_1)$ as described in Section~\ref{sec:geom} with $BCDE \subseteq L$, we require that $\sigma_L$ is $(B,C,D,E)$-extendible.  While all the state data may arise from restrictions of a global state, we do not assume this. 

We have already defined the information convex for embedded regions in Definition~\ref{def:ICS}.  Here we generalize this notion to immersed regions. First we formalize the notion of a spin system on the immersed region. Recall we have spins located at points $x \in J \subset \RR^d$, with associated Hilbert space $\calH_x$, and each finite ball contains finitely many spins.
\begin{Definition}[Hilbert spaces on immersed regions] \label{def:Hilbert-space-immersed}
Consider an immersion $f \colon M \to \RR^d$ of a compact manifold $M$ with boundary. We define a spin system on the immersed region $(M,f)$ with Hilbert space 
\begin{align}
\calH_{M,f}=\bigotimes_{x \in J\cap f(M)} \bigotimes_{\wt{x} \in f^{-1}(x)} \calH_x \,.
\end{align} 
\end{Definition}
\noindent That is, we take an isomorphic copy of $\calH_x$ for each point in the fiber $f^{-1}(x)$.
Because $M$ is compact, $\calH_{M,f}$ contains finitely many spins.  Our constructions will be essentially invariant under diffeomorphisms of $M$.  That is, if $\phi : M' \to M$ is a diffeomorphism, then $\IC(M, f)$ and $\IC(M',f \circ \phi)$ are canonically isomorphic; they only differ by how the local Hilbert spaces are labeled by points in the domain.

We want a notion of states on immersed regions that are consistent with reference patches.
\begin{Definition}[Consistent states on immersed regions] \label{def:immersed-consistent}
For immersed region $f \colon M \to \RR^d$, we say that $\rho \in \St_{M,f}$ is \emph{$r_1$-consistent} when $\rho|_{\wt{L}}=\sigma_{L}$ for all $r_1$-balls $L \subset \RR^d$ and all $\wt{L} \subseteq M$ such that $f$ maps $\wt{L}$ homeomorphically onto $L$.
\end{Definition}
That is, we require $\rho$ to be consistent with all admissible copies of $r_1$-balls, in the language of Definition~\ref{def:admissible}. 

To define the information convex set of the immersed region $(M,f)$, we assume a boundary curvature bound $-(2r_1)^{-1} g \le K$, so that we can define the extended immersion $f_{+2r_1} : M^{+2r_1} \to \RR^d$, using the outer collar of Definition~\ref{def:outer-collar}.
\begin{Definition}\label{def:ICS-immersed}
The \emph{information convex set} of an immersed region $f : M \to \RR^d$ is the set $\IC(M,f)$ of states $\rho \in\St_{M,f}$ that admit an extension $\rho_+ \in\St_{M^{+2r_1},f_{+2r_1}}$ on the extended immersed region $(M^{+2r_1},f_{+2r_1})$ that is $r_1$-consistent (Definition~\ref{def:immersed-consistent}).
\end{Definition}
This definition coincides with Definition~\ref{def:ICS} when $f$ is an embedding, in which case we continue to write $\IC(Y)$ for the embedded image region $Y=f(M)$.  For immersed regions, we keep the notation $\IC(M,f)$. By construction, $\IC(M,f)$ is a convex subset of $\St_{M,f}$.  By the characterization of admissible balls in Proposition~\ref{prop:admissible-ball}, the $r_1$-consistency of $\rho_+$ in Definition~\ref{def:ICS-immersed} implies that for all $\wt{x} \in M^{+r_1}$, $\rho_+|_{\wt{L}} = \sigma_L$ for $\wt{L} = \Ba_{r_1}(\wt{x})$ and $L=\Ba_{r_1}(f_{+2r_1}(\wt{x}))$. 

\begin{Lemma}[Restriction to embedded subregions]
\label{lem:restriction_embedded}
Let $h:P\to\RR^d$ be an immersed region for which $\IC(P,h)$ is defined, and let $X\subseteq P$ be a compact codimension-zero submanifold with $X^{+2r_1}\subseteq P^{+2r_1}$, on which $h_{+2r_1}$ is injective. Then $h(X)$ is an embedded region, and $\tau|_X\in\IC(h(X))$ for every $\tau\in\IC(P,h)$.
\end{Lemma}
\begin{proof}
Let $\hat{\tau}$ be an $r_1$-consistent extension of $\tau$ on $P^{+2r_1}$. Its restriction to $X^{+2r_1}$ is $r_1$-consistent, and through the injective map $h_{+2r_1}$ it is a state on $h_{+2r_1}(X^{+2r_1})=\Ba_{2r_1}(h(X))$. Every $r_1$-ball $L$ meeting $h(X)$ has its center within distance $r_1$ of $h(X)$, so by Proposition~\ref{prop:admissible-ball} it is the image of an admissible copy in $X^{+2r_1}$; thus $\hat{\tau}|_L=\sigma_L$ for all $L\in\Gamma_1(h(X))$, which is Definition~\ref{def:ICS}.
\end{proof}

For the isomorphism theorem, we treat regions with further regularity, controlled by length scale $r_2$.  Our results hold when $r_1 \le c_2 r_2$ for an unspecified constant $c_2>0$ that depends only on spatial dimension.  Its value is determined by geometric considerations in the proofs of Theorem~\ref{thm:geom_circuit} and Proposition~\ref{prop:macro-extend}.  Note we use a single scale $r_2$ to define both (a) the radius of balls for data patches and (b) the curvature bounds below, but there is no special significance in having taken these equal; while the scales are related,  the results would hold \textit{a fortiori} with larger data patches or smoother regions. 
 
\begin{Definition}[Tame regions and deformations] \label{def:macro-region}
For $d$-manifold $M$ with boundary, we say $f : M \to \RR^d$ is an \emph{$r$-tame region} when $f|_{\partial M}$ is an $r$-tame hypersurface (Section~\ref{sec:macro}) and $M$ admits an inner $r$-collar.
We say $F : M \times [0,1] \to \RR^d$ with $f_t = F(\cdot,t)$ is a \emph{$r$-tame deformation} from region $f_0$ to $f_1$ when $F$ is a regular homotopy such that $f_t$ is $r$-tame for all $t \in [0,1]$.  If unspecified, $r=r_2$ by default. 
\end{Definition} 
\noindent When we deform regions, we do not require their boundary is fixed.   

Now we can state the isomorphism theorem.  Informally, when we smoothly deform one region to another, we obtain a quantum channel between the associated information convex sets, which yields an isomorphism.  The isomorphism is canonical, in the sense that it only depends on the homotopy class of the path of deformations between regions. Again $M$ is a compact $d$-manifold with boundary, and the following holds for $r_0  \ll r_1 \ll r_2$, i.e., for $r_0 \le c_1 r_1$ and $r_1 \le c_2 r_2$ for constants $c_1, c_2>0$ depending only on spatial dimension. 
\begin{Theorem}[Isomorphism theorem]\label{thm:isomorphism}
Let $F\colon M\times[0,1]\to\RR^d$ be a tame deformation of regions, as in Definition~\ref{def:macro-region}.
Then there exists an assignment $F \mapsto \Phi^F$ for all such $F$, where $\Phi^F$ is a quantum channel
\[
\Phi^F \colon \IC(M,f_0)\to\IC(M,f_1)
\]
that depends only on the homotopy class of $F$ and satisfies 
\[
\Phi^{F_2 \circ F_1} = \Phi^{F_2} \circ \Phi^{F_1}
\]
where $F_2 \circ F_1$ denotes the composition of homotopies, with $\Phi^\id=\id$.
In particular, $\IC(M,f_0)$ and $\IC(M,f_1)$ are isomorphic as convex sets.  
\end{Theorem}
\noindent The final line follows because each homotopy $F$ has an inverse $\bar{F}$ given by reversing $t \in [0,1]$, so that $\Phi^{\bar{F}} \circ \Phi^F = \id_{\IC(M,f_0)}$ and $\Phi^F \circ \Phi^{\bar{F}} = \id_{\IC(M,f_1)}$. Note $\IC(M,f)$ also only depends on $f$ up to pre-composition with diffeomorphisms on $M$.

\begin{Remark} \label{rmk:scale-homotopy}
Rather than demand $f_t$ has curvature bounds and collars for all $t$, we could assume these properties only for $f_0$ and $f_1$.  Then we could apply Theorem~\ref{thm:isomorphism} to the modified homotopy $\wt{F} = S_2 \circ F' \circ S_1 : M\times[0,1]\to\RR^d$, where $\wt{F}$ satisfies $\wt{f}_0=f_0$ and $\wt{f}_1 = f_1$, but now $\wt{f}_t$ has appropriate curvature bounds and collars for all $t$.  Here $\circ$ refers to composition of homotopies,  $S_1$ is the homotopy that uniformly scales the image of $f_0$ by some large constant $c$, $F'$ is the scaled analog of $F$, and $S_2$ is the homotopy that uniformly scales the image by $c^{-1}$.  The constant $c$ may depend on $F$, and this construction relies on the compactness of $M$ and the smoothness of $F$. The same works relative to a boundary component. If $F$ fixes a collar of a boundary component, say $N$, and moves points only within a ball $\Ba_{R}(p)$, one may scale $\Ba_R(p)$ alone by a large factor, keeping it away from $N$, and interpolate back to the identity in between, obtaining a deformation with a better tameness parameter. 
\end{Remark}

\begin{Remark}
We may rephrase Theorem \ref{thm:isomorphism} categorically. Let $\mathrm{Imm}(M,\mathbb{R}^d)$ denote the space of immersions $f : M \to \mathbb{R}^d$, up to pre-composition with diffeomorphisms of $M$, and let
\(
\Pi_1(\mathrm{Imm}(M,\mathbb{R}^d))
\)
be its fundamental groupoid, whose objects are tame immersions and whose morphisms are homotopy classes (with endpoints fixed) of regular homotopies between them. Then Theorem~\ref{thm:isomorphism} says that
\[
\IC : \Pi_1(\mathrm{Imm}(M,\mathbb{R}^d)) \longrightarrow \mathrm{QState}
\]
is a functor, where $\mathrm{QState}$ is the category whose objects are convex sets of density matrices and whose morphisms are quantum channels, i.e., completely positive trace-preserving maps.  For $f : M \to \RR^d$  smooth but not tame, we do not define a state space $\IC(M,f)\subset \St_{\calH_{M,f}}$.  However, abstractly, we can still associate $f$ with the state space $\IC(M, f')$ where $f' \colon x \mapsto \lambda f(x)$ scales $\RR^d$ by sufficiently large $\lambda$ such that $f'$ is tame.  This data does not depend on $\lambda$, up to canonical isomorphism.
\end{Remark}

\begin{Remark}[Finite domains]\label{rmk:finite-domains}
For finite-volume applications, one should impose the local data and axioms only for configurations contained in reference patches lying in the interior of a chosen domain $\Omega \subset \RR^d$; the isomorphism theorem then applies to deformations through regions that remain sufficiently far from the boundary of $\Omega$. 
 Note that in principle, one could impose extendibility for reference patches that only partially overlap $\Omega$, even if all the spins are contained within $\Omega$. This makes sense formally because empty regions are assigned the trivial Hilbert space $\CC$, so the axioms remain well-defined even when some of the regions involved contain no spins. However, in that case, by Theorem~\ref{thm:isomorphism}, any bounded region deformable into the empty exterior of $\Omega$ would have trivial information convex, ultimately due to the extendibility constraints imposed near $\partial \Omega$. 
\end{Remark}

We postpone the \hyperlink{proof:isomorphism}{proof of Theorem \ref{thm:isomorphism}} while developing a few tools. 

\begin{Definition}[From geometric circuits to circuits of channels] \label{def:geom-circ-to-channel} 
Consider a geometric circuit as in Definition~\ref{def:geom_gate}, transforming region $Y_0$ to $Y_n$ using $n$ layers.  Assume each gate $G_{ij}$ in layer $i=1,\ldots,n$ is associated to a configuration $(B_{ij},C_{ij},D_{ij},E_{ij})$ on which the reference state is locally extendible, with associated recovery map $R_{B_{ij} E_{ij} \to B_{ij} C_{ij}}.$ Denote $\wt{A}_i = \cup_j A_{ij}$ for $A \in \{B,C,D,E\}$, and denote $R_{\wt{B}_i \wt{E}_i \to \wt{B}_i \wt{C}_i} = \bigotimes_j R_{B_{ij} E_{ij} \to B_{ij} C_{ij}}$.  For each layer $i$, define channel 
\begin{align} \label{eq:channel-gate}
R_i = \Tr_{Y'_i \setminus Y_i}\circ  R_{\wt{B}_i \wt{E}_i \to \wt{B}_i \wt{C}_i} \circ \Tr_{Y_{i-1} \cap (\wt{C}_i \wt{D}_i)}.
\end{align}
where $Y'_i =( Y_{i-1}   \setminus (\wt{D}_i \wt{E}_i))  \cup \wt{B}_i \wt{C}_i$. Here the extendibility map $R_{\wt{B}_i \wt{E}_i \to \wt{B}_i \wt{C}_i}$ is understood to be tensored with the identity on $Y_{i-1} \setminus \wt{B}_i \wt{E}_i \wt{C}_i \wt{D}_i$, so that $R_i \colon \St_{Y_{i-1}}  \to \St_{Y_i}$.  Leaving the identities implicit, we define the channel associated to the entire circuit as 
\begin{align}
R_{Y_0 \to Y_n} = R_n \circ \ldots \circ R_1.
\end{align}
\end{Definition}

Below we use the notation $\partial M^{[-a,b)}$ to denote the tubular neighborhood of $\partial M$ consisting of a (closed) inner $a$-collar and (open) outer $b$-collar; see Definitions~\ref{def:inner-collar} and \ref{def:outer-collar}. We also denote $M^{+b} = M \cup \partial M^{[0,b)}$ and $M^{-a} = M \setminus \partial M^{[-a,0)}$.   Following Remark~\ref{rmk:geom_circuit_collar}, sometimes we work with an immersed oriented hypersurface $N$ and tubular neighborhoods $N^{[-a,b)}$, defined analogously. In the case $N = \partial M$, with $M$ admitting an inner $a$-collar, we can identify $N^{[-a,b)} = \partial M^{[-a,b)}$. 

\begin{Proposition}[Recovery from erased collar] \label{prop:collar-recovery}
Let $f : N \to \RR^d$ be an immersed oriented hypersurface (Remark~\ref{rmk:geom_circuit_collar}) with curvature bound $-(2r_1)^{-1}g\le K\le r_1^{-1}g$.
Define adjacent shells $B,C,D$, with $B= N^{[-r_1,0)}, C = N^{[0,r_1)}, D = N^{[r_1,2r_1)}$.  There exists quantum channel $R_{B \to BC}$ such that, for arbitrary subsystem $A$, for all $\rho \in \St_{ABCD}$ such that $\rho|_{BCD}$ is $r_1$-consistent, $\rho|_{ABC} = \id_A \otimes R_{B \to BC} [ \rho|_{AB} ]$.  
\end{Proposition}
\noindent To emphasize, given consistency on $BCD$, the restriction of $\rho$ to $ABC$ is uniquely determined by its further restriction to $AB$. The basic principle is to iterate the extendibility property of Eq.~\eqref{eq:loc_extend}.
\begin{proof} Consider the geometric circuit of Theorem~\ref{thm:geom_circuit} (in the collar form of Remark~\ref{rmk:geom_circuit_collar}) that transforms $Y_0 = B = N^{[-r_1,0)} $ to $Y_n = BC = N^{[-r_1,r_1)}$, using $n$ layers, with the $i$'th layer transforming $Y_{i-1}$ to $Y_i$.  Let $R_{B \to BC}$ be the associated quantum channel, as in Definition~\ref{def:geom-circ-to-channel}, composed of layers $R_{B \to BC} = R_n \circ \ldots \circ R_1$.
Let $\tau_0 = \rho|_{AB}$ and define the sequence of states $\tau_i = \id_A \otimes R_i[\tau_{i-1}]$ for $i=1,\ldots,n$.  We prove by induction that $\tau_i = \rho|_{AY_i}$.  The $i=0$ case follows by definition. Take the inductive hypothesis $\tau_i = \rho|_{AY_i}$.  
Note the channel $R_i$ of Eq.~\eqref{eq:channel-gate} is composed of erasures and $(B,C,D,E)$-extendibility maps. These $BCDE$ are supported in admissible $r_1$-balls in $N^{[-r_1,2r_1)}$, and $\rho|_{BCD}$ is $r_1$-consistent, hence $\rho$ is $(B,C,D,E)$-extendible. 
Then
\begin{align} \label{eq:channel-circ-1}
 \tau_{i+1} = \id_A \otimes R_{i+1}[\tau_i] = \id_A \otimes R_{i+1}[\rho|_{AY_i}] = \rho|_{AY_{i+1}},
\end{align}
where the final equality follows from the extendibility of $\rho$ and the defining property~\eqref{eq:loc_extend}.  Finally, $\tau_n = \rho|_{AY_n}= \rho|_{ABC}$ yields the desired result.
\end{proof}

Now we can show $\IC(M,f_1)$ has at most one state when the tame embedded topological ball $f_1 \colon M \to \RR^d$ is isotopic to a geometric ball.%
\footnote{We do not address the separate question of whether every tame ball is isotopic to a geometric ball, to avoid possible pathologies with exotic balls in higher dimension.} 
Later we find that $\IC(M,f_1)$ for such a ball contains exactly one state, but for now we show a more limited statement.
\begin{Proposition}[Uniqueness on balls]\label{prop:ball-unique-0}
Let $F : M \times [0,1] \to \RR^d$ be an isotopy of a topological ball $M$ such that $f_0(M)$ is a geometric $r_1$-ball.  Assume $f_t(M)$ has boundary curvature bound $-(2r_1)^{-1}g\le K\le r_1^{-1}g$ with embedded outer $2r_1$-collar and inner $r_1$-collar, and
 $(f_t)_{+2r_1}(M^{+2r_1}) \subset (f_1)_{+2r_1}(M^{+2r_1})$ for all $t \in[0,1]$.  Then $\IC(M,f_1)$ contains at most one state.
\end{Proposition}
\noindent Roughly speaking, we demand the embedded ball $f_1(M)$ is isotopic to a geometric $r_1$-ball via an isotopy contained in an $2r_1$-neighborhood of the former.  
\begin{proof}
Let $Y = f_1(M)$ and $Y^+ = (f_1)_{+2r_1}(M^{+2r_1})$. Suppose $\rho \in \IC(M,f_1)$, witnessed by an $r_1$-consistent extension $\rho_+ \in \St_{Y^+}$.
By compactness, choose $0 = t_0 < \cdots < t_n = 1$ such that $Y_{i+1} := f_{t_{i+1}}(M) \subset \Ba_{r_1}(Y_i)$ for each $i$, where $Y_i := f_{t_i}(M)$. At each step, apply Proposition~\ref{prop:collar-recovery} with $N = \partial Y_i$, taking $A = Y_i^{-r_1}$, $B = (\partial Y_i)^{[-r_1,0)}$, $C = (\partial Y_i)^{[0,r_1)}$, $D = (\partial Y_i)^{[r_1,2r_1)}$. The hypothesis $(f_{t_i})_{+2r_1}(M^{+2r_1}) \subset Y^+$ ensures that $BCD \subset Y^+$, so $\rho_+|_{BCD}$ is $r_1$-consistent. Proposition~\ref{prop:collar-recovery} then gives
\begin{align}
\rho_+|_{Y_i \cup BC} = \id_A \otimes R_{B \to BC}[\rho_+|_{Y_i}].
\end{align}
where $R_{B \to BC}$ depends only on the local data and not on $\rho$.  Since $Y_{i+1} \subset \Ba_{r_1}(Y_i) = Y_i \cup BC$, restricting yields $\rho_+|_{Y_{i+1}} = \Phi_i[\rho_+|_{Y_i}]$ for some channel $\Phi_i$ independent of $\rho$.
Composing, 
\begin{equation}
\rho_+|_Y = \Phi_{n-1} \circ \cdots \circ \Phi_0[\rho_+|_{Y_0}]. 
\end{equation}
Since $Y_0$ is a geometric $r_1$-ball, $r_1$-consistency forces $\rho_+|_{Y_0} = \sigma_L$.  The right-hand side is then independent of $\rho$, so $\IC(M,f_1)$ contains at most one state.
\end{proof}

Below we work with an oriented immersed hypersurface $f : N \to \RR^d$, as in Remark~\ref{rmk:geom_circuit_collar}.  In practice this will often arise as the boundary $N= \partial M$ of a tame region (Definition~\ref{def:macro-region}).
\begin{Proposition}[Extending consistent states]\label{prop:macro-extend}
Let $f : N \to \RR^d$ be an oriented immersed hypersurface with boundary curvature $-r_2^{-1} g\le K \le r_2^{-1} g$.  Define adjacent shells $A, B,C$, with $A=N^{[-r_2,-2r_1)}, B=N^{[-2r_1,r_1)}, C=N^{[r_1,2r_1)}$.   Then there exists a  channel $R_{B \to BC}$ such that for all $\rho \in \St_{AB}$ that are $r_1$-consistent on $AB$, $\id_{A} \otimes R_{B \to BC}[\rho]$ is $r_1$-consistent on $ABC$.   
\end{Proposition}
\noindent First we sketch the idea, taking $N=\partial M$ for concreteness. The channel $R_{B \to BC}$ is obtained via the circuit of Definition~\ref{def:geom-circ-to-channel}. To see that an $r_1$-consistent state $\rho$ remains consistent after applying a single gate (a single channel $R_{B_i E_i \to B_i C_i}$), note that the restriction of $\rho$ to an $r_1$-ball containing $B_i C_i D_i E_i$ is consistent with a reference patch. Then the extendibility property~\eqref{eq:loc_extend} guarantees the state after applying $R_{B_i E_i \to B_i C_i}$ is consistent with the \textit{same} patch.  However, $R_{B_i E_i \to B_i C_i}$ does not preserve consistency with patches that overlap $C_i$ but do not contain $B_iC_iD_iE_i$. Then naively, the radius of consistency shrinks slightly with each extension.  The argument therefore requires more delicacy.  First, on the interior, Proposition~\ref{prop:ball-unique-0} allows one to upgrade $r_1$-consistency to consistency with larger data patches.  Then, to show consistency on $r_1$-balls near the boundary, we trace their lightcone under the circuit into the interior, where it lands within a data patch.  This argument is illustrated in Fig.~\ref{fig:extend-with-cone}.
\begin{proof} 
We work with extended immersion $f_{(-r_2,r_2)} : N^{(-r_2,r_2)} \to \RR^d$.
Take $\rho \in \St_{A B}$ that is $r_1$-consistent.  Subdivide $B = B_1 B_2$ into shells $B_1 = N^{[-2r_1, -r_1)}$, $B_2 = N^{[-r_1,r_1)}$, so that we have adjacent shells $A,B_1,B_2,C$.
We define $R_{B \to BC} = R'_{B_1 \to BC} \circ \Tr_{B_2}$ with channel $R'_{B_1 \to BC}$ defined as follows. By applying Theorem~\ref{thm:geom_circuit} (and Remark~\ref{rmk:geom_circuit_collar}) three times and concatenating circuits, 
there exists a  constant-depth geometric circuit mapping $B_1 \to BC$.  (The upper boundary of $B_1$, at height $-r_1$, advances distance $3r_1$ to height $2r_1$.)
Let $R'$ be the channel associated to this circuit, as in Definition~\ref{def:geom-circ-to-channel}. 

The argument below matches the schematic in Fig.~\ref{fig:extend-with-cone}, where $AB_1$ here plays the role of $X$ in the figure.
We want to show $\rho' := R_{B \to BC}[\rho] = \id_A \otimes R'_{B_1 \to BC}[\rho|_{AB_1}] \in \St_{ABC}$ is $r_1$-consistent on $ABC$. Consider any admissible copy $W \subseteq ABC$ of some $r_1$-ball $L \subset \RR^d$ (the reference patch in Fig.~\ref{fig:extend-with-cone}).  We must show $\rho'|_{W} = \sigma_L$.  The nontrivial case is when $W$ overlaps $BC$. 

Let $W^{C}$ denote the causal cone of $W$ with respect to the circuit (Remark~\ref{rmk:causal_cone}), i.e., the union of supports of all gates that influence $W$.  Because the circuit has constant depth, $\mathrm{diam}(W^{C}) < c_{\mathrm{cone}} r_1$ for a constant $c_{\mathrm{cone}}$ depending only on that depth, hence only on spatial dimension. Let $W^{B} =  W^{C} \cap (AB_1)$ be the causal base, the portion of the input $AB_1$ that influences $W$.  

Choose a topological ball $V \subset AB$ containing $W^B$ (the ``extended causal base'' in Fig.~\ref{fig:extend-with-cone}) with $\Ba_{2r_1}(V) \subset AB$.   We choose $V$ so that its embedding satisfies the hypotheses of Proposition~\ref{prop:ball-unique-0}, which demands smoothness at scale $r_1$ and the existence of an isotopy to some geometric $r_1$-ball in $V$, with isotopy contained in $V$.  Because $W^B$ has diameter $\le c_{\mathrm{cone}} r_1$, such a ball $V \supset W^B$ exists when $r_1 \le c_2 r_2$, for sufficiently small universal constant $c_2$.  We may choose $V$ with boundary radius of curvature comfortably between $r_1$ and $r_2$.  

By Proposition~\ref{prop:ball-unique-0}, $\IC(V)$ contains at most one state.  Since $\Ba_{2r_1}(V) \subset AB$ and $\rho$ is $r_1$-consistent on $AB$, we have $\rho|_{V} \in \IC(V)$.  Because $\Ba_{2r_1}(V)$ is admissible and has radius $< r_2$, there exists a data patch $\sigma_K$ for an $r_2$-ball $K \subset \RR^d$ with $K\supset f_{(-r_2,r_2)}(\Ba_{2r_1}(V))$.  Then $\sigma_K|_{V} \in \IC(V)$.  By uniqueness, $\rho|_{V} = \sigma_K|_{V}$, and in particular $\rho|_{W^B} = \sigma_K|_{W^B}$.

Finally, choose an $r_2$-ball $K'$ containing the image of $W^C$ (the ``data patch'' in Fig.~\ref{fig:extend-with-cone}).  This is possible because $\mathrm{diam}(W^C) < c_{\mathrm{cone}} r_1 \ll r_2$.  The gates in the causal cone form a circuit mapping $W^B$ to $W$; let $R_{W^B \to W}$ denote the corresponding channel.  Then
\begin{align} \label{eq:macro-extend-1}
\rho'|_{W} = R_{W^{B} \to W}[\rho|_{W^{B}}] = R_{W^{B} \to W}[\sigma_{K}|_{W^{B}}]  = R_{W^{B} \to W}[\sigma_{K'}|_{W^{B}}] = \sigma_{K'}|_W = \sigma_L,
\end{align}
where the third equality uses the consistency of data patches, the fourth because $\sigma_{K'}$ is $r_1$-consistent and hence extendible on each gate in the causal cone, and the final equality is by consistency of data and reference patches.
\end{proof}

Let us sketch the basic principle behind the proof of Theorem~\ref{thm:isomorphism}. Take the special case of embedded regions $X, X' \subset \RR^d$ with $X \subseteq X'$.  We want channels $\Phi : \IC(X) \to \IC(X')$ and $\Phi' : \IC(X') \to \IC(X)$ that compose both ways to the identity.  Take $\Phi'$ as the restriction map $\Tr_{X' \setminus X}$, and define $\Phi$ via Proposition~\ref{prop:macro-extend}, which allows us to extend consistent states to consistent states on larger regions.  Loosely speaking, $\Phi' \circ \Phi = \id_{\IC(X)}$ because extension followed by erasure yields the identity, while $\Phi \circ \Phi' = \id_{\IC(X')}$ follows from Proposition~\ref{prop:collar-recovery}, because the erasure can be recovered. 

To prove Theorem~\ref{thm:isomorphism}, we will need to consider nearby immersions $f, f' : M \to \RR^d$.  Say $\|f-f'\|_{C^0} <  r$, so that $f'(M) \subseteq \Ba_r(f(M))$.  For the case of embeddings, it follows $\Alg_{f'(M)} \subseteq \Alg_{\Ba_r(f(M))}$, and we obtain a restriction map on states $\IC(M^{+r},f_{+r}) \to \IC(M,f')$. However, for the case of immersions, the inclusion of image sets $f'(M) \subseteq \Ba_r(f(M))$  does not immediately yield a relation between $\calH_{M,f}$ and $\calH_{M,f'}$, because the Hilbert space may contain multiple copies of each spin in the image (Definition~\ref{def:Hilbert-space-immersed}).  Instead, we start with the canonical inclusion 
\begin{equation} \label{eq:MMtilde-inclusion}
\Alg_{\wt{M},\wt{f}} \subseteq \Alg_{M^{+r},f_{+r}}
\end{equation}
for submanifold $\wt{M} \subset M^{+r}$ with $\wt{f}=f_{+r}|_{\wt{M}}$.  We can express nearby immersions $f, f'$ roughly in this form.  With the hypotheses of  Proposition~\ref{prop:small-diffeomorphism}, we find diffeomorphism $\phi : M \to \wt{M} \subset M^{+r}$ such that  
\begin{equation}
f' =\wt{f} \circ \phi, \qquad \wt{f} =  f_{+r}|_{\wt{M}}.
\end{equation} 
 Note that in general, for diffeomorphism $\phi : M \to \wt{M}$ and immersion $\wt{f} : \wt{M} \to \RR^d$,
\begin{equation}
\calH_{\wt{M}, \wt{f}} \cong \calH_{\phi^{-1}(\wt{M}), \wt{f} \circ \phi}
\end{equation}
because the construction of the Hilbert space is invariant (up to re-labeling) after diffeomorphisms on the domain. Combined with~\eqref{eq:MMtilde-inclusion} we obtain
\begin{equation}
\calH_{M,f'} =   \calH_{\phi^{-1}(\wt{M}), \wt{f} \circ \phi} \cong \calH_{\wt{M},\wt{f}}  \subset \calH_{M^{+r},f_{+r}}.
\end{equation} 
Let us denote the above map
\begin{equation}
\iota_{f' \to f_{+r}} \colon \calH_{M,f'} \to \calH_{M^{+r},f_{+r}}.
\end{equation}
It behaves nicely under composition for nearby immersions $f,f',f'' : M \to \RR^d$,
\begin{equation}
\iota_{f'' \to f_{+2r}} = \iota_{f'_{+r} \to f_{+2r}}  \circ \iota_{f'' \to f'_{+r}},
\end{equation}
using the fact that the corresponding diffeomorphisms $\phi$ in Proposition~\ref{prop:small-diffeomorphism} compose properly, as in Eq.~\eqref{eq:phi-compose}.

We immediately obtain a restriction map on states, $\St_{\calH_{M^{+r},f_{+r}}} \to \St_{\calH_{M,f'}}$.
We also obtain a restriction map on information convex sets,
\begin{align} \label{eq:restrict-ICS}
\mathrm{Res}^\IC_{f_{+r} \to f'} \colon \IC(M^{+r},f_{+r}) \to \IC(M,f').
\end{align}
To see this, note we can reformulate the target as $\IC(M,f') \cong \IC(\wt{M},\wt{f})$, because the definition of the information convex set is invariant under diffeomorphisms on $M$.  Then using the inclusion $\wt{M}^{+2r_1} \subset M^{+(r+2r_1)}$, we can straightforwardly restrict $r_1$-consistent states on $(M^{+(r+2r_1)},f_{+(r+2r_1)})$ to $r_1$-consistent states on $(\wt{M}^{+2r_1},f_{+(r+2r_1)}|_{\wt{M}^{+2r_1}})$.

With these ingredients, we turn to the proof of Theorem~\ref{thm:isomorphism}. Recall that given a regular homotopy
$F\colon M\times[0,1]\to\RR^d$ of immersions $f_t=F(\cdot,t)$, each with boundary curvature bound $|K|\le r_2^{-1}$ and inner $r_2$-collar, we seek a quantum channel $ \Phi^F \colon \IC(M,f_0)\to\IC(M,f_1)$, such that the assignment $F \mapsto \Phi^F$ respects composition and only depends on the homotopy class of $F$.

\begin{proof}[\textbf{Proof of Theorem~\ref{thm:isomorphism}}] 
\hypertarget{proof:isomorphism}{}

First we treat some properties of nearby immersions $f , f' : M \to \RR^d$, with induced metrics $g_f, g_{f'}$ on $M^{+r_1}$, satisfying 
\begin{align} \label{eq:ffdist}
\| f_{+r_1}- f'_{+r_1} \|_{C^0} < \delta, 
\qquad
g_f\, c^{-2} \;\le\; g_{f'} \;\le\; g_f\, c^{2},
\end{align}
fixing constants $c=2$ and $\delta < r_1/(12c) = r_1/24$. The discussion at Eq.~\eqref{eq:restrict-ICS} yields the canonical map
\[
\mathrm{Res}^\IC_{f_{+r_1} \to f'} \colon \IC(M^{+r_1},f_{+r_1}) \to \IC(M,f').
\]
Consider the collar-extension channel $R^{(f)}_{M\to M^{+r_1}}:\ \IC(M,f)\longrightarrow \IC(M^{+r_1},f_{+r_1})$
constructed by Proposition~\ref{prop:macro-extend}, applied three times, to the hypersurfaces $\partial M^{-r_1}$, $\partial M$ and $\partial M^{+r_1}$ in turn, followed by a partial trace, and define
\begin{align}\label{eq:nearby-Phi}
\Phi_{f\to f'}\ =\ \mathrm{Res}^\IC_{f_{+r_1} \to f'}\circ R^{(f)}_{M\to M^{+r_1}} \, \colon \, \IC(M,f) \to \IC(M,f').
\end{align}
Each $\Phi_{f \to f'}$ acts as the identity on $M^{-3r_1}$, because $R^{(f)}_{M\to M^{+r_1}}$ acts only outside $M^{-3r_1}$.  

By Proposition~\ref{prop:collar-recovery}, states $\rho \in \IC(M,f)$ are uniquely determined by their restriction $\rho|_{M^{-r_1}}$.  Iterating this claim, $\rho$ is determined uniquely by its restriction $\rho|_{M^{-r}}$, for $r$ sufficiently smaller than $r_2$. 
Defining $\Phi_{f' \to f}$ analogously to $\Phi_{f\to f'}$, we find
\begin{align}  \label{eq:ff-id}
\Phi_{f' \to f} \circ \Phi_{f\to f'} = \id_{\IC(M,f)},
\end{align}
because  $\Phi_{f' \to f} \circ \Phi_{f\to f'}$ acts as the identity on $M^{-3r_1}$, so $\Phi_{f' \to f} \circ \Phi_{f\to f'}[\rho]$ and $\rho$ have the same restriction to $M^{-3r_1}$, hence they are equal.
By the same reasoning, 
\begin{align} \label{eq:fff-id}
\Phi_{f'' \to f} \circ \Phi_{f'\to f''} \circ \Phi_{f\to f'} = \id_{\IC(M,f)},
\end{align}
for any immersion $f''$ close to both $f$ and $f'$ in the sense of Eq.~\eqref{eq:ffdist}.

Now we consider the extended immersions $(f_t)_{+r_1}$. By smoothness of $F$ and compactness of $M$, there exist some $n \in \NN$ and times $t_0 < \ldots < t_n$ with $t_0=0, t_n=1$, such that, for $i = 0,\ldots,n-1$,
\begin{align} \label{eq:ffdist-t}
\| (f_{t_{i+1}})_{+r_1}- (f_{t_i})_{+r_1} \|_{C^0} < \delta, \qquad g_{f_{t_{i+1}}}\, c^{-2} \;\le\; g_{f_{t_{i}}} \;\le\; g_{f_{t_{i+1}}}\, c^{2},
\end{align}

We assemble maps $\Phi_{f \to f'}$ constructed by Eq.~\eqref{eq:nearby-Phi} into the total channel
\[ 
\Phi^F =\ \Phi_{f_{t_{n-1}}\to f_{t_n}}\circ\cdots\circ \Phi_{f_{t_0}\to f_{t_1}} \, \colon \, \IC(M,f_0) \to \IC(M,f_1).
\]
This $\Phi^F$ is independent of the choice of times $\{t_i\}_i$, so long as the time-intervals are sufficiently small that Eq.~\eqref{eq:ffdist-t} remains satisfied.  To see this, we can use Eq.~\eqref{eq:fff-id} to insert or remove time points.
Our construction also directly yields
\[
\Phi^{F_2 \circ F_1} = \Phi^{F_2} \circ \Phi^{F_1}
\]
where $F_2 \circ F_1$ denotes the composition of homotopies, as well as $\Phi^{\id}=\id$. To see $\Phi^F$ only depends on the homotopy class of $F$, consider $\mathcal{F} : M \times [0,1]^2 \to \RR^d$, with $F_t = \mathcal{F}(\cdot,\cdot,t)$.  Then $\Phi^{F_0} = \Phi^{F_1}$ because we can discretize $[0,1]^2$ into a square grid satisfying Eq.~\eqref{eq:ffdist} for neighboring points, then use Eq.~\eqref{eq:fff-id} to see that the associated diagram commutes.

Finally, for $F : M \times [0,1] \to \RR^d$, consider $\bar{F}$ with $t \in [0,1]$ reversed.  Then $\Phi^{\bar{F}}  \circ\Phi^F  = \id$ and $\Phi^F  \circ \Phi^{\bar{F}}  = \id$  because $\bar{F}  \circ F$ and $F \circ \bar{F}$ are both homotopic to a constant homotopy. Alternatively, the isomorphism follows directly from Eq.~\eqref{eq:ff-id}; in particular, $\Phi^F$ is an isomorphism of convex sets.
\end{proof}

\begin{Remark}[Decomposition of the isomorphism theorem]\label{rmk:iso-steps}
The channel $\Phi^F$ is a composition of quantum channels, each of which extends the region along its boundary by at most $r_1/6$. It uses the channel of Proposition~\ref{prop:macro-extend}, and then applies a partial trace to a region within distance $r_1/6$ of the original. In particular, the region at any step lies within distance $r_1$ of the region at the preceding step, and the causal base (Remark~\ref{rmk:causal_cone}) of a subregion at one step lies within distance $c_{\mathrm{step}}r_1$ of that subregion at the preceding step, where $c_{\mathrm{step}}:=3c_{\mathrm{cone}}+1$. We use this decomposition in Section~\ref{sec:rev_extendibility_merging}.
\end{Remark}

Now we can verify that the information convex set of a topological ball is trivial.   Essentially, this follows from the isomorphism theorem and the fact that the information convex set of an $r_1$-ball has a single state, given by restricting from a data patch.
\begin{Proposition} \label{prop:ball-unique}
Let $f : M \to \RR^d$ be a tame immersed region, with $M$ a topological ball, and assume $f$ is regular homotopic to a geometric ball.  Then $\IC(M,f)$ has exactly one state. 
\end{Proposition}
\begin{proof}
By the assumption that $f$ is regular homotopic to a geometric ball, by Remark~\ref{rmk:scale-homotopy}, we can moreover find a tame deformation from $f$ to an embedding $f'$ where $f'(M)$ is a geometric ball of radius $r = r_2$. By the isomorphism theorem~\ref{thm:isomorphism}, $\IC(M,f) \cong \IC(M,f')$ where $f' : M \to \RR^d$ embeds a geometric ball of radius $r = r_2$.  By Proposition~\ref{prop:ball-unique-0}, $\IC(M,f')$ contains at most one state.  A data patch $\sigma_K$ for $r_2$-ball $K \supseteq f'(M)$ can be extended to an $r_1$-consistent state on a larger ball by Proposition~\ref{prop:macro-extend}, hence furnishing a single state for $\IC(M,f')$.
\end{proof}

We can formulate a variant of Theorem~\ref{thm:isomorphism} in which some
boundary components of a tame immersed region are held fixed.  This generalization
will be used later in Lemma~\ref{lem:extendibility-from-annulus}.  For
simplicity of notation, we assume the fixed boundary component is embedded and
that there are only two boundary components, although both restrictions can be
relaxed.

Let \(f:M\to\RR^d\) be tame, with
\[
\partial M=N_1\sqcup N_2,
\]
and suppose \(f|_{N_1}\) is an embedding.  Let
\[
\widetilde B_f:=N_1^{[-r_2,0]}\subset M,
\qquad
B:=f(\widetilde B_f)\subset\RR^d
\]
be the corresponding embedded \(r_2\)-collar, and assume that \(B\) has tame \textit{boundary}.
\footnote{The requirement that the \(r_2\)-collar \(B\) has
tame boundary is stronger than the requirement that \(f\) be tame, although it
follows if \(f\) is \(c r_2\)-tame for sufficiently large universal constant \(c\).}
If \(f':M\to\RR^d\) is another tame immersion with
\(f'|_{N_1}=f|_{N_1}\) and the same outward normal along \(N_1\), then the collar determined by \(f'\) has the same
embedded image \(B\).  

Recall $M^{+r}$ denotes the manifold obtained by adjoining an outer $r$-collar to $M$. However, in the fixed-boundary setting, when $M$ has a distinguished fixed boundary component $N_1$, we use $M^{+r}$ to denote the manifold obtained by adjoining a collar only along the non-fixed component $N_2$, and $f_{+r}$ denotes the corresponding extension of $f$.  Let $A$ be an
abstract subsystem, and choose $\tau \in \St_{AB}$ which admits an extension to $A M^{+2r_1}$ that is $r_1$-consistent on $M^{+2r_1}$. 
 Then we can define the information convex set \textit{relative}
to the fixed state $\tau$ on the boundary collar and ancilla by
\begin{equation}\label{eq:IC-rel}
\ICrel(M,f',\tau) = \bigl\{ \rho|_{AM} \in \St_{AM,f'} \,: \rho \in
\St_{A M^{+2r_1},(f')_{+2r_1}} \text{ s.t. } 
\rho|_{AB}=\tau \text{ and } \rho|_{M^{+2r_1}} \text{is $r_1$-consistent} \bigr\}.
\end{equation}
That is, $\ICrel(M,f',\tau)$ consists of states on $AM$ that match $\tau$ on $AB$ and admit an $r_1$-consistent extension on $A M^{+2r_1}$ in this fixed-boundary sense. 

We obtain a variant of Theorem~\ref{thm:isomorphism} in this setting.
\begin{Theorem}[Fixed-boundary isomorphism theorem] \label{thm:isomorphism-fixed-boundary} 
For relative information convex sets $\ICrel(M,f',\tau)$ as above, the
conclusion of Theorem~\ref{thm:isomorphism} remains valid under tame
deformations $F : M \times [0,1] \to \RR^d$ such that
\[
f_t|_{N_1} = f_0|_{N_1}
\qquad\text{for all } t\in[0,1].
\]
\end{Theorem}
\noindent 
The proof of Theorem~\ref{thm:isomorphism} directly adapts to
Theorem~\ref{thm:isomorphism-fixed-boundary}.  The key is that
Proposition~\ref{prop:macro-extend} and
Proposition~\ref{prop:collar-recovery} are already formulated to work with a
single boundary component.  Proposition~\ref{prop:collar-recovery} allows an
ancilla $A$, and so does Proposition~\ref{prop:macro-extend}, since its channel
acts only on $B$.  The definition of
$\ICrel(M,f',\tau)$ supplies the compatible $r_1$-consistent extensions needed
to apply them.
As an analog of Proposition~\ref{prop:ball-unique}, we have:
\begin{Corollary}[Uniqueness for annuli with fixed collar state] \label{cor:fixed-boundary-annulus}
In the setting of Theorem~\ref{thm:isomorphism-fixed-boundary}, suppose $M$
is an annulus and $f : M \to \RR^d$ is tamely deformable, fixing $N_1$, to a
tame immersion $g:M\to\RR^d$ with $g(M)=B'$, where $B'\supseteq B$ is the embedded $2r_2$-collar of $N_1$. Then $\ICrel(M,f,\tau)$ is a singleton.
\end{Corollary}
\noindent 
Indeed, by Theorem~\ref{thm:isomorphism-fixed-boundary}, $\ICrel(M,f,\tau)$ is isomorphic to $\ICrel(M,g,\tau)$.  The restriction to $AM$ of the assumed $r_1$-consistent extension of $\tau$ belongs to $\ICrel(M,f,\tau)$, so this set is nonempty, and hence so is $\ICrel(M,g,\tau)$.
Moreover, every $\rho\in\ICrel(M,g,\tau)$ satisfies \(\rho|_{AB}=\tau\), and its $r_1$-consistent extension is determined on $B'\setminus B$ by $\tau$: applying Proposition~\ref{prop:collar-recovery} to the parallel hypersurfaces of $N_1$ at successive depths $r_2, r_2+r_1,\dots$ recovers $\rho$ on the next strip of width $r_1$ from the previous one, by channels that do not depend on $\rho$. Hence $\rho$ is determined by $\tau$ on all of $AB'=AM$. 
Therefore $\ICrel(M,g,\tau)$ has a single element, and so does $\ICrel(M,f,\tau)$.

\section{Extendibility at larger scales}
\label{sec:extension}

In Section~\ref{sec:isomorphism}, our starting point was extendible local data, without assuming the existence of a consistent global state.  Below we show a unique such state exists on $\RR^d$ (Proposition~\ref{prop:global-state}).  The purpose of this section is then to demonstrate various properties of global states that satisfy local extendibility:
\begin{enumerate}
\item Local extendibility at some scale implies extendibility at all sufficiently larger scales, for standard micro-geometries (Theorem~\ref{thm:extend-A1'}) and for more general smooth shapes (Theorem~\ref{thm:large-smooth-extendibility}).
\item After a local circuit, an extendible state remains extendible at sufficiently larger scales (Theorem~\ref{thm:circuit-invariance-global}). Thus locally extendible states form a circuit-invariant class (Corollary~\ref{cor:circ-invariant-class}).
\item For an extendible state, local purifiability for a single patch implies local purifiability everywhere at larger scales (Theorem~\ref{thm:extend-A0}).
\item An extendible state admits a decomposition on every finite ball, as a mixture of states that are also locally purifiable and mutually locally orthogonal (Theorem~\ref{thm:decomposition-extendible}).
\end{enumerate}

Recall we work with extendible local data at scale $(r_0,r_1)$ with $r_0 \ll r_1$, meaning $r_0 \le c_1 r_1$ for a constant $c_1>0$ depending only on spatial dimension. We refer to the standard configurations $\Lambda(r_0,r_1)$ of Definition~\ref{def:Lambda-micro}. As before, we assume local data covering all of $\RR^d$, though the constructions below are purely local. In particular, all results could be reformulated for local data covering only a sufficiently large domain $\Omega \subset \RR^d$, as in Remark~\ref{rmk:finite-domains}, and the conclusions would then hold on the interior of $\Omega$.  Nonetheless, we retain the formulation over all of $\RR^d$ because it is often more convenient.

In our setting, it is easy to discuss states on infinite systems.  Recall that we work with a spin system on $\RR^d$, namely a family $(\calH_x : x \in J)$ where $J \subseteq \RR^d$ has finite intersection with any ball. In general, we define a state $\sigma$ on this infinite spin system as an assignment of mutually consistent states $\sigma|_X$ to every finite region $X$; equivalently, $\sigma$ is a positive normalized linear functional on the algebra of finitely supported operators.

\begin{Proposition}[Global states from local data] \label{prop:global-state}
Given extendible local data at scales $(r_0,r_1)$, there exists a unique global state $\sigma$ on $\RR^d$ consistent with the reference patches.
\end{Proposition}
\noindent Note, however, that $\sigma$ need not be consistent with the data patches near their boundaries.
\begin{proof}
For every finite region $X \subset \RR^d$, choose a geometric ball $B \supset X$, let $\rho_B$ denote the unique state in $\IC(B)$ from Proposition~\ref{prop:ball-unique}, and define \(\rho_X := \rho_B|_X\).
The uniqueness of $\IC(B)$ on balls implies that $\rho_X$ is independent of the choice of $B$ and that the family $\{ \rho_X \colon X \subset \RR^d\}$ is consistent under restriction, hence determines a global state $\sigma$.  The same uniqueness also implies that $\sigma$ is consistent with the reference patches and that $\sigma$ is the unique such state.
\end{proof}

From here on, $\sigma$ denotes the global state of Proposition~\ref{prop:global-state}, which is locally extendible on standard configurations $\Lambda(r_0,r_1)$.  Equivalently, the corresponding reference and data patches form extendible local data.  All information convex sets below are understood with respect to these reference patches.

\newcommand{\extendAlargerScalesStatement}{For every ratio
\(\alpha\in(0,c_1],\) 
there exists a scale factor \(s^{(\alpha)}>0\) such that when $\sigma$ is extendible for standard configurations \(\Lambda(r_0,r_1)\), it is also extendible for configurations \(\Lambda(r_0',r_1')\) when  \[ r_1' \ge s^{(\alpha)} r_1 , \qquad r_0' \ge r_1' \alpha.\]}  

We now describe how extendibility of the global state $\sigma$ at small scales implies extendibility at larger scales.  Recall the universal constant $c_1>0$ such that $r_0 < c_1 r_1$  suffices for valid extendible local data.

\begin{Theorem}[Extendibility on larger scales] \label{thm:extend-A1'}
\extendAlargerScalesStatement
\end{Theorem}
\noindent The \hyperref[proof:extend-A1p]{proof} is postponed until after the geometric preparations below.

\begin{Remark}[Alternative micro-geometries]
Although Theorem~\ref{thm:extend-A1'} assumes extendibility for standard
micro-geometries $\Lambda(r_0,r_1)$, the argument only uses the conclusion of the fixed-boundary isomorphism theorem~\ref{thm:isomorphism-fixed-boundary}.  This shows that the standard
micro-geometries are not an unduly restrictive choice: any alternative class of
micro-geometries rich enough to establish Theorem~\ref{thm:isomorphism-fixed-boundary} would likewise yield
extendibility on sufficiently scaled standard micro-geometries.
\end{Remark}

Note that Theorem~\ref{thm:extend-A1'} demonstrates extendibility at larger scales with ratio \(\alpha'=r_0'/r_1'\) that can be taken smaller  than the original ratio $\alpha=r_0/r_1$.  This will be a crucial feature for Theorem~\ref{thm:circuit-invariance-global}. 

Given extension maps $R_{BE \to BC}$ for regions $(B,C,D,E)$ at scale $(r_0,r_1)$, Theorem~\ref{thm:extend-A1'} constructs recovery maps $R_{B'E' \to B'C'}$ for regions $(B',C',D',E')$ at scale $(r_0',r_1')$.  A straightforward strategy would be to compose $R_{B'E' \to B'C'}$ directly from a circuit of channels $R_{BE \to BC}$ on smaller regions. However, the depth of this circuit scales like $r'_1/r_1$, and $E$ has scale $r_0$, so the accumulated loss of small regions $E$ results in the loss of a region $E'$ of scale $r'_0 \gtrsim r_0 r'_1/r_1$. Then $r'_0 / r'_1 \gtrsim r_0/r_1$, which is unsuitable: we would have reproduced extendibility at a larger scale but with worse ratio $r'_0/r'_1$.

Rather than directly construct  $R_{B'E' \to B'C'}$, we show extendibility on $(B',C',D',E')$ by showing the existence of an immersed state there, as in Theorem~\ref{thm:immersion1}.  This immersed state is created using the tools of Section~\ref{sec:macro} and is therefore a macro-geometry of scale $r_2$ for some $r_2 \gg r_1$. 
 This only requires $r'_0 \gg r_2 \gg r_1$, so the lower bound on suitable $r'_0$ no longer grows linearly with $r_1'$.  Thus for each fixed thinness ratio \(\alpha=r_0'/r_1'\), one may take \(r_0'\) sufficiently large and obtain extendibility on scale \((r_0',r_1')\) with \(r_1'=r_0'/\alpha\).

The above result only demonstrates extendibility on larger scaled versions of the standard configurations, or scaled ``micro-geometries.''  Such $(B,C,D,E)$ are defined by the recipe in Section~\ref{sec:micro} built on a sphere embedding, and they are geometrically constrained: the hemispheres defining $BE$ and $DE$ are somewhat flat, due to the Lipschitz condition in Eq.~\eqref{fmp_Lip}.  (That is, the ball $BCDE$ is constrained to be slightly pancake-shaped.)  Meanwhile, we can also demonstrate extendibility for configurations $(B,C,D,E)$ with the same topology but allowing more general geometrical shapes, as long as they are sufficiently smooth.  Such a result appears in Theorem~\ref{thm:large-smooth-extendibility}.

\begin{figure}[t]
    \centering
    \input{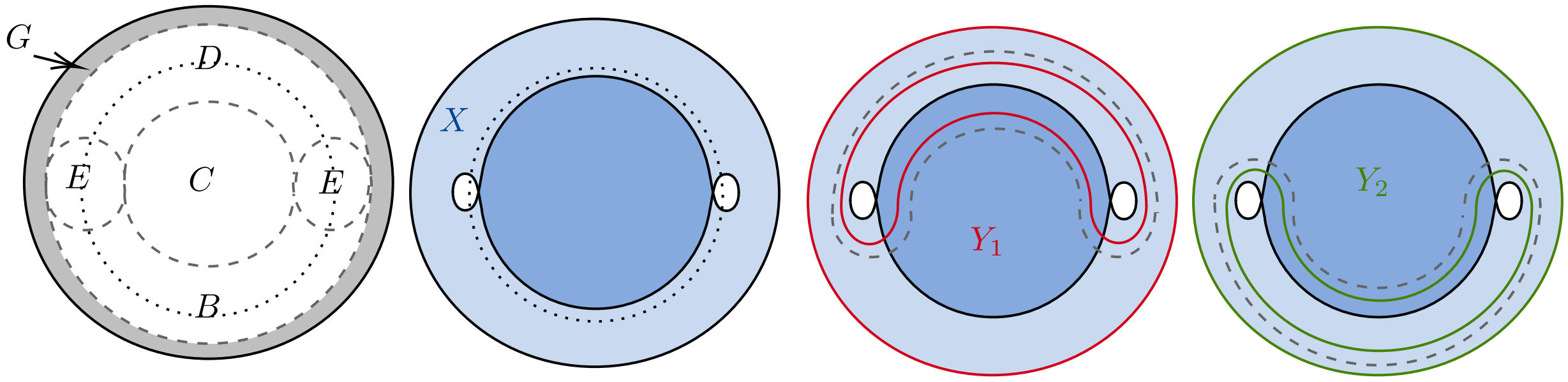}
    \caption{A typical setup for Lemma~\ref{lem:extendibility-from-annulus} in \(\RR^2\).  The dashed curves mark regions \((B,C,D,E)\) associated to an embedded sphere (dotted curve), with \(G\) an embedded annulus and \(X_0=GBCDE\) an embedded ball.  The blue region is an immersed annulus \(X\) containing two overlapping lobes (darker blue), which contain two overlapping copies \(C_1\) and \(C_2\) of \(C\), attached to \(B\) and \(D\), respectively.  The red and green-outlined regions are embedded annuli \(Y_1\) and \(Y_2\).  As embedded regions, we arrange \(GBC \subset Y_1\) and \(GDC \subset Y_2\).  Inside the immersed annulus \(X\), the embedded regions \(GBC\) and \(GDC\) are realized as \(GBC_1\) and \(GDC_2\).}
\label{fig:smooth-immersed-2}
\end{figure}

We begin by sketching Lemma~\ref{lem:extendibility-from-annulus}; see Fig.~\ref{fig:smooth-immersed-2}.   Our goal is to prove extendibility for regions $(B,C,D,E)$ associated to an embedded sphere, in the sense of Definition~\ref{def:BCDE-region-recipe}.  Let $G$ be an annulus%
\footnote{We use ``annulus'' and ``spherical shell'' synonymously to indicate a region diffeomorphic to $S^{d-1} \times [0,1]$. Likewise, from here on a ball is a region diffeomorphic to the closed disk $D^d$.} surrounding the ball $BCDE$, chosen so that \(X_0=GBCDE\) is a tame ball.  Let \(\bar{\sigma}\in\St_{AGBCDE}\) be a pure state extending \(\sigma|_{GBCDE}\) using ancilla $A$.  Treating \(\bar{\sigma}\) as a purification of \(\sigma|_{BCDE}\) with ancilla $AG$, Theorem~\ref{thm:immersion1} reduces extendibility of \(\sigma|_{BCDE}\) to the existence of a state \(\rho \in \St_{AGBC_1C_2D}\) such that \(\rho|_{AGBC_1}=\bar{\sigma}|_{AGBC}\) and \(\rho|_{AGDC_2}=\bar{\sigma}|_{AGDC}\), where $C_1$ and $C_2$ are abstract copies of $C$.

To argue for such a state $\rho$, let us momentarily ignore issues of tameness, and also arrange that $B$ and $D$ do not touch, say by enlarging $E$ slightly.  Then $GBD$ is an annulus.  Glue two copies $C_1$ and $C_2$ of $C$ to $GBD$, with $C_1$ glued along \(\partial B\cap\partial C\) and $C_2$ glued along \(\partial D\cap\partial C\).  The result is an immersed annulus \(\wt X=GBC_1C_2D\), with overlapping flaps $C_1$ and $C_2$ attached to $B$ and $D$, respectively.  Since \(\wt X\) is regular homotopic to $G$ through a homotopy fixing the outer boundary, we hope to obtain a state on \(A\wt X\) by acting on \(\bar{\sigma}|_{AG}\) with a sequence of recovery maps and erasures, to grow and reshape the inner boundary.  This is achieved by the fixed-boundary isomorphism theorem~\ref{thm:isomorphism-fixed-boundary}.  Because $GBC_1$ and $GDC_2$ are embedded annuli inside \(\wt X\) with shared outer boundary, the same theorem should yield $\rho|_{AGBC_1}=\bar{\sigma}|_{AGBC}$ and $\rho|_{AGDC_2}=\bar{\sigma}|_{AGDC}$, implying extendibility by Theorem~\ref{thm:immersion1}.

To legitimize the application of Theorem~\ref{thm:isomorphism-fixed-boundary}, we must construct tame annuli to serve as proxies for $\wt X$, $GBC_1$, and $GDC_2$.  The existence of such tame regions is essentially the hypothesis of Lemma~\ref{lem:extendibility-from-annulus} encapsulated in Definition~\ref{def:good-triple}, which allows $(B,C,D,E)$ themselves to be non-smooth.  In particular, we ask for a tame immersed annulus $X$ and tame embedded annuli $Y_1,Y_2$, all sharing outer boundary $\partial X_0$, such that \(X \supset \wt X\), \(Y_1 \supset GBC\), and \(Y_2 \supset GDC\).

\begin{Definition}\label{def:good-triple}
Let $X_0=GBCDE\subset\RR^d$ be an embedded ball, decomposed into disjoint embedded subregions $G,B,C,D,E$. Let $W\subset X_0$ be the inner $2r_2$-collar of $\partial X_0$, and assume $W\subset G$. We consider triples $(X,Y_1,Y_2)$ of annuli with outer boundary $\partial X_0$, where $X$ is immersed and $Y_1,Y_2\subseteq X_0$ are embedded. We say that such a triple is \emph{good for} $(B,C,D,E)$ if the following conditions are met.
\begin{enumerate}
\item \emph{Immersed annulus condition}: $X$ is a tame immersed annulus $f_X\colon M\to\RR^d$ with a distinguished boundary component $N\subset\partial M$ such that $f_X|_N$ is an embedding onto $\partial X_0$.
\item \emph{Embedded annulus condition}: $Y_i=f_X(M_i)$ for a subannulus $M_i\subset M$ containing $N$ on which $f_X$ is an embedding, with $GBC\subset Y_1$, $GDC\subset Y_2$, and $M_i^{+2r_1}\subseteq M^{+2r_1}$ for the thickening along the inner boundary. The preimages of $C$ in $M_1$ and in $M_2$ are disjoint, while those of $G$ coincide.
\item \emph{Regularity condition}: $W,Y_1,Y_2$ have tame boundary and embedded outer $2r_1$-collars, and $X,Y_1,Y_2$ are tamely deformable to $W$, fixing $\partial X_0$.
\end{enumerate}
\end{Definition}

\begin{Lemma}[Extendibility on larger regions using immersions]
\label{lem:extendibility-from-annulus}
Let $\sigma$ be a global state that is extendible on standard configurations $\Lambda(r_0,r_1)$, and let $(X,Y_1,Y_2)$ be good for $(B,C,D,E)$, with $X_0=GBCDE$ as in Definition~\ref{def:good-triple}. Then $\sigma|_{X_0}$ is $(B,C,D,E)$-extendible.
\end{Lemma}

\begin{proof}
    Let $\overline{\sigma} \in \St_{\bar{A}BCDE}$ be a purification of $\sigma|_{BCDE}$, where $\bar{A}$ is a purifying system. We aim to show the existence of some state $\rho \in \St_{\bar{A}BC_1C_2D}$ satisfying the following:
    \begin{equation}
\rho|_{\bar{A}BC_1} = \overline{\sigma}|_{\bar{A}BC}, \qquad \rho|_{\bar{A}C_2D} = \overline{\sigma}|_{\bar{A}CD}.
\label{eq:extend_large_region_sufficient_condition}
    \end{equation}
    Then Theorem~\ref{thm:immersion1} establishes the claim. 

    We first define an information convex set relative to some fixed boundary state. Let $\rho_0=\sigma|_{X_0} \in \IC(X_0)$ and choose a purification $\bar{\rho}_0 \in \St_{AX_0}$. Note this state can be also viewed as a purification of $\sigma|_{BCDE}$ discussed above, with the purifying system $\bar{A}$ identified with $AG$. Let $\tau = \bar{\rho}_0 |_{AW'}$, where $W'\subset W$ is the inner $r_2$ collar of $\partial X_0$.
Let \(W^{+2r_1} \subset X_0 \) be a thickening of $W$ only along the inner (non-fixed) boundary of \(W\). The restriction \(\bar{\rho}_0|_{A W^{+2r_1}}\) is then an $r_1$-consistent extension of $\tau$. Thus we can define an information convex set relative to the fixed boundary $\partial X_0$ and state $\tau$, in the sense of Eq.~\eqref{eq:IC-rel}.

Let $M_1, M_2 \subset M$ be the subannuli in Definition~\ref{def:good-triple}. Since \(f_X|_{M_i}\) is an embedding onto \(Y_i\), we can view states on \(Y_i\) as states on \(M_i\). In particular, we use the following short-hand notation.
\begin{equation}
\ICrel(X):=\ICrel(M,f_X,\tau),
\qquad
\ICrel(Y_i):=\ICrel(M_i,f_X|_{M_i},\tau).
\end{equation}
Let $g\colon M\to\RR^d$ be the endpoint of the deformation of $X$ to $W$. Then \(\bar{\rho}_0|_{AW}\in\ICrel(M,g,\tau)\), so $\ICrel(X)$ is nonempty by Theorem~\ref{thm:isomorphism-fixed-boundary}. By Corollary~\ref{cor:fixed-boundary-annulus}, the information convex sets above are all singletons. We will denote the unique state in $\ICrel(X)$ as \(\rho_X \in \ICrel(X)\).

For each \(i=1,2\), we compare two elements of \(\ICrel(Y_i)\).  First, note \(\rho_X|_{A M_i} \in \ICrel(Y_i)\) because any witness for \(\rho_X\in\ICrel(X)\) restricts to \(M_i\).  Second, the inner boundary of $Y_i$ lies in the interior of $X_0$, so its embedded outer $2r_1$-collar, which we denote $Y_i^{+2r_1}$, is contained in $X_0$. In particular, the restriction \(\bar{\rho}_0|_{A Y_i^{+2r_1}}\) supplies a compatible \(r_1\)-consistent extension of \(\bar{\rho}_0|_{A Y_i}\), and its restriction to \(AW'\) is \(\tau\).  Hence \(\bar{\rho}_0|_{A Y_i}\in \ICrel(Y_i)\).  Since \(\ICrel(Y_i)\) is a singleton, these two states agree:
\begin{equation} \label{eq:AM-AYi}
\rho_X|_{A M_i}=\bar{\rho}_0|_{A Y_i}.
\end{equation}

Define $Z_1:=(f_X|_{M_1})^{-1}(GBC)$ and $Z_2:=(f_X|_{M_2})^{-1}(GDC)$. We denote these regions as \(Z_1=GBC_1\) and \(Z_2=GDC_2\), where \(C_1\subset M_1\) and \(C_2\subset M_2\) are the two disjoint subsets mapped onto \(C\) from the embedded annulus condition of Definition~\ref{def:good-triple}. Using this notation and restricting Eq.~\eqref{eq:AM-AYi} above to \(Z_1\) and \(Z_2\), we have
\begin{equation}
\rho_X|_{AGBC_1}=\bar{\rho}_0|_{AGBC},
\qquad
\rho_X|_{AGDC_2}=\bar{\rho}_0|_{AGDC}.
\end{equation}
The restriction of $\rho_X$ to $AGBC_1C_2D$ satisfies Eq.~\eqref{eq:extend_large_region_sufficient_condition}, using $\bar{A}=AG$. Thus the claim follows.
\end{proof}

\begin{figure}[t]
\centering
\begin{tabular}{cc}
\begin{tikzpicture}[line width=0.75pt, scale=0.62]
\pgfmathsetmacro{\unitR}{2.55}
\pgfmathsetmacro{\alphaDraw}{0.285}
\pgfmathsetmacro{\tubeR}{\alphaDraw*\unitR}
\pgfmathsetmacro{\coreR}{\unitR-\tubeR}
\pgfmathsetmacro{\outerR}{\unitR+\tubeR}
\pgfmathsetmacro{\collarR}{\outerR+\tubeR}
\fill[fill={rgb, 255:red, 191; green, 191; blue, 191}] (0,0) circle (\collarR);
\fill[fill=tikzbluelight] (0,0) circle (\outerR);
\fill[fill=tikzblue] (0,0) circle (\coreR);
\draw[] (0,0) circle (\collarR);
\draw[dashed] (0,0) circle (\outerR);
\draw[dashed] (0,0) circle (\coreR);
\draw[dashed] (\unitR,0) circle (\tubeR);
\draw[dashed] (-\unitR,0) circle (\tubeR);
\draw[black!60, dash pattern=on 0.1pt off 4.6pt, line width=1.1pt] (0,0) circle (\unitR);
\node[] () at (0,0) {$C$};
\node[] () at (0,\unitR) {$D$};
\node[] () at (0,-\unitR) {$B$};
\node[] () at (\unitR,0) {$E$};
\node[] () at (-\unitR,0) {$E$};
\node[] () at (45:{\outerR+\tubeR/2}) {$G$};
\end{tikzpicture} & \scalebox{0.62}{\begin{tikzpicture}[line cap=round, line join=round]
\pgfmathsetmacro{\unitR}{2.55}
\pgfmathsetmacro{\alphaDraw}{0.48}
\pgfmathsetmacro{\tubeR}{\alphaDraw*\unitR}
\pgfmathsetmacro{\coreR}{\unitR-\tubeR}
\pgfmathsetmacro{\outerR}{\unitR+\tubeR}

\pgfmathsetmacro{\gammaCapCenter}{\unitR}
\pgfmathsetmacro{\gammaCapR}{\tubeR/4}
\pgfmathsetmacro{\gammaOuterX}{\gammaCapCenter+\gammaCapR}
\pgfmathsetmacro{\gammaMidR}{\gammaCapCenter-\gammaCapR}

\pgfmathsetmacro{\sigmaCapCenter}{\unitR}
\pgfmathsetmacro{\sigmaCapR}{\tubeR/2}
\pgfmathsetmacro{\sigmaOuterR}{\sigmaCapCenter+\sigmaCapR}
\pgfmathsetmacro{\sigmaInnerR}{\sigmaCapCenter-\sigmaCapR}

\def\upperTubePath{%
  (\outerR,0)
  arc[start angle=0,end angle=-180,radius=\tubeR]
  arc[start angle=0,end angle=180,radius=\coreR]
  arc[start angle=0,end angle=-180,radius=\tubeR]
  arc[start angle=180,end angle=0,radius=\outerR]
  -- cycle}

\def\gammaDraftPath{%
  (-\gammaOuterX,0)
  arc[start angle=180,end angle=0,radius=\gammaCapR]
  arc[start angle=180,end angle=360,radius=\gammaMidR]
  arc[start angle=180,end angle=0,radius=\gammaCapR]
  arc[start angle=0,end angle=-180,radius=\gammaCapR]
  arc[start angle=0,end angle=180,radius=\gammaMidR]
  arc[start angle=0,end angle=-180,radius=\gammaCapR]
  -- cycle}

\def\sigmaDraftPath{%
  (\sigmaOuterR,0)
  arc[start angle=0,end angle=-180,radius=\sigmaCapR]
  arc[start angle=0,end angle=180,radius=\sigmaInnerR]
  arc[start angle=0,end angle=-180,radius=\sigmaCapR]
  arc[start angle=180,end angle=0,radius=\sigmaOuterR]
  -- cycle}

\path[use as bounding box] (-4.2,-4.2) rectangle (4.2,4.2);

\draw[black, line cap=butt, dash pattern=on 6pt off 3.6pt, line width=0.65pt] \upperTubePath;
\begin{scope}[yscale=-1]
  \draw[black, line cap=butt, dash pattern=on 6pt off 3.6pt, line width=0.65pt] \upperTubePath;
\end{scope}
\draw[black!60, dash pattern=on 0.1pt off 4.6pt, line width=1.1pt] (0,0) circle (\unitR);

\draw[blue!65!black, line width=1.35pt] \gammaDraftPath;
\draw[red!70!black, line width=1.25pt] \sigmaDraftPath;
\end{tikzpicture}}\\
(a) & (b)
\end{tabular}
\caption{\textbf{(a)} The configuration $(B,C,D,E)=\lambda(a,f)$ of the unit sphere (dotted) and the surrounding annulus $G=X_0\setminus BCDE$ in the plane; $E$ is the $a$-neighborhood of the equator, and $B$ and $D$ are the remaining parts of the $a$-neighborhoods of the two hemispheres. \textbf{(b)} Curves used in the proof of~Lemma~\ref{lem:standard-unit-witness}, drawn for the unit sphere, for which the signed distance $t$ to the sphere used in the proof is the distance to the center minus one. The underlying unit sphere is dotted, and the associated \((B,C,D,E)\) are outlined with dashed black curves.  The blue curve is the inner boundary of \(X\) and the red curve that of \(Y_1\), both before the smoothing at their joins; inside the blue curve, \(X\) covers \(C'\) twice, as \(C'_1\) and \(C'_2\), and omits \(E'=T_{a/4}\), for the fixed choice of \(a\) under discussion.  The omitted inner boundary of \(Y_2\) is obtained from that of \(Y_1\) by reflection across the horizontal axis.}
\label{fig:standard-unit-witness}
\end{figure}
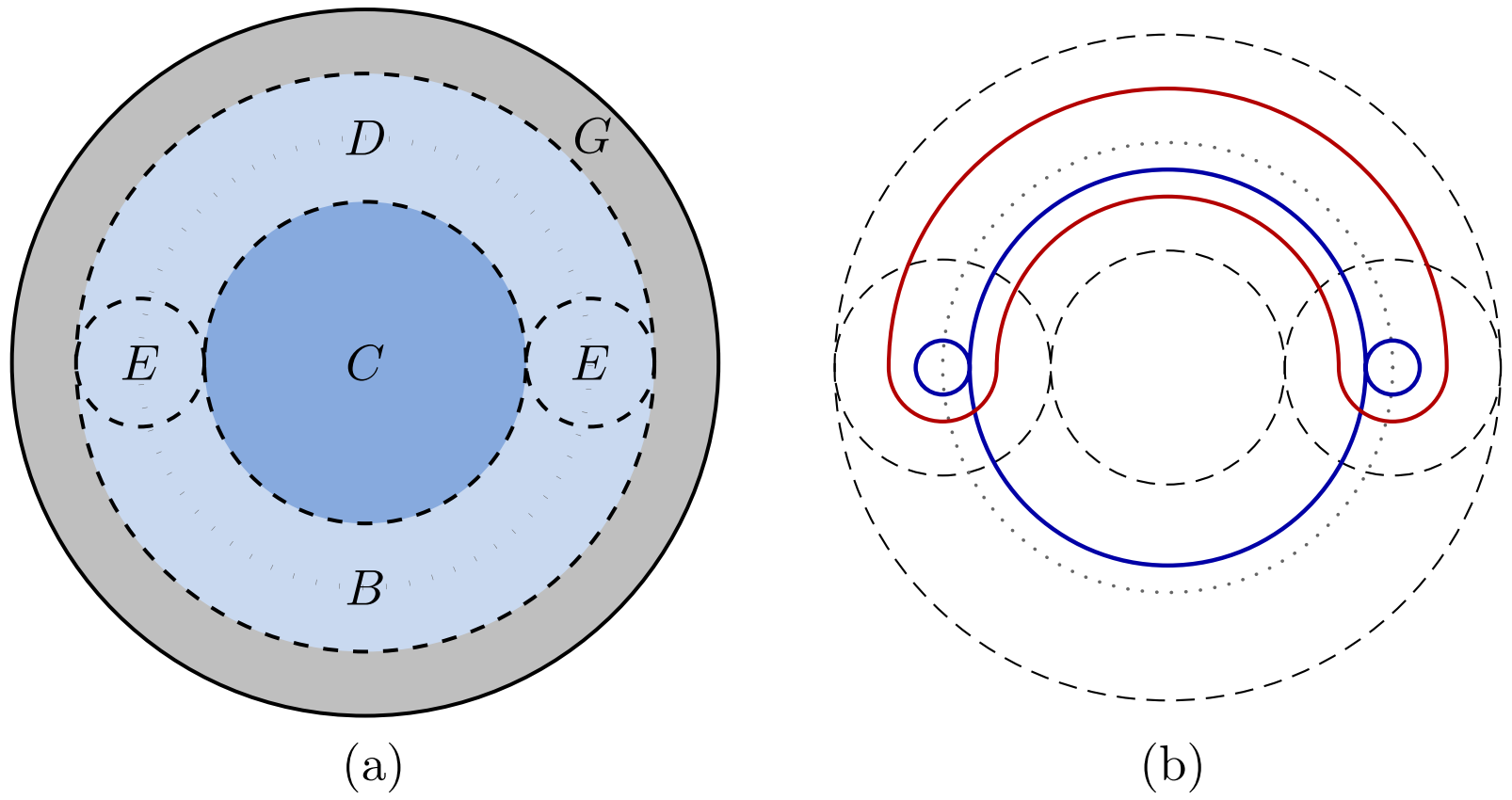

We now construct regions satisfying the hypotheses of Lemma~\ref{lem:extendibility-from-annulus} for a tame embedded sphere $f$ and $(B,C,D,E)=\lambda(a,f)$, when the thickness $a$ is small compared to the scale of tameness. Fig.~\ref{fig:standard-unit-witness} illustrates the construction for the unit sphere. We also need a notion of tameness for the equator of a sphere, which is a submanifold of codimension two. Let $\Gamma\subset\RR^d$ be a closed submanifold of codimension two. We say that $\Gamma$ is \emph{$r$-tame} if every point within distance $r$ of $\Gamma$ has a unique nearest point on $\Gamma$. This implies that $\Gamma$ has curvature at most $1/r$. The codimension-two submanifold we consider will be the equator of a sphere throughout this section.

\begin{Lemma}[Annular immersed regions for tame spheres]
\label{lem:standard-unit-witness}
There exist constants $\alpha_0,k>0$ depending only on $d$ such that the following holds. Let $F\colon D^d\to\RR^d$ be a smooth embedding whose boundary sphere $f:=F|_{\partial D^d}$ is an $r$-tame embedded hypersurface with $r$-tame equator $f(\partial S_-)$. Let $(B,C,D,E)=\lambda(a,f)$ with $r_2/k\le a\le\alpha_0r$. Then there exist an embedded ball $X_0=GBCDE$ and a triple $(X,Y_1,Y_2)$ good for $(B,C,D,E)$, in the sense of Definition~\ref{def:good-triple}.
\end{Lemma}

\begin{proof}

    Recall $(B,C,D,E)=\lambda(a,f)$ is
    \begin{equation}
    BE= \Ba_a(f(S_-)), \qquad DE= \Ba_a(f(S_+)), \qquad E= (BE)\cap (DE), \qquad C= \mathrm{Int}(S^{d-1},f)\setminus \Ba_a(f(S^{d-1})).
    \end{equation}
    We choose the triple as follows. Let $K:=F(D^d)$. To start with, we will choose $X_0$ to be a round ball of radius $R$ centered at a point of $K$ and containing $\Ba_a(BCDE)$. The radius $R$ will be chosen to be sufficiently large, with the precise condition deferred to the end of the proof. For $0<a'\le a$ write $V_\pm(a'):=X_0\setminus\Ba_{a'}(f(S_\pm))$, so that $V_\pm(a') \subseteq V_\pm(a'')$ for $0<a'' \le a'$. The embedded annuli are chosen as follows:
    \begin{equation}
    Y_1:= V_+(a/2) = X_0 \setminus \Ba_{a/2}(f(S_+)), \qquad Y_2:= V_-(a/2) = X_0\setminus \Ba_{a/2}(f(S_-)).
    \end{equation}
    For $X$, let $(B', C', D', E') := \lambda(a/4,f)$ and $G' := X_0 \setminus \Ba_{a/4}(K)$ so that $X_0=G'B'C'D'E'$. Since $\Ba_{a/4}(K)$ is a ball (see below), $G'$ is an annulus. Then $V_+(a/4)=G'B'C'$ and $V_-(a/4)=G'C'D'$. We set $X:= G'B'C_1'C_2'D'$, the immersed annulus obtained by gluing these two along $G'$, duplicating $C'$ into copies $C_1'$ and $C_2'$. Since $G'$ and $C'$ are separated by the shell $B'E'D'$, they are non-adjacent, so this gluing yields a manifold; the preimages of $C$ in the two halves lie in the disjoint copies $C_1'$, $C_2'$, while those of $G$ both lie in $G'$ and so coincide.

    Note this settles the setting of Definition~\ref{def:good-triple}. With $G:=X_0\setminus BCDE$, the region $X_0=GBCDE$ is an embedded ball. Since $a\ge2r_2$ (see below), the containment $X_0\supset\Ba_a(BCDE)$ already gives $W\subset G$ (Definition~\ref{def:good-triple}). It remains to show that $\Ba_{a/2}(f(S_\mp))$ are balls contained in its interior. Let $0<a'\le a$, so that $a'\le r/2$ since $a\le\alpha_0r$ with $\alpha_0\le1/2$. Since $f$ and its equator are $r$-tame, $\Ba_{a'}(f(S_\pm))$ is a tubular neighborhood of the disk $f(S_\pm)$, hence a ball. Taking $a'=a/2$, it is contained in the interior of $X_0$, since $\Ba_{a/2}(f(S_\mp))\subset\Ba_a(BCDE)\subset X_0$. A smoothly embedded copy of $D^d$ in the interior of a connected $d$-manifold can be deformed to any other by an ambient isotopy; in particular $\Ba_{a/2}(f(S_\mp))$ are ambient isotopic, within the interior of $X_0$, to round balls. Hence $Y_1$ and $Y_2$ are embedded annuli contained in $X_0$ with outer boundary $\partial X_0$, and it remains to prove the three conditions in Definition~\ref{def:good-triple}.

    Let $k>0$ be a sufficiently small constant depending only on $d$. In particular $k\le1/2$ so that $a\ge r_2/k\ge2r_2$. In particular, $a\gg r_1$. Tameness of $X$, $Y_1$ and $Y_2$ at scale $r_2$ will follow as soon as their boundaries have curvature $O(1/a)$ and they admit inner collars of depth proportional to $a$. We construct them with $C^{1,1}$ boundary of curvature $O(1/a)$ and inner collars of depth proportional to $a$, and then smooth the boundaries.

    Let $n$ be the outward unit normal of $f$ and $\hat{f}(x,t):=f(x)+t\,n(x)$ the normal map of Section~\ref{sec:macro}; since $f$ is $r$-tame this is a diffeomorphism of $S^{d-1}\times(-r,r)$ onto the points within $r$ of $f(S^{d-1})$, under which $t$ is the signed distance to $f(S^{d-1})$, positive outside $K$. Write $f_t:=\hat f(\cdot,t)$ for the parallel hypersurface at distance $t$, of curvature at most $2/r$ for $|t|\le r/2$. Write $T_w:=\Ba_w(f(\partial S_-))$ for the tube of radius $w$ about the equator; for $w\le r/2$ its boundary is smooth of curvature at most $2/w$ and meets $f_{-w}$ and $f_w$ tangentially along their equators $\hat f(\partial S_-\times\{\mp w\})$. Since $t$ is $1$-Lipschitz, $T_w$ lies in the region $|t|<w$. For $0<a'\le a$, $\Ba_{a'}(K)=K\cup\hat f(S^{d-1}\times[0,a'))$ is a ball.

    For $0<a'\le a$, the neighborhood $\Ba_{a'}(f(S_\pm))$ is the union of a normal neighborhood of the hemisphere and the tube of the same radius. A point within $a'$ of $f(S_+)$ has its nearest point on $f(S_+)$ in the interior or on the equator, so $\Ba_{a'}(f(S_+))=\hat f(S_+\times(-a',a'))\cup T_{a'}$. Its boundary is the union of the parts of $f_{a'}$ and $f_{-a'}$ over $S_+$ with the part of $\partial T_{a'}$ over $S_-$, the three joining tangentially along the equators of $f_{\pm a'}$; see Fig.~\ref{fig:standard-unit-witness}. The same holds with $S_+$ and $S_-$ exchanged, and the two neighborhoods intersect in $T_{a'}$, so $E'=T_{a/4}$.

    We first verify the immersed annulus condition. Let $M$ be the manifold obtained by gluing the copies $G'B'C'_1$ and $G'C'_2D'$ of $V_+(a/4)$ and $V_-(a/4)$ along $G'$. Let $f_X\colon M\to\RR^d$ be the immersion which is the inclusion on each copy, so that $X$ is the immersed region $(M,f_X)$. The copy $N\subset M$ of $\partial X_0$ is a boundary component of $M$, embedded by $f_X$ onto $\partial X_0$. Moreover, $M$ is an annulus: it is obtained from the annulus $G'$ by attaching in turn the balls $B'C'_1$ and $C'_2D'$ along disks in its inner boundary sphere $f_{a/4}$, which leaves its topology unchanged. Indeed, the equator $\hat f(\partial S_-\times\{a/4\})$ of $f_{a/4}=\partial\Ba_{a/4}(K)$ divides it into the two disks $Q_\mp:=\hat f(S_\mp\times\{a/4\})$, along which $B'C'_1$ and $C'_2D'$ meet $G'$. The balls $D'E'=\Ba_{a/4}(f(S_+))$ and $B'E'=\Ba_{a/4}(f(S_-))$ have $C^1$ boundary and meet $f_{a/4}$ in the disks $Q_+$ and $Q_-$, so their complements $B'C'$ and $C'D'$ in the ball $\Ba_{a/4}(K)=B'C'D'E'$ are balls.

    It remains to show that $\partial X$ has curvature $O(1/a)$ and $X$ an inner collar of depth proportional to $a$, and likewise for $Y_1$ and $Y_2$. The inner boundary of $X$ consists of the closures of the two disks $\partial\Ba_{a/4}(f(S_\pm))\cap\Ba_{a/4}(K)$, glued along their common boundary, the equator of $f_{a/4}$. Each consists of a piece of $f_{-a/4}$ and a piece of $\partial T_{a/4}$, of curvature at most $2/r\le2/a$ and $8/a$ respectively, meeting tangentially along the equator of $f_{-a/4}$. Along that equator each disk is a piece of the tube boundary $\partial T_{a/4}$, one over $S_-$ and one over $S_+$, so the inner boundary of $X$ is smooth there. Hence $\partial X$ is $C^{1,1}$ with curvature $O(1/a)$. Its inner collar has depth at least $a/2$: since every point within $r$ of $f(S_\pm)$ has a unique nearest point on it, the normal rays from $\partial\Ba_{a/4}(f(S_\pm))$ pointing away from $f(S_\pm)$ do not cross within distance $r-a/4\ge a/2$, and they remain in $X$. The same holds for $Y_1$ and $Y_2$, whose inner boundaries $\partial\Ba_{a/2}(f(S_\pm))$ consist of pieces of $f_{\pm a/2}$ and $\partial T_{a/2}$, of curvature at most $2/a$ and $4/a$, meeting tangentially along the equators of $f_{\pm a/2}$.

    We now verify the embedded annulus condition. By the symmetry exchanging $S_+$ and $S_-$, it suffices to treat $Y_1$ and $M_1$ only. Since $G$, $C$ and $B$ all lie at distance at least $a$ from $f(S_+)$ and $X_0\setminus Y_1=\Ba_{a/2}(f(S_+))$, $GBC\subset Y_1$. Moreover, $Y_1$ is the image of an embedded annulus $M_1\subset M$. Since $f_X$ is injective on the copy $G'B'C'_1$ of $V_+(a/4)$ and $Y_1=V_+(a/2)\subset V_+(a/4)$, the set $M_1:=f_X^{-1}(Y_1)\cap G'B'C'_1$ is a subannulus of $M$ containing $N$ with $Y_1=f_X(M_1)$. In particular, $f_X$ embeds $M_1$. Moreover, $M_1^{+2r_1}\subseteq M^{+2r_1}$. Indeed, the inner boundary of $Y_1$ lies at distance $a/2$ from $f(S_+)$, while $f_X(G'B'C'_1)=V_+(a/4)$ extends to distance $a/4$. Since $a\gg r_1$, the $2r_1$-thickening of $M_1$ stays in $G'B'C'_1$, where $f_X$ is injective; hence $M_1^{+2r_1}\subseteq M^{+2r_1}$. The conditions on the preimages of $C$ and $G$ follow from the construction.

    We can now smooth the boundaries. The inner boundaries of $X$, $Y_1$ and $Y_2$ are $C^{1,1}$ with curvature $O(1/a)$, their pieces meeting tangentially along the equators of $f_{-a/4}$ and $f_{\pm a/2}$. We may smooth each around the equators, keeping the curvature $O(1/a)$ (Section~\ref{sec:macro}). Taking the smoothing $C^1$-small and within a small fraction of $a$ of the equators, the normal maps stay injective to a depth proportional to $a$, and each boundary moves by a small fraction of $a$. The containments $GBC\subset Y_1$ and $GDC\subset Y_2$ start with a margin of $a/2$ and therefore survive. Also, the thickenings of $M_1$ and $M_2$ of depth $a/4$ retain a depth proportional to $a$, exceeding $2r_1$ as $a\gg r_1$. The attachments defining $M$, hence its topology, remain unchanged. The curvature radii and collar depths obtained are proportional to $a$ with constants depending only on $d$. So for $k$ small enough, $X$, $Y_1$ and $Y_2$ are $r_2$-tame.

    Finally, we verify the regularity condition. $W$ is the inner $2r_2$-collar of the round sphere $\partial X_0$, hence $r_2$-tame, so $W$, $Y_1$ and $Y_2$ have tame boundary. Each has an embedded outer $2r_1$-collar: outside $\partial X_0$ and inside the hole of $W$ the normal map is injective to depth $r_2\ge2r_1$, while into the hole of $Y_i$ it is injective to a depth proportional to $a$, since $\partial\Ba_{a/2}(f(S_\pm))$ has focal distance $a/2$ into the ball and the smoothing keeps curvature $O(1/a)$; this exceeds $2r_1$ as $r_1\ll r_2\le a/2$. It remains to show that $X$, $Y_1$ and $Y_2$ are tamely deformable to $W$ while fixing $\partial X_0$. Each of $Y_1$, $Y_2$ and $G'$ is $X_0$ with an embedded ball removed from its interior, and so is $W$, the removed ball being concentric. For each, an ambient isotopy of $X_0$ confined to the concentric ball of radius $R_0:=2\,\mathrm{diam}(K)$ deforms the removed ball to the concentric round ball of radius $r_2$; enlarging this ball to radius $R-2r_2$ then reaches $W$ through round shells, which are $r_2$-tame throughout. For $X$, first retract $B'C'_1$ and $D'C'_2$ into collars of the disks $Q_-$ and $Q_+$ along which they are attached, which deforms $X$ to $G'$ within $\Ba_{a/2}(K)$, inside the concentric ball of radius $R_0$; then continue with the deformation of $G'$. The confined stages of the deformation move points only within the concentric ball of radius $R_0$ and do not depend on $R$. For $R$ large enough, we can make them tame while fixing $\partial X_0$ (Remark~\ref{rmk:scale-homotopy}); we fix such an $R\ge R_0$. The inward normal segments of length $r$ lie in $K$, so $\mathrm{diam}(K)\ge r\ge2a$, and $R\ge R_0$ then gives $X_0\supset\Ba_a(BCDE)=\Ba_{2a}(K)$.
\end{proof}

The next result shows extendibility for regions
\((B,C,D,E)=\lambda(a,f)\) associated to an embedded sphere whenever \(f\) is
sufficiently large and smooth.  

\begin{Theorem}[Extendibility for large smooth regions]
\label{thm:large-smooth-extendibility}
There are constants $a_0,s_0>0$, depending only on $d$, such that the following holds. Let $F\colon D^d\to\RR^d$ be a smooth embedding whose boundary sphere $f:=F|_{\partial D^d}$ is an $r$-tame embedded hypersurface with $r$-tame equator $f(\partial S_-)$, and let $(B,C,D,E)=\lambda(a,f)$ with $a\ge a_0r_2$ and $r\ge s_0a$. Then $\sigma$ is $(B,C,D,E)$-extendible.
\end{Theorem}
\begin{proof}
By Proposition~\ref{prop:g_mono}, $\lambda(a',f)\cge\lambda(a,f)$ for $a'\le a$, so it suffices to treat $a=a_0r_2$ with $a_0:=1/k$, where $k$ is the constant of Lemma~\ref{lem:standard-unit-witness}. Then $ka=r_2$, and $a\le\alpha_0r$ if $s_0\ge1/\alpha_0$. Lemma~\ref{lem:standard-unit-witness} then provides $X_0$ and a triple $(X,Y_1,Y_2)$ good for $(B,C,D,E)$, and one can apply Lemma~\ref{lem:extendibility-from-annulus}.
\end{proof}

\noindent To apply the above to scaled micro-geometries, it will be helpful to replace them with smoothed versions. Earlier in Section~\ref{sec:micro} we described how to build micro-geometries using smooth hemisphere embeddings, but here we moreover use a smooth embedding of the full sphere.

\begin{Proposition}[Smooth replacements for micro-geometries]
\label{prop:smooth-replacement}
There is a constant $c_{\mathrm{sm}}>0$, depending only on $d$, such that the following holds. Let $(B,C,D,E)=\lambda(r_0,f)$ with $f\in\Delta(r_1)$ and $r_0\ll r_1$. Then there is a smooth embedding $\tilde F\colon D^d\to\RR^d$, with $\tilde f:=\tilde F|_{\partial D^d}$, such that $\tilde f$ is a $(c_{\mathrm{sm}}r_0)$-tame embedded hypersurface with $(c_{\mathrm{sm}}r_0)$-tame equator $\tilde f(\partial S_-)$, and for every $a\in(0,c_{\mathrm{sm}}r_0]$ the configuration $(\tilde B,\tilde C,\tilde D,\tilde E):=\lambda(a,\tilde f)$ satisfies $(\tilde B,\tilde C,\tilde D,\tilde E)\cge(B,C,D,E)$, with $\tilde C$, $\tilde B\tilde C$, $\tilde C\tilde D$, and $\tilde B\tilde C\tilde D\tilde E$ embedded topological balls.
\end{Proposition}
\begin{proof}

We use the coordinates of Section~\ref{sec:micro}, rescaled to $r=r_1/6=1$, so that $r_0\ll1$. After the isometry $\wt v^{-1}$ of Eq.~\eqref{gmp_def}, the two hemispheres are graphs,
\[
f(S_\mp)=\{(z,h_\mp(z)):\,z\in\overline Z\}\subset\RR^{d-1}\oplus\RR,
\]
where $Z=g_\mp^\parallel(\Ba_1(0))$ and $h_\mp\colon\RR^{d-1}\to\RR$ are the $1$-Lipschitz functions of Eq.~\eqref{hmp_def}, which satisfy $h_-<h_+$ on $Z$ and $h_-=h_+$ outside $Z$ by Eq.~\eqref{hmp_prop}. Thus $f(S^{d-1})$ bounds the region between the two graphs, whose thickness vanishes along the equator over $\partial Z$.

Fix $b:=r_0/4$, and let $\delta,\eta\le b/8$ be sufficiently small multiples of $r_0$. We mollify $h_\mp$ and $g^\parallel_-$ at scale $\delta$ with the kernel $\ph_\delta$ of Lemma~\ref{lem:smoothing}, taking $\ph$ even:
\begin{equation}
\tilde h_\mp:=\ph_\delta*h_\mp,\qquad \tilde g^\parallel:=\ph_\delta*g_-^\parallel.
\end{equation}
By Lemma~\ref{lem:smoothing}, $\tilde h_\mp$ and $\tilde g^\parallel$ are smooth, with $|\tilde h_\mp-h_\mp|\le\delta$, $|\nabla\tilde h_\mp|\le1$, and second derivatives $O(1/\delta)$; likewise $\KLip(\tilde g^\parallel-\id)\le1/2$ and, as $\ph_\delta*\id=\id$ for even $\ph$, $|\tilde g^\parallel-g_-^\parallel|\le\delta\,\KLip(g_-^\parallel-\id)\le\delta/2$. The last bound gives $\|D\tilde g^\parallel-\one\|\le1/2$, so $D\tilde g^\parallel$ is invertible everywhere, and $\tilde g^\parallel$ is bijective by Lemma~\ref{lem:Lipschitz}; hence $\tilde g^\parallel$ is a diffeomorphism of $\RR^{d-1}$, with second derivatives $O(1/\delta)$. Let $\tilde Z:=\tilde g^\parallel(\bar\Ba_1(0))$, a smooth closed $(d-1)$-ball whose boundary has curvature $O(1/\delta)$ and admits unique nearest points within a distance proportional to $\delta$. Since $\partial\tilde Z$ lies within $\delta$ of $\partial Z$, where $h_-=h_+$, and the $h_\mp$ are $1$-Lipschitz, we have $|\tilde h_+-\tilde h_-|\le4\delta$ on $\partial\tilde Z$.

Following Fig.~\ref{fig:gmp}(b), we thicken the region between the two mollified graphs by $b$:
\begin{equation}
\tilde K_0:=\{(z,y):\,z\in\tilde Z,\ \tilde h_-(z)-b\le y\le\tilde h_+(z)+b\}.
\end{equation}
Its top and bottom are graphs of the smooth functions $\tilde h_++b$ and $\tilde h_--b$, whose gradients have norm at most $1$ and whose second derivatives are $O(1/\delta)$. Its side is the vertical cylinder over $\partial\tilde Z$, of height $\tilde h_+-\tilde h_-+2b\ge2b-4\delta\ge b$, meeting the top and the bottom transversally along two smooth $(d-2)$-spheres. Let $\tilde K$ be obtained by rounding these two edges within distance $\eta$ of them. Since $2\eta<b$, the two roundings do not meet and therefore $\tilde K$ is a smooth ball.

We now verify the tameness conditions in the statement. $\partial\tilde K$ has curvature $O(1/\eta)$, with the normal map of Section~\ref{sec:macro} injective to a depth proportional to $\eta$. We choose the equator $\tilde\Gamma$ to be the graph of $\tfrac12(\tilde h_++\tilde h_-)$ over $\partial\tilde Z$; it lies on $\partial\tilde K$, midway between the two rounded edges. It is a smooth $(d-2)$-sphere with curvature $O(1/\delta)$ and unique nearest points within a distance proportional to $\delta$. Choose a diffeomorphism $\tilde F\colon D^d\to\tilde K$ whose restriction $\tilde f$ maps $S_+$ onto the part of $\partial\tilde K$ above $\tilde\Gamma$; this is possible since both are smooth disks in a sphere. Hence $\tilde f$ and its equator are $(2c_{\mathrm{sm}}r_0)$-tame for $c_{\mathrm{sm}}$ sufficiently small, so that every $a\le c_{\mathrm{sm}}r_0$ is at most half their tameness scale.

For the inclusions, we apply Lemma~\ref{lem:ggtilde} to $\tilde f$ and $f$. This requires a map $u$, which we now construct. Parametrize $S_\mp$ by $x\in D^{d-1}$ as in Section~\ref{sec:micro}. Then $f(x)$ lies over $g^\parallel_\mp(x)$. Likewise, $\tilde f(x)$ lies over $\tilde g^\parallel(x)$, away from the side and the rounded edges. Let $u_\mp(x)$ be the parameter of the point of $f(S_\mp)$ over $\tilde g^\parallel(x)$, namely $u_\mp(x):=(g^\parallel_\mp)^{-1}(\tilde g^\parallel(x))$. This point may fall outside $D^{d-1}$, but only within $\delta$ of its boundary. In that case, we project it radially back. Within $\delta$ of the equator, we interpolate linearly to the identity, so that $u_-$ and $u_+$ agree there. Thus $u$ is continuous, maps $S_\mp$ into $S_\mp$, and fixes the equator. In particular, $u$ is homotopic to the identity.

Every point $\tilde f(x)$ lies within $b+O(\delta)+\eta$ of $f(u(x))$. Indeed, the two points lie over base points within $O(\delta)$ of each other, $\tilde h_\mp$ differs from $h_\mp$ by at most $\delta$, and the rounding moves points by at most $\eta$. Hence, for $\delta$ and $\eta$ sufficiently small multiples of $r_0$, $\lambda(a,\tilde f)\cge\lambda(a+b+O(\delta)+\eta,f)\cge\lambda(r_0,f)$ for all $a\le r_0/2$, by Lemma~\ref{lem:ggtilde} and Proposition~\ref{prop:g_mono}.

Finally, we verify that $\tilde C$, $\tilde B\tilde C$, $\tilde C\tilde D$ and $\tilde B\tilde C\tilde D\tilde E$ are balls. Since $\Ba_a(\partial\tilde K)$ is a tubular neighborhood, both $\tilde B\tilde C\tilde D\tilde E=\Ba_a(\tilde K)$ and $\tilde C$ are balls. For $\tilde B\tilde C=\Ba_a(\tilde K)\setminus\Ba_a(\tilde f(S_+))$, note that $\Ba_a(\tilde f(S_+))$ is bounded by the two parallel copies of $\tilde f(S_+)$ at distance $a$ and by the tube of radius $a$ about the equator, since $a$ is below the tameness scale of $\tilde f$ and its equator. Removing it from $\Ba_a(\tilde K)$ leaves a region bounded below by the parallel copy of $\tilde f(S_-)$ at distance $a$ outside $\tilde K$, above by the copy of $\tilde f(S_+)$ at depth $a$ inside $\tilde K$, and near the equator by the lower half of the tube. Every vertical line meets this region in an interval, over a disk of the base, so it is a ball. The same argument holds for $\tilde C\tilde D$.
\end{proof}

\phantomsection\label{proof:extend-A1p}
\begin{TheoremRestated}[Extendibility on larger scales]
\extendAlargerScalesStatement
\end{TheoremRestated}
\begin{proof}[Proof of Theorem \ref{thm:extend-A1'}]
We prove the extendibility of configurations in $\Lambda(r_0',r_1')$ from that of smooth configurations, using Proposition~\ref{prop:smooth-replacement}, and then apply Theorem~\ref{thm:large-smooth-extendibility}. We work at the scale $r_2:=r_1/c_2$. Let $a_0,s_0$ be the constants of Theorem~\ref{thm:large-smooth-extendibility} and $c_{\mathrm{sm}}$ that of Proposition~\ref{prop:smooth-replacement}, and set $s^{(\alpha)}:=\frac{s_0a_0}{c_{\mathrm{sm}}\,\alpha\,c_2}$. Let $(B,C,D,E)\in\Lambda(r_0',r_1')$ with $r_1'\ge s^{(\alpha)}r_1$ and $r_0'\ge\alpha r_1'$. Recall that $r_0'\le c_1r_1'$, as $\Lambda(r_0',r_1')$ requires.

We apply Proposition~\ref{prop:smooth-replacement} at the scale $r_0'$, with $a:=c_{\mathrm{sm}}r_0'/s_0$. Note that $a\le c_{\mathrm{sm}}r_0'$, since $s_0\ge1/\alpha_0\ge2$. This gives a configuration $(\tilde B,\tilde C,\tilde D,\tilde E)=\lambda(a,\tilde f)\cge(B,C,D,E)$, where $\tilde f$ and its equator are $r$-tame for $r:=c_{\mathrm{sm}}r_0'$. We now check the hypotheses of Theorem~\ref{thm:large-smooth-extendibility}. First, $r=s_0a$ by the choice of $a$. Second, $a\ge a_0r_2$, because $r_0'\ge\alpha r_1'\ge\alpha s^{(\alpha)}r_1=s_0a_0r_2/c_{\mathrm{sm}}$. Hence $\sigma$ is $(\tilde B,\tilde C,\tilde D,\tilde E)$-extendible. By Proposition~\ref{prop:g_mono}, $\sigma$ is $(B,C,D,E)$-extendible.
\end{proof}

\begin{Theorem}[Purifiability on larger scales] \label{thm:extend-A0}
Consider a state $\sigma$ extendible at scales $(r_0,r_1)$.  For some fixed
disjoint regions $(B,C)$, assume $\sigma|_{BC}$ is $(B,C)$-purifiable, where
\(BC\) is a topological ball and \(C\) is a $2r_2$-tame ball.
Then \(\sigma\) is \((B',C')\)-purifiable for any disjoint regions
\((B',C')\) such that \(B'C'\) is a topological ball, \(C'\) is a tame
ball, and \((C')^{+2r_1}\) is a tame ball contained in \(B'C'\).
\end{Theorem}
\noindent We first prove the case of concentric round balls \(C\subseteq C'\) with \(B'C'=BC\), and then
reduce the general case to that one.
\begin{proof}
We first treat the case \(B'C'=BC\) and \(C^{+2r_2}\subseteq C'\), where \(C\) and \(C'\) are concentric round balls and \(C\) has radius at least \(2r_2\).  Set \(Y:=BC\), and
let \(\bar{\sigma}_{AY}\) be a purification of \(\sigma|_Y\).  Since
\(\sigma|_Y\) is \((Y\setminus C,C)\)-purifiable, we have
\(\bar{\sigma}_{AC}=\bar{\sigma}_A\otimes \bar{\sigma}_C\).

Let \(C^-\) be the set of points of \(C\) at distance at least \(r_2\) from \(\partial C\). It is a round ball of radius at least \(r_2\), and \(K:=C\setminus C^-\) is its outer \(r_2\)-collar, to which the fixed-boundary isomorphism theorem applies. Set \(X:=C'\setminus C^-\).  Let
\(\widetilde A:=AC^-\) and \(\tau:=\bar{\sigma}_{\widetilde A K}\).  Because
\(\bar{\sigma}_{AC}\) is a product, \(\tau\) factors as
\(\tau=\bar{\sigma}_A\otimes\bar{\sigma}_{C^-K}\).

Now consider the relative information convex set \(\ICrel(X,\tau)\).  The
actual state \(\bar{\sigma}_{\widetilde A X}\) lies in \(\ICrel(X,\tau)\):
its witness is the actual restriction
\(\bar{\sigma}_{\widetilde A X^{+2r_1}}\), which is
available because \(X^{+2r_1}\subset (C')^{+2r_1}\subset Y\).  The product state
\(\bar{\sigma}_A\otimes \bar{\sigma}_{C^-X}\) also lies in \(\ICrel(X,\tau)\),
witnessed by
\(\bar{\sigma}_A\otimes \bar{\sigma}_{C^-X^{+2r_1}}\).

Since $X$ is an annulus tamely deformable, fixing the inner collar $K$, to the $2r_2$-collar annulus, Corollary~\ref{cor:fixed-boundary-annulus} implies that $\ICrel(X,\tau)$ is a singleton.  Therefore
\[
\bar{\sigma}_{\widetilde A X}
=
\bar{\sigma}_A\otimes \bar{\sigma}_{C^-X}.
\]
Because \(C'=C^- \cup X\), this is exactly
\(\bar{\sigma}_{AC'}=\bar{\sigma}_A\otimes \bar{\sigma}_{C'}\).  Hence
\(\sigma|_Y\) is \((Y\setminus C',C')\)-purifiable in this case.

We next show that $B$ can be reduced, while the decoupled ball $C$ remains fixed. If \(V\subseteq BC\) is a tame ball with \(C^{+2r_1}\subseteq V\), then \(\sigma\) is \((V\setminus C,C)\)-purifiable. The argument is basically the one above with the roles of the two boundaries exchanged. Let \(Y\supseteq (BC)^{+2r_2}\) be a sufficiently large round ball and \(\bar{\sigma}_{AY}\) a purification of \(\sigma|_Y\). Then \(\bar{\sigma}\) also purifies \(\sigma|_{BC}\) with purifying system \(A(Y\setminus BC)\), yielding
\[
\bar{\sigma}_{A(Y\setminus BC)C}=\bar{\sigma}_{A(Y\setminus BC)}\otimes\bar{\sigma}_{C}.
\]
Set \(Z:=Y\setminus V\), an annulus containing \(Y\setminus BC\), and consider \(\ICrel(Z,\tau)\) with the outer boundary \(\partial Y\) fixed, ancilla \(AC\), and \(\tau:=\bar{\sigma}_{AC\,\partial Y^{[-r_2,0]}}\). The actual state \(\bar{\sigma}_{ACZ}\) lies in \(\ICrel(Z,\tau)\), witnessed by \(\bar{\sigma}_{ACZ^{+2r_1}}\); here \(Z^{+2r_1}\) is disjoint from \(C\), because \(C^{+2r_1}\subseteq V\). The product state \(\bar{\sigma}_{AZ}\otimes\bar{\sigma}_{C}\) also lies in \(\ICrel(Z,\tau)\), witnessed by \(\bar{\sigma}_{AZ^{+2r_1}}\otimes\bar{\sigma}_{C}\); its restriction to the ancilla and the fixed collar is \(\tau\) since \(\partial Y^{[-r_2,0]}\subseteq Y\setminus BC\). The annulus \(Z\) can be tamely deformed to the \(2r_2\)-collar of \(\partial Y\) while fixing \(\partial Y\). Indeed, an isotopy of \(V\) to a round ball can be made tame since \(Y\) is large (Remark~\ref{rmk:scale-homotopy}). Hence Corollary~\ref{cor:fixed-boundary-annulus} implies that \(\ICrel(Z,\tau)\) is a singleton. In particular, \(\bar{\sigma}_{ACZ}=\bar{\sigma}_{AZ}\otimes\bar{\sigma}_{C}\). As \(AZ\) purifies \(\sigma|_{V}\), this is \((V\setminus C,C)\)-purifiability.

Now consider general disjoint regions \(B',C'\) as in the statement. Since \(C\) is \(2r_2\)-tame, it contains a round ball \(D\) of radius \(2r_2\). Choose a
sufficiently large round ball \(C''\) concentric with \(D\) and containing both \(C^{+2r_2}\) and \(C'\).  Then choose a
topological ball \(X\) containing both ambient regions \(BC\) and \(B'C'\), and
large enough that \((C'')^{+2r_1} \subset X\).  By monotonicity of local
purifiability from Section~\ref{subsec:circuit_invariance},
\(\sigma|_X\) is \((X\setminus C,C)\)-purifiable and hence \((X\setminus D,D)\)-purifiable. Applying the case already
proved inside the ambient ball \(X\), with \(D\) in place of \(C\) and \(C''\) in place of \(C'\), we find
that \(\sigma|_X\) is \((X\setminus C'',C'')\)-purifiable.  Since
\(C' \subset C''\), another application of monotonicity gives
\((X\setminus C',C')\)-purifiability. Finally, let $V:=(C')^{+2r_1}$.
Reducing the ambient ball from \(X\) to \(V\) as above, and then enlarging it to \(B'C'\) by monotonicity, we conclude that \(\sigma\) is \((B',C')\)-purifiable.
\end{proof}

\begin{Theorem}[Circuit invariance for extendible states] \label{thm:circuit-invariance-global}
Let \(\sigma\) be extendible on scale \((r_0,r_1)\) for some \(r_0 \ll r_1\), and set \(\alpha:=\frac{r_0}{r_1}\).  Then there exists a constant \(c^{(\alpha)}>0\), depending only on \(\alpha\), such that for every circuit \(U\) with light-cone radius \(r_U\), the state \(U\sigma U^\dagger\) is extendible on every standard configuration \(\Lambda(r_0',r_1')\) satisfying
\[
\frac{r_0'}{r_1'}=\frac{r_0}{r_1},
\qquad
r_0' \ge \max\!\bigl(c^{(\alpha)}r_0,6r_U\bigr).
\]
\end{Theorem}
One may take \(c^{(\alpha)}:=s^{(\alpha/2)}\), where \(s^{(\alpha/2)}\) is the threshold factor furnished by Theorem~\ref{thm:extend-A1'} for the ratio \(\alpha/2\).
Thus \(U\sigma U^\dagger\) is extendible on sufficiently larger scales $(r_0', r_1')$ with the same ratio \(r_0/r_1\) as the original one. 
\begin{proof}
Set \(c^{(\alpha)}:=s^{(\alpha/2)}\).
Fix a pair \((r_0',r_1')\) satisfying
\[
\frac{r_0'}{r_1'}=\frac{r_0}{r_1},
\qquad
r_0' \ge \max\!\bigl(c^{(\alpha)}r_0,6r_U\bigr).
\]
Set \(\bar r_0:=\frac{r_0'}{2}\).  Then \(\frac{\bar r_0}{r_1'}=\frac{r_0'}{2r_1'}=\frac{\alpha}{2}\).  Also, the bound \(r_0'\ge c^{(\alpha)}r_0=s^{(\alpha/2)}r_0\) implies \(\bar r_0=\frac{r_0'}{2}\ge \frac{1}{2}s^{(\alpha/2)}r_0=\frac{\alpha}{2}\, s^{(\alpha/2)} r_1\).  Since \(\bar r_0/r_1'=\alpha/2\), this is equivalent to \(r_1'\ge s^{(\alpha/2)} r_1\).
Therefore Theorem~\ref{thm:extend-A1'} implies that \(\sigma\) is extendible on scale \((\bar r_0,r_1')\).

Applying Theorem~\ref{thm:circuit-invariance} with light-cone radius \(r_U\), we obtain that \(U\sigma U^\dagger\) is extendible on scale
\[
(\bar r_0+3r_U,\; r_1').
\]
Since \(r_0' \ge 6r_U\), we have \(\bar r_0=\frac{r_0'}{2}\ge 3r_U\), and hence \(\bar r_0+3r_U \le 2\bar r_0 = r_0'\).
By Proposition~\ref{prop:g_mono}, extendibility on \((\bar r_0+3r_U,r_1')\) implies extendibility on \((r_0',r_1')\).
\end{proof}
\noindent Then  we can formalize the sense in which the class of locally extendible states is circuit-invariant.
\begin{Corollary}[Circuit-invariant class] \label{cor:circ-invariant-class}
Let \(\mathcal C\) denote the class of states \(\sigma\) that are extendible
for standard configurations \(\Lambda(r_0,r_1)\) for some pair of scales
\(r_0 \ll r_1\), possibly depending on \(\sigma\).
Then \(\mathcal C\) is closed under finite-depth local circuits.
\end{Corollary}

Below we learn that locally extendible states are not significantly more general than states that are both locally extendible and locally purifiable: the former are simply mixtures of the latter.
\begin{Theorem}[Decomposition into locally purifiable states]
\label{thm:decomposition-extendible}
Let $\sigma$ be extendible on scale $(r_0,r_1)$, and let $W$ be a finite ball.  Then $\sigma|_W$ admits a decomposition $\sigma|_W = \sum_i p_i \sigma^i$ for $p_i>0$ such that the states $\sigma^i$ are extendible on scale $(r_0,r_1)$, locally purifiable in the sense of the conclusion of Theorem~\ref{thm:extend-A0}, and mutually locally orthogonal, i.e., $\supp(\sigma^i|_L) \perp \supp(\sigma^j|_L)$ for $i \neq j$, for reference patches.
\end{Theorem}
This demonstrates a sense in which local purifiability is a secondary axiom.  For instance, as a corollary, the classification of locally extendible states under constant-depth circuits reduces to the classification of states that are both locally extendible and purifiable.  As another corollary, if a locally extendible state satisfies $I(X:Y)=0$ for a fixed pair of sufficiently large and separated regions $X,Y$, then it is also locally purifiable.%
\footnote{We expect that in the approximate setting, where extendibility holds
only up to a small error (the subject of ongoing work), most of our results
admit close analogs, without dependence on the finite local Hilbert space
dimension.  However, we caution that
Theorem~\ref{thm:decomposition-extendible} may require more substantial
modification, for instance through passage to the ``subspace constraints''
mentioned in Section~\ref{sec:formal-intro}.}
\begin{proof}
Choose a larger geometric ball $X$ containing $W$,
set $r=8r_2$, and define concentric annuli $B_2 = \partial X^{[0,r)}$ and
$B_1 = \partial X^{[r,2r)}$.  Let $Y:=B_1 B_2 X$, which is again a ball, and let $Y_+$ be a ball containing $Y^{+2r_2}$. Let
$\bar{\sigma} \in \St_{AY_+}$ be a purification of $\sigma|_{Y_+}$ with ancilla $A$, and set $\tilde A:=A(Y_+\setminus Y)$, which purifies $\sigma|_Y$.
Using Proposition~\ref{prop:collar-recovery} and
Proposition~\ref{prop:macro-extend}, there exists a channel
$R_{B_1 \to B_1 B_2}$ extending the inner boundary of $B_1$ such that
\(\id_{\tilde A} \otimes R_{B_1 \to B_1 B_2}[\bar{\sigma}|_{\tilde AB_1}] = \bar{\sigma}_{\tilde AB_1 B_2}\).
Hence \(I(\tilde A:B_2|B_1)_{\bar{\sigma}}=0\).  Let
\(\rho := \bar{\sigma}|_{\tilde AB_1B_2}\).  By the structure
theorem on Markov states \cite{hayden2004structure}, there exist orthogonal
projections $P_i \in \Alg_{B_1}$ such that \(\rho = \sum_i p_i \rho^i\), with
\(\rho^i = p_i^{-1} P_i \rho P_i\), \(p_i = \Tr(\rho P_i) > 0\), and
\(I(\tilde A:B_2)_{\rho^i}=0\) for each $i$.

Define \(\bar{\sigma}^i := p_i^{-1} P_i \bar{\sigma} P_i\), and let
\(\sigma^i := \bar{\sigma}^i|_{Y_+}\), \(\sigma_Y^i := \sigma^i|_Y\) and \(\sigma_W^i := \sigma_Y^i|_W\).  Since
$P_i$ acts only on $B_1$ and $W \subset X$ is disjoint from $B_1$, tracing out
$Y_+ \setminus W$ gives
\[
\sigma|_W = \sum_i p_i \sigma_W^i.
\]

Choose a $2r_2$-tame ball \(C \subset B_2\), for instance a round ball of radius $3r_2$.  Since
\(\bar{\sigma}^i|_{\tilde AB_1B_2}=\rho^i\) and \(I(\tilde A:B_2)_{\rho^i}=0\), monotonicity
of mutual entropy gives \(I(\tilde A:C)_{\bar{\sigma}^i}=0\).  Since \(\bar{\sigma}^i\) purifies \(\sigma^i|_Y\) with purifying system \(\tilde A\), the state \(\sigma^i\) is
\((Y\setminus C,C)\)-purifiable.  On the other hand, for every $O(1)$-size extendibility
patch \(L \subset Y_+\), the restriction \(\sigma^i|_L\) is supported on
\(\supp(\sigma|_L)\), because \(p_i \sigma^i|_L\) is a positive summand of
\(\sigma|_L\).  Therefore Proposition~\ref{prop:extendibility-mixture} implies
that each \(\sigma^i\) is locally extendible on $Y_+$.  By
Theorem~\ref{thm:extend-A0}, applied within $Y_+\supseteq Y^{+2r_2}$, each \(\sigma^i\) is then locally purifiable on
$Y$ in the sense of the conclusion of that theorem, and in particular on all
sufficiently large balls contained in $W$.

Next we show the \(\sigma_W^i\) are mutually locally orthogonal.  The states
\(\sigma_Y^i\) are mutually orthogonal on \(Y\), since the projections \(P_i\)
are mutually orthogonal on the subsystem \(B_1 \subset Y\).  Let \(L \subset W\)
be a reference patch.  Because each \(\sigma_Y^i\) is locally extendible, and
the same local extension channels for \(\sigma\) also work for every
\(\sigma_Y^i\), the proof of Proposition~\ref{prop:collar-recovery} gives channels
\(R_{L \to Y}\) and \(R_{Y \to L}\), independent of \(i\), such that
\[
R_{L \to Y}[\sigma_Y^i|_L] = \sigma_Y^i,
\qquad
R_{Y \to L}[\sigma_Y^i] = \sigma_Y^i|_L.
\]
If the states \(\sigma_Y^i|_L\) and \(\sigma_Y^j|_L\) had nonzero fidelity for
some \(i \neq j\), then after applying \(R_{L \to Y}\) the resulting states
\(\sigma_Y^i\) and \(\sigma_Y^j\) would also have nonzero fidelity, since
fidelity cannot decrease under a quantum channel.  This contradicts their
orthogonality on \(Y\).  Hence \(\supp(\sigma_W^i|_L) \perp \supp(\sigma_W^j|_L)\)
for \(i \neq j\).

\end{proof}

\section{Boundary gluing} \label{sec:cmi}

In this section, we show that elements of information convex sets over different regions can be merged to an element of the information convex set on a glued region. This discussion hinges on properties of quantum conditional mutual information.
\begin{Definition}
    Let $\rho\in \St_{ABC}$. Its conditional mutual information is
    \begin{equation}
        I(A:C|B)_{\rho} := S(\rho|_{AB}) + S(\rho|_{BC}) - S(\rho|_B) - S(\rho|_{ABC}),
    \end{equation}
    where $S(\rho) = -\Tr(\rho \log \rho)$.
\end{Definition}
Conditional mutual information satisfies the following well-known properties. First, it is non-increasing under a channel acting on $A$ or $C$. Also, $I(A:C|B)_{\rho}=0$ if and only if there is a channel $\Phi: \St_B \to \St_{BC}$ such that $\rho|_{ABC} = \id_A \otimes \Phi[\rho|_{AB}]$~\cite{petz1988sufficiency}. Under this condition, we say $\rho|_{ABC}$ is \emph{Markov}. Lastly, the following lemma was proven in Ref.~\cite{kato2016information}, which will play an important role.
\begin{Lemma}[Merging lemma]
    \label{lem:merging_lemma}
    Let $\rho \in \St_{ABC}$ and $\rho' \in \St_{BCD}$, where $\rho|_{BC}=\rho'|_{BC}$ and $I(A:C|B)_{\rho} = I(B:D|C)_{\rho'}=0$. There exists $\tau\in \St_{ABCD}$ such that $\tau|_{ABC} = \rho$, $\tau|_{BCD} = \rho'$, and $I(A:CD|B)_{\tau} = I(AB:D|C)_{\tau}=0$.
\end{Lemma}

Conditional mutual information also obeys the following bound, which we use in Section~\ref{sec:rev_extendibility_merging} to find a density matrix that is nearly Markov. For disjoint regions $\{X_i\}_{i=1}^n$ we write $X^{(k)}= \cup_{i=1}^k X_i$, a convention we apply to other indexed families as well.
\begin{Lemma}\label{lem:existence_small_cmi}
Let $Y$ be a region disjoint from $X^{(n)}$ and let $\rho\in \St_{X^{(n)}Y}$. There exists $k\le n$ such that
\begin{equation}
I(Y:X_k|X^{(k-1)})_{\rho} \le \frac{2S(\rho|_Y)}{n}.
\end{equation}
\end{Lemma}
\begin{proof}
    Note $I(Y:X_k|X^{(k-1)})_{\rho} = I(Y:X^{(k)})_\rho - I(Y:X^{(k-1)})_\rho$. Since 
    \begin{equation}
    \sum_{k=1}^n I(Y:X_k|X^{(k-1)})_\rho = I(Y:X^{(n)})_\rho \le 2S(\rho|_Y)
    \end{equation}
    and $I(Y:X_k|X^{(k-1)})_\rho\ge 0$ by strong subadditivity~\cite{lieb1973proof}, we get
    \begin{equation}
    \min_{k\le n} I(Y:X_k|X^{(k-1)})_\rho \le \frac{2S(\rho|_Y)}{n}.
    \end{equation}
\end{proof}

Let us first define what it means to glue two regions. For embedded regions that share a boundary, this entails simply taking their union. For immersed regions, we can generalize this definition as follows. 
\begin{Definition}[Glued immersion]
    \label{def:glued_immersion}
    Let $f:M\to \mathbb{R}^d$ and $f': M' \to \mathbb{R}^d$ be proper immersions. Partition $\partial M$ and $\partial M'$ into disjoint closed subsets:
    \begin{equation}
        \partial M = N_1 \sqcup N_2, \quad \partial M' = N_1' \sqcup N_2'.
    \end{equation}
    Let $\varphi:N_1 \to N_1'$ be a diffeomorphism satisfying the following.
    \begin{enumerate}
        \item $f' \circ \varphi = f|_{N_1}$.
        \item Let $n(x)$ and $n'(x')$ be outward unit normal vectors of $f$ and $f'$ at $x\in \partial M$ and $x' \in \partial M'$, respectively. For every $x\in N_1$, $n(x)= -n'(\varphi(x))$.
    \end{enumerate}
    A $\varphi$-gluing of $f$ and $f'$ is an immersion $F:M \cup_{\varphi} M' \to \mathbb{R}^d$, where
    \begin{equation}
    M\cup_{\varphi} M' := (M\cup M') / (x\sim \varphi(x)\,\, \forall x \in N_1)
    \end{equation}
    and $F$ is induced by $f$ and $f'$ on $M\cup_{\varphi} M'$.
\end{Definition}
Note that $F$ is well defined, with $F|_M=f$ and $F|_{M'}=f'$. The glued region inherits smoothness and tameness from its two halves.
\begin{Lemma}[Gluing tame regions]
\label{lem:gluing_tame}
Let $f:M\to\RR^d$ and $f':M'\to\RR^d$ be $r$-tame immersed regions and $\varphi:N_1\to N_1'$ a diffeomorphism as in Definition~\ref{def:glued_immersion}. Then $M\cup_\varphi M'$ admits a smooth structure for which the $\varphi$-gluing $F$ is a smooth immersion, and $(M\cup_\varphi M',F)$ is an $r$-tame immersed region.
\end{Lemma}
\begin{proof}
Let $u:N_1\times(-r,0]\to M$ and $u':N_1'\times(-r,0]\to M'$ be the inner collars along $N_1$ and $N_1'$, and define $v:N_1\times(-r,r)\to M\cup_\varphi M'$ by $v(x,t)=u(x,t)$ for $t\le0$ and $v(x,t)=u'(\varphi(x),-t)$ for $t\ge0$. By the two conditions on $\varphi$, namely $f'\circ\varphi=f|_{N_1}$ and $n'\circ\varphi=-n$, we have $F(v(x,t))=f(x)+t\ts n(x)$ on all of $N_1\times(-r,r)$. We give $M\cup_\varphi M'$ the smooth structure generated by the charts of $M\setminus N_1$, those of $M'\setminus N_1'$, and $v$; the transition maps are $u^{-1}$ on $t<0$ and $(\varphi^{-1}\times(t\mapsto-t))\circ u'^{-1}$ on $t>0$, hence smooth. In this structure $F$ is smooth, and it is an immersion because $F\circ v$ is the normal map of $N_1$, whose differential is nondegenerate on $N_1\times(-r,r)$ since $u$ and $u'$ are diffeomorphisms onto their images. The boundary of $M\cup_\varphi M'$ is $N_2\sqcup N_2'$, which is $r$-tame, and the inner $r$-collars of $M$ along $N_2$ and of $M'$ along $N_2'$ lie in the respective pieces and combine to an inner $r$-collar of the glued region; so $F$ is $r$-tame.
\end{proof}

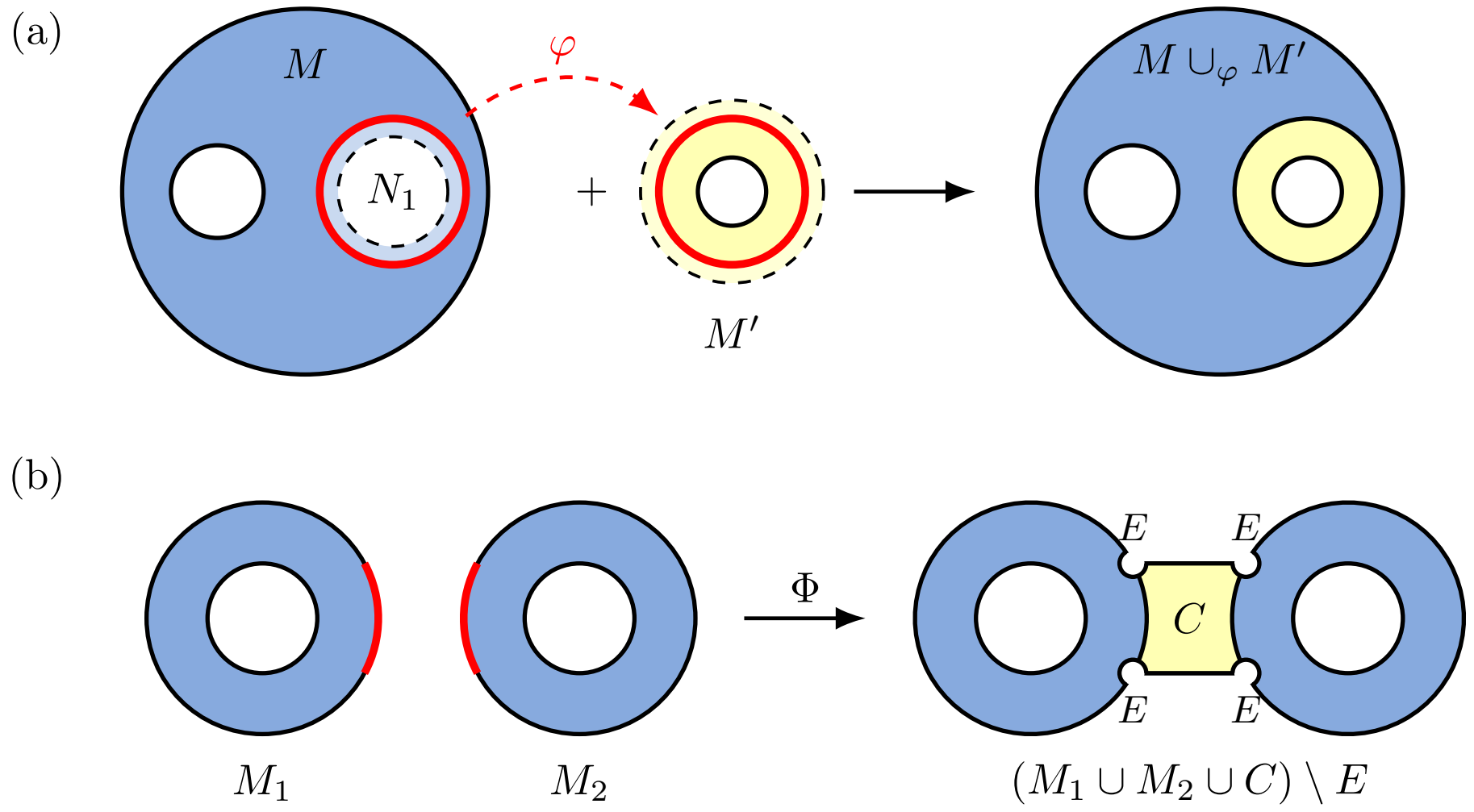
\begin{figure}[t]
    \centering
\begin{tikzpicture}[
    line width=1pt,
    reg/.style={fill=tikzblue},
    new/.style={fill=my_color1},
    col/.style={fill=tikzbluelight},
    colp/.style={fill=my_color1!55!white},
    hole/.style={fill=white},
    glue/.style={draw=red, line width=1.8pt},
    collaredge/.style={dashed, line width=0.7pt},
    mapto/.style={-{Latex[length=2.5mm]}, line width=0.9pt},
]

\begin{scope}

  \node at (-2.2,1.3) {(a)};

  \begin{scope}
    \draw[reg]  (0,0) circle (1.5);
    \draw[hole] (-0.72,0) circle (0.38);
    \path[col]  ( 0.72,0) circle (0.6);
    \draw[hole, collaredge] (0.72,0) circle (0.45);
    \draw[glue] ( 0.72,0) circle (0.6);
    \node at (0.72,0) {$N_1$};
    \node at (0,1.05) {$M$};
  \end{scope}

  \node at (2.35,0) {$+$};

  \begin{scope}[shift={(3.5,0)}]
    \path[colp] (0,0) circle (0.75);
    \draw[collaredge] (0,0) circle (0.75);
    \draw[new]  (0,0) circle (0.6);
    \draw[hole] (0,0) circle (0.28);
    \draw[glue] (0,0) circle (0.6);
    \node at (0,-1.15) {$M'$};
  \end{scope}

  \draw[mapto, dashed, red]
    (1.32,0.62) to[out=40,in=140] node[midway, above] {$\varphi$} (2.9,0.62);

  \draw[mapto] (4.5,0) -- (5.5,0);

  \begin{scope}[shift={(7.5,0)}]
    \draw[reg]  (0,0) circle (1.5);
    \draw[hole] (-0.72,0) circle (0.38);
    \draw[new]  ( 0.72,0) circle (0.6);
    \draw[hole] ( 0.72,0) circle (0.28);
    \node at (0,1.05) {$M\cup_{\varphi}M'$};
  \end{scope}

\end{scope}

%
\begin{scope}[yshift=-3.5cm, xshift=-0.35cm]

  \node at (-1.85,1.15) {(b)};

  \foreach \x in {0, 2.6} {
    \begin{scope}[shift={(\x,0)}]
      \draw[reg]  (0,0) circle (0.95);
      \draw[hole] (0,0) circle (0.45);
    \end{scope}
  }
  \draw[glue] (28.27371:0.95) arc (28.27371:-28.27371:0.95);
  \draw[glue] ([shift={(151.72629:0.95)}]2.6,0) arc (151.72629:208.27371:0.95);
  \node at (0,-1.35)   {$M_1$};
  \node at (2.6,-1.35) {$M_2$};

  \draw[mapto] (3.95,0) -- node[midway, above] {$\Phi$} (4.95,0);

  \begin{scope}[shift={(6.3,0)}]
    \path[reg, draw=black]
      (28.27371:0.95) arc (28.27371:331.72629:0.95)
      -- ([shift={(208.27371:0.95)}]2.6,0) arc (208.27371:511.72629:0.95)
      -- cycle;
    \path[new, draw=black]
      (0.83666,-0.45) -- (1.76334,-0.45) arc (208.27371:151.72629:0.95)
      -- (0.83666, 0.45) arc (28.27371:-28.27371:0.95) -- cycle;
    \draw[hole] (0,0)   circle (0.45);
    \draw[hole] (2.6,0) circle (0.45);
    \foreach \p in {(0.83666,0.45), (0.83666,-0.45), (1.76334,0.45), (1.76334,-0.45)}
      { \fill[white] \p circle (0.095); }
    \begin{scope}
      \clip
        (28.27371:0.95) arc (28.27371:331.72629:0.95)
        -- ([shift={(208.27371:0.95)}]2.6,0) arc (208.27371:511.72629:0.95)
        -- cycle;
      \foreach \p in {(0.83666,0.45), (0.83666,-0.45), (1.76334,0.45), (1.76334,-0.45)}
        { \draw \p circle (0.11); }
    \end{scope}
    \node at (1.3,0) {$C$};
    \foreach \p in {(0.83666,0.75), (0.83666,-0.75), (1.76334,0.75), (1.76334,-0.75)}
      { \node[font=\small] at \p {$E$}; }
    \node at (1.3,-1.35) {$(M_1\cup M_2\cup C)\setminus E$};
  \end{scope}

\end{scope}

\end{tikzpicture}
    \caption{Comparison of different merging processes; in each case the merging takes place over the part of the boundary marked in red. \textbf{(a)} Merging density matrices over the full boundary (Theorem~\ref{thm:boundary_gluing}). The boundary component $N_1$ of the two-holed disk $M$ is identified by $\varphi$ with the outer boundary of the annulus $M'$, and the two density matrices are merged along all of $N_1$. \textbf{(b)} Merging density matrices over strict subset of a connected boundary component (Section~\ref{sec:rev_extendibility_merging}). A merging map $\Phi$ joins two annuli along arcs of their outer boundaries, creating the region $C$ and yielding a two-holed disk. \label{fig:boundary_gluing}}
\end{figure}

\begin{Theorem}[Boundary gluing]
    \label{thm:boundary_gluing}
    Let $f:M\to \mathbb{R}^d$ and $f':M'\to \mathbb{R}^d$ be tame immersed regions satisfying the conditions in Definition~\ref{def:glued_immersion}.\footnote{We assume that no spin lies on the gluing surface $F(N_1)$. Since each ball contains finitely many spins, this can be arranged by displacing the surface slightly.} Let $\rho \in \IC(M, f)$ and $\rho' \in \IC(M', f')$, and let $\hat{\rho}$ and $\hat{\rho}'$ be $r_1$-consistent extensions of $\rho$ and $\rho'$ on $M^{+2r_1}$ and ${M'}^{+2r_1}$. Suppose that\footnote{Note $N_1^{[-r_1, r_1]} = {N_1'}^{[-r_1, r_1]}$.}
    \begin{equation}
    \label{eq:gluing_collar_match}
        \hat{\rho}|_{F(N_1^{[-r_1, r_1)})} = \hat{\rho}'|_{F(N_1^{[-r_1, r_1)})}.
    \end{equation}
    Then there exists $\tau\in \IC(M\cup_{\varphi} M', F)$ such that $\tau|_{f(M)}=\rho|_M$, $\tau|_{f'(M')}=\rho'|_{M'}$. Furthermore, there are recovery maps $\Phi: \St_{F(N_1^{[-r_1, 0)})} \to \St_{F(M'\cup N_1^{[-r_1,0))}}$ and $\Phi': \St_{F({N_1'}^{[-r_1,0)})} \to \St_{F(M\cup {N_1'}^{[-r_1,0)})}$, depending only on $\rho'$ and on $\rho$, respectively, such that
    \begin{equation}
        \tau = \id_{F(M\setminus N_1^{[-r_1,0)})} \otimes \Phi[\rho|_M] = \id_{F(M'\setminus {N_1'}^{[-r_1,0)})} \otimes \Phi'[\rho'|_{M'}].
    \end{equation}
\end{Theorem}
\noindent 
We note that the condition~\eqref{eq:gluing_collar_match} does not depend on the choice of the extensions, and therefore this is a condition on $\rho$ and $\rho'$ alone. Indeed, $\hat{\rho}$ is $r_1$-consistent on $N_1^{[-r_1,2r_1)}$, so Proposition~\ref{prop:collar-recovery}, applied to $N_1$ with the outward normal of $M$, determines $\hat{\rho}|_{N_1^{[-r_1,r_1)}}$ from $\rho|_{N_1^{[-r_1,0)}}$ by a channel that does not depend on the state. The same holds for $\hat{\rho}'$, with the outward normal of $M'$. Thus Eq.~\eqref{eq:gluing_collar_match} says that the extension of $\rho$ across $N_1$ agrees with $\rho'$, or equivalently that the extension of $\rho'$ across $N_1$ agrees with $\rho$.
\begin{proof}
    We aim to glue the reduced density matrices of these states over $f_{+2r_1}(M^{+2r_1} \setminus N_1^{[r_1, 2r_1)})$ and $f'_{+2r_1}({M'}^{+2r_1} \setminus {N_1'}^{(r_1, 2r_1]})$, respectively. Let us denote the partitions of these regions as $ABC$ and $BCD$, where
    \begin{equation}
        B= N_1^{[-r_1,0)} = {N_1'}^{(0, r_1]}, \quad C= N_1^{[0, r_1)} = {N_1'}^{(-r_1, 0]}.
    \end{equation}
    $A$ and $D$ can be obtained by subtracting these regions. 

    By Proposition~\ref{prop:collar-recovery}, $I(A:C|B)_{\hat{\rho}} = I(D:B|C)_{\hat{\rho}'}=0$. The states $\hat{\rho}|_{ABC}$ and $\hat{\rho}'|_{BCD}$ agree on $BC$ by Eq.~\eqref{eq:gluing_collar_match}. From Lemma~\ref{lem:merging_lemma}, it follows that there is $\hat{\tau} \in \St_{ABCD}$ satisfying $\hat{\tau}|_{ABC} = \hat{\rho}|_{ABC}$ and $\hat{\tau}|_{BCD} = \hat{\rho}'|_{BCD}$. In particular, $\hat{\tau}$ is $r_1$-consistent. Since $ABCD = F_{+2r_1}((M\cup_{\varphi}M')^{+2r_1})$, $\hat{\tau}$ is an $r_1$-consistent extension of its reduced density matrix on $F(M\cup_{\varphi}M')$. Thus $\tau:=\hat{\tau}|_{F(M\cup_{\varphi}M')} \in \IC(M\cup_{\varphi}M', F)$.

    Moreover, from the same lemma we also get $I(A:CD|B)_{\hat{\tau}} = I(D:AB|C)_{\hat{\tau}}=0$. After tracing out the outer $2r_1$-collars, we get $I(A':CD'|B)_{\tau} = I(D':A'B|C)_{\tau}=0$, where $A'B= f(M)$ and $CD'=f'(M')$. From this, the existence of the recovery maps follows~\cite{petz1988sufficiency}. Concretely, $\Phi$ is the Petz recovery map of $\hat{\rho}'|_{BCD'}$, which is determined by $\rho'$ through Proposition~\ref{prop:collar-recovery}, as noted after the statement. Hence $\Phi$ depends only on $\rho'$, and likewise $\Phi'$ depends only on $\rho$.
\end{proof}

The same proof applies to the elements of the relative information convex set~\eqref{eq:IC-rel} as well. Below we use the same setup for $f:M\to \mathbb{R}^d$ and $f':M'\to \mathbb{R}^d$. 
\begin{Corollary}[Boundary gluing for relative information convex sets]
    \label{cor:boundary_gluing_rel}
    Let $\rho \in \ICrel(M, f, \widetilde{\rho})$, where $\widetilde{\rho}\in \St_{P N_2^{[-r_2,0]}}$ for an abstract system $P$ and a boundary component $N_2\subseteq\partial M$ disjoint from $N_1$, and let $\rho' \in \IC(M', f')$. Suppose that the $r_1$-consistent extensions of $\rho$ and $\rho'$ provided by Eq.~\eqref{eq:IC-rel} satisfy Eq.~\eqref{eq:gluing_collar_match}. Consider the following state:
    \begin{equation}
    \tau:=(\id_{P F(M\setminus N_1^{[-r_1,0)})}\otimes\Phi)[\rho|_{PM}],
    \end{equation}
    where $\Phi$ the recovery map of Theorem~\ref{thm:boundary_gluing}. This state belongs to $\ICrel(M\cup_{\varphi}M', F, \widetilde{\rho})$ and satisfies $\tau|_{P f(M)}=\rho|_{PM}$ and $\tau|_{f'(M')}=\rho'|_{M'}$. The same holds when the abstract system is attached to $M'$ instead of $M$.
\end{Corollary}
\begin{proof}
The proof of Theorem~\ref{thm:boundary_gluing} applies with $P$ adjoined to $A$. Indeed, Eq.~\eqref{eq:IC-rel} provides an $r_1$-consistent extension $\hat{\rho}$ that includes $P$; Proposition~\ref{prop:collar-recovery} holds for an arbitrary adjoined system, so $I(PA:C|B)_{\hat{\rho}}=I(D:B|C)_{\hat{\rho}'}=0$; and Lemma~\ref{lem:merging_lemma} yields $\hat{\tau}$ with $\hat{\tau}|_{PABC}=\hat{\rho}|_{PABC}$ and $\hat{\tau}|_{BCD}=\hat{\rho}'|_{BCD}$, together with the recovery maps. The restriction of $\hat{\tau}$ to $PN_2^{[-r_2,0]}$ is that of $\hat{\rho}$, namely $\widetilde{\rho}$.
\end{proof}

Theorem~\ref{thm:boundary_gluing} and its version for relative information convex set (Corollary~\ref{cor:boundary_gluing_rel}) let us merge elements of information convex sets; see Fig.~\ref{fig:boundary_gluing}. However, these statements allow the merging only along a closed boundary component. We will sometimes need to merge the density matrices over strict subsets of a connected component. We describe such a method in Section~\ref{sec:rev_extendibility_merging}.

\section{Reverse extendibility and merging}
\label{sec:rev_extendibility_merging}

One might ask how one can generalize Theorem~\ref{thm:boundary_gluing} to the setup in which the gluing occurs along a subset of the boundary. To that end, it will be helpful to introduce closely related notions of local extendibility. These will be useful for proving such a statement, but also for more advanced arguments that will appear in our future work. One such condition is ``reverse'' local extendibility. This condition is analogous to local extendibility, but with the roles of regions $A$ and $C$ switched. Throughout this section we assume $d\ge2$.

\begin{Definition}[Reverse local extendibility] \label{def:rev_loc_extend}
Let $B$, $C$, $D$, $E$ be disjoint regions, $\sigma$ a state on some region containing $BCDE$, and $\overline{\sigma} \in \St_{ABCDE}$ a purification of $\sigma|_{BCDE}$ using some abstract system $A$. The state $\sigma$ is said to be \emph{$(B,C,D,E)$-reverse extendible} if there exists a quantum channel $R_{BE\to BA}$ such that
\begin{equation}\label{eq:rev_loc_extend}
(\id_C \otimes R_{BE\to BA})[\overline{\sigma}|_{BCE}]=\overline{\sigma}|_{ABC}.
\end{equation}
\end{Definition}

States that are reverse extendible satisfy properties analogous to extendible states.  
\begin{Corollary}
\label{thm:immersion2}
    Let $\overline{\sigma}\in\St_{ABCDE}$ be a purification of $\sigma|_{BCDE}$ with abstract system $A$, as in Definition~\ref{def:rev_loc_extend}. Then $\sigma$ is $(B,C,D,E)$-reverse extendible if and only if there is a state $\rho\in\St_{A_1A_2BCD}$, with $A_1,A_2$ copies of $A$, such that $\rho|_{A_1BC} = \overline{\sigma}|_{ABC}$ and $\rho|_{A_2CD} = \overline{\sigma}|_{ACD}$.
\end{Corollary}
\noindent
Viewed as a statement about the pure state $\overline{\sigma}$, this is Theorem~\ref{thm:immersion1} with the labels $A$ and $C$ switched, so the proof is not repeated. Neither condition depends on the choice of purification, as any two purifications of $\sigma|_{BCDE}$ are related by a channel on the purifying system.

\begin{Proposition}[Reverse geometric monotonicity]\label{prop:reverse-g_mono}
Suppose that
\begin{equation}\label{eq:g_mono_reverse}
B'E'\supseteq BE,\qquad D'E'\supseteq DE,\qquad
C'\subseteq C,\qquad B'C'D'E'\supseteq BCDE.
\end{equation}
Then any $(B,C,D,E)$-reverse extendible state $\sigma$ on a region containing $B'C'D'E'$ is also $(B',C',D',E')$-reverse extendible.
\end{Proposition}
\noindent This could be proven directly in the vein of Proposition~\ref{prop:g_mono}, but we instead give a proof that applies Proposition~\ref{prop:g_mono} itself, to directly illustrate the symmetry between local extendibility and reverse local extendibility.
\begin{proof}
Let $\overline{\sigma}'$ be a purification of $\sigma|_{B'C'D'E'}$ with an abstract system $A'$, and let $G=B'C'D'E'\setminus BCDE$. The last condition in \eqref{eq:g_mono_reverse} implies that $B'C'D'E'=GBCDE$. Therefore, $\overline{\sigma}'$ is also a purification of $\sigma|_{BCDE}$ with system $A=A'G$.

Because $\sigma$ is $(B,C,D,E)$-reverse extendible, the pure state $\overline{\sigma}'$ is equivalently $(B,A'G,D,E)$-extendible, with $C$ playing the role of the purifying system. We now apply Proposition~\ref{prop:g_mono} to $\overline{\sigma}'$, with $A'G$ and $A'$ playing the roles of $C$ and $C'$. The first two inclusions in \eqref{g_mono} are unchanged, and clearly $A'\subseteq A'G$. It remains to check that
\[
B'A'D'E'\supseteq BA'GDE.
\]
Indeed, $B,D,E\subseteq B'D'E'$ by the first two inclusions of \eqref{eq:g_mono_reverse}. Also, $G\subseteq B'D'E'$: by definition, $G\subseteq B'C'D'E'$, while $C'\subseteq C\subseteq BCDE$, so $G\cap C'=\emptyset$. Thus $G$ cannot meet the $C'$ part of $B'C'D'E'$, and hence $G\subseteq B'D'E'$. Therefore, $BA'GDE\subseteq B'A'D'E'$.

By Proposition~\ref{prop:g_mono}, the pure state $\overline{\sigma}'$ is $(B',A',D',E')$-extendible. Switching back the roles of $A'$ and $C'$, this is exactly $(B',C',D',E')$-reverse extendibility of $\sigma$.
\end{proof}

Despite these similarities, for a fixed configuration, extendibility and reverse extendibility are generally inequivalent. For instance, consider the following counterexample. Let $\rho_{ABCDE} = \rho_{ACDE} \otimes \rho_B$, where both $\rho_{ACDE}$ and $\rho_B$ are pure. Let
\begin{equation}
\rho_{ACE} = \frac{1}{2} \sum_{a\in \{0,1\}} \ket{a}_A\bra{a} \otimes \mathcal{N}_p(\ket{a}_C\bra{a}) \otimes \mathcal{N}_p(\ket{a}_E\bra{a}),
\end{equation}
where $\mathcal{N}_p(\cdot) = (1-p) (\cdot) + p X (\cdot )X$, $p \in (0, \frac{1}{2})$, and $X$ is the Pauli $X$ matrix. It is easy to see that there is a recovery map from $BE$ to $BC$. On the other hand, the recovery map from $BE$ to $BA$ cannot exist. The proof follows from the data processing inequality. Because $I(C:BE) < I(C:AB)$, and mutual information is nonincreasing under channels on one of the subsystems, the recovery map from $BE$ to $BA$ cannot recover $\rho_{ABC}$. 

A closely related concept is \emph{merging}, which entails extending a density matrix to a larger region, but with a weaker requirement than extendibility. Let $B$, $C$, $E$ be disjoint regions and $\sigma_{ABCE} \in \St_{ABCE}$. We may demand
\begin{equation}
    \label{eq:merging}
(\id_A\otimes(\Tr_C  \circ R_{BE\to BC}))[\sigma|_{ABE}] = \sigma|_{AB}, \quad
R_{BE\to BC} [\sigma|_{BE}] = \sigma|_{BC},
\end{equation}
which is weaker than Definition~\ref{def:loc_extend}. An example of the merging process is shown in Fig.~\ref{fig:attaching_example}(a-c), wherein one begins with a union of two annuli and obtain a two-holed ball via a channel localized to some region. In this example, $A$ can be viewed as part of the original annuli that lie outside of this region.

In a future work, we will use reverse extendibility for the standard configurations (more precisely, for appropriately scaled-up regions) and merging operations. It is thus natural to ask if those statements require genuinely new assumptions or if they follow from local extendibility alone. As we prove below, local extendibility on the standard configuration set does imply reverse extendibility at a larger scale. It also lets us attach a bridge to a region, which is the tool from which merging operations are built.

\begin{figure}[t]
\begin{tikzpicture}[line width=1pt, scale=0.5]
\draw[fill=tikzblue] (-3,0) circle (2cm);
\draw[fill=tikzblue] ( 3,0) circle (2cm);
\begin{scope}[xshift=-3cm]
\draw[] (-90:1.2cm) arc (-90:90:1.2cm);
\draw[] (90:1.2cm) arc (90:270:0.3cm);
\draw[] (-90:1.2cm) arc (270:90:0.3cm);
\end{scope}
\begin{scope}[xshift=3cm, xscale=-1]
\draw[] (-90:1.2cm) arc (-90:90:1.2cm);
\draw[] (90:1.2cm) arc (90:270:0.3cm);
\draw[] (-90:1.2cm) arc (270:90:0.3cm);
\end{scope}
\draw[fill=tikzblue, draw=none] (-3cm, 0.6cm) -- (3cm, 0.6cm) -- (3cm, -0.6cm) -- (-3cm, -0.6cm) -- cycle;
\draw[] (-3cm, 0.6cm) -- (3cm, 0.6cm);
\draw[] (-3cm, -0.6cm) -- (3cm, -0.6cm);
\draw[fill=white] (-3cm, 0.6cm) arc (-90:-270:0.3cm) arc (90:30:1.2cm) -- cycle;
\draw[fill=white] (-3cm, -0.6cm) arc (90:270:0.3cm) arc (270:330:1.2cm) -- cycle;
\begin{scope}[xscale=-1]
\draw[fill=white] (-3cm, 0.6cm) arc (-90:-270:0.3cm) arc (90:30:1.2cm) -- cycle;
\draw[fill=white] (-3cm, -0.6cm) arc (90:270:0.3cm) arc (270:330:1.2cm) -- cycle;
\end{scope}

\node[] () at (0, -3cm) {(a)};

\begin{scope}[xshift=11cm, xslant=0.3, scale=0.5, yshift=8cm]

\begin{scope}[yshift= 1.5cm]
\draw[fill=tikzblue] (3,-2.5) -- ++ (5, 0) -- ++ (0, 5) -- ++ (-5 ,0) -- cycle;
\draw[fill=red, draw=red] (5.5,0) circle (2cm);
\draw[fill=white] (5.5,0) circle (1.8cm);
\draw[dotted] (2, 0) -- ++ (-1, 0);
\draw[dotted] (9, 0) -- ++ (1, 0);
\draw[dotted] (5.5, 3) -- ++ (0, 1);
\draw[dotted] (5.5, -3) -- ++ (0, -1);
\end{scope}

\begin{scope}[yshift= -8.5 cm, xshift=1.5cm ]
\draw[fill=tikzblue] (3,-2.5) -- ++ (5, 0) -- ++ (0, 5) -- ++ (-5 ,0) -- cycle;
\draw[fill=red, draw=red] (5.5,0) circle (2cm);
\draw[fill=white] (5.5,0) circle (1.8cm);
\draw[dotted] (2, 0) -- ++ (-1, 0);
\draw[dotted] (9, 0) -- ++ (1, 0);
\draw[dotted] (5.5, 3) -- ++ (0, 1);
\draw[dotted] (5.5, -3) -- ++ (0, -1);
\end{scope}

\draw[dashed] (-4, 2) -- (5.5, 3.5);
\draw[dashed] (-4, -2) -- (5.5, -0.5);

\draw[dashed] (-2, -5) -- (7, -6.5);
\draw[dashed] (-2, -9) -- (7, -10.5);

\draw[fill=red, draw=red] (-4,0) circle (2cm);
\draw[fill=tikzblue] (-4,0) circle (1.8cm);
\begin{scope}[xshift=-4cm]
\draw[] (-90:1.2cm) arc (-90:90:1.2cm);
\draw[] (90:1.2cm) arc (90:270:0.3cm);
\draw[] (-90:1.2cm) arc (270:90:0.3cm);

\draw[fill=white] (120:1.2cm) arc (120:-120:1.2cm) arc (-120:-270:0.3cm) -- ++ (0, 0.96cm) arc (270:120:0.3cm);

\begin{scope}[yshift=-7cm, xshift=2cm, xscale=-1]
\draw[fill=red, draw=red] ( 0,0) circle (2cm);
\draw[fill=tikzblue] ( 0,0) circle (1.8cm);
\draw[] (-90:1.2cm) arc (-90:90:1.2cm);
\draw[] (90:1.2cm) arc (90:270:0.3cm);
\draw[] (-90:1.2cm) arc (270:90:0.3cm);

\draw[fill=white] (120:1.2cm) arc (120:-120:1.2cm) arc (-120:-270:0.3cm) -- ++ (0, 0.96cm) arc (270:120:0.3cm);
\end{scope}
\draw[fill=tikzblue, draw=none] (-0.6, 0.5196) arc (240:270:0.3cm) -- ++ (2.9cm, -7cm) arc (270:300:0.3cm)
  -- (2.6, -7.5196) arc (-300:-270:0.3cm) -- ++ (-2.9cm, 7cm) arc (-270:-240:0.3cm)
  -- cycle;
\draw[] (120:1.2cm) arc (120:270:0.3cm) -- ++ (2.9cm, -7cm) arc (270:420:0.3cm);
\draw[] (-120:1.2cm) arc (-120:-270:0.3cm) -- ++ (2.9cm, -7cm) arc (-270:-420:0.3cm);

\draw[fill=tikzblue, draw=none] (240:1.8cm) arc (240:360:1.8cm)  -- (360:1.2cm) arc (360:240:1.2cm) -- cycle;
\draw[fill=red, draw=red] (240:2cm) arc (240:360:2cm)  -- (360:1.8cm) arc (360:240:1.8cm) -- cycle;
\draw[] (220:1.8cm) arc (220:380:1.8cm);
\draw[] (240:1.2cm) arc (240:360:1.2cm);

\end{scope}
\end{scope}

\node[] () at (12cm, -3cm) {(b)};

\begin{scope}[xshift=24cm, xslant=0.3, scale=0.5]
\begin{scope}[yshift=6cm]
\draw[fill=tikzblue, draw=none] (-6, -9) -- ++ (14.5,0) -- ++ (0,6) -- ++ (-14.5, 0) -- cycle;

\draw[dotted] (-7, -6) -- ++ (-1, 0);
\draw[dotted] (9.5, -6) -- ++ (1, 0);
\draw[dotted] (1.25, -9.5) -- ++ (0, -1);
\draw[dotted] (1.25, -2.5) -- ++ (0, 1);

\draw[dashed] (-4.5, -6) -- (-6, 0);
\draw[dashed] (-0.5, -6) -- (-2, 0);

\draw[dashed] (7.5, -6) -- (6, 0);
\draw[dashed] (3.5, -6) -- (2, 0);

\draw[fill=red, draw=red] (-4,0) circle (2cm);
\draw[fill=red, draw=red] ( 4,0) circle (2cm);
\draw[fill=tikzblue] (-4,0) circle (1.8cm);
\draw[fill=tikzblue] ( 4,0) circle (1.8cm);

\begin{scope}[xshift=-4cm]
\draw[] (-90:1.2cm) arc (-90:90:1.2cm);
\draw[] (90:1.2cm) arc (90:270:0.3cm);
\draw[] (-90:1.2cm) arc (270:90:0.3cm);
\end{scope}
\begin{scope}[xshift=4cm, xscale=-1]
\draw[] (-90:1.2cm) arc (-90:90:1.2cm);
\draw[] (90:1.2cm) arc (90:270:0.3cm);
\draw[] (-90:1.2cm) arc (270:90:0.3cm);
\end{scope}
\draw[fill=tikzblue, draw=none] (-4cm, 0.6cm) -- (4cm, 0.6cm) -- (4cm, -0.6cm) -- (-4cm, -0.6cm) -- cycle;
\draw[] (-4cm, 0.6cm) -- (4cm, 0.6cm);
\draw[] (-4cm, -0.6cm) -- (4cm, -0.6cm);
\draw[fill=white] (-4cm, 0.6cm) arc (-90:-270:0.3cm) arc (90:30:1.2cm) -- cycle;
\draw[fill=white] (-4cm, -0.6cm) arc (90:270:0.3cm) arc (270:330:1.2cm) -- cycle;
\begin{scope}[xscale=-1]
\draw[fill=white] (-4cm, 0.6cm) arc (-90:-270:0.3cm) arc (90:30:1.2cm) -- cycle;
\draw[fill=white] (-4cm, -0.6cm) arc (90:270:0.3cm) arc (270:330:1.2cm) -- cycle;
\end{scope}

\draw[fill=red, draw=red] (-2.5,-6) circle (2cm);
\draw[fill=red, draw=red] ( 5.5,-6) circle (2cm);
\draw[fill=white,draw=none] (-2.5,-6) circle (1.8cm);
\draw[fill=white,draw=none] ( 5.5,-6) circle (1.8cm);

\end{scope}

\end{scope}

\node[] () at (24cm, -3cm) {(c)};

\end{tikzpicture}

\caption{\textbf{(a)} Bridge region. \textbf{(b)} The two annuli of the bridge can be glued to different punctured spaces, creating a state whose existence implies reverse extendibility (Corollary~\ref{thm:immersion2}). \textbf{(c)} The bridge can be glued to a region along its two outer boundaries by a quantum channel (Theorem~\ref{thm:merging_thm}). \label{fig:bridge_applications}}
\end{figure}
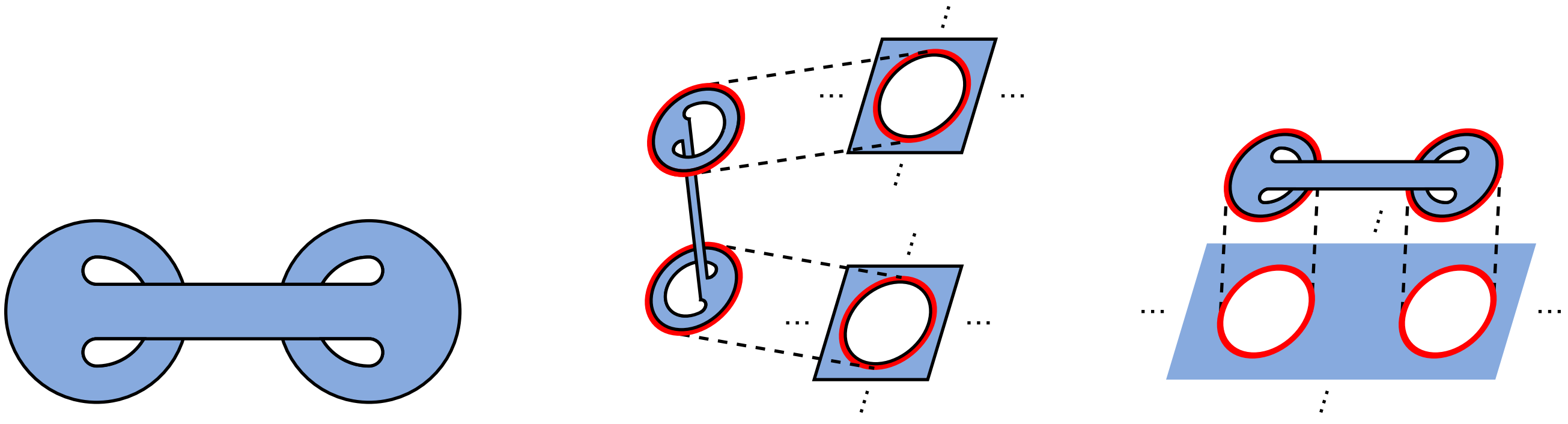

Here is a sketch of the argument. We first define a ``bridge'' region, which is a union of two disjoint annuli $A_1$ and $A_2$ connected by a tube through their inner boundaries. This is an immersed region, which we denote as $g:W\to \RR^d$; see Fig.~\ref{fig:bridge_applications}(a). We will define convex subsets of $\Sigma(W, g)$ consisting of states whose reduced density matrices on the annuli $A_1, A_2\subset W$ match those of the global state $\sigma$ defined in Proposition~\ref{prop:global-state}. This set, denoted as $\Sigma_{1,1}(W,g)$, shall be defined more precisely in Section~\ref{subsec:bridge}. In particular, we show they are nonempty in Section~\ref{subsec:rev_extend_merging}.

If such sets are nonempty, we can use an element of those sets to (i) create a state on an immersed region that implies reverse extendibility and (ii)attach a bridge to a region (Section~\ref{subsec:rev_extend_merging}). For the reverse extendibility, this is achieved by gluing an element of $\Sigma_{1,1}(W,g)$ to two independent punctured spaces and using Corollary~\ref{thm:immersion2}; see Fig.~\ref{fig:bridge_applications}(b). For merging, we can glue an element to a two-punctured space and remove the overlapping regions by a partial trace (Fig.~\ref{fig:bridge_applications}(c)). Note the gluing procedure in the latter case requires the reduced density matrix of the union of the annuli to be consistent with $\sigma$. This is stronger than the constraint put on $\Sigma_{1,1}(W,g)$, which only requires the reduced density matrix of each annulus to be consistent with $\sigma$ individually. 

\subsection{Bridge: overview}
\label{subsec:bridge}

We first formally define the bridge. Let $W$ be the smooth region obtained from two disjoint annuli $A_1,A_2$ by attaching the two end disks of a tube $T$ to disks in their respective inner boundary spheres. Topologically, the tube is $D^{d-1}\times[0,1]$. As a smooth manifold with boundary, $W$ is a closed $d$-ball with the interiors of two disjoint balls removed, since it is a boundary connected sum of two annuli along their inner spheres~\cite[Chapter~8, Theorem~3.1]{hirsch1976}. An immersion $g:W\to\RR^d$ that embeds each of the three pieces defines a bridge. Different pieces may have overlapping images. More generally, any region regular homotopic to such a region is called a bridge. As before, we primarily focus on tame bridges.

We start with a few remarks. One may consider a seemingly different type of region in which the two annuli are connected by their outer boundaries. Because such a region can be embedded and the aforementioned bridge region is immersed, one might think that these two types of regions are topologically inequivalent. However, a counterintuitive fact is that they are nonetheless regular homotopic to each other in some dimensions, for instance, in $d=3$~\cite{smale1959}. For this reason, we will consider a more refined notion we refer to as the \emph{standard bridge}. To that end, note that a bridge has three boundary components. Two of them are the outer spheres of the two annuli which we refer to as \emph{outer boundaries}, and we denote them as $N_1$ and $N_2$. The remaining one is called the \emph{inner boundary}, formed from the inner boundary spheres of the annuli and the surface of the tube connecting them. We write $W=A_1A_2T$ for the decomposition of a bridge into its two annuli and the tube, where $N_i$ is the outer sphere of $A_i$. We say the outer boundary $N_i$ is \emph{standard} if $g|_{N_i}$ is a tame embedded hypersurface (Section~\ref{sec:macro}) bounding a ball $F_i$ and $g$ maps a neighborhood of $N_i$ in $W$ into $F_i$. We say the annulus $A_i$ is \emph{standard} if the immersion $g:W\to\RR^d$ embeds it as a tame annulus $F_i\setminus\Int(I_i)$, where $I_i\subset F_i$ are tame balls and $F_i$ is bounded by $g(N_i)$. A bridge is \emph{standard} if both annuli are standard.

For standard tame bridges, we define a subset of its information convex set. For a subregion $Y$ of an immersed region on which the immersion is injective, we write $\sigma|_Y$ for the pullback along the immersion of the reduced density matrix of the global state $\sigma$ of Proposition~\ref{prop:global-state} on the image of $Y$. Let $K_i\subseteq A_i$ be the inner $r_2/2$-collar of $N_i$. We single out the states of $\Sigma(W,g)$ that agree with the global state on each collar:
\begin{equation}
\Sigma_{1,1}(W,g):= \{\rho \in \Sigma(W,g): \rho|_{K_1}= \sigma|_{K_1},\ \rho|_{K_2}= \sigma|_{K_2} \}.
\label{eq:bridge_ic_vacuum}
\end{equation}
Here $\sigma|_{K_i}$ is the reduced density matrix of the unique element of $\Sigma(F_i)$ (Proposition~\ref{prop:ball-unique}). Physically, the subscripts $1$ refer to the superselection sectors of the two annuli, which are the vacuum~\cite{shi2020fusion}. When $g$ is injective on $K_1K_2$ (for instance, when $F_1\cap F_2=\varnothing$), we can consider the following related subset:
\begin{equation}
\Sigma_{11}(W,g):= \{\rho \in \Sigma(W,g): \rho|_{K_1K_2}= \sigma|_{K_1K_2} \}.
\label{eq:bridge_ic_vacuum_joint}
\end{equation}
Clearly $\Sigma_{11}(W,g)\subseteq\Sigma_{1,1}(W,g)$.

Our main objective is to show that $\Sigma_{1,1}(W,g)$ is nonempty for any standard tame bridge. To that end, we note that it suffices to show the nonemptiness of some distinguished standard tame bridge. Let us begin with a geometric lemma.
\begin{Lemma}[Deforming bridges]
\label{lem:deform_bridge}
Any two standard tame bridges admit a tame deformation to one another along which the outer boundaries remain standard. If the balls $F_1$ and $F_2$ of both bridges are at distance at least $\delta\ge0$, the deformation can be chosen so that they remain at distance at least $\delta$ throughout.
\end{Lemma}
\begin{proof}
    It suffices to construct a regular homotopy between two bridges $g:W\to\RR^d$ and $g':W'\to\RR^d$ through immersions that embed each $N_i$ as the boundary of a smoothly embedded ball and map a neighborhood of $N_i$ into that ball. Then the tameness can be ensured by scaling up the regions; see Remark~\ref{rmk:scale-homotopy}.
    
    We now construct the regular homotopy. Let $K_i\subseteq A_i$ and $K_i'\subseteq A_i'$ be collars of the outer boundaries. Any two smoothly embedded balls are isotopic, and so are their collars. Such an isotopy from the ball of $g$ to that of $g'$ identifies $K_i$ with $K_i'$, and it deforms $g|_{K_i}$ to $g'|_{K_i}$ through embeddings onto collars of balls. The identification extends over the annuli. Since the disks along which the two tubes are attached are isotopic~\cite[Chapter~8, Theorem~3.1]{hirsch1976}, it extends to a diffeomorphism from $W$ to $W'$.
    
    The deformation of the collars extends to a regular homotopy of $W$ starting from $g$. Indeed, $W$ is obtained from $K_1\sqcup K_2$ by thickening the collars to $A_1\sqcup A_2$ and attaching the tube, which is a handle of index one. Since the index is less than $d$, the restriction of immersions of $W$ to $A_1\sqcup A_2$ has the covering homotopy property~\cite[Proposition~1.15]{weiss_immersion}. The regular homotopy ends at an immersion $g''$ that agrees with $g'$ on $K_1\sqcup K_2$. Deforming it further with $K_1\sqcup K_2$ fixed, we may assume that $g''$ agrees with $g'$ on $A_1\sqcup A_2$. We identify the tube with $D^{d-1}\times[0,1]$ and refer to the image of $\{0\}\times[0,1]$ as its {\emph{core}}. Along the core, the derivatives of $g''$ and $g'$ are two paths in $GL(d,\RR)$ with the same endpoints. They differ by an element of $\pi_1(SO(d))$, the twist. If the twist is trivial, $g''$ can be deformed to $g'$ with the annuli fixed~\cite[Theorem~1.3 and Proposition~1.15]{weiss_immersion}. A full rotation of $K_1$, inserted into the deformation of $K_1$, changes the twist by a generator. Sufficiently many such rotations remove the twist.

    Suppose now that the balls of both bridges are at distance at least $\delta$. The pairs of tame balls at distance at least $\delta$ form a connected space.  Indeed, we shrink each ball inside itself, move the small balls while keeping them at least $\delta$ apart, and expand each within its target ball.  Choosing the deformation of the collars along such a path, with the rotations performed while the ball of $K_1$ is small and round, we can keep the balls at distance at least $\delta$ throughout; scaling up preserves this as well.
\end{proof}

\begin{Proposition}
\label{prop:transport}
Let $F:W\times[0,1]\to\RR^d$ be a tame deformation from $(W,g_0)$ to $(W,g_1)$ along which the outer boundaries remain standard, where $g_0$ and $g_1$ are standard. Then $\Phi^F$ in Theorem~\ref{thm:isomorphism} maps $\Sigma_{1,1}(W,g_0)$ to $\Sigma_{1,1}(W,g_1)$. If moreover the balls $F_1$ and $F_2$ remain at distance $\gg r_1$ throughout, $\Phi^F$ also maps $\Sigma_{11}(W,g_0)$ to $\Sigma_{11}(W,g_1)$.
\end{Proposition}
\begin{proof}
    Note $\Phi^F$ is a composition of channels, each extending the region by a geometric circuit along its boundary and then applying a partial trace. Let $\rho_t$ be the state at step $t$ and $\hat{\rho}_t$ its $r_1$-consistent extension. Let $K_1$ at step $t$ denote the inner $r_2/2$-collar of $N_1$ at that step. We refine the subdivision in Eq.~\eqref{eq:ffdist-t} so that the inward normal parametrizations of $K_1$ at consecutive steps differ by less than $r_1$; this does not change $\Phi^F$. We show by induction that $\rho_t|_{K_1}=\sigma|_{K_1}$; the case of $K_2$ is identical, and the base case is the hypothesis $\rho_0\in\Sigma_{1,1}(W,g_0)$.

    The causal base (Section~\ref{subsec:iso_sketch}) of $K_1$ at step $t+1$ lies within distance $2c_{\mathrm{step}}r_1$ of $K_1$ at step $t$ (Remark~\ref{rmk:iso-steps}). As such, it lies in the inner $(r_2/2+2c_{\mathrm{step}}r_1)$-collar of $N_1$ together with the outer collar of $W$ along $N_1$. This region is embedded: $N_1$ is standard and $W$ is tame, so the inner $r_2$-collar of $N_1$ is mapped onto the inner $r_2$-collar of $\partial F_1$, and $c_{\mathrm{step}}r_1\ll r_2$.

    Applying Proposition~\ref{prop:collar-recovery} to the hypersurfaces parallel to $N_1$ at depth $r_2/2+s$, for $s=0,r_1,2r_1,\dots$, and then outward from $N_1$, we find that $\hat{\rho}_t$ on this region is determined by $\rho_t|_{K_1}$ through channels that do not depend on the state. These channels map $\sigma|_{K_1}$ to $\sigma$ by Lemma~\ref{lem:extension_reference}, so $\hat{\rho}_t=\sigma$ on the causal base. The gates of the causal cone form a geometric circuit from the base to $K_1$ at step $t+1$, and the cone is embedded; by the same lemma, $\rho_{t+1}|_{K_1}=\sigma|_{K_1}$.

    For the joint condition, we apply the same argument to $K_1\sqcup K_2$. The two causal cones have disjoint images when $F_1$ and $F_2$ are at distance $\gg r_1$, as they lie in the $c_{\mathrm{step}}r_1$-neighborhoods of $F_1$ and $F_2$; so their union is embedded, and Lemma~\ref{lem:extension_reference} applies to it.

\end{proof}

\begin{Lemma}
    \label{lem:extension_reference}
Consider a geometric circuit whose gates are copies of $(BCDE, BE, BC)$ for configurations in $\Lambda(r_0,r_1)$, transforming $Y$ to $Y'$. Denote the resulting channel as $R_{Y\to Y'}$. Then
\[
R_{Y\to Y'}[\sigma|_Y] = \sigma|_{Y'}.
\] 
\end{Lemma}
\begin{proof}
The claim follows from the induction over gates as in Eq.~\eqref{eq:channel-circ-1}. Each gate's $BCDE$ lies in an $r_1$-ball $L$, and $\sigma|_L=\sigma_L$ is a reference patch, hence $(B,C,D,E)$-extendible. Taking $A=Y\setminus BCDE$, the global state $\sigma$ is an extension of $\sigma_L|_{BCDE}$, so Eq.~\eqref{eq:loc_extend} yields the claim.
\end{proof}

Thus it suffices to establish the nonemptiness of $\Sigma_{1,1}(W_0,g_0)$ for some distinguished standard tame bridge. We will in fact prove a stronger statement by showing that $\Sigma_{11}(W_0, g_0)$ is nonempty. This is done by constructing elements of the \emph{approximate information convex set}; see Definition~\ref{def:ICS-immersed} for comparison.
\begin{Definition}
    \label{def:approximate_ic}
The $\eps$-\emph{approximate information convex set} of a region $Y$ is the set $\Sigma_{\eps}(Y)$ of states $\rho\in\St_Y$ that admit an extension $\rho_+\in \St_{Y^+}$ satisfying $\| (\rho_+)|_L - \sigma_L \|_1 \le \eps$ for all $L\in\Gamma_1(Y)$.
\end{Definition}
\noindent
This definition straightforwardly generalizes to the immersed regions (Definition~\ref{def:ICS-immersed}), which we would denote as $\Sigma_{\eps}(M,f)$ for an immersion $f:M\to \RR^d$. For a standard tame bridge $g:W\to\RR^d$ with $F_1\cap F_2=\varnothing$ we further set
\begin{equation}
\Sigma_{11,\eps}(W,g) := \{ \rho \in \Sigma_{\eps}(W,g): \|\rho|_{K_1K_2} - \sigma|_{K_1K_2} \|_1\le \eps \}.
\end{equation}
We will construct states $\rho_n$ with $(\rho_n)|_{W_0}\in \Sigma_{11,\eps_n}(W_0,g_0)$ and $\eps_n\downarrow 0$; the nonemptiness of $\Sigma_{11}(W_0,g_0)$ then follows by compactness.

We build $\rho_n$ on a bridge by gluing the two annuli to a tube (which is topologically a ball); see Fig.~\ref{fig:bridge_geometry}(a). Using Lemma~\ref{lem:existence_small_cmi}, one can show that for a sufficiently long tube its conditional mutual information can be made arbitrarily small. Thus one may hope to use the Fawzi-Renner theorem~\cite{fawzi2015quantum} to glue the tube to the two annuli. 

To make this approach work, two issues need to be resolved. First, Lemma~\ref{lem:existence_small_cmi} only guarantees small conditional mutual information for some partition which we cannot choose. For instance, the subsystem that has small conditional mutual information may be too short to glue with the annuli. Thus it is necessary to elongate the tube whilst ensuring the small conditional mutual information condition remains intact. Secondly, the compactness argument for establishing the nonemptiness of $\Sigma_{11}(W_0,g_0)$ requires $W_0$ to be a fixed region. Thus we need to apply a channel that shortens the arbitrarily long tube to a tube of fixed length, obtaining a fixed-size bridge. These are dealt with by the channels $\mathcal{E}_k$ and $\mathcal{S}_n$ constructed in Section~\ref{subsec:geometric_details}; see Fig.~\ref{fig:bridge_geometry}(b).

Both channels are constructed from geometric circuits and partial traces. One complication is that the tube is only locally tame (Definition~\ref{def:locally_tame}), so the isomorphism theorem (Theorem~\ref{thm:isomorphism}) does not apply. Nonetheless, the states we transform are reduced density matrices of the global state $\sigma$ (Proposition~\ref{prop:global-state}), and Lemma~\ref{lem:extension_reference} applies to them with no tameness requirement. Thus we conclude that the density matrices obtained this way are again reduced density matrices of the same global state. This is enough to show that $\Sigma_{11}(W_0,g_0)$ is nonempty.

The construction of these channels are geometric in nature, which will be the focus on Section~\ref{subsec:geometric_details}.

\subsection{Geometric details}
\label{subsec:geometric_details}

Throughout this section we work with a particular geometry we refer to as the \emph{hairpin geometry}. It will be used to define the specific bridge we consider as well as the geometric circuits transforming it. Generally, we can specify a region $\mathcal{C}_{\gamma, R}$ in terms of a curve $\gamma: [0,L] \to \mathbb{R}^d$, parametrized by arclength, and a thickness function $R(t)\ge 0$:
\begin{equation}
    \label{eq:cylinder}
\mathcal{C}_{\gamma, R}|_{[a,b]}:= \{\gamma(t)+ v| v\perp \dot{\gamma}(t), |v|\le R(t), t\in [a,b] \}.
\end{equation}
(If $[a,b]=[0,L]$, we simply denote this as $\mathcal{C}_{\gamma, R}$.) We refer to a region of this form as a \emph{tube}. We can also define its boundary hypersurface as 
\begin{equation}
    \label{eq:cylinder_boundary}
\partial \mathcal{C}_{\gamma, R}|_{(a,b)}:= \{\gamma(t)+ v| v\perp \dot{\gamma}(t), |v|= R(t), t\in (a,b) \}.
\end{equation}
The hairpin geometry is obtained from a curve of the following form:
\begin{equation}
\gamma(t) =
\begin{cases}
\gamma_1(t) & \quad 0\le t <\ell, \\
\gamma_2(t-\ell) & \quad \ell \le  t\le \ell+s_{\rm arc}, \\
\gamma_3(t-\ell-s_{\rm arc}) & \quad \ell+s_{\rm arc} < t \le L,
\end{cases}
\end{equation}
where
\begin{equation}
\begin{aligned}
\gamma_1(t) &= (t, r, 0, \ldots, 0), \\
\gamma_2(t) &= (\ell, 0, \ldots, 0) + r\beta(t/r), \\
\gamma_3(t) &= (\ell-t, -r, 0, \ldots ,0).
\end{aligned}
\end{equation}
Here $\beta$ is a fixed smooth embedded curve in the $(x_1,x_2)$-plane from $(0,1)$ to $(0,-1)$, parametrized by arclength. It lies in $\{x_1\ge 0\}$ and is straight and parallel to the $x_1$-axis near its endpoints, e.g., a semicircle which is smoothed where it meets $\gamma_1$ and $\gamma_3$.  The curve $\gamma$ is parametrized by arclength, with total length $L=2\ell+s_{\rm arc}$, where $s_{\rm arc}$ is the length of $\gamma_2$. We refer to the images of $\gamma_1$ and $\gamma_3$ as the \emph{arms} and the image of $\gamma_2$ as the \emph{arc}. The arms have length $\ell$ and separation $2r$. The curve $\gamma$ has an embedded normal tube of radius proportional to $r$, independent of $\ell$. We take the thickness $R$ to be constant on the arc and at most $w$, with $w\ll r$, so that this normal tube contains $\mathcal{C}_{\gamma,R}$ and the normal disks over the curved part of the arc lie in $\{x_1>\ell\}$.

This tube is partitioned into
\begin{equation}
C= \mathcal{C}_{\gamma, R}|_{\left[0, \delta\right)} \sqcup \mathcal{C}_{\gamma, R}|_{\left(L-\delta, L \right]}, \quad \overline{X}_n = \mathcal{C}_{\gamma, R}\setminus C, \qquad \delta:=\frac{\ell}{n+1}.
\end{equation}
$\overline{X}_n$ is further partitioned into $\overline{X}_n= Y_nZ_1\cdots Z_n$, where
\begin{equation}
Z_k = \mathcal{C}_{\gamma, R}|_{\left[\ell-k\delta,\, \ell-(k-1)\delta\right)} \sqcup \mathcal{C}_{\gamma, R}|_{\left(L-\ell+(k-1)\delta,\, L-\ell+k\delta\right]}, \quad Y_n= \overline{X}_n \setminus Z^{(n)}.
\end{equation}
We assume $R(t)\gg r_1$ for all $t$. For the initial bridge, we take the thickness to be constant, $R(t)=w/2$ with $r_2\ll w\ll r$, and set $\ell=(n+1)w$, so that $\delta=w$; thickness functions varying with $t$ will describe the intermediate regions arising below. We do not specify the geometry of $\bar{A}$, aside from demanding a few conditions. $\bar{A}$ is partitioned into $\bar{A}=AS$, where $A$ consists of two disjoint tame annuli, whose balls $F_1$ and $F_2$ obtained from the outer boundary are at distance $\gg r_1$, and $S$ consists of the two initial segments of the tube, each attached to the inner boundary of an annulus and crossing it. We choose the two initial segments disjoint from each other so that, together with the hairpin, they form an embedded tube. Thus $A$ is embedded, while $\bar{A}$ is immersed. We also assume that $\bar{A}$ lies in $x_1<0$ and that $\overline{W}_n=\bar{A}C\overline{X}_n$ is $r'$-tame for some $r'\gg r_2$.

\begin{figure}[t]
    \begin{tikzpicture}[line width=1pt, scale=0.315,
                        every node/.style={font=\small}]

\draw[fill=tikzblue] (-3,0) circle (2cm);
\begin{scope}[xshift=-3cm]
\draw[] (-90:1.2cm) arc (-90:90:1.2cm);
\draw[] (90:1.2cm) arc (90:270:0.3cm);
\draw[] (-90:1.2cm) arc (270:90:0.3cm);
\end{scope}
\draw[fill=tikzblue, draw=none] (-3cm, 0.6cm) -- (7.5cm, 0.6cm) -- (7.5cm, -0.6cm) -- (-3cm, -0.6cm) -- cycle;
\draw[] (-3cm, 0.6cm) -- (7.5cm, 0.6cm);
\draw[] (-3cm, -0.6cm) -- (7.5cm, -0.6cm);

\foreach \x in {1, ..., 5}
{
\draw[dotted] (1.5cm * \x, 0.6cm ) -- ++ (0, -1.2cm);
}

\draw[fill=white] (-3cm, 0.6cm) arc (-90:-270:0.3cm) arc (90:30:1.2cm) -- cycle;
\draw[fill=white] (-3cm, -0.6cm) arc (90:270:0.3cm) arc (270:330:1.2cm) -- cycle;

\begin{scope}[yshift= -6cm]
\draw[fill=tikzblue] (-3,0) circle (2cm);
\begin{scope}[xshift=-3cm]
\draw[] (-90:1.2cm) arc (-90:90:1.2cm);
\draw[] (90:1.2cm) arc (90:270:0.3cm);
\draw[] (-90:1.2cm) arc (270:90:0.3cm);
\end{scope}
\draw[fill=tikzblue, draw=none] (-3cm, 0.6cm) -- (7.5cm, 0.6cm) -- (7.5cm, -0.6cm) -- (-3cm, -0.6cm) -- cycle;
\draw[] (-3cm, 0.6cm) -- (7.5cm, 0.6cm);
\draw[] (-3cm, -0.6cm) -- (7.5cm, -0.6cm);
\draw[fill=white] (-3cm, 0.6cm) arc (-90:-270:0.3cm) arc (90:30:1.2cm) -- cycle;
\draw[fill=white] (-3cm, -0.6cm) arc (90:270:0.3cm) arc (270:330:1.2cm) -- cycle;

\foreach \x in {1, ..., 5}
{
\draw[dotted] (1.5cm * \x, 0.6cm ) -- ++ (0, -1.2cm);
}
\end{scope}

\begin{scope}[xshift=4.5cm]
\draw[fill=tikzblue, draw=none] (3cm, 0.6cm) arc (90:-90:3.6cm) -- ++ (0, 1.2cm) arc (-90:90:2.4cm) -- cycle;
\draw[] (3cm, 0.6cm) arc (90:-90:3.6cm);
\draw[] (3cm, -0.6cm) arc (90:-90:2.4cm);
\node[] () at (6cm, -3cm) {$Y_4$};
\end{scope}

\draw[dashed] (0, -8.4cm) -- (0, 2.4cm);
\node[below] () at (0, -8.4cm)  {$x_1=0$};

\node[] () at (-3cm, 0) {$\bar{A}$};
\node[] () at (-3cm, -6cm) {$\bar{A}$};

\node[] () at (0.75cm, 0) {$C$};
\node[] () at (0.75cm, -6cm) {$C$};

\node[] () at (2.25cm, 0) {$Z_4$};
\node[] () at (2.25cm, -6cm) {$Z_4$};

\node[] () at (3.75cm, 0) {$Z_3$};
\node[] () at (3.75cm, -6cm) {$Z_3$};

\node[] () at (5.25cm, 0) {$Z_2$};
\node[] () at (5.25cm, -6cm) {$Z_2$};

\node[] () at (6.75cm, 0) {$Z_1$};
\node[] () at (6.75cm, -6cm) {$Z_1$};

\draw[Stealth-Stealth] (0cm, 1.2cm) -- (7.5cm, 1.2cm) node [midway, above] {$\ell$};

\draw[Stealth-Stealth] (7.5cm, -3cm) -- ++ (2.1213cm, 2.1213cm) node[midway, above] {$r$};

\node[] () at (3.05, -10.4) {(a)};

\def\hairpin#1#2{(#1,0.6) -- (#2,0.6) arc (90:-90:3.6) -- (#1,-6.6)
   -- (#1,-5.4) -- (#2,-5.4) arc (-90:90:2.4) -- (#1,-0.6) -- cycle}

\begin{scope}[xshift=20.6cm]

  \begin{scope}[xshift=-8cm]
    \fill[tikzblue] \hairpin{3}{7.5};
    \draw \hairpin{3}{7.5};
    \foreach \x in {4.5, 6, 7.5} {
      \draw[dotted] (\x, 0.6) -- (\x, -0.6);
      \draw[dotted] (\x, -5.4) -- (\x, -6.6);
    }
    \foreach \x/\lb in {3.75/{Z_3}, 5.25/{Z_2}, 6.75/{Z_1}} {
      \node[] () at (\x, 0) {$\lb$};
      \node[] () at (\x, -6) {$\lb$};
    }
    \node[] () at (10.5, -3) {$Y_4$};
  \end{scope}

  \draw[-{Latex[length=2.5mm]}, line width=0.9pt] (3.7, -3) -- (5.7, -3)
    node[midway, above] {$\mathcal{E}_k$};

  \begin{scope}[xshift=6.3cm]
    \fill[tikzblue] \hairpin{0}{7.5};
    \draw \hairpin{0}{7.5};
    \foreach \x in {1.5, 3, 4.5, 6, 7.5} {
      \draw[dotted] (\x, 0.6) -- (\x, -0.6);
      \draw[dotted] (\x, -5.4) -- (\x, -6.6);
    }
    \foreach \p in {(3.75,0.6), (3.75,-0.6), (3.75,-5.4), (3.75,-6.6)}
      { \fill[white] \p circle (0.20); }
    \begin{scope}
      \clip \hairpin{0}{7.5};
      \foreach \p in {(3.75,0.6), (3.75,-0.6), (3.75,-5.4), (3.75,-6.6)}
        { \draw \p circle (0.24); }
    \end{scope}
    \foreach \x/\lb in {0.75/{C}, 2.25/{Z_4}, 3.75/{Z_3}, 5.25/{Z_2}, 6.75/{Z_1}} {
      \node[] () at (\x, 0) {$\lb$};
      \node[] () at (\x, -6) {$\lb$};
    }
    \node[] () at (10.5, -3) {$Y_4$};
  \end{scope}

  \draw[-{Latex[length=2.5mm]}, line width=0.9pt] (18, -3) -- (20, -3)
    node[midway, above] {$\mathcal{S}_n$};

  \begin{scope}[xshift=20.6cm]
    \path[fill=tikzblue,draw]
      (0,0.6) -- (1.875,0.6) .. controls (1.95,0.6) and (1.99,0.48) .. (2.065,0.48)
      -- (4.5,0.48) arc (90:-90:3.48) -- (2.065,-6.48)
      .. controls (1.99,-6.48) and (1.95,-6.6) .. (1.875,-6.6) -- (0,-6.6)
      -- (0,-5.4) -- (1.875,-5.4) .. controls (1.95,-5.4) and (1.99,-5.52) .. (2.065,-5.52)
      -- (4.5,-5.52) arc (-90:90:2.52) -- (2.065,-0.48)
      .. controls (1.99,-0.48) and (1.95,-0.6) .. (1.875,-0.6) -- (0,-0.6) -- cycle;
    \draw[dotted] (1.5, 0.6) -- (1.5,-0.6) (1.5,-5.4) -- (1.5,-6.6);
    \node[] () at (0.75, 0) {$C$};
    \node[] () at (0.75, -6) {$C$};
    \node[] () at (7.5, -3) {$X$};
  \end{scope}

  \node[] () at (10.35, -10.4) {(b)};
\end{scope}
    
    \end{tikzpicture}

    \caption{\textbf{(a)} The geometric description of the bridge for $n=4$. \textbf{(b)} Schematic descriptions of the channels $\mathcal{E}_k$ and $\mathcal{S}_n$. Here $k=3$ and therefore $\mathcal{E}_k$ acts only on $Z_3$. $\mathcal{S}_n$ transforms the long tube to a shorter tube. \label{fig:bridge_geometry} }

\end{figure}
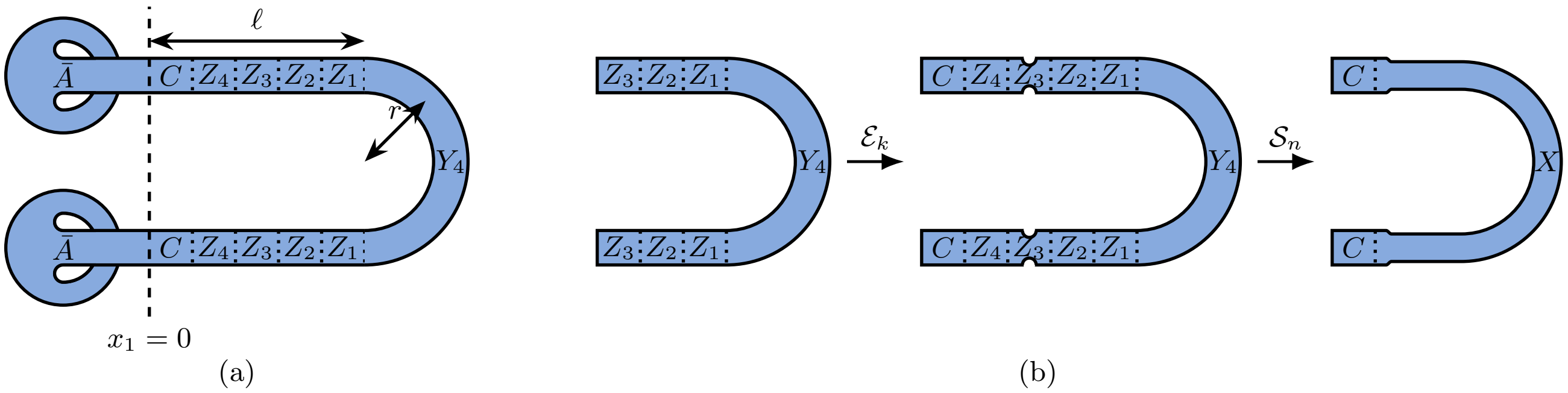

Now we construct the two channels using this geometry. Let $k$ be an integer such that $I(Y_n:Z_{k}|Z^{(k-1)})_{\sigma} \le O(1/n)$ (Lemma~\ref{lem:existence_small_cmi}). We aim to construct two channels: (i) $\mathcal{E}_k$, which extends $\sigma|_{Y_n Z^{(k)}}$ to $\sigma|_{Y_nZ^{(n)}C\setminus E_k} $ and (ii) $\mathcal{S}_n$, which maps $\sigma|_{Y_nZ^{(n)}C\setminus E_k} $ to the reduced density matrix on a shorter tube (Fig.~\ref{fig:bridge_geometry}(b)). Both channels can be realized by a sequence of geometric circuits and partial traces.

\begin{figure}[h]
    \centering
    \includegraphics{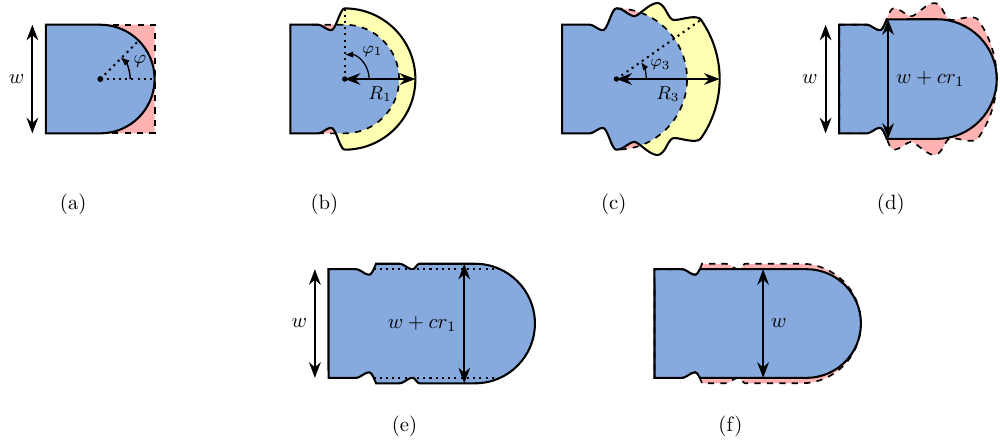}%
    \caption{\textbf{(a)} After tracing out the red region, $Z_k$ becomes a spherical cap attached to a tube. The black dot is the origin of the coordinate system used in this analysis. \textbf{(b,c)} Using Theorem~\ref{thm:geom_circuit_local}, the tame subregion of the hypersurface advances (yellow), leaving an erased region (red) behind. \textbf{(d)} After tracing out an excess region, we obtain a tube with a diameter expanded by $cr_1$. \textbf{(e)} The steps from (b) to (d) can be repeated indefinitely while ensuring the diameter never dips below $w$ (the dotted line) except at the initially removed region. \textbf{(f)} Tracing out the excess region, we obtain an elongated tube whose length can be made arbitrarily long. \label{fig:elongation}}
\end{figure}

We first describe the sequence of geometric circuits and partial traces underlying $\mathcal{E}_k$. The two components of $Z_k$ are treated identically; we place the origin at the center of one of them, orient the $x_1$-axis along $\gamma$ so that $C$ lies in the positive direction, and write $x=(x_1,x_\perp)$, with $\varphi=\cos^{-1}(x_1/\|x\|)\in[0,\pi]$ the polar angle. First, we smoothen $Z_k$ by tracing out $\{\|x\|> w/2,\ \varphi\le\pi/2\}$; see Fig.~\ref{fig:elongation}(a). The intersection of the remaining region with $\{x_1\ge 0,\ \|x\|\le w\}$ is the half-ball $\bar{\Ba}_{w/2}(0)\cap\{x_1\ge 0\}$, the rest of the region lying in $\{x_1<0\}$ or at distance at least $2r-w/2>w$ from the origin. The region is then elongated by repeatedly using the following lemma.

\begin{Lemma}\label{lem:elongation_step}
Let $u'=w/2+cr_1/2$, and let $M$ be a compact region such that $M\cap\{x_1\ge 0,\ \|x\|\le w\}$ is the half-ball $\bar{\Ba}_u(0)\cap\{x_1\ge 0\}$ for some $u\in[w/2,u']$. There is a constant $c'>0$ such that $M$ can be transformed into
\begin{equation}\label{eq:elongation_step}
M'=(M\cup G)\setminus E,\qquad
G=\bar{\Ba}_{u'}\bigl(\{0\le x_1\le c'w,\ x_\perp=0\}\bigr)\cap\{x_1\ge 0\},
\end{equation}
for some $E\subseteq\{0\le x_1\le 10r_1,\ u-cr_1/2<|x_\perp|\le w\}$, by a sequence of constant-depth geometric circuits with gates as described in Theorem~\ref{thm:geom_circuit_local}, supported in $\{x_1\ge -2r_1\}$, followed by a partial trace.
\end{Lemma}

\begin{proof}
The circuits are obtained by iterating Theorem~\ref{thm:geom_circuit_local} along a sequence of spherical caps,
\begin{equation}\label{eq:elong_caps}
N_m=\{\|x\|=R_m,\ \varphi\le\varphi_m\},\qquad R_m=u+cr_1m,\quad
\varphi_m=\frac{\pi}{2}-\frac{(8m+1)r_1}{w},
\end{equation}
for $0\le m<m^*=\lfloor cw/(2048\ts r_1)\rfloor$, with $\widetilde{N}_m=\{\|x\|=R_m,\ \varphi<\varphi_m-7r_1/w\}$; see Fig.~\ref{fig:elongation}(b,c). At $m=0$, the region contains the half-ball, whose radial segments provide the inner collar; thus $M$ is $(u/2)$-tame along $N_0$. Since $\varphi_0<\pi/2$, $N_0$ lies in the smooth part of $\partial M$. At every step, $\dist_{N_m}(\widetilde{N}_m,\partial N_m)\ge 7r_1R_m/w\ge 3.5\ts r_1>3r_1$, as the theorem requires. Moreover, $N_{m+1}\subseteq\widetilde{N}_m\times\{cr_1\}$ is a round sphere of radius $R_{m+1}\ge u$ whose inward radial segments of length $u/2$ lie in the region, away from the removals; so the region remains $(u/2)$-tame along $N_{m+1}$. By the first part of Theorem~\ref{thm:geom_circuit_local}, the $m$-th application adds a region containing the shell $\{R_m<\|x\|\le R_{m+1},\ \varphi\le\varphi_{m+1}\}$, and its gates lie in admissible copies of balls of radius $r_1$ centered on $N_m\subseteq\{x_1\ge 0\}$, hence in $\{x_1\ge -2r_1\}$.

By the second part of the theorem, points are removed only near where the cap ends, namely in
\begin{equation}
V_m=\{R_m-cr_1/3<\|x\|\le R_m,\ \ \varphi_m-7r_1/w\le\varphi\le\varphi_m\}.
\end{equation}
For $x\in V_m$, the bounds $R_m\le w$, $\sin t\le t$, and $\cos t\ge 1-t^2/2$ give
\begin{equation}\label{eq:erosion_bounds}
x_1\le (8m+8)r_1,\qquad
|x_\perp|\ \ge\ u+cr_1\Bigl(m-\frac{1}{3}\Bigr)-\frac{32r_1^2(m+1)^2}{w}.
\end{equation}
Since $w\gg r_1$, $V_0$ lies in the set specified for $E$. For $0<m<m^*$, each $V_m$ lies in $\{\|x\|>u,\ |x_\perp|>u'\}$. In particular, the removals intersect $M\cup G$ only at $m=0$, within the set specified for $E$.

For $c'=c^2/4096$, the region obtained after $m^*$ steps contains every point of $G$ outside the set specified for $E$; the concluding partial trace then yields $M'$ exactly. To see this, let $x$ be such a point and $\varrho=\|x\|$. If $\varrho\le u$, then $x\in M$. Otherwise $x_1\ge 10r_1$, since $x_1\le 10r_1$ would force $|x_\perp|\le u-cr_1/2$ and hence $\varrho<u$. Since $x\in G$ lies within distance $u'$ of the segment from $0$ to $(c'w,0,\ldots,0)$, the triangle inequality gives $\varrho\le c'w+u'\le w$. So we have $R_m<\varrho\le R_{m+1}$ for some $m\ge 0$ with $m\le c'w/(cr_1)+1<m^*$. (The last inequality follows from the choice of $c'$ and $w\gg r_1$.) For $m=0$, $x_1/\varrho\ge 10r_1/w$; for $m\ge 1$, $\varrho-u'\ge cr_1(m-1/2)\ge cr_1m/2$ and $\varrho+u'\ge w$, so that
\begin{equation}
x_1\ \ge\ \sqrt{(\varrho-u')(\varrho+u')}\ \ge\ \sqrt{cr_1mw/2}\ \ge\ (8m+9)r_1.
\end{equation}
In either case $\cos\varphi(x)=x_1/\varrho\ge\sin\bigl((8m+9)r_1/w\bigr)=\cos\varphi_{m+1}$, so $x$ belongs to the shell added at the $m$-th step.
\end{proof}

We apply Lemma~\ref{lem:elongation_step} repeatedly, first to the smoothed region with $u=w/2$ and then translating the origin by $c'w$ along the $x_1$-axis after each application, with $u=u'$. In the first step, a region in $|x_\perp|>w/2-cr_1/2$ is removed. Afterwards, the radius of the half-ball is increased to $u'>w/2$, ensuring the erased regions in these steps remain at $|x_\perp|>w/2$; see Fig.~\ref{fig:elongation}(d,e). In the end, the resulting region contains $Y_nZ^{(n)}C\setminus E_k$ for some
\begin{equation}\label{eq:Ek_shape}
E_k\subseteq\{x\in Z_k:\ 0\le x_1\le 10r_1,\ |x_\perp|>w/2-cr_1/2\}.
\end{equation}
(The initially traced-out part of $Z_k$ is recreated by $G$ except within this set.) A final partial trace reduces the region to $Y_nZ^{(n)}C\setminus E_k$ exactly; see Fig.~\ref{fig:elongation}(f).

Throughout this process, every gate acts on systems contained in $Z_k$ or in regions created by preceding gates, so the composition of these circuits and partial traces defines a channel $\mathcal{E}_k:\St_{Z_k}\to\St_{Z_k\cdots Z_nC\setminus E_k}$ acting nontrivially only on $Z_k$. By Lemma~\ref{lem:extension_reference}, each such circuit (as well as each partial trace) maps a reduced density matrix of $\sigma$ to a reduced density matrix. Therefore,
\begin{equation}\label{eq:Ek_extension}
(\id_{Y_nZ^{(k-1)}}\otimes\mathcal{E}_k)[\sigma|_{Y_nZ^{(k)}}]=\sigma|_{Y_nZ^{(n)}C\setminus E_k}.
\end{equation}

We now turn to the construction of $\mathcal{S}_n$, which retracts the arc of the hairpin; see Fig.~\ref{fig:bridge_geometry}(b). For $\lambda\in[3w,\ell]$, let $\gamma^{(\lambda)}$ denote the hairpin curve with the arms shortened to length $\lambda$, so that $\gamma^{(\ell)}=\gamma$. Note that every point of $\gamma^{(\lambda-s)}$ lies within distance $s$ of $\gamma^{(\lambda)}$. Each application of Lemma~\ref{lem:retraction_step} transforms the tube around $\gamma^{(\lambda)}$ into the tube around $\gamma^{(\lambda-\tau)}$ for a retraction $\tau$ of order $cw$; iterating moves the arc to $\lambda=3w$, and a final partial trace yields a tube of fixed size. 

Throughout this discussion we refer to a thickness function, regarded as a function of the arclength, as a \emph{profile}. The conditions stated in terms of $x_1$ are imposed on both arms. Let $p$ be an integer in $[256/c,512/c]$, and let $\tau=w/p$ and $\Delta=cr_1/2$. Let $h:\RR\to[0,1]$ be a smooth nondecreasing function that equals $0$ on $(-\infty,1/4]$ and $1$ on $[3/4,\infty)$. Let $\chi$ be a smooth nondecreasing function of $x_1/w$ that equals $0$ for $x_1\le5w/4$ and $1$ for $x_1\ge11w/8$. We consider the following profiles on the arms:
\begin{equation}\label{eq:retract_profile}
R_\lambda(x_1)=w/2-\Delta\chi(x_1)-\Delta h\bigl((x_1-q_\lambda)/\tau\bigr),\qquad q_\lambda:=\lambda-3w/16.
\end{equation}
On the arc, $R_\lambda=w/2-2\Delta$. (This is also the value of $R_\lambda$ near the end of each arm, since $\lambda\ge3w$.) Let $T_\lambda:=\mathcal C_{\gamma^{(\lambda)},R_\lambda}$. For $w\gg r_1$, the profiles satisfy
\begin{equation}\label{eq:profile_cond}
w/2-2\Delta\le R_\lambda\le w/2,\qquad |R_\lambda'|\le\sqrt{cr_1/w},\qquad |R_\lambda''|\le 1/w
\end{equation}
uniformly in $\lambda$. Indeed, the first and second derivatives of $R_\lambda$ are $O(\Delta/\tau)$ and $O(\Delta/\tau^2)$, respectively.

The hypersurface $\partial T_\lambda$ of Eq.~\eqref{eq:cylinder_boundary} has an embedded normal collar of depth $w/10$ on both sides, with principal curvatures at most $10/w$. Indeed, consider the hypersurface obtained by displacing $\partial T_\lambda$ by $s$ along the outward unit normals, where $|s|\le w/10$. Over an arm, the displaced point lies at distance $R_\lambda+s/\sqrt{1+(R_\lambda')^2}>0$ from the arm, and its first coordinate is $x_1-sR_\lambda'/\sqrt{1+(R_\lambda')^2}$. The latter is increasing in $x_1$ by Eq.~\eqref{eq:profile_cond}. Thus each displaced hypersurface is a graph over the arm, and at a fixed first coordinate its distance from the arm increases with $s$. Hence these graphs are disjoint. Over the arc, the displaced hypersurfaces are the boundaries of the tubes of thickness $w/2-2\Delta+s$. They lie in the embedded normal tube of $\gamma^{(\lambda)}$ and agree with the graphs near the ends of the arms. The same formulas bound the principal curvatures.

\begin{figure}[t]
    \centering
    \includegraphics{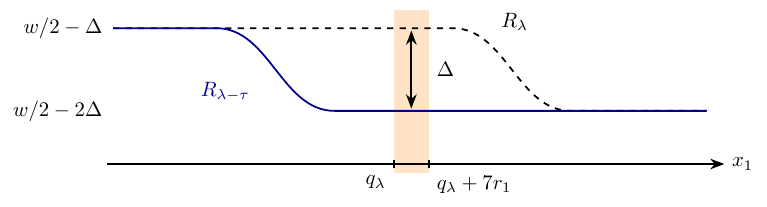}%
    \caption{The profiles $R_\lambda$ (dashed) and $R_{\lambda-\tau}$ (solid) on an arm, for $x_1\ge11w/8$. On the shaded interval they differ by $\Delta$.\label{fig:retraction_step}}
\end{figure}

\begin{figure}[tp]
    \centering
    \includegraphics{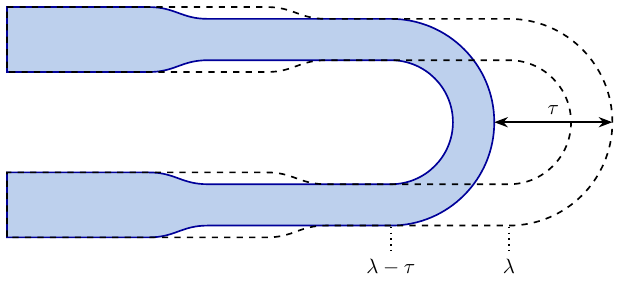}%
    \caption{The tubes $T_\lambda$ (dashed outline) and $T_{\lambda-\tau}$ (blue). Translating the arc of $T_{\lambda-\tau}$ by $\tau$ gives the arc of $T_\lambda$.\label{fig:retraction}}
\end{figure}

\begin{Lemma}\label{lem:retraction_step}
Let $3w+\tau\le\lambda\le\ell$. The tube $T_\lambda$ can be transformed into $T_{\lambda-\tau}$ by a sequence of constant-depth geometric circuits with gates as described in Theorem~\ref{thm:geom_circuit_local}, supported in $\{x_1\ge\lambda-w/2\}$, followed by a partial trace; see Fig.~\ref{fig:retraction}.
\end{Lemma}

\begin{proof}

The circuits are obtained by iterating Theorem~\ref{thm:geom_circuit_local} along a sequence of hypersurfaces enclosing the end of the tube. The relevant hypersurfaces are defined as follows:
\begin{equation}\label{eq:retract_surfaces}
N_m=\bigl(\partial T_\lambda\times\{cr_1m\}\bigr)\cap\{x_1\ge q_\lambda+8r_1m\},
\end{equation}
where $0\le m<m^*=\lceil 2\tau/(cr_1)\rceil$. (Here $\times\{t\}$ denotes displacement by $t$ along the outward unit normals.) Since $cr_1m^*\le 3\tau\ll w$, all displacements $cr_1m$ with $m<m^*$ stay within the collar of depth $w/10$. We apply the theorem once along each $N_m$, growing the region over the truncated hypersurface $\widetilde{N}_m=N_m\cap\{x_1>q_\lambda+8r_1m+7r_1\}$ in the normal direction.

For each $N_m$ and $\widetilde{N}_m$, we now verify the hypotheses of Theorem~\ref{thm:geom_circuit_local}. For $m=0$, the tube $T_\lambda$ is $(w/10)$-tame along $N_0$. At every step, $\widetilde{N}_m$ is separated from $\partial N_m$ by $7r_1$ in the $x_1$-coordinate, so $\dist_{N_m}(\widetilde{N}_m,\partial N_m)\ge 7r_1>3r_1$, as the theorem requires. Moreover, each application of the theorem leaves the region tame along $\widetilde{N}_m\times\{cr_1\}$, at a scale reduced by $r_1$; since $m^*r_1\le w/128+r_1$, the scale remains $\gg r_1$ throughout. This covers $N_{m+1}$: displacements along the normals compose, and each changes $x_1$ by at most $cr_1<r_1$, so $N_{m+1}\subseteq\widetilde{N}_m\times\{cr_1\}$.

After applying the theorem $m^*$ times, we will obtain a region containing $T_{\lambda-\tau}$. To that end, we first specify what each application of the theorem adds and removes. At the $m$-th step, the added region contains $\widetilde{N}_m\times(0,cr_1]$; see the first part of the theorem. Note that all gates lie in admissible copies of balls of radius $r_1$ centered on $N_m$, hence in $\{x_1\ge\lambda-w/2\}$. The removed region lies within normal distance $cr_1/3$ below $N_m$, over $\{q_\lambda+8r_1m\le x_1\le q_\lambda+8r_1m+7r_1\}$.

First, consider the part of $T_{\lambda-\tau}$ over the arms of $\gamma^{(\lambda-\tau)}$. It lies in $T_\lambda$, since $R_{\lambda-\tau}\le R_\lambda$ there by the monotonicity of $h$. It is also not removed. Indeed, the removal at $m=0$ occurs over $q_\lambda\le x_1\le q_\lambda+7r_1$, where $\chi=1$ and $7r_1<\tau/4$. On this interval $R_\lambda=w/2-\Delta$ and $R_{\lambda-\tau}=w/2-2\Delta$; see Fig.~\ref{fig:retraction_step}. So $T_{\lambda-\tau}$ lies at depth at least $\Delta$ below $\partial T_\lambda$ there, while the removal reaches only depth $cr_1/3<\Delta$. The removals at $m\ge1$ lie outside $T_\lambda$.

Second, consider the part of $T_{\lambda-\tau}$ over the arc of $\gamma^{(\lambda-\tau)}$. It lies in $\{x_1\ge\lambda-\tau\}$. Let $x$ be a point of this part that is not in $T_\lambda$. The translate of $x$ by $\tau$ in the $x_1$-direction lies in $T_\lambda$, since the two arcs have the same profile. Hence the shortest segment from $x$ to $T_\lambda$ is normal to $\partial T_\lambda$, has length at most $\tau$, and stays in $\{x_1\ge\lambda-2\tau\}$. Since $8r_1m^*\le w/16+8r_1$, we have $q_\lambda+8r_1m^*+7r_1<\lambda-3\tau$. Thus the segment lies over every $\widetilde{N}_m$, and the $m^*$ applications add the normal layers along it up to height $m^*cr_1\ge2\tau$. So $x$ is added at some step. No point of this part is removed, since every removal occurs over $x_1<\lambda-3\tau$. The concluding partial trace then yields $T_{\lambda-\tau}$ exactly.

\end{proof}

We are now in a position to define $\mathcal{S}_n$ for $n\ge2$. In the beginning, a partial trace reduces the region $Y_nZ^{(n)}C\setminus E_k$ to the tube $T_\ell$. Indeed, since $k\le n$, Eq.~\eqref{eq:Ek_shape} implies that every point of $E_k$ has $x_1\ge3w/2-10r_1>11w/8$ and lies at distance greater than $w/2-\Delta$ from $\gamma$. On the other hand, $R_\ell\le w/2-\Delta$ for $x_1\ge11w/8$. Hence $T_\ell$ is disjoint from $E_k$ for every $k$. Since $\ell=(n+1)w$, applying Lemma~\ref{lem:retraction_step} $(n-2)p$ times then yields $T_{3w}$.

A final partial trace yields $X:=\mathcal C_{\gamma^{(3w)},R_X}$, where $R_X=w/2-2\Delta\chi$ on the arms and $R_X=w/2-2\Delta$ on the arc. Indeed, $R_X\le R_{3w}$, since $\chi=1$ wherever $h\bigl((x_1-q_{3w})/\tau\bigr)>0$. Note that $X$ is independent of $n$ and $k$, and its thickness is $w/2$ for $x_1\le5w/4$. Its boundary has a normal collar of depth $w/10$ except where it meets $A$, by the argument given for $T_\lambda$.

Throughout this process, every gate acts on systems contained in $Y_nZ^{(n)}\setminus E_k$ or in regions created by preceding gates; in particular, $C$ is never acted upon, as all gates are supported in $\{x_1\ge 2w\}$ and every partial trace leaves the profile equal to $w/2$ on $\{x_1\le 5w/4\}$. The composition therefore defines a channel $\mathcal{S}_n:\St_{Y_nZ^{(n)}C\setminus E_k}\to\St_{X}$. By Lemma~\ref{lem:extension_reference}, each circuit maps a reduced density matrix of $\sigma$ to a reduced density matrix, as does each partial trace; hence
\begin{equation}\label{eq:Sn_map}
\mathcal{S}_n[\sigma|_{Y_nZ^{(n)}C\setminus E_k}]=\sigma|_{X}.
\end{equation}

\begin{Remark}
    \label{rmk:tame_bridge_sitting}
Applying $\mathcal{E}_k$, gluing the density matrix of $\bar{A}C$ using the Fawzi-Renner theorem~\cite{fawzi2015quantum}, and applying $\mathcal{S}_n$, we obtain a density matrix on some bridge. Moreover, this bridge is thick enough that there is a standard tame bridge sitting inside this region with room to spare: it can be chosen so that its $2r_1$-neighborhood is still contained in that region.
\end{Remark}

\subsection{Reverse extendibility and merging}
\label{subsec:rev_extend_merging}

With the geometric details in Section~\ref{subsec:geometric_details}, we can now prove the main result of this section. In the following discussion $g_0:W_0 \to \RR^d$ should be understood as the standard tame bridge mentioned in Remark~\ref{rmk:tame_bridge_sitting}.
\begin{Theorem}
    \label{thm:bridge_nonempty}
Let $g:W\to\RR^d$ be a standard tame bridge. Then $\Sigma_{1,1}(W,g)$ is nonempty. If moreover the balls $F_1$ and $F_2$ of $W$ are at distance $\gg r_1$, then $\Sigma_{11}(W,g)$ is nonempty.
\end{Theorem}

\begin{proof}
    Let $\rho_n$ be the density matrix constructed in Section~\ref{subsec:geometric_details} for each $n\ge2$. We aim to show that $(\rho_n)|_{W_0}\in\Sigma_{11,\eps_n}(W_0,g_0)$ and $\eps_n\downarrow 0$, and then conclude by compactness.

    By Lemma~\ref{lem:existence_small_cmi}, there is $k\le n$ with $I(Y_n:Z_k|Z^{(k-1)})_\sigma\le 2S(\sigma|_{Y_n})/n=O(1/n)$, since $Y_n$ is of a fixed size and therefore $\dim\calH_{Y_n}$ is uniformly bounded for all $n$. By Eq.~\eqref{eq:Ek_extension}, the channel $\mathcal{E}_k$, acting nontrivially only on $Z_k$, extends $\sigma|_{Y_nZ^{(k)}}$ to $\sigma|_{Y_nZ^{(n)}C\setminus E_k}$; monotonicity of the conditional mutual information then gives $I(Y_n:Z_k\cdots Z_nC\setminus E_k|Z^{(k-1)})_\sigma=O(1/n)$. From this bound we obtain the state on the long bridge, writing $\widetilde{X}_n = Z^{(n)}Y_n\setminus E_k$. By the Fawzi--Renner theorem~\cite{fawzi2015quantum}, there is some channel $\Phi: \St_{Z^{(k-1)}} \to \St_{Z^{(k-1)}Y_n}$ such that 
\begin{equation}
F(\sigma|_{C\overline{X}_n \setminus E_k}, \id_{CZ^{(n)} \setminus (Z^{(k-1)}E_k)}\otimes \Phi[\sigma|_{CZ^{(n)} \setminus E_k}]) \ge 1- O(1/n). \label{eq:fidelity_bound_rho_tilde}
\end{equation}

Let us define a state $\omega_n$ on $\bar{A}CZ^{(n)}$. Let $Y_0$ be the union of the two annuli $A_1$, $A_2$ with short initial segments of the tubes inside the balls $I_1$, $I_2$ so that this region is embedded. Then $\sigma|_{Y_0}\in\Sigma(Y_0)$. Let $\hat{Z}^{(n)}$ be $Z^{(n)}$ with a smooth cap of length at least $2r_1$ attached at each of the two ends adjoining the arc $Y_n$, so that $\bar{A}C\hat{Z}^{(n)}$ is tame. Let $F$ be a tame deformation from $Y_0$ to $\bar{A}C\hat{Z}^{(n)}$ that extends the tubes and fixes the outer boundaries, and set $\omega_n:=\Tr_{\hat{Z}^{(n)}\setminus Z^{(n)}}\Phi^F[\sigma|_{Y_0}]$. Then $\Phi^F[\sigma|_{Y_0}]\in\Sigma(\bar{A}C\hat{Z}^{(n)})$ by Theorem~\ref{thm:isomorphism}, so $\omega_n$ is $r_1$-consistent on $\bar{A}CZ^{(n)}$. Moreover, $\omega_n|_{K_1K_2}=\sigma|_{K_1K_2}$; the proof of Proposition~\ref{prop:transport} applies, since it did not require the two tubes to be joined. The same argument gives $\omega_n|_{CZ^{(n)}}=\sigma|_{CZ^{(n)}}$, jointly on the two parallel segments of the hairpin; their causal cones remain disjoint, since the arms lie at distance $2r\gg r_1$.

Let $\widetilde{\rho}_n:= \Tr_{E_k}[\id_{\bar{A}CZ^{(n)} \setminus Z^{(k-1)}}\otimes \Phi[\omega_n]]$. From the support of $\Phi$ it follows that $(\widetilde{\rho}_n)|_{\bar{A}C} =\omega_n|_{\bar{A}C}$, which together with $\omega_n|_{CZ^{(n)}}=\sigma|_{CZ^{(n)}}$ and Eq.~\eqref{eq:fidelity_bound_rho_tilde} gives
\begin{equation}
    (\widetilde{\rho}_n)|_{\bar{A}C} = \omega_n|_{\bar{A}C}, \quad F\left((\widetilde{\rho}_n)|_{C\widetilde{X}_n}, \sigma|_{C\widetilde{X}_n}\right) \ge 1- O(1/n).
    \label{eq:cond_rho_n}
\end{equation}

To shorten the bridge to a fixed size, set $\rho_n:=(\id_{\bar{A}}\otimes\mathcal{S}_n)[\widetilde{\rho}_n]$, a state on $\bar{A}X$. Since $\mathcal{S}_n$ acts nontrivially only on $Y_nZ^{(n)}\setminus E_k$, we get $(\rho_n)|_{\bar{A}C}=(\widetilde{\rho}_n)|_{\bar{A}C}=\omega_n|_{\bar{A}C}$. Since the fidelity is nondecreasing under channels, Eqs.~\eqref{eq:cond_rho_n} and~\eqref{eq:Sn_map} give $F\bigl((\rho_n)|_{X},\,\sigma|_{X}\bigr)\ge 1-O(1/n)$.

The state $(\rho_n)|_{W_0}$ belongs to $\Sigma_{11,\eps_n}(W_0,g_0)$ with $\eps_n=O(n^{-1/2})$, the extension being $(\rho_n)|_{W_0^+}$. This is defined because $W_0^{+2r_1}\subseteq\bar{A}X$ (Remark~\ref{rmk:tame_bridge_sitting}). Indeed, every patch $L\in\Gamma_1(W_0)$ lies in $\bar{A}C$ or in $X$, since the tubes $C\subseteq X$ separate $\bar{A}$ from the rest of $X$ at distance $\gg r_1$. On $\bar{A}C$, $\rho_n$ agrees with $\omega_n$, which is $r_1$-consistent. On $X$, the marginals of $\rho_n$ and $\sigma$ are $O(n^{-1/2})$-close in trace distance. Moreover, the condition on the two collars holds exactly since $\omega_n|_{K_1K_2}=\sigma|_{K_1K_2}$ and the channels above do not act on $A$. 

Since the state space $\St_{W_0^+}$ of the immersed region $W_0^+$ is compact, the extensions $(\rho_n)|_{W_0^+}$ have a limit point; as $\eps_n\downarrow 0$, it satisfies every local consistency condition exactly, and its restriction to $W_0$ belongs to $\Sigma_{11}(W_0,g_0)$, which is therefore nonempty. Moreover, $\Sigma_{1,1}(W_0,g_0)\supseteq\Sigma_{11}(W_0,g_0)$ is also nonempty. The first claim then follows from Lemma~\ref{lem:deform_bridge} and Proposition~\ref{prop:transport}.

For the second claim, note that the balls of $W_0$ are at distance $\gg r_1$ by the conditions on $\bar{A}$. By Lemma~\ref{lem:deform_bridge}, there is a tame deformation from $W_0$ to $W$ along which the outer boundaries remain standard and the balls stay at distance $\gg r_1$ throughout. Proposition~\ref{prop:transport} then transports $\Sigma_{11}(W_0,g_0)$ to $\Sigma_{11}(W,g)$.
\end{proof}

\begin{Proposition}
    \label{prop:reverse_extendibility_from_bridge}
    Let $BCDE$ be a tame ball and $A$ a tame annulus with inner boundary $\partial(BCDE)$. Let $W=A_1A_2T$ be a tame bridge where $T:=BCD$ and  $A_1,A_2$ are two copies of $A$, attached to $T$ along the faces of $B$ and of $D$ on $\partial(BCDE)$. Let $g:W\to\RR^d$ be the identity on $T$ and maps each $A_i$ to $A$. If $A_1BC$ and $A_2CD$ are tame annuli with embedded outer $2r_1$-collars, $\sigma$ is $(B,C,D,E)$-reverse extendible.
\end{Proposition}
\begin{proof}
We first note that the bridge is standard. Indeed, each $A_i$ is embedded as $A=F\setminus\Int(BCDE)$, where $F:=ABCDE$ is a tame ball. Since $W$ is also tame, $\Sigma_{1,1}(W,g)$ is nonempty (Theorem~\ref{thm:bridge_nonempty}).

The main idea is to glue a purification of the outside to each of $A_1$ and $A_2$, and then to identify the two resulting halves with the reference state. Let $\omega\in\Sigma_{1,1}(W,g)$ so that $\omega$ agrees with the global state $\sigma$ on the collars $K_1$ and $K_2$. Since $A$ is tame, it contains an inner $r_2$-collar along its outer boundary. The gluing below takes place at depth $r_2/4$ (in the coordinate normal to the outer boundary, pointing inward), with collars of width $R:=r_2/8$ on each side. These collars lie in the inner $r_2$-collar of $A$. Throughout the proof, $R$ plays the role of the macroscopic scale, which is allowed because $r_1\ll R$.

Let $A'$ be an annulus whose inner boundary is the outer boundary of $A$ and whose outer boundary is a sufficiently large round sphere, and let $Q:=AA'BCDE$. Let $\overline{\sigma}$ be a purification of $\sigma|_{Q^{+2r_1}}$, and let $P$ denote the purifying system together with the collar $Q^{+2r_1}\setminus Q$. We now glue $\omega$ with two copies of $\overline{\sigma}$ on $A'P$ together with the part of $A$ outside the respective gluing sphere, one near the outer boundary of $A_1$ and the other near the outer boundary of $A_2$. We use Corollary~\ref{cor:boundary_gluing_rel}, with $P_i$ as the abstract system attached along the outer $R$-collar of $A_i'$. The gluing is performed along the sphere $S_i\subset A_i$ at depth $r_2/4$ below the outer boundary of $A_i$. It is $R$-tame, being a parallel hypersurface of an $r_2$-tame one. The part of $A_i$ outside $S_i$ is supplied by $\overline{\sigma}$. The two states agree on the collar $S_i^{[-r_1,r_1]}$, because this collar lies in $K_i$, where $\omega$ agrees with $\sigma$; here $\omega$ and $\overline{\sigma}$ themselves serve as the extensions in Eq.~\eqref{eq:gluing_collar_match}, both being defined and $r_1$-consistent across $S_i$.

We denote the resulting density matrix by $\widetilde{\sigma}$.
We claim that
\begin{equation}
\label{eq:rev_extendibility_after_gluing}
\widetilde{\sigma}|_{A_1A_1'P_1BC}=\overline{\sigma}|_{AA'PBC},\qquad \widetilde{\sigma}|_{A_2A_2'P_2CD}=\overline{\sigma}|_{AA'PCD},
\end{equation}
where $A_i$ is identified with $A$. Consider the first equality. The glued density matrix is obtained from $\omega$ by applying a channel near $A_1$ and a channel near $A_2$, which act on disjoint regions. Tracing out $A_2A_2'P_2$ removes the second channel. Thus $\widetilde{\sigma}|_{A_1A_1'P_1BCD}$ is obtained by gluing $\omega$ and $\overline{\sigma}|_{A_1'P_1}$ along $S_1$ only. By Corollary~\ref{cor:boundary_gluing_rel}, the state $\widetilde{\sigma}|_{A_1A_1'P_1BC}$ belongs to the information convex set of the annulus $A_1A_1'BC$ relative to $P_1$, with the outer boundary of $A_1'$ fixed. So does $\overline{\sigma}|_{AA'PBC}$, with $A_1$ identified with $A$. By Corollary~\ref{cor:fixed-boundary-annulus}, this set is a singleton: an ambient isotopy of $Q$ fixing $\partial Q$ takes $DE$ to a round ball, and it can be made tame since $Q$ is large (Remark~\ref{rmk:scale-homotopy}). Hence the two states agree, which is the first equality. The second equality follows in the same way, with the annulus $A_2A_2'CD$.
Finally, Eq.~\eqref{eq:rev_extendibility_after_gluing} says that $\widetilde{\sigma}$ is a state on two copies of $AA'P$ together with $BCD$ whose reduced density matrices on $A_1A_1'P_1BC$ and on $A_2A_2'P_2CD$ are those of the purification $\overline{\sigma}$ of $\sigma|_{BCDE}$ with purifying system $AA'P$. By Corollary~\ref{thm:immersion2}, $\sigma$ is $(B,C,D,E)$-reverse extendible. The choice of purification does not matter, because any two purifications of $\sigma|_{BCDE}$ are related by a channel on the purifying system.
\end{proof}

\begin{Theorem}[Reverse extendibility for scaled standard configurations]
\label{thm:extend-reverse-A1p}
Let $\sigma$ be the global state of Proposition~\ref{prop:global-state}, and suppose that $\sigma$ is extendible on the standard configurations $\Lambda(r_0,r_1)$. Then there is a constant $s_*>0$, depending only on $d$ and $r_2/r_0$, such that $\sigma$ is $(B,C,D,E)$-reverse extendible for every $(B,C,D,E)\in\Lambda(sr_0,sr_1)$ with $s\ge s_*$.
\end{Theorem}
\begin{proof}
Let $(B,C,D,E)\in\Lambda(sr_0,sr_1)$. We apply Proposition~\ref{prop:smooth-replacement} at the scale $sr_0$. Let $\tilde F$ and $\tilde f$ be the resulting maps and let 
\begin{equation}
\tilde{K} := \tilde{F}(D^d),\quad \tilde{\Gamma} := \tilde{f}(\partial S_-).
\end{equation}
Let $a:=c_{\mathrm{sm}}sr_0/4$ and choose $s_*$ so that $a\ge c\,r_2$ for a sufficiently large constant $c$ depending only on $d$.

We aim to use Proposition~\ref{prop:reverse_extendibility_from_bridge} but it cannot be applied to $\lambda(a,\tilde f)$ directly, since the faces of $B$ and $D$ along which the annuli are attached would touch, and the bridge would not be tame. We therefore thicken $E$ as follows. Let $\tilde E$ be the open $a$-neighborhood of the set of points of $\partial\tilde K$ within distance $a/2$ of $\tilde\Gamma$. Its boundary is smoothed by removing material within some distance (say, $a/32$) of the four $(d-2)$-spheres along which its curvature jumps. Then define the regions as follows:
\begin{equation}
\tilde B:=\Ba_a(\tilde f(S_-))\setminus\tilde E,\qquad \tilde D:=\Ba_a(\tilde f(S_+))\setminus\tilde E,\qquad \tilde C:=\tilde K\setminus\Ba_a(\partial\tilde K).
\end{equation}
These regions are disjoint, because $\Ba_a(\tilde f(S_-))\cap\Ba_a(\tilde f(S_+))=\Ba_a(\tilde\Gamma)\subseteq\tilde E$. Moreover, by construction and by Proposition~\ref{prop:smooth-replacement} with $2a$ in place of $a$, we get
\begin{equation}
\tilde B\tilde E\subseteq\Ba_{2a}(\tilde f(S_-))\subseteq BE,\quad
\tilde D\tilde E\subseteq\Ba_{2a}(\tilde f(S_+))\subseteq DE,\quad
\tilde C\supseteq\tilde K\setminus\Ba_{2a}(\partial\tilde K)\supseteq C.
\end{equation}
Moreover, $\tilde B\tilde C\tilde D\tilde E=\Ba_a(\tilde K)\subseteq BCDE$. Thus they satisfy $(\tilde B,\tilde C,\tilde D,\tilde E)\cge(B,C,D,E)$.

Now let $A$ be a tame annulus with inner boundary $\partial(\tilde B\tilde C\tilde D\tilde E)$, for instance its outer $2r_2$-collar, and let $W=A_1A_2\tilde B\tilde C\tilde D$ be the bridge of Proposition~\ref{prop:reverse_extendibility_from_bridge}, with $A_1$ and $A_2$ attached along the faces of $\tilde B$ and of $\tilde D$. The inner boundary of $W$ consists of $\partial\tilde E\cap\partial(\tilde B\tilde C\tilde D)$ and the unattached parts of the inner boundaries of $A_1$ and $A_2$, which join it smoothly along the two outer edges of $\tilde E$. It is a smooth sphere with curvature $O(1/a)$, and $W$ has an inner $r_2$-collar along it, since $a\ge c\,r_2$. Hence $W$ is tame. Likewise, $A_1\tilde B\tilde C$ and $A_2\tilde C\tilde D$ are tame annuli with embedded outer $2r_1$-collars: their holes $\tilde D\tilde E$ and $\tilde B\tilde E$ are balls of thickness $2a$ whose boundaries are smooth with curvature $O(1/a)$. The constants here depend only on $d$. Indeed, before the smoothing, $\tilde E$, $\tilde D\tilde E$, and $\tilde B\tilde E$ are the $a$-neighborhoods of the sets of points of $\partial\tilde K$ within distance $a/2$ of $\tilde\Gamma$, $\tilde f(S_+)$, and $\tilde f(S_-)$, respectively. Since $\tilde f$ and its equator are $4a$-tame, every point within distance $2a$ of such a set has a unique nearest point on it. Hence the normal rays from the boundary of its $a$-neighborhood do not cross within distance $a$ on either side, and the smoothing retains a depth proportional to $a$. By Proposition~\ref{prop:reverse_extendibility_from_bridge}, $\sigma$ is $(\tilde B,\tilde C,\tilde D,\tilde E)$-reverse extendible, and by Proposition~\ref{prop:reverse-g_mono}, $\sigma$ is $(B,C,D,E)$-reverse extendible.
\end{proof}

To summarize, starting from the local extendibility axiom at scale $r_1$, the reverse extendibility holds at some larger scale. Because the local extendibility also holds at this larger scale (Theorem~\ref{thm:extend-A1'}), all these properties hold at a common scale. Moreover, the reverse extendibility ought to remain circuit-invariant in the sense described in Section~\ref{subsec:circuit_invariance}. Indeed, because the local extendibility axiom is circuit-invariant, the reverse extendibility will continue to hold at some proportionately larger scale. Nonetheless, we note that these statements can be proved without going through this indirect route.

\begin{Theorem}[Circuit-invariance of reverse extendibility]\label{thm:reverse-circuit-invariance}
    Consider disjoint regions $(B,C,D,E)$ and, separately, disjoint regions $(B',C',D',E')$.  Consider a unitary circuit $U$ composed of $\ell$ layers of gates, commuting within each layer, each gate with support of diameter at most $s$.  Define light-cone radius $r := \ell s$.  Assume $(BE)^{+2r}  \subseteq B'E'$, $(DE)^{+2r} \subseteq D'E'$, $C' \subseteq C$, $(BCDE)^{+2r} \subseteq B'C'D'E'$, and $C^{+s} \subseteq BCDE$. Let $\sigma$ be a state on $(B'C'D'E')^{+r}$ that is $(B,C,D,E)$-reverse extendible, and let $\sigma' = U \sigma U^\dagger$.  Then $\sigma'$ is $(B',C',D',E')$-reverse extendible.
    \end{Theorem}
\begin{proof}
The proof is identical to that of Theorem~\ref{thm:circuit-invariance}, using Proposition~\ref{prop:reverse-g_mono} in place of Proposition~\ref{prop:g_mono}. 
\end{proof}

\begin{figure}[t]
\centering
\begin{tikzpicture}[line width=1pt, scale=0.42]
\pgfmathsetmacro{\cx}{3.5}   
\pgfmathsetmacro{\hx}{1.6}   
\pgfmathsetmacro{\rh}{0.55}  
\pgfmathsetmacro{\ri}{0.33}  
\pgfmathsetmacro{\w}{0.17}   
\pgfmathsetmacro{\fl}{0.08}  
\begin{scope}
\foreach \s in {-1,1}{
  \draw[fill=tikzblue] (\s*\cx,0) circle (2.6cm);
  \draw[fill=white] (\s*\cx,0) circle (1.2cm);
  \draw[fill=white, draw=red] (\s*\hx,0) circle (\rh cm);
}
\node[red] at (-\hx-0.4,-1) {$N_1$};
\node[red] at (\hx+0.4,-1) {$N_2$};
\node at (0,-3.4) {(a)};
\end{scope}
\begin{scope}[xshift=13.2cm]
\foreach \s in {-1,1}{
  \draw[fill=tikzblue] (\s*\cx,0) circle (2.6cm);
  \draw[fill=white] (\s*\cx,0) circle (1.2cm);
  \draw[red] (\s*\hx,0) circle (\rh cm);
}
\begin{scope}[xshift=-\hx cm]
\draw[] (-90:\ri cm) arc (-90:90:\ri cm);
\draw[] (90:\ri cm) arc (90:270:\fl cm);
\draw[] (-90:\ri cm) arc (270:90:\fl cm);
\end{scope}
\begin{scope}[xshift=\hx cm, xscale=-1]
\draw[] (-90:\ri cm) arc (-90:90:\ri cm);
\draw[] (90:\ri cm) arc (90:270:\fl cm);
\draw[] (-90:\ri cm) arc (270:90:\fl cm);
\end{scope}
\draw[fill=tikzblue, draw=none] (-\hx cm, \w cm) -- (\hx cm, \w cm) -- (\hx cm, -\w cm) -- (-\hx cm, -\w cm) -- cycle;
\draw[] (-\hx cm, \w cm) -- (\hx cm, \w cm);
\draw[] (-\hx cm, -\w cm) -- (\hx cm, -\w cm);
\foreach \sc in {1,-1}{
  \begin{scope}[xscale=\sc]
  \draw[fill=white] (-\hx cm, \w cm) arc (-90:-270:\fl cm) arc (90:30:\ri cm) -- cycle;
  \draw[fill=white] (-\hx cm, -\w cm) arc (90:270:\fl cm) arc (270:330:\ri cm) -- cycle;
  \end{scope}
}
\foreach \s in {-1,1}{ \draw[line width=0.4pt, dash pattern=on 0.6pt off 0.9pt] (\s*\hx,-\w) -- (\s*\hx,\w); }
\node at (-\hx-0.4,-1) {$A_1$};
\node at (\hx+0.4,-1) {$A_2$};
\node at (0,0.7) {$T$};
\node at (0,-3.4) {(b)};
\end{scope}
\pgfmathsetmacro{\m}{\ri+0.1}                     
\pgfmathsetmacro{\xo}{-\cx+sqrt(2.6*2.6-\m*\m)}   
\pgfmathsetmacro{\tho}{atan2(\m,\xo+\cx)}         
\begin{scope}[xshift=26.4cm]
\foreach \sc in {1,-1}{
  \begin{scope}[xscale=\sc]
  \fill[tikzblue, rounded corners=2.5pt] (-\hx+0.4,\w) -- (-\hx,\w) -- (-\hx,\m) -- (\xo,\m) arc (\tho:360-\tho:2.6cm) -- (-\hx,-\m) -- (-\hx,-\w) -- (-\hx+0.4,-\w) [sharp corners] -- cycle;
  \draw[rounded corners=2.5pt] (-\hx+0.4,\w) -- (-\hx,\w) -- (-\hx,\m) -- (\xo,\m) arc (\tho:360-\tho:2.6cm) -- (-\hx,-\m) -- (-\hx,-\w) -- (-\hx+0.4,-\w);
  \draw[fill=white] (-\cx,0) circle (1.2cm);
  \end{scope}
}
\draw[fill=tikzblue, draw=none] (-\hx cm, \w cm) -- (\hx cm, \w cm) -- (\hx cm, -\w cm) -- (-\hx cm, -\w cm) -- cycle;
\draw[] (-\hx cm, \w cm) -- (\hx cm, \w cm);
\draw[] (-\hx cm, -\w cm) -- (\hx cm, -\w cm);
\node at (0,-3.4) {(c)};
\end{scope}
\end{tikzpicture}
\caption{An example application of Theorem~\ref{thm:merging_thm}. \textbf{(a)} The region $M$ is the union of two annuli $M_1$ and $M_2$, each with a ball removed; the boundaries $N_1$ and $N_2$ of the removed balls are shown in red. \textbf{(b)} A small bridge is glued to $M$ along $N_1\sqcup N_2$. Its annuli $A_1$, $A_2$ fill part of the holes, and its tube $T$ passes over $M_1$ and $M_2$. The inner and outer boundaries of the two annuli are unchanged. \textbf{(c)} Some of the materials of $A_1$, $A_2$, $M_1$, and $M_2$ along the tube, on both sides of it, are erased. The resulting region is embedded, forming a ball with two holes.}
\label{fig:attaching_example}
\end{figure}
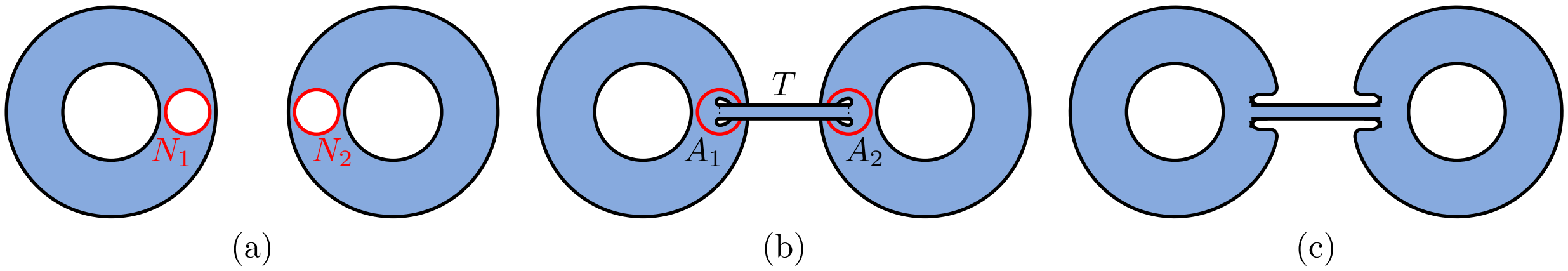

We now turn to the merging operation~\eqref{eq:merging}. The following is the setup one should have in mind. One is given a region $K=ABCE$, and the goal is to extend a density matrix in $\Sigma(ABE)$ to that of $\Sigma(ABC)$ by a channel acting on $BE$. The input region of the channel $BE$ is a union of two balls and the channel may change the topology of the region, as in Fig.~\ref{fig:attaching_example}. 

The merging map can be constructed by gluing the two outer boundaries of a bridge to two holes; see Theorem~\ref{thm:boundary_gluing} and Corollary~\ref{cor:boundary_gluing_rel}. More precisely, one would first apply a partial trace to the hole, after which the gluing occurs. After the gluing, we obtain an immersed region; see Fig.~\ref{fig:bridge_applications}(c). By removing the overlapping region and its neighborhood appropriately, we realize the merging map.

There is one subtle difference between merging and reverse extendibility. In the latter case, we glued the boundaries of the bridge along two independent elements of the relative information convex set of an annulus, one at a time. For merging, we glue along both boundaries simultaneously. During this process, note Corollary~\ref{cor:boundary_gluing_rel} requires that the extensions of the density matrices across the gluing boundary should match. Since the density matrix over the two balls that constitute $BE$ is inherited from the global state $\sigma$, we must use the element of $\Sigma_{11}(W)$, not $\Sigma_{1,1}(W)$. Indeed, elements in the former set are included in the latter, but the opposite may not be true in general.

Here is our setup. Let $f:M\to\RR^d$ be a tame immersed region and let $N=N_1\sqcup N_2\subseteq\partial M$ be two of its boundary components. Let $g:W\to\RR^d$ be a standard tame bridge, $W=A_1A_2T$, whose balls $F_1$ and $F_2$ are at distance $\gg r_1$. We glue its outer boundaries to $N_1$ and $N_2$ by a diffeomorphism $\varphi$ as in Definition~\ref{def:glued_immersion}, leaving the inner boundary of the bridge free. Let $M':=M\cup_\varphi W$ and let $F:M'\to\RR^d$ be the glued immersion, which is tame by Lemma~\ref{lem:gluing_tame}. We identify $N$ with its image in $M'$ and orient it by the unit normal pointing into $W$. Then $N\subset\Int(M')$, and $\partial M'$ consists of $\partial M\setminus N$ and the inner boundary of the bridge. 
\begin{Theorem}[Attaching a bridge]
\label{thm:merging_thm}
In the setup above, let $\rho\in\Sigma(M,f)$ satisfy
\begin{equation}
\label{eq:vacuum_seam}
\hat{\rho}|_{N^{(0,r_1]}}=\sigma|_{F(N^{(0,r_1]})}
\end{equation}
for some $r_1$-consistent extension $\hat{\rho}$ of $\rho$ on $M^{+2r_1}$. Let $\omega\in\Sigma_{11}(W,g)$. Then there is a channel $\Phi_\omega:\St_{N^{(-r_1,0]}}\to\St_{N^{(-r_1,0]}\cup W}$, depending only on $\omega$, such that
\begin{equation}
\label{eq:merged_state}
\rho':=\bigl(\id_{M\setminus N^{(-r_1,0]}}\otimes\Phi_\omega\bigr)[\rho]
\end{equation}
belongs to $\Sigma(M',F)$ and satisfies $\rho'|_M=\rho$ and $\rho'|_W=\omega$. The same holds for the relative information convex sets~\eqref{eq:IC-rel} with fixed boundary components in $\partial M\setminus N$, the abstract system being kept alongside $M$.
\end{Theorem}
\noindent In Eq.~\eqref{eq:vacuum_seam}, the region $N^{(0,r_1]}$ is the inner $r_1$-collar of $W$ along $N$, on which $F$ is injective: it lies in the annuli $A_1\sqcup A_2$, which $g$ embeds into the disjoint balls bounded by $g(N_1)$ and $g(N_2)$. The condition does not depend on the choice of the extension, as noted after Theorem~\ref{thm:boundary_gluing}.

\begin{proof}
We reduce the statement to Theorem~\ref{thm:boundary_gluing}, applied to $f:M\to\RR^d$ and $g:W\to\RR^d$ with $N$ in place of $N_1$. Note nothing in that theorem requires the glued boundary to be connected, so we can glue both components of $N$ at once. It remains to verify the hypothesis~\eqref{eq:gluing_collar_match}. Let $B:=N^{(-r_1,0]}$ and $C:=N^{(0,r_1]}$ be the two halves of the collar $N^{(-r_1,r_1]}$, and let $\hat{\rho}$ and $\hat{\omega}$ be $r_1$-consistent extensions of $\rho$ and $\omega$ on $M^{+2r_1}$ and $W^{+2r_1}$.

We show that $\hat{\rho}$ and $\hat{\omega}$ agree on $BC$. They agree on $C$, because $C\subseteq K_1\sqcup K_2$, so that $\omega|_C=\sigma|_{F(C)}=\hat{\rho}|_C$ by $\omega\in\Sigma_{11}(W,g)$ and Eq.~\eqref{eq:vacuum_seam}. For $B$, we use Proposition~\ref{prop:collar-recovery} with $N$ oriented into $M$. It determines the restriction of each state to $CB$ from its restriction to $C$ by one and the same channel, so the two states agree on $BC$ as well.

Theorem~\ref{thm:boundary_gluing} now yields $\rho'\in\Sigma(M',F)$ with $\rho'|_M=\rho$ and $\rho'|_W=\omega$, of the form $\rho'=\id_{M\setminus B}\otimes\Phi[\rho]$ with $\Phi$ depending only on $\omega$; this is $\Phi_\omega$. For relative information convex sets, we use Corollary~\ref{cor:boundary_gluing_rel} instead, with the abstract system adjoined to $M$.
\end{proof}

Theorem~\ref{thm:merging_thm} can be used to merge density matrices of two regions, changing the topology of the underlying region. For instance, in Fig.~\ref{fig:attaching_example}(b) we see the merging process joining two disjoint annuli. After applying some partial traces, we obtain a two-holed disk (Fig.~\ref{fig:attaching_example}(c)). The glued state lies in the information convex set of the immersed region, and the partial trace leaves an element of the information convex set of the resulting embedded region by Lemma~\ref{lem:restriction_embedded}. The channels that map Fig.~\ref{fig:attaching_example}(a) to (c) are local to some region, and map elements of the information convex set to an element of the information convex set with a different topology. Thus this can be viewed as a version of the merging theorem introduced in Ref.~\cite{shi2020fusion}, which serves a similar purpose. Of course, unlike the method in Ref.~\cite{shi2020fusion}, ours is robust to shallow quantum circuits. However, this theorem can be also used for a different purpose, for instance, for creating an overpass as in Fig.~\ref{fig:bridge_applications}(c).

More formally, we can prove the existence of the merging operation as in Eq.~\eqref{eq:merging} for the following regions. Similar to Section~\ref{sec:micro}, we define a set of standard geometries. To that end, fix the two caps $S_1=\{x\in S^{d-1}:\,x_d\ge 1/2\}$ and $S_2=\{x\in S^{d-1}:\,x_d\le -1/2\}$ of the unit sphere, and let $S_0=S^{d-1}\setminus(S_1\cup S_2)$ be the band between them. Given an embedding $f\colon S^{d-1}\to\RR^d$ and a thickness $a>0$, we set
\begin{equation}\label{eq:merging_config}
BE=\Ba_a\bigl(f(S_1\cup S_2)\bigr),\qquad DE=\Ba_a\bigl(f(S_0)\bigr),
\end{equation}
which define $E=BE\cap DE$, $B=BE\setminus E$ and $D=DE\setminus E$. We choose $C=\mathrm{Int}(S^{d-1},f)\setminus\Ba_a(f(S^{d-1}))$ as the inside of the sphere, as in Definition~\ref{def:BCDE-region-recipe}. We write $\lambda_{\mathrm{m}}(a,f)=(B,C,D,E)$. Let $\Delta_{\mathrm{m}}(s)$ be the set of boundary spheres $f=F|_{\partial D^d}$ of smooth embeddings $F\colon D^d\to\RR^d$ such that $f(S^{d-1})$ is an $s$-tame embedded hypersurface, $f(\partial S_1)$ and $f(\partial S_2)$ are $s$-tame, and $f(S_1)$ and $f(S_2)$ are at distance at least $s$.
For $a\le s/640$, the \emph{standard set of merging configurations} is
\begin{equation}\label{std_Lambda_merging}
\Lambda_{\mathrm{m}}(a,s)=\{\lambda_{\mathrm{m}}(a,f):\, f\in\Delta_{\mathrm{m}}(s)\}.
\end{equation}
\begin{Theorem}[Merging for standard merging configurations]
\label{thm:merging_standard}
Let $\sigma$ be the global state of Proposition~\ref{prop:global-state}. There is a constant $c_{\mathrm{m}}>0$, depending only on $d$, such that for every $(B,C,D,E)\in\Lambda_{\mathrm{m}}(a,s)$ with $a\ge c_{\mathrm{m}}r_2$ there is a channel $R_{BE\to BC}$ satisfying
\begin{equation}\label{eq:merging_standard}
(\id_A\otimes(\Tr_C\circ R_{BE\to BC}))[\overline{\sigma}|_{ABE}]=\overline{\sigma}|_{AB},\qquad R_{BE\to BC}[\sigma|_{BE}]=\sigma|_{BC}
\end{equation}
for every purification $\overline{\sigma}\in\St_{ABCDE}$ of $\sigma|_{BCDE}$.
\end{Theorem}
Let us first sketch the proof. The main idea is shown in Fig.~\ref{fig:bridge_applications}(c). Recall that $BE$ is a thickening of two caps, say $(BE)_1$ and $(BE)_2$. We first remove a ball $F_i$ from $(BE)_i$ and attach a standard tame bridge along $\partial F_1\sqcup\partial F_2$ by Theorem~\ref{thm:merging_thm}. The annuli of the bridge refill most of the two thickenings, and its tube covers $C$ plus the part of $B$ adjacent to $C$. We can then trace out $E$ and every part of $B$ that is covered twice, keeping only the copy in the tube. Over $BC$, this leaves a chain of three regions: the outer layer of $B$, the annuli, and the tube. The state on $BC$ is the unique state of a ball. The state on $AB$ is the unique state of an annulus with the purifying system attached along its outer boundary. Thus Proposition~\ref{prop:ball-unique} and Corollary~\ref{cor:fixed-boundary-annulus} establish the claim.

\begin{proof}
Let $(BE)_i:=\Ba_a(f(S_i))$, so that $BE=(BE)_1\sqcup(BE)_2$. Throughout, all lengths other than $r_1$ and $r_2$ are assumed to be fixed multiples of $a$. We will take $c_{\mathrm{m}}$ to be large enough so that each of them exceeds $r_2$ by any factor the argument requires. Edges of the regions below are rounded at a scale small compared to $a$; we omit those details.

We define a standard bridge $g: W\to \RR^d$, $W=A_1A_2T$, with $g$ acting as the identity on each piece. We demand the following properties. First, $F_i\subset (BE)_i$ is a tame ball with distance from $\partial(BE)_i$ sufficiently large, so that the gluing procedure in Theorem~\ref{thm:merging_thm} can be realized by a channel acting within $BE$. Second, $E\supset F_1F_2\setminus g(W)$. Third, $\partial g(T)$ meets $\overline{B}$ only in $g(\delta_1\sqcup\delta_2)$, where $\delta_i\subset\partial I_i$ is the disk along which $T$ is attached to $A_i$. Fourth, $C\subseteq g(T)\subseteq BCDE$. It is straightforward to show that there is a standard bridge with a smooth boundary that satisfies these conditions. Then the tameness can be ensured by scaling up.

The merging map is a composition of the partial trace of $F_1F_2$, boundary gluing using Theorem~\ref{thm:merging_thm} along $\partial F_1 \sqcup \partial F_2$, and a partial trace. The gluing uses a fixed $\omega\in\Sigma_{11}(W,g)$, which is nonempty by Theorem~\ref{thm:bridge_nonempty} since $F_1$ and $F_2$ are at distance at least $s-2a\gg r_1$. The last partial trace is over $E$ and over $B':=B\cap g(T)$, the original copy of the part of $B$ covered by the tube. 

As noted after Definition~\ref{def:loc_extend}, it suffices to verify Eq.~\eqref{eq:merging_standard} for one purification. Let $Y$ be a sufficiently large round ball containing $BCDE$, let $\bar{A}:=Y\setminus BCDE$, and let $\overline{\sigma}\in\St_{PY}$ be a purification of $\sigma|_Y$ with an abstract system $P$, so that $A=P\bar{A}$. Let $M:=Y\setminus F_1F_2$. The state $\overline{\sigma}|_{PM}$ belongs to the relative information convex set of $M$ with $P$ attached along $\partial Y$ and the state $\widetilde{\rho}:=\overline{\sigma}|_{P\,\partial Y^{[-r_2,0]}}$ fixed. (Note here $\overline{\sigma}$ is the $r_1$-consistent extension, so Eq.~\eqref{eq:vacuum_seam} holds.) Theorem~\ref{thm:merging_thm} gives $\rho'\in\ICrel(M',F,\widetilde{\rho})$ with $\rho'|_{PM}=\overline{\sigma}|_{PM}$ and $\rho'|_W=\omega$, obtained from $\overline{\sigma}|_{PM}$ by the gluing channel. Let $\hat B\hat C\subset M'$ be the copy of $BC$ left after the last partial trace, namely $B\setminus B'$ in $M\cup A_1A_2$ together with the copy of $B'\cup C$ in $T$. Then $(\id_A\otimes R_{BE\to BC})[\overline{\sigma}|_{ABE}]=\rho'|_{A\hat B\hat C}$, because the gluing channel acts within $BE$. It remains to show $\rho'|_{\hat B\hat C}=\sigma|_{BC}$ and $\rho'|_{A\hat B}=\overline{\sigma}|_{AB}$.

Note that $BC$ and $CDE$ are not tame: $\partial(BE)$ and $\partial(DE)$ are tangent at distance $a$ above and below the edges of the caps; this gives $BC$ cusps outside the sphere $f(S^{d-1})$ and $CDE$ cusps inside it. As such, we reason about the tame balls $X\supset BC$ and $G\subset CDE$ instead, which are defined as follows. Let $X$ be $BC$ with a small neighborhood of its cusps in $E$ added, and let $G$ be $CDE$ with a small neighborhood of its cusps removed. Since the cusps of $BC$ lie in $M$ outside $g(T)$ and those of $CDE$ lie inside $g(T)$, the region $M'$ contains copies $\hat X\supseteq\hat B\hat C$ of $X$ and $\hat Z\supseteq\bar{A}\hat B$ of the annulus $Y\setminus G$, obtained by adding points of $M$ and of $T$, respectively.

The map $F$ is injective on $\hat X$ and $\hat Z$, and by the margins in the choice of $T$, $F_{+2r_1}$ is injective on their $2r_1$-neighborhoods in $M'^{+2r_1}$. By Lemma~\ref{lem:restriction_embedded}, $\rho'|_{\hat X}\in\Sigma(X)$, which contains $\sigma|_X$ and is a singleton by Proposition~\ref{prop:ball-unique}; hence $\rho'|_{\hat B\hat C}=\sigma|_{BC}$. Likewise, restricting the $r_1$-consistent extension of $\rho'$ shows that $\rho'|_{P\hat Z}$ belongs to the relative information convex set of $Y\setminus G$ with $P$ attached along $\partial Y$ and $\widetilde{\rho}$ fixed, and so does $\overline{\sigma}|_{P(Y\setminus G)}$. This set is a singleton by Corollary~\ref{cor:fixed-boundary-annulus}: an ambient isotopy of $Y$ fixing $\partial Y$ takes $G$ to a round ball, and it can be made tame since $Y$ is large (Remark~\ref{rmk:scale-homotopy}); hence $\rho'|_{A\hat B}=\overline{\sigma}|_{AB}$.
\end{proof}

\subsection{Manifolds}
\label{subsec:manifolds}

We end this section with a few remarks. First, one might wonder whether the proof of Proposition~\ref{prop:reverse_extendibility_from_bridge} needed the entirety of $\mathbb{R}^d$. One might instead hope to prove the same statements in a local way, i.e., proving reverse extendibility within each region from a finite set of local extendibility conditions. In the proof we needed a family of hairpin geometries (Section~\ref{subsec:geometric_details}) whose length can grow arbitrarily. Such a geometry can be embedded into an immersed region defined as follows. Consider the curve $\gamma(t)=(R\cos t, R \sin t, 0, \ldots, 0)$ for $t\in[0,t_*]$. We can thicken this curve along the direction normal to the tangent vector by $w$, which becomes a tame immersed tube if $R\gg w \gg r_2$, whose image on $\RR^d$ is of a fixed size. Thus the hairpin geometry of arbitrary length can be embedded in this immersed region by choosing a sufficiently large $t_*$. To make this argument work, other parts of the paper, e.g., the isomorphism theorem (Theorem~\ref{thm:isomorphism}), should also be modified to a local version. We leave such a treatment for future work.

\begin{figure}[t]
    \centering
    \includegraphics[width=0.95\textwidth]{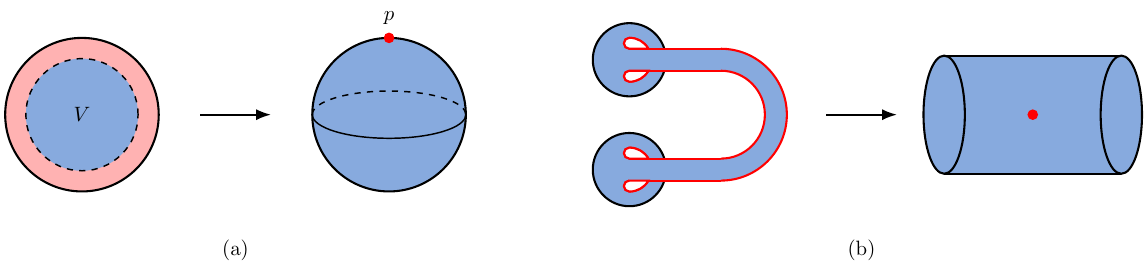}%
    \caption{\textbf{(a)} Compactifying an inner collar of a ball to a point $p$, we obtain a sphere. \textbf{(b)} Compactifying the red inner boundary of the bridge, we obtain a cylinder; its two black boundaries descend from the outer boundaries of the annuli.  \label{fig:compactification}}
\end{figure}

Second, the results in this section can be used to define a closed manifold on which the local extendibility, reverse extendibility, and merging operation hold everywhere. For instance, let $V$ be a tame ball. We can view its inner collar $\partial V^{[-r_2, 0)}$ and the purification of $V$ as a single point, say $p$, similar to the one-point compactification; see Fig.~\ref{fig:compactification}(a). For the regions that do not contain $p$, the aforementioned conditions hold by the results established so far. For the regions that do contain $p$, the extendibility/reverse extendibility follows from the reverse extendibility/extendibility on a region that does not contain $p$. 

Moreover, given two manifolds, one can connect them by attaching a bridge between the two. The inner collar of the inner boundary of the bridge can be identified with a point, after which we obtain a cylinder; see Fig.~\ref{fig:compactification}(b). This way, one can attach handles, creating various closed manifolds. To see that a region containing the compactified point satisfies the axioms, one can make the following observation. Consider a bridge region immersed on a sphere, wherein the sphere was obtained by the aforementioned method. On the sphere, the bridge is regular homotopic to a ball with two punctures, wherein the boundaries of the two punctures correspond to the two outer boundaries of the bridge. After compactifying these punctures to a point, we obtain a ball. This ball can be completed into a sphere by identifying the remaining boundary to a point. Notably, this boundary is the one deformed from the inner boundary of the initial bridge. Thus we can apply the previous argument for extendibility/reverse extendibility/merging for the regions containing the compactified point. 

\section{Extendibility and local topological quantum order}
\label{sec:extendibility_ltqo}

One may wonder if the local extendibility condition follows from a more fundamental condition. In this section, we prove that for the special case of commuting Hamiltonians, they follow from a variant of a condition known as \emph{local topological quantum order}. The latter was used in proving the stability of gap for frustration-free Hamiltonians~\cite{michalakis2013stability}. 

Without loss of generality, consider a local frustration-free Hamiltonian $H= \sum_X (I- Q_X)$, summed over a set of $O(1)$-size ball $X\subset \mathbb{R}^d$, with projections $Q_X$ local to $X$. Recall a Hamiltonian is called \emph{frustration-free} when every ground state is also a ground state of each individual term.  Then each $Q_X$ stabilizes the ground state. We denote the projection onto  $\{|\psi\rangle_A \in \mathcal{H}_A \,:\, (1-Q_X) |\psi\rangle_A= 0\; \forall X \subset A \}$ as $Q_A$; this is the projector onto the space of states on $A$ with zero energy. In particular, we say a state $\rho$ is \emph{zero-energy on} $A$ if $Q_A \rho=\rho$.

We now introduce the local topological quantum order (LTQO) condition and its variant. We use the following notation. Define the canonical expansion of a region:
\begin{equation}
\Gamma_H(A) := \{X: Q_X\neq I, X\cap A \neq \emptyset  \}, \quad
 A_H^+ := A\cup \bigcup_{X \in \Gamma_H(A)} X.
\end{equation}
We set $r_H$ as the infimum of $x>0$ such that $A^{+x} \supseteq A_H^+$ for every region $A$, where $A^{+x}=\Ba_x(A)$. We also define $A^{[x, y)} := \Ba_y(A) \setminus \Ba_x(A)$, where $y>x\ge 0$ and $\Ba_0(A):=A$. A restriction of the Hamiltonian to a region is denoted as $H_A = \sum_{X\subseteq A} (I-Q_X)$. 

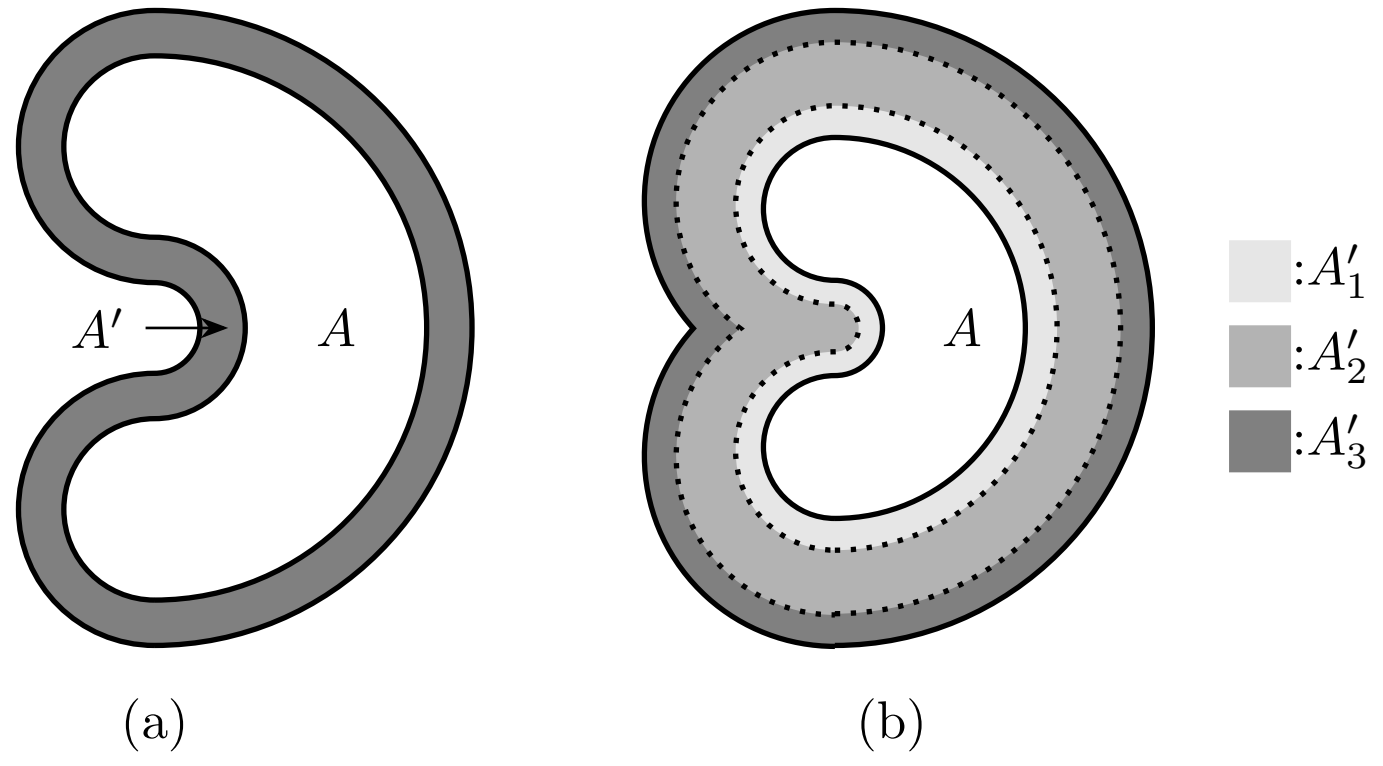
\begin{figure}[t]
 \centering
 \begin{tikzpicture}[line width=1pt, scale=1.5]
 \draw[fill=black!50!white] (0, -1.4cm) arc (-90:90:1.4cm) arc (90:270:0.6cm) arc (90:-90:0.2cm) arc (90:270:0.6cm);
 \draw[fill=white] (0, -1.2cm) arc (-90:90:1.2cm) arc (90:270:0.4cm) arc (90:-90:0.4cm) arc (90:270:0.4cm);
 
 \node[] () at (0.8cm, 0) {$A$};
  \node[] (ap) at (-0.25cm, 0) {$A'$};
  \draw[-Stealth, line width=0.5pt] (ap) -- (0.325cm, 0);

 \node[] () at (0, -1.75cm) {(a)};
 
 \begin{scope}[xshift= 3cm]

\begin{scope}[scale=0.7]
\draw[fill=black!50!white] (0, -2cm) arc (-90:90:2cm) arc (90:222:1.2cm) arc (138:270:1.2cm);
 \draw[fill=black!30!white, dotted] (0, -1.8cm) arc (-90:90:1.8cm) arc (90:233.5:1cm) arc (126.5:270:1cm);
 \draw[fill=black!10!white, dotted] (0, -1.4cm) arc (-90:90:1.4cm) arc (90:270:0.625cm) arc (90:-90:0.15cm) arc (90:270:0.625cm);
 \draw[fill=white] (0, -1.2cm) arc (-90:90:1.2cm) arc (90:270:0.45cm) arc (90:-90:0.3cm) arc (90:270:0.45cm);
 
 \node[] () at (0.8cm, 0) {$A$};
\end{scope}

\begin{scope}[xshift= 1.75cm, yshift=-0.25cm, scale=0.5]
\filldraw[black!10!white] (0, 0.75cm) --++ (0.5cm, 0) -- ++ (0, 0.5cm) -- ++ (-0.5cm, 0) -- cycle;
\filldraw[black!30!white] (0, 0) --++ (0.5cm, 0) -- ++ (0, 0.5cm) -- ++ (-0.5cm, 0) -- cycle;
\filldraw[black!50!white] (0, -0.75cm) --++ (0.5cm, 0) -- ++ (0, 0.5cm) -- ++ (-0.5cm, 0) -- cycle;

\node[] () at (0.875cm, 1cm) {:$A_1'$};
\node[] () at (0.875cm, 0.25cm) {:$A_2'$};
\node[] () at (0.875cm, -0.5cm) {:$A_3'$};
\end{scope}
 \node[] () at (0.25cm, -1.75cm) {(b)};
 \end{scope}

 \end{tikzpicture}
 \caption{\textbf{(a)} An exemplary region used in the formulation of the local topological quantum order. Here $A$ is chosen as a topological ball and $A'=A^{[0,\ell)}$ for some $\ell> 0$. \textbf{(b)} Regions used in the formulation of the spherical local topological quantum order. Generally the thickness of $A_1'$ and $A_3'$ is chosen to be the same, say $\ell$. The thickness of $A_2'$ is chosen to be $L$, which is often larger than $\ell$.}
 \label{fig:ltqo_sltqo_figs}
 \end{figure}

\begin{Definition}[Local topological quantum order]
\label{definition:ltqo}
Let $A$ and $A'$ be disjoint regions. $H$ satisfies $(A, A')$-local topological quantum order if ground states of $H_{AA'}$ are indistinguishable over $A$.
\end{Definition}
For the standard choice, we let $A$ be an open set and choose $A'=A^{[0,\ell)}$. In this case, we say $H$ satisfies $(A,\ell)$-\emph{local topological quantum order}.%
\footnote{If $A$ is a smooth embedding of a $d$-dimensional manifold, $A'$ is its outer $\ell$-collar (assuming it exists).} One might say: ``all zero-energy states on $A^{+\ell}$ look identical on $A$." The original definition of LTQO~\cite{michalakis2013stability} allows the reduced states to be approximately indistinguishable, but we do not consider such cases. More importantly, we envision imposing the LTQO condition on a set of tame topological balls; see Fig.~\ref{fig:ltqo_sltqo_figs}(a). This is different from the usual formulation, which chooses $A$ and $AA'$ as square-shaped regions~\cite{michalakis2013stability}. The fracton models satisfy the LTQO condition for such regions, but not generally for non-convex balls. 
We also introduce a notion of spherical LTQO.  
\begin{Definition}[Spherical local topological quantum order]\label{definition:spherical_ltqo}
Let $A,A'\subset X$ be disjoint regions. Let $A'=A_1'A_2'A_3'$ be a partition of $A'$. $H_X$ satisfies $(A, A_1', A_2', A_3')$-spherical local topological quantum order if every ground state of $H_X$ satisfies the following condition. Let $\rho_{X}$ be a ground state and $\rho_{AA_1'A_2'A_3'}$ be its reduced state. Denote its purification as  $\bar{\rho}_{AA_1'A_2'A_3'Y}$. For any state $\sigma_{A_1'A_2'A_3'Y}$ that (i) is zero-energy on $A_1'A_2'A_3'$ and (ii) $\sigma_{A_3'Y} = \bar{\rho}_{A_3'Y}$,
\begin{equation}
\sigma_{A_2'A_3'Y} = \bar{\rho}_{A_2'A_3'Y}. \label{eq:sltqo_equality}
\end{equation}
\end{Definition}
Note as usual the choice of purification used in the definition is unimportant: if it is satisfied for one purification, it is satisfied for all. We also note that, unlike the definition of LTQO, spherical LTQO may depend on the region $X$ that includes $AA'$.

For the standard choice, we let $A$ be an open set and choose $A'$ as $A_1'= A^{[0,\ell)}$, $A_2' = A^{[\ell, \ell+L)}$, and $A_3'=A^{[\ell+L, 2\ell+L)}$. In this case, we say $H_X$ satisfies $(A,\ell,L)$-\emph{spherical LTQO}. (For the intended application, $L$ will be taken to be $c\ell$ for some sufficiently large constant $c$.) In principle, one may consider more generally $A_1'=A^{[0,\ell_1)}$, $A_2'=A^{[\ell_1,\ell_1+\ell_2)}$, $A_3'=A^{[\ell_1+\ell_2,\ell_1+\ell_2+\ell_3)}$ for arbitrary $\ell_1,\ell_2,\ell_3>0$. However, the spherical LTQO condition satisfies a monotonicity property (Proposition~\ref{prop:mono_SLTQO}). Thus the more general formulation implies the standard choice for $\ell = \max(\ell_1,\ell_3)$ and $L=\ell_2 - |\ell_3-\ell_1|$, provided $\ell_2>|\ell_3-\ell_1|$. Moreover, for proving extendibility/reverse extendibility, it suffices to only consider the standard choice.

\begin{Proposition}
\label{prop:mono_SLTQO} 
If a Hamiltonian satisfies $(A,A_1',A_2',A_3')$-spherical LTQO, it also satisfies $(A, \wt{A}_1', \wt{A}_2', \wt{A}_3')$-spherical LTQO for 
\begin{equation}
\wt{A}_1'\supseteq A_1', \quad \wt{A}_3' \supseteq A_3', \quad \wt{A}'=A',
\end{equation}
where $A'=A_1'A_2'A_3'$ and $\wt{A}' = \wt{A}_1'\wt{A}_2'\wt{A}_3'$.
\end{Proposition}
\begin{proof}

Let $\rho_{A\wt{A}'}$ be a reduced state of a ground state of the Hamiltonian and denote its purification as $\bar{\rho}_{A\wt{A}' Y}$, which is also a purification of $\rho_{AA'}$. Consider a state $\sigma_{\wt{A}'Y}$ that is (i) zero-energy on $\wt{A'}$ and (ii) $\sigma_{\wt{A}_3'Y}=\bar{\rho}_{\wt{A}_3'Y}$. Since $\wt{A}'=A'$, it is also zero-energy on $A'$. Since $\wt{A}_3'\supseteq A_3'$, $\sigma_{A_3'Y}=\bar{\rho}_{A_3'Y}$. Thus the $(A, A_1',A_2',A_3')$-spherical LTQO condition implies $\sigma_{A_2'A_3'Y}=\bar{\rho}_{A_2'A_3'Y}$. Since $\wt{A}_2'\wt{A}_3'=\wt{A}'\setminus \wt{A}_1'\subseteq A'\setminus A_1' = A_2'A_3'$, it follows that $\sigma_{\wt{A}_2'\wt{A}_3'Y}=\bar{\rho}_{\wt{A}_2'\wt{A}_3'Y}$, proving the claim.
\end{proof}

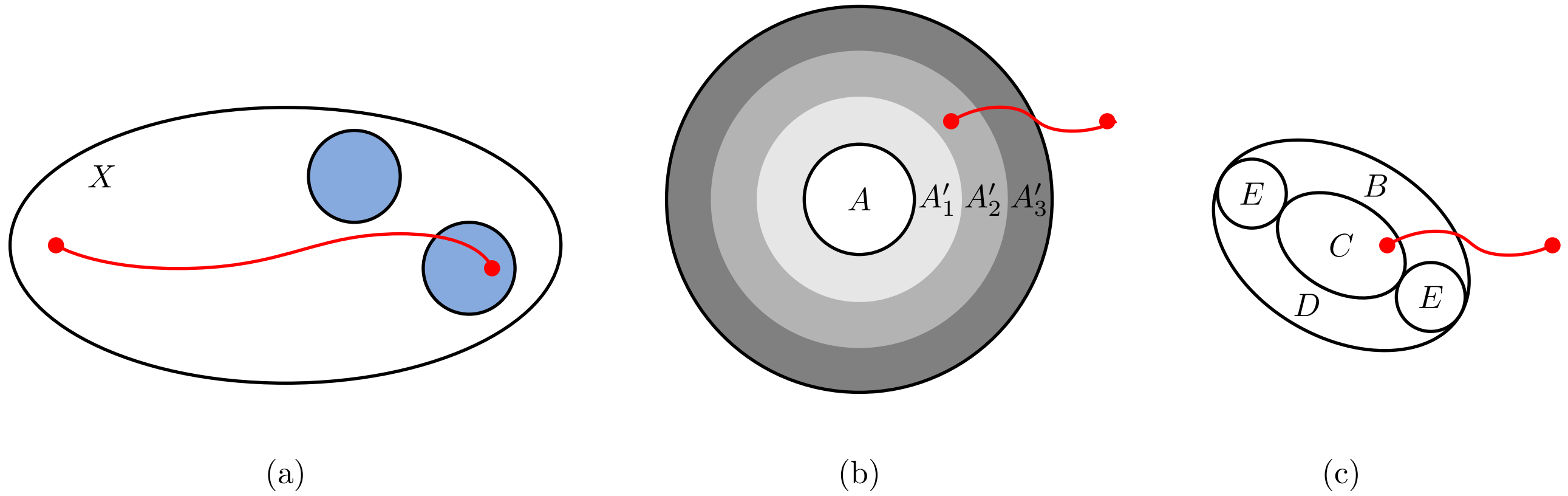
\begin{figure}[t]
    \centering
    \begin{tikzpicture}[line width=1pt]
    \draw[] (0,0) ellipse (3cm and 1.5cm);
    \node[] () at (-2cm, 0.75cm) {$X$};
    \draw[fill=tikzblue] (0.75cm, 0.75cm) circle (0.5cm); 
    \draw[fill=tikzblue] (2cm, -0.25cm) circle (0.5cm); 
    \filldraw[red] (-2.5cm, 0) circle (2pt);
    \filldraw[red] (2.25cm, -0.25cm) circle (2pt);
    \draw[red] plot [smooth, tension=1] coordinates {(-2.5cm, 0) (-1cm, -0.25cm) (1.25cm, 0.125cm) (2.25cm, -0.25cm)};
    \node[] () at (0, -2.5cm) {(a)};
    
    \begin{scope}[xshift=6.25cm, yshift=0.5cm]
        \draw[fill=black!50!white] (0,0) circle (2.1cm);
        \filldraw[black!30!white] (0,0) circle (1.6cm);
        \filldraw[black!10!white] (0,0) circle (1.1cm);
     \draw[fill=white] (0,0) circle (0.6cm);
        \node[] () at (0,0) {$A$};
        \node[] (a1p) at (0.85cm, 0) {$A_1'$};
        \node[] (a2p) at (1.35cm, 0) {$A_2'$};
        \node[] (a3p) at (1.85cm, 0) {$A_3'$};
        \filldraw[red] (1, 0.85cm) circle (2pt);
        \filldraw[red] (2.7cm, 0.85cm) circle (2pt);
        \begin{scope}[xshift=0.5cm, yshift=0.85cm]
        \draw[red] plot [smooth, tension=1] coordinates {(0.5cm, 0) (1.1cm, 0.15cm) (1.7cm, -0.1cm) (2.3cm, 0)};
        \end{scope}

        \node[] () at (0, -3cm) {(b)};
    \end{scope}
    
    \begin{scope}[xshift=11.5cm]
    \begin{scope}[rotate=-30]
    \draw[] (0,0) ellipse (1.5cm and 1cm);
    \draw[] (0,0) ellipse (0.75cm and 0.5cm);
    \draw[] (-1.125cm, 0) circle (0.375cm);
    \draw[] (1.125cm, 0) circle (0.375cm);
    \node[] () at (0,0) {$C$};
    \node[] () at (0, 0.75cm) {$B$};
    \node[] () at (0, -0.75cm) {$D$};
    \node[] () at (-1.125cm, 0) {$E$};
    \node[] () at (1.125cm, 0) {$E$};
    \end{scope}
        \filldraw[red] (0.5cm, 0) circle (2pt);
        \filldraw[red] (2.3cm, 0) circle (2pt);
        \draw[red] plot [smooth, tension=1] coordinates {(0.5cm, 0) (1.1cm, 0.15cm) (1.7cm, -0.1cm) (2.3cm, 0)};
         \node[] () at (0, -2.5cm) {(c)};
    \end{scope}
    \end{tikzpicture}
    \caption{\textbf{(a)} Two non-abelian anyons (red dots) are separated by a large distance. While LTQO holds for all small topological balls (blue), it fails for the large ball $X$, due to multiple fusion outcomes.  \textbf{(b)} We illustrate the violation of spherical LTQO on a small ball $A$ surrounded by $A'$, with a single non-abelian anyon in $A'$.  \textbf{(c)} The partial recovery property also fails for a small ball surrounding a non-abelian anyon.
 }
    \label{fig:SLTQO-example}
\end{figure}

To illustrate the importance of spherical LTQO, we demonstrate an example with the following simultaneous properties: LTQO holds on small topological balls, LTQO fails for a large ball, and both spherical LTQO and local extendibility fail for a small ball. This example consists of two distant non-abelian anyons, explained in Fig.~\ref{fig:SLTQO-example}. We conclude that LTQO on $O(1)$-size balls guarantees neither LTQO on larger scales nor local extendibility.  On the other hand, we find LTQO and spherical LTQO together on small balls are enough to guarantee both local extendibility and LTQO on all scales (Theorem~\ref{thm:LTQO_local_extendibility}, and Proposition~\ref{prop:ball-unique-0}).

Spherical LTQO may appear mysterious compared to LTQO. However, they are in fact related. For instance, on a sphere, LTQO  implies spherical LTQO; see Lemma~\ref{lemma:ltqo_to_sltqo}. There is also a sort of converse: LTQO and spherical LTQO on a ball implies the existence of a modified Hamiltonian defined on a sphere satisfying the LTQO condition on a complementary region; see Section~\ref{subsec:sphere_completion} for details. 

Later in Section~\ref{subsec:recovery_from_ltqo}, $A, \ell,$ and $L$ will be related to the lengths associated with the standard configuration set (Section~\ref{sec:micro}). Specifically, we will impose a set of LTQO/spherical LTQO conditions with the following choice:
\begin{equation}
    A = \Ba_{a}(f_+(D^{d-1})), \quad \ell, a = O(r_0), \quad L=O(r_1), \label{eq:translation_btwn_lengths}
\end{equation}
from which the local extendibility for the standard configuration set follows, and hence its reverse version at a larger scale (Theorem~\ref{thm:extend-reverse-A1p}). Since $r_1 \gg r_0$, $L\gg \ell$ for the intended application.

In the discussions below we assume $A$ is an open set and use a short-hand notation $A^{[x, \infty)}:=\mathbb{R}^d\setminus A^{+x}$. In particular, we choose $B=A^{[L+\ell, \infty)}$ and assume $L/2>\ell> r_H$.
\begin{Lemma}
 \label{lemma:ltqo_to_sltqo} 
 If $H$ satisfies $(A^{+L}, \ell)$-LTQO and $(A^{[\ell, \infty)},A^{[0, \ell)})$-LTQO, $H_{A^{+(L+\ell)}}$ satisfies $(A, \ell, L-2\ell)$-spherical LTQO.
\end{Lemma}
 \begin{proof}
Following the notation of Definition~\ref{definition:spherical_ltqo}, let $\rho_{A^{+(L+\ell)}}$ be an arbitrary ground state of $H_{A^{+(L+\ell)}}$. By the $(A^{+L},\ell)$-LTQO condition, $\rho_{A^{+L}}=\psi_{A^{+L}}$, where $\psi$ is a pure ground state of $H$. Thus we can choose the purification of $\rho_{A^{+L}}$ as $\psi,$ setting the purifying system as $Y=A^{[L, \infty)}$.

Let $A'=A_1'A_2'A_3'$, where $A_1'=A^{[0, \ell)}$, $A_2'=A^{[\ell, L-\ell)}$, $A_3' = A^{[L-\ell,L)}$. Without loss of generality, let $\sigma_{A'Y}$ be any state that is (i) zero-energy on $A_1'A_2'A_3'$ and (ii) $\sigma_{A_3'Y} = \psi_{A_3'Y}$. The former and the latter condition implies that $\sigma_{A'Y}$ is zero-energy on $A_1'A_2'A_3'=A^{[0,L)}$ and $A_3'Y=A^{[L-\ell, \infty)}$, respectively. Because $\ell > r_H$, every local term of the Hamiltonian either belongs to the former or the latter. Thus $\sigma_{A'Y}$ is zero-energy on $A'Y = A^{[0,\infty)}$. By the $(A^{[\ell, \infty)},A^{[0, \ell)})$-LTQO condition, its reduced state on $A^{[\ell, \infty)}=A_2'A_3'Y$ is equal to that of $\psi$, proving the spherical LTQO condition.
 \end{proof}
 One can envision a similar argument on a general manifold, a sphere in particular. In that setup, one can view $A$ and $B$ as two ball-like regions. Thus imposing LTQO on two balls would imply a spherical LTQO condition on a ball. Conversely, one might wonder if spherical LTQO implies LTQO on a complementary region. This is generally false. However, one can instead prove a different statement that will be useful later: if LTQO holds on one region, spherical LTQO on its complementary region can increase the size of the region on which LTQO holds.

\begin{Lemma}
\label{lemma:sltqo_to_larger_ltqo} Let $H$ be a frustration-free Hamiltonian satisfying the following conditions. 
    \begin{enumerate}
        \item $H_{A^{+(L+\ell)}}$ satisfies $(A, \ell, L-2\ell)$-spherical LTQO.
        \item $H$ satisfies $(A^{+L},\ell)$-LTQO. 
    \end{enumerate}
If $H$ satisfies $(A^{[L-\ell, \infty)}, A^{[L-2\ell, L-\ell)})$-LTQO, it satisfies $(A^{[\ell,\infty)}, A^{[0, \ell)})$-LTQO.
\end{Lemma}

\begin{proof}

Without loss of generality, consider a pair of ground states $\tau$ and $\tau'$ of $H_{A^{[0, \infty)}}$. We aim to show that they are indistinguishable on $A^{[\ell, \infty)}$.

Consider a ground state of $H_{A^{+(L+\ell)}}$, denoted as $\rho_{A^{+(L+\ell)}}$. By the $(A^{+L},\ell)$-LTQO condition $\rho$ is indistinguishable from any ground state of $H$ on $A^{+L}$. Let $\psi$ be a pure ground state of $H$. We choose the purification of $\rho_{A^{+L}}$ as $\psi$. The purifying system is thus $Y=A^{[L,\infty)}$.

By the $(A^{[L-\ell, \infty)}, A^{[L-2\ell, L-\ell)})$-LTQO condition, $\psi$, $\tau$, and $\tau'$ are indistinguishable on $A^{[L-\ell, \infty)}$. We now aim to apply the spherical LTQO condition. Note the relevant subsystems are $A_1'=A^{[0,\ell)}$, $A_2'=A^{[\ell, L-\ell)}$, $A_3'=A^{[L-\ell, L)}$. In particular, $A_3'Y = A^{[L-\ell, \infty)}$ and as such $\psi_{A_3'Y}=\tau_{A_3'Y}=\tau_{A_3'Y}'$. Moreover, $\tau$ and $\tau'$ are zero-energy on $A'=A^{[0, L)}$ by construction. Thus by the spherical LTQO condition $\tau_{A_2'A_3'Y}=\tau'_{A_2'A_3'Y}=\psi_{A_2'A_3'Y}$, which implies they are indistinguishable on $A^{[\ell, \infty)}$.

\end{proof}

We note that the proof of Lemma~\ref{lemma:ltqo_to_sltqo} and~\ref{lemma:sltqo_to_larger_ltqo} applies to all  regions. However, in Section~\ref{subsec:sphere_completion} and~\ref{subsec:recovery_from_ltqo} we will focus on the case where $AA'$ is bounded (so that the associated Hilbert space is finite-dimensional). Furthermore, in those sections, the local terms of the Hamiltonian supported on $A^{[0,\infty)}$ will have finite number of terms acting on a finite-dimensional Hilbert space. Thus in that setup the state can be also viewed as a density matrix.

\subsection{Spherical completion}
\label{subsec:sphere_completion}

In this section, we show that LTQO and spherical LTQO condition implies the existence of a certain Hamiltonian we call as \emph{spherical completion}. We start with the formal definition.

\begin{Definition}
\label{def:spherical_completion}
    Let $H_A$ be a frustration-free Hamiltonian and let $A'\subseteq A$. A $(B,A')$-spherical completion of $H_{A}$ is a Hamiltonian of the form $H'=H_A+H_{A'B}$ satisfying these properties:
    \begin{enumerate}
        \item $H'$ is frustration-free and has a unique ground state.
        \item $H_{A'B} = I-Q_{A'B}$, where $Q_{A'B}$ is a projection.
    \end{enumerate}
     If $H_A$ is a local commuting Hamiltonian and $H_{A'B}$ commutes with these terms, $H'$ is a commuting spherical completion.
\end{Definition}
\noindent
For the standard choice, we let $A=\wt{A} A'$, where $A'= \wt{A}^{[0, r)}$ for some $r>0$. If $A$ is a ball and $A'$ is its inner collar, we can view the auxiliary system $B$ as a ball that completes a sphere, hence the name. 

We remark that a spherical completion can be always promoted to a commuting spherical completion if the Hamiltonian consists of commuting local terms.
\begin{Proposition}
\label{prop:spherical_completion_commuting_upgrade}
    Let $H$ be a frustration-free local commuting Hamiltonian. Let $A=\wt{A}^{+r}$ and $r_H<x<r$. If $H_A$ admits a $(B, \wt{A}^{[x, r)})$-spherical completion, it also admits a $(B, \wt{A}^{[x-r_H, r)})$-commuting spherical completion.
\end{Proposition}
\begin{proof}
    Let $H'=H_A + (I-Q_{\wt{A}^{[x,r)} B})$ be a $(B,\wt{A}^{[x,r)} )$-spherical completion and write $Q:=Q_{\wt{A}^{[x,r)}B}$. One can replace $Q_{\wt{A}^{[x,r)}B}$ by the projection to the kernel of $I-(Q_{\wt{A}^{[x,r)}B} + Q')/2$ without changing the ground state, where $Q'$ is a product of local projectors supported on $A$ acting nontrivially on $\wt{A}^{[x,r)}B$. In particular, $Q'$ is supported on $\wt{A}^{[x-r_H, r)}$. Let this projection be $\wh{Q}:=Q_{\wt{A}^{[x-r_H, r)}B}= \lim_{n\to \infty} ((Q + Q')/2)^n$. By construction, the resulting Hamiltonian is frustration-free and has a unique ground state. Furthermore, it is supported on $\wt{A}^{[x-r_H, r)}B$.

    Now we prove that $\wh{Q}$ commutes with the local terms in $H_A$. Note first that $\wh{Q}$ is the projection onto $\supp Q\cap\supp Q'$. Let $X$ be the support of a local term. If $X$ meets $\wt{A}^{[x,r)}$, then $Q_X$ is one of the factors of $Q'$, so $\supp\wh{Q}\subseteq\supp Q_X$ and hence $Q_X\wh{Q}=\wh{Q}=\wh{Q}Q_X$. Otherwise $X$ is disjoint from $\wt{A}^{[x,r)}B$, so $Q_X$ commutes with $Q$; it also commutes with $Q'$, since $H$ is commuting. Hence $Q_X$ commutes with $(Q+Q')/2$ and therefore with $\wh{Q}$.

\end{proof}

We now show that LTQO and spherical LTQO on a set of subsystems implies the existence of a spherical completion. We first show that spherical completion follows from LTQO conditions alone (Proposition~\ref{prop:nested_ltqo_sphere_completion}). However, for deriving extendibility/local extendibility, we later require a spherical completion that satisfies the LTQO condition on a larger region (Section~\ref{subsec:recovery_from_ltqo}). By additionally assuming a spherical LTQO condition, this can be ensured (Proposition~\ref{prop:spherical_completion_large_ltqo}). In this discussion, we assume the region $AA'$ is bounded; in particular $\mathcal{H}_{AA'}$ is finite-dimensional.

\begin{Proposition}[Spherical completion from LTQO on nested balls] Let $A$ be a bounded region and let $A_{i+1}=A_i^{+\ell_{i+1}}$ for $i\in \{0,1,2,3\}$, $A_0:=A$. If $H$ satisfies $(A_i,\ell_{i+1})$-LTQO for $i\in \{0,1,2,3\}$, $H_{A^{+(\ell_1+\ell_2+\ell_3)}}$ admits a $(B,A^{[0,\ell_1+\ell_2+\ell_3)})$-spherical completion $H'$. Furthermore, $H'$ satisfies $(A^{[\ell_1,\ell_1+\ell_2+\ell_3)}B, A^{[0,\ell_1)})$-LTQO.
\label{prop:nested_ltqo_sphere_completion}
\end{Proposition}

The proof of Proposition~\ref{prop:nested_ltqo_sphere_completion} is based on the relations between local purifiability and LTQO. We briefly discuss these first.

\begin{Theorem}
\label{thm:LTQO_purifiability}
Consider a frustration-free Hamiltonian satisfying $(A,A')$-LTQO. Every ground state $\sigma$ is $(A', A)$-purifiable.
\end{Theorem}
\begin{proof}
By Proposition~\ref{prop:supp}, it suffices to show that for every state $\tau_{AA'}$ with $\supp(\tau_{AA'}) \subset \supp(\sigma_{AA'})$ its reduced density matrix on $A$ must be identical.  Note $\tau_{AA'}$ satisfying $\supp(\tau_{AA'}) \subset \supp(\sigma_{AA'})$ must have zero energy with respect to Hamiltonian terms on $AA'$. Hence by LTQO, the claim follows.
\end{proof}
\begin{Theorem}
    \label{thm:purifiability_LTQO}
    Let $\rho$ be a state that is $(A', A)$-purifiable. There is a Hamiltonian supported on $AA'$ that satisfies $(A,A')$-LTQO and has $\rho_{AA'}$ as a ground state.
\end{Theorem}
\begin{proof}
Because of the condition (2) in the definition of local purifiability (Definition~\ref{def:loc_puri}), there is a unitary isomorphism from $\mathcal{H}_{A'}$ to $(\mathcal{H}_{A_1'}\otimes \mathcal{H}_{A_2'})\oplus \mathcal{H}_{{A'}^{\perp}}$ that takes $\rho_{AA'}$ to $\rho_{A_1'}' \otimes \rho_{AA_2'}'$, where $\rho_{AA_2'}'$ is a pure state. Denote the projection onto this pure state conjugated by the unitary isomorphism as $Q_{AA'}$. Then the Hamiltonian is $I-Q_{AA'}$. By construction, $Q_{AA'}\rho_{AA'}=\rho_{AA'}$. Note the reduced density matrix of any elements in the kernel of this Hamiltonian is $\rho_A$, because the isomorphism acts only on $A'$.
\end{proof}

\begin{proof}[Proof of Proposition~\ref{prop:nested_ltqo_sphere_completion}]

Because of the $(A^{+(\ell_1+\ell_2+\ell_3)}, \ell_4)$-LTQO condition, every ground state of $H_{A^{+(\ell_1+\ell_2+\ell_3+\ell_4)}}$ has the same reduced density matrix on $A^{+(\ell_1+\ell_2+\ell_3)}$. Denote the reduced density matrix as $\rho_{A^{+(\ell_1+\ell_2+\ell_3)}}$ and its purification as $\psi_{A^{+(\ell_1+\ell_2+\ell_3)}B}$. 

Let $A_i':=A_i\setminus A_{i-1}$. From the LTQO conditions, it follows that $\psi$ is $(A_{i+1}',A_i)$-purifiable for $i\in\{0,1,2 \}$ (Theorem~\ref{thm:LTQO_purifiability}).  Note any tripartite finite-dimensional pure state over $UVW$ is $(V,U)$-purifiable if and only if it is $(V,W)$-purifiable. Thus, writing $L_i:=\sum_{j=1}^i \ell_j$ with $L_0:=0$, the state $\psi$ is $(A^{[L_{i-1},L_i)},\, A^{[L_i,\infty)})$-purifiable for $i\in\{1,2,3\}$, where 
\begin{equation}
    A^{[x, \infty)} := A^{[x, \ell_1+\ell_2+\ell_3)}B \label{eq:purification_convention}
\end{equation}
for any $0\le x \le \ell_1+\ell_2+\ell_3$. Note the Hilbert space associated with $A^{[x,\infty)}$ is finite-dimensional because both $A^{[x, \ell_1+\ell_2+\ell_3)}$ and $B$ are finite-dimensional.

Let $H_{A^{[0, \infty)}}$ be the Hamiltonian introduced in the proof of Theorem~\ref{thm:purifiability_LTQO}, based on the $(A^{[0,\ell_1)}, A^{[\ell_1,\infty)})$-purifiability condition. By construction, $(H_{A^{+(\ell_1+\ell_2+\ell_3)}} + H_{A^{[0, \infty)}})\psi=0$. Thus this Hamiltonian has at least one zero energy state and satisfies the $(A^{[\ell_1,\infty)}, A^{[0,\ell_1)})$-LTQO. 

We now show that $\psi$ is the unique ground state of this Hamiltonian. Let $\rho$ and $\rho'$ be two ground states. Note the elements of the kernel of $H_{A^{+(\ell_1+\ell_2+\ell_3)}} + H_{A^{[0, \infty)}}$ are indistinguishable on $AA_1'A_2'$ and $BA_3'A_2'$. Furthermore, they are  $(A_2',AA_1')$-purifiable and $(A_2',BA_3')$-purifiable. Any such state can be reconstructed from the reduced density matrix on $AA_1'A_2'$ by applying a Petz map from $A_2'$ to $BA_3'A_2'$. Because the reduced density matrices are equal on both systems, $\rho$ and $\rho'$ must be equal globally. Thus the ground state is unique.
\end{proof}

We are now in a position to prove that a finite set of LTQO and spherical LTQO condition implies the existence of a spherical completion that satisfies a certain LTQO condition. The existence of this spherical completion is shown to later imply extendibility and reverse extendibility (Section~\ref{subsec:recovery_from_ltqo}). For the discussions below, we use the following short-hand notation:
\begin{equation}
        A^{[x, \infty)}:= A^{[x, L+\ell)}B
\end{equation}
for any $0\le x \le L+\ell$. In particular, $B$ shall be chosen as a purification of $A^{+(L+\ell)}$.

\begin{Proposition}
\label{prop:spherical_completion_large_ltqo}
    Let $A$ be a bounded region and let $L>2\ell$. If $H_{A^{+(L+\ell)}}$ satisfies $(A, \ell, L-2\ell)$-spherical LTQO and $H$ satisfies $(A^{+x},\ell)$-LTQO for $x\in \{L-2\ell, L-\ell, L, L+\ell\}$, $H_{A^{+(L+\ell)}}$ admits a $(B,A^{[L-2\ell, L+\ell)})$-spherical completion satisfying $(A^{[\ell,\infty)}, A^{[0, \ell)})$-LTQO.
\end{Proposition}
\begin{proof}
Applying the second assumption to Proposition~\ref{prop:nested_ltqo_sphere_completion}, it follows that $H_{A^{+(L+\ell)}}$ admits a $(B,A^{[L-2\ell, L+\ell)})$-spherical completion satisfying $(A^{[L-\ell, L+\ell)}B,A^{[L-2\ell,L-\ell)})$-LTQO, which is equivalent to the $(A^{[L-\ell, \infty)},A^{[L-2\ell,L-\ell)})$-LTQO condition. 

Let us denote the spherical completion as $H'= H_{A^{+(L+\ell)}} + H_{A^{[L-2\ell,\infty)}}'$. Because the additional term $H_{A^{[L-2\ell,\infty)}}'$ does not contain terms supported on $A^{+(L+\ell)}$, $H_{A^{+(L+\ell)}}'=H_{A^{+(L+\ell)}}$. As such, $H_{A^{+(L+\ell)}}'$ satisfies $(A,\ell, L-2\ell)$-spherical LTQO. Moreover, $H$, and hence $H'$, satisfies $(A^{+L},\ell)$-LTQO condition thanks to the second assumption. These two, together with $(A^{[L-\ell, \infty)},A^{[L-2\ell,L-\ell)})$-LTQO condition lets us apply Lemma~\ref{lemma:sltqo_to_larger_ltqo} verbatim, proving the claim.
\end{proof}

Moreover, we can upgrade the spherical completion to a commuting spherical completion. This can be done by modifying the nonlocal term appearing in the spherical completion using the method in the proof of Proposition~\ref{prop:spherical_completion_commuting_upgrade}.
\begin{Corollary}
\label{corollary:commuting_case}
    Let $H$ be a local commuting Hamiltonian. Let $A$ be a bounded region and let $L>3\ell$ and $\ell>r_H$. If $H_{A^{+(L+\ell)}}$ satisfies $(A, \ell, L-2\ell)$-spherical LTQO and $H$ satisfies $(A^{+x},\ell)$-LTQO for $x\in \{L-2\ell, L-\ell, L, L+\ell\}$, $H_{A^{+(L+\ell)}}$ admits a $(B,A^{[L-2\ell-r_H, L+\ell)})$-commuting spherical completion satisfying $(A^{[\ell,\infty)}, A^{[0, \ell)})$-LTQO.
\end{Corollary}
Note all the assumptions are imposed on a finite region $A^{+(L+2\ell)}$. From this, we deduced the existence of a frustration-free commuting Hamiltonian on $A^{+(L+\ell)}B$. This Hamiltonian contains a single non-local term supported on $A^{[L-2\ell-r_H, \infty)}\subseteq A^{[L-3\ell, \infty)}$. Moreover, the $(A^{[\ell,\infty)}, A^{[0,\ell)})$-LTQO implies $(A^{[\ell',\infty)}, A^{[0,\ell')})$-LTQO for any $\ell' \ge \ell$ because of the monotonicity property (Proposition~\ref{prop:mono_LTQO}). We consolidate pertinent properties of this Hamiltonian as follows.
\begin{Corollary}
    \label{corollary:spherical_completion_consolidated}
    Let $L>3\ell$. Under the assumptions of Corollary~\ref{corollary:commuting_case}, $H_{A^{+(L'+\ell')}}$ admits a $(B, A^{[L', L'+\ell')})$ commuting spherical completion that satisfies $(A^{[\ell',\infty)}, A^{[0,\ell')})$-LTQO, where $L'=L-3\ell$, $\ell'=4\ell$.
\end{Corollary}

\begin{Proposition}
\label{prop:mono_LTQO}
If a Hamiltonian satisfies $(A,A')$-LTQO, it also satisfies $(\wt{A}, \wt{A}')$-LTQO for
\begin{equation}
\wt{A}\subseteq A, \quad\wt{A}\wt{A}' \supset AA'.
\end{equation}
\end{Proposition}
\begin{proof}
Let $\rho_{\wt{A} \wt{A}'}$ be a ground state of $H_{\wt{A} \wt{A}'}$, i.e., $\Tr(\rho_{\wt{A} \wt{A}'}H_{\wt{A} \wt{A}'})=0$. Since $H_{\wt{A} \wt{A}'}\ge H_{AA'}$, it follows that $\Tr(\rho_{\wt{A} \wt{A}'} H_{AA'})=0$. Thus $\Tr_{\wt{A}\wt{A}' \setminus (AA')}(\rho_{\wt{A} \wt{A}'})$ is a ground state of $H_{AA'}$. 

Now consider two ground states of $H_{\wt{A} \wt{A}'}$. By the discussion in the previous paragraph, their reduced density matrices on $AA'$ are ground states of $H_{AA'}$. Thus $(A,A')$-LTQO implies their reduced density matrices on $A$ must be identical. Taking the partial trace on  $A\setminus \wt{A}$, the claim follows.
\end{proof}

The commuting spherical completion appearing in Corollary~\ref{corollary:spherical_completion_consolidated} will be used to construct the recovery map in Section~\ref{subsec:recovery_from_ltqo}. In the setup described in Eq.~\eqref{eq:translation_btwn_lengths}, we will get a commuting spherical completion with $L' = O(r_1)$ and $\ell = O(r_0)$.

\subsection{Recovery maps}
\label{subsec:recovery_from_ltqo}

We now show that LTQO and spherical LTQO implies extendibility. For this discussion, we specialize to commuting Hamiltonians. Recalling  $H = \sum_X(I - Q_X)$, we also use 
\begin{align} \label{eq:ham_proj_prod}
P_A = \prod_{X \subset A} Q_X,
\end{align}
where $X$ runs over supports of Hamiltonian terms, to denote the product of projections supported in $A$.  The following lemma will be our main tool for defining recovery maps using LTQO. 
\begin{Lemma} \label{lemma:ABC_recovery} 
    Consider commuting, frustration-free Hamiltonian $H$. Let $A,B,C$ be disjoint regions such that for any local term supported on $X\subseteq ABC$, it satisfies either $X\subseteq AB$ or $X\subseteq BC$. Let $R_{B \to BC}$ be the Petz recovery map associated to the state $(\Tr P_{BC})^{-1} P_{BC}$, given by the quantum channel
\begin{align}  \label{eq:RBBC}
    R_{B \to BC}[\lambda_B] 
    &= P_{BC} \Omega_B^{-1/2} W_B \lambda_B W_B \Omega_B^{-1/2}  P_{BC} + (\dim\calH_C)^{-1} \mathds{1}_C W_B^\perp \lambda_B W_B^\perp,
\end{align}
where $\Omega_B = \Tr_C P_{BC}$ and $W_B$ is the projection onto $\supp(\Omega_B)$. Then for any state $\sigma_{AB}$ such that $P_{AB} \sigma_{AB} =\sigma_{AB}$ and $\supp(\sigma_B) \subset \supp(\Omega_B)$ the state $\sigma'_{ABC} = R_{B \to BC}[\sigma_{AB}]$ satisfies 
\begin{align}  \label{eq:P_ABC_preserve}
    P_{ABC} \sigma'_{ABC} = \sigma'_{ABC}.
\end{align}
\end{Lemma}
Here the map $R_{B \to BC}$ is the usual Petz map for state $(\Tr P_{BC})^{-1} P_{BC}$ when restricted to inputs supported on subspace $\supp[\Omega_B]$, where $\Omega_B^{-1/2}$ is then well-defined.  The second term in \eqref{eq:RBBC} has been added explicitly so that $R_{B \to BC}$ is a CPTP map on the whole domain.
\begin{proof}
We consider  $\sigma'_{ABC} = R_{B \to BC}[\sigma_{AB}]$ given by Eq.~\eqref{eq:RBBC}, for a state $\sigma_{AB}$ such that $P_{AB} \sigma_{AB} =\sigma_{AB}$ and $\supp(\sigma_B) \subset \supp(\Omega_B)$. The latter implies the second term vanishes on the RHS of \eqref{eq:RBBC}.  (We used notation $W_B^\perp=(I-W_B)$ for projection $W_B$.)
    To see Eq.~\eqref{eq:P_ABC_preserve},
 note $P_{ABC}=P_{AB}P_{BC}$ because of the condition on the support of the local terms. Therefore, it is sufficient to show  $P_{AB} \sigma'_{ABC}=\sigma'_{ABC}$ and $P_{BC} \sigma'_{ABC}=\sigma'_{ABC}$.  
 The latter follows immediately from the form of $\eqref{eq:RBBC}$.  To see $P_{AB} \sigma'_{ABC}=\sigma'_{ABC}$, first note $[P_{AB},P_{BC}]=0$, so $[P_{AB},\Omega_B]=0$ where $\Omega_B=\Tr_C P_{BC}$, hence also $[P_{AB},\Omega_B^{-1/2}]=0$.  Then $P_{AB} \sigma'_{ABC}=\sigma'_{ABC}$ also follows immediately from the form of $\eqref{eq:RBBC}$, by commuting $P_{AB}$ through and using the assumption that $P_{AB} \sigma_{AB}=\sigma_{AB}$.
\end{proof}

This lemma works independent of the topologies of the regions, insofar as the combinatorial condition on the support of the local terms are satisfied. However, unlike the usual recovery maps, it does not guarantee $\sigma_{AB}'= \sigma_{AB}$. All it can guarantee is that the state obtained from an application of $R_{B\to BC}$ is \emph{some} state which is zero-energy on $ABC$. Nevertheless, with the LTQO condition the zero-energy condition can be upgraded to a statement about extendibility. Recall that $(V,V')$-LTQO condition implies that any zero-energy state on $VV'$ looks identical on $V$. Thus if $ABC=VV'$ satisfies the LTQO condition, the restriction of the extended state to $V$ is the ground state reduced density matrix. A similar argument can be applied to density matrices that include purifications, using the commuting spherical completion (Definition~\ref{def:spherical_completion}).

Now we are in a position to prove the local extendibility condition for the standard configuration set $\Lambda (r_0, r_1)= \{\lambda(r_0, f): f\in \Delta(r_1) \}$ (Section~\ref{sec:micro}) from a set of LTQO and spherical LTQO conditions. Without loss of generality, consider a configuration $(B,C,D,E) \in \Lambda(r_0, r_1)$:
\begin{equation}
    BE = \Ba_{r_0}(f_-(D^{d-1})), \quad DE= \Ba_{r_0}(f_+(D^{d-1})), \label{eq:be_de_def}
\end{equation}
where $r_0\le r_1/2$ and $C$ is the part of $\mathbb{R}^d\setminus (BE\cup DE)$ that lies inside $f(S^{d-1})$.

Below, we show that $(B,C,D,E)$-extendibility follows from a finite set of LTQO and spherical LTQO conditions. To that end, we consider a configuration $(\wt{B}, \wt{C}, \wt{D}, \wt{E})\cge (B,C,D,E)$, chosen as follows:
\begin{equation}
    \wt{B}\wt{E} = \Ba_a(f_-(D^{d-1})), \quad \wt{D}\wt{E} = \Ba_a(f_+(D^{d-1})),
\end{equation}
and $\wt{C}$ is the inside part of $\mathbb{R}^d \setminus (\wt{B}\wt{E} \cup \wt{D}\wt{E})$ (Section~\ref{sec:micro}). We consider $X= (\wt{D}\wt{E})^{+(L+a)}$ and make the following assumptions. 
\begin{enumerate}
    \item $H_X$ satisfies $((\wt{D}\wt{E})^{+L}, a)$-LTQO. \label{assumption:1} 
    \item $H_X$ admits a $(Y, (\wt{D}\wt{E})^{[L, L+a)})$-commuting spherical completion satisfying $((\wt{D}\wt{E})^{[a,\infty)} ,(\wt{D}\wt{E})^{[0,a)})$-LTQO. (Here $(\wt{D}\wt{E})^{[x,\infty)}:= (\wt{D}\wt{E})^{[x, L+a)}Y$ for $ 0\le x\le L+a$)\label{assumption:2} 
    \item Geometric assumptions: (i) $L$ is chosen to be large enough so that $(\wt{D}\wt{E})^{+L}\supseteq BCDE$ and (ii) $\Ba_a(\wt{D}\wt{E})\subseteq DE$ and $a > r_H$. \label{assumption:3}
\end{enumerate}
Note the assumptions about the Hamiltonian can be checked locally from $H_{X^{+a/4}}$ and follow from a finite set of LTQO and spherical LTQO conditions (Corollary~\ref{corollary:spherical_completion_consolidated}).

\begin{Theorem}
Under the assumptions~\eqref{assumption:1}~\eqref{assumption:2}, and~\eqref{assumption:3}, every ground state of $H$ satisfies $(B, C, D, E)$-extendibility.
\label{thm:LTQO_local_extendibility}
\end{Theorem}
\begin{proof}
Since $BCDE \subseteq (\wt{D}\wt{E})^{+L}$, \eqref{assumption:1} implies that the reduced density matrix over $BCDE$ is indistinguishable for every ground state of $H$ as well as that of the spherical completion. We denote this reduced density matrix as $\rho_{BCDE}$. Since $(\wt{B},\wt{C},\wt{D},\wt{E})\cge (B,C,D,E)$, $\wt{B}\wt{C}\wt{D}\wt{E} \subseteq BCDE$, so the indistinguishability condition also holds for $\wt{B}\wt{C}\wt{D}\wt{E}$. 

Below, we consider two different purifications with respect to these regions. Let $\psi$ be the ground state of the spherical completion of $H_X$. For $BCDE$ and $\wt{B}\wt{C}\wt{D}\wt{E}$, the respective purifications become $A=YX\setminus (BCDE)$ and $\wt{A}=YX\setminus (\wt{B}\wt{C}\wt{D}\wt{E})$. By the discussion of the previous paragraph, $\psi_{BCDE} = \rho_{BCDE}$ and also $\psi_{\wt{B}\wt{C}\wt{D}\wt{E}} = \rho_{\wt{B}\wt{C}\wt{D}\wt{E}}$. 

We apply the recovery map of Lemma~\ref{lemma:ABC_recovery} from $\wt{B}\wt{E}$ to $\wt{B}\wt{C}\wt{E}$, where the input state is $\psi_{\wt{A}\wt{B}\wt{E}}$. Because $a>r_H$, for any support $Z\subseteq \wt{A}\wt{B}\wt{C}\wt{E}$ of a term of the spherical completion, either $Z\subseteq \wt{A}\wt{B}\wt{E}$ or $Z\subseteq \wt{B}\wt{C}\wt{E}$. Thus we have a zero-energy state on $\wt{A}\wt{B}\wt{C}\wt{E}$. Consider a restriction of the zero-energy state on $\wt{A}\wt{B}\wt{C}\wt{E}$ to $ABC$. (Since $DE\supseteq \wt{D}\wt{E}$, $ABC\subseteq \wt{A}\wt{B}\wt{C}$.) Because $ABC = XY \setminus DE \subseteq XY\setminus \Ba_a(\wt{D}\wt{E}) = (\wt{D}\wt{E})^{[a, \infty)}$ by~\eqref{assumption:3}, \eqref{assumption:2} implies that the resulting state is $\psi_{ABC}$.

Now we can formally view the recovery map from $\wt{B}\wt{E}$ to $\wt{B}\wt{C}\wt{E}$, followed by the partial trace, as a map from $BE$ to $BC$. Thus $\psi$ is $(B,C,D,E)$-extendible and so is $\rho_{BCDE}$.
\end{proof}

When choosing the regions involved in the LTQO/spherical LTQO condition, it is desirable to ensure they are topological balls. This can be ensured in the following way. Given $f\in \Delta(r_1)$, one can choose $a$ sufficiently small so that $(\wt{D}\wt{E})^{+a}$ is a ball. Maps in $\Delta(r_1)$ may not be smooth, so we define the regions using the smooth replacement $\tilde f$ of Proposition~\ref{prop:smooth-replacement} in place of $f$. It is $(c_{\mathrm{sm}}r_0)$-tame, and choosing $2a\le c_{\mathrm{sm}}r_0$ places $a$ below the curvature radius. The proposition, applied with $2a$ in place of $a$, gives $\Ba_a(\wt{D}\wt{E})=\Ba_{2a}(\tilde f(S_+))\subseteq DE$, as required in~\eqref{assumption:3}, and $(\wt{B},\wt{C},\wt{D},\wt{E})\cge(B,C,D,E)$ still holds. Assuming $L\gg a$, there will be also regions of the form of $(\wt{D}\wt{E})^{+(L+O(a))}$. Choosing $L=cr_1$ for some sufficiently large constant $c$, one can ensure this is also topologically a ball. 

To summarize, by imposing the LTQO/spherical LTQO conditions over a finite set of regions for each configuration, we proved local extendibility of the standard configuration set. The regions satisfy the conditions for the lengths specified in Eq.~\eqref{eq:translation_btwn_lengths}, assuming $r_H \ll r_0 \ll r_1$. We note that the reverse extendibility then follows at a larger scale by Theorem~\ref{thm:extend-reverse-A1p}, and so does the merging operation (Section~\ref{sec:rev_extendibility_merging}).

\subsection{Preparing ground states with sequential circuits}
\label{subsec:sequential_circuit}

For quantum computing and quantum simulation, one may want to prepare a ground state using a quantum circuit. With the results in Section~\ref{subsec:recovery_from_ltqo}, one can show that LTQO conditions imply that a ground state can be prepared using a circuit whose depth and number of gates that are both linear in volume. The circuit is sequential in the sense of ~\cite{chen2024sequential}, with each qudit only acted on by $O(1)$ gates. It uses ancillas in addition to the qudits on which the ground state is defined.  

Ref.~\cite{kim2024learning} presented results about learning a sequential circuit to prepare a ground state, given copies of the quantum state, and assuming local extendibility. The results in this section are different in that we focus on commuting Hamiltonians and do not assume local extendibility, but rather LTQO or variants thereof.

First we formulate a result with a weaker assumption than LTQO. Specifically, we consider a tame topological ball $\wt{B}=B^{+\ell}$ and $\wh{B}\supset \wt{B}$ and posit that
\begin{equation}
\supp [\Tr_{\wt{B}\setminus B} P_{\wt{B}}] = \supp [\Tr_{\wh{B} \setminus B} P_{\wh{B}}]. \label{eq:supps_assumption_v2}
\end{equation}
While Eq.~\eqref{eq:supps_assumption_v2} follows from $(B, \ell)$-LTQO, it is not directly related to the ground state degeneracy or reduced density matrix. In fact, it even allows $H=0$. Nevertheless, the following lemma lets us create \textit{some} ground state of $H$ on a large ball from this assumption.

\begin{Lemma}\label{lemma:general-growth_v2}
Let $H$ be a local, frustration-free, commuting projector Hamiltonian and $\sigma_X$ be a ground state of $H_X$. Consider a partition  $X=X'BB'E$ and let $C$ be disjoint from $X$. If $\wt{B}=B^{+\ell}\subseteq X$ contains $B'$ and $\wh{B}=\wt{B}C$ satisfies Eq.~\eqref{eq:supps_assumption_v2} and every local term in $H_{X'BB'C}$ is supported either on $BB'C$ or $X'B$, there is a channel $R_{BB'E \to BB'C}$ such that $R_{BB'E \to BB'C}(\sigma_X)$ is a ground state of $H_{X'BB'C}$.
\end{Lemma}
\begin{proof}
The channel $R_{BB'E \to BB'C}$ can be constructed by erasing $B'E$ and then applying the recovery channel from Lemma~\ref{lemma:ABC_recovery}. (More precisely,  $C$ in Lemma~\ref{lemma:ABC_recovery} should be replaced by $B'C$ here.) In order to apply this lemma, one should show that $P_{X'B}\sigma_{X'B}=\sigma_{X'B}$, which follows from the assumption. It also requires $\supp[\sigma_B] \subset \supp[\Tr_{B'C} P_{BB'C}]$, which follows from the following chain of argument:
\begin{equation}
\supp[\sigma_{B}] \subset \supp[\Tr_{\wt{B} \setminus B} P_{\wt{B}}] =\supp [\Tr_{\wh{B}\setminus B} P_{\wh{B}}]\subset \supp[\Tr_{B'C} P_{BB'C}].
\end{equation}
The first inclusion follows from $\supp[\sigma_{\wt{B}}] \subset \supp[P_{\wt{B}}]$ because $\wt{B}\subset X$ and $P_X\sigma_X =\sigma_X$. The next equality follows from Eq.~\eqref{eq:supps_assumption_v2}. The second inclusion follows from $B'C\subset \wh{B}$.
\end{proof}

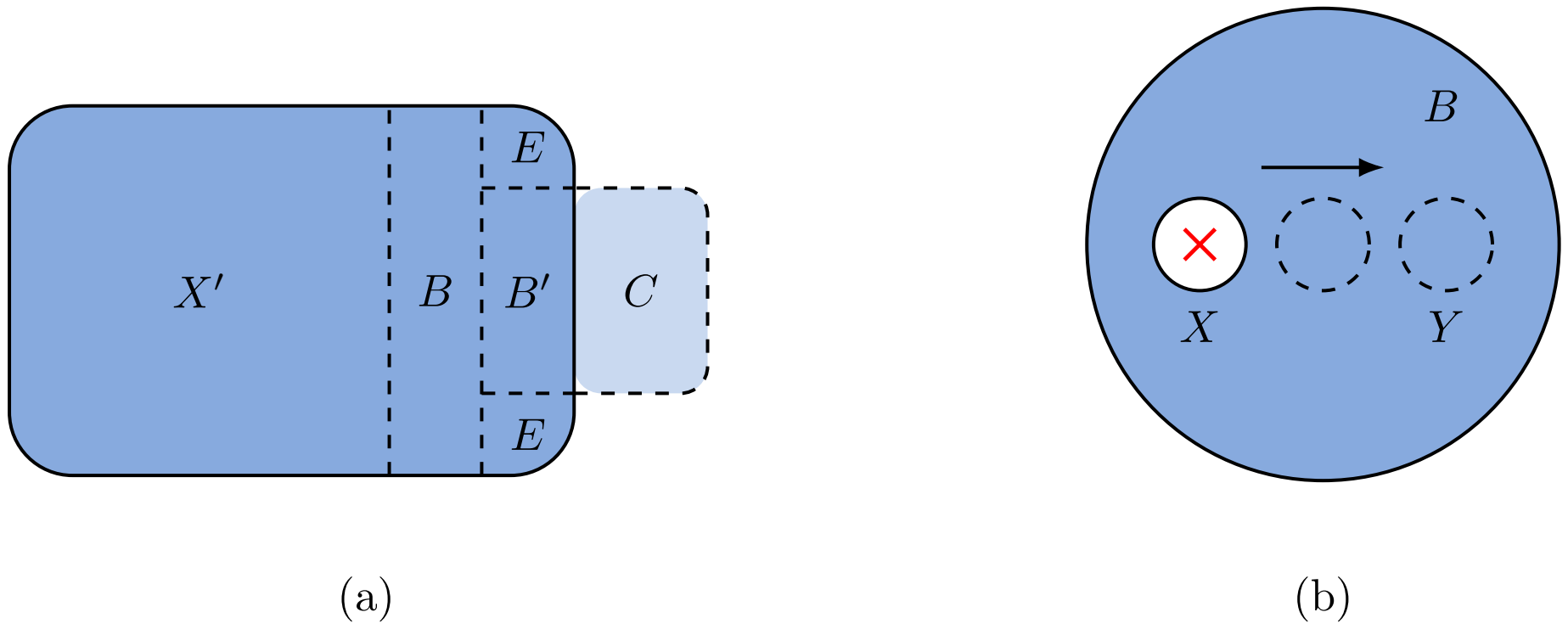
\begin{figure}[t]
    \centering
\begin{tabular}{c@{\hspace{8em}}c}
\begin{tikzpicture}[line width=0.75pt, scale=0.8, baseline={(0,1.8)}]
\pgfmathsetmacro{\xl}{0}   \pgfmathsetmacro{\xb}{4.6}  \pgfmathsetmacro{\xr}{5.5}
\pgfmathsetmacro{\yb}{0}   \pgfmathsetmacro{\yt}{3.6}
\pgfmathsetmacro{\ye}{0.8}
\fill[tikzblue, rounded corners=14pt] (\xl,\yb) rectangle (\xr,\yt);
\fill[tikzbluelight, rounded corners=6pt] (\xr,\ye) rectangle (\xr+1.3,\yt-\ye);
\draw[dashed, rounded corners=6pt] (\xr,\ye) -- (\xr+1.3,\ye) -- (\xr+1.3,\yt-\ye) -- (\xr,\yt-\ye);
\draw[rounded corners=14pt] (\xl,\yb) rectangle (\xr,\yt);
\draw[dashed] (\xb-0.9,\yb) -- (\xb-0.9,\yt);
\draw[dashed] (\xb,\yb) -- (\xb,\yt);
\draw[dashed] (\xb,\ye) -- (\xr,\ye);
\draw[dashed] (\xb,\yt-\ye) -- (\xr,\yt-\ye);
\node at ({(\xl+\xb-0.9)/2},{(\yb+\yt)/2}) {$X'$};
\node at ({\xb-0.45},{(\yb+\yt)/2}) {$B$};
\node at ({(\xb+\xr)/2},{(\yb+\yt)/2}) {$B'$};
\node at ({(\xb+\xr)/2},{\ye/2}) {$E$};
\node at ({(\xb+\xr)/2},{\yt-\ye/2}) {$E$};
\node at (\xr+0.65,{(\yb+\yt)/2}) {$C$};
\end{tikzpicture} & \begin{tikzpicture}[line width=0.75pt, scale=0.8, baseline={(0,0)}]
\fill[tikzblue] (0,0) circle (2.3);
\draw (0,0) circle (2.3);
\fill[white] (-1.2,0) circle (0.45);
\draw (-1.2,0) circle (0.45);
\draw[red, line width=0.9pt] (-1.35,-0.15) -- (-1.05,0.15);
\draw[red, line width=0.9pt] (-1.35,0.15) -- (-1.05,-0.15);
\draw[dashed] (0,0) circle (0.45);
\draw[dashed] (1.2,0) circle (0.45);
\draw[-{Latex[length=2mm]}] (-0.6,0.75) -- (0.6,0.75);
\node at (-1.2,-0.8) {$X$};
\node at (1.2,-0.8) {$Y$};
\node at (1.15,1.35) {$B$};
\end{tikzpicture}\\ \noalign{\vskip 20pt}
(a) & (b)
\end{tabular}
    \caption{\textbf{(a)} We illustrate the incremental growth step in the proof of Lemma \ref{lemma:general-growth_v2}. We expand state $\sigma_X$ on $X=X' B B' E$ to state $\sigma_{XC\setminus E}$  via channel acting on $B B'E \to B B'C.$ Region $X'$ can be arbitrarily large, while $BB'CE$ is $O(1)$-size, so the channel is local. \textbf{(b)} Moving an excitation (red cross), as in Proposition~\ref{prop:mobility}: the hole $X$ of the annulus $B\setminus X$ is deformed to $Y$ through the dashed positions, and the excitation moves with it.}
    \label{fig:general-growth}
\end{figure}  

In order to use Lemma~\ref{lemma:general-growth_v2} to build a state preparation circuit, one can choose the regions as in Fig.~\ref{fig:general-growth}, extending a ground state on $X$ to a ground state on $XC\setminus E$. By demanding Eq.~\eqref{eq:supps_assumption_v2} for $O(1)$-size balls, one can apply the recovery map repeatedly (Fig.~\ref{fig:growth-step}(c)), creating a ground state on a ball of radius $r$ using a circuit of $O(r)$ depth. The existence of ground state preparation circuit also follows from an alternative set of assumptions, which we state informally below. Both statements concern commuting and frustration-free Hamiltonian.
\begin{enumerate}
\item Since LTQO implies Eq.~\eqref{eq:supps_assumption_v2}, LTQOs on $O(1)$-size balls imply that a ground state on a ball $X$ of radius $r$ can be prepared in $O(r)$-depth sequential circuit.
\item LTQO and spherical LTQO on $O(1)$-size balls imply local extendibility (Theorem~\ref{thm:LTQO_local_extendibility}). Thus the reduced density matrix of the ground state on a ball $X$ of radius $r$ can be prepared in $O(r)$-depth sequential circuit.
\end{enumerate}

\section{Mobility of excitations}
\label{sec:mobility}

Localized excitations are called mobile when they are able to move through the system (without creating further excitations).  This motion may occur via the natural dynamics of the Hamiltonian or via local operations implemented externally.  These two viewpoints are related, because one can imagine adding local Hamiltonian terms that implement the desired local operations via dynamics.

A hallmark of fracton systems~\cite{Haah2011} is that some excitations are either immobile or at least have restricted mobility; they are unable to move in all directions.  This is also a feature of spatial inhomogeneities such as domain walls in a traditional topological phase: not all excitations can pass through the domain wall.

Here we show that states satisfying local purifiability and local extendibility do have mobile excitations.  In the entanglement bootstrap program, mobility of excitations may be shown through the development of string operators \cite{shi2020fusion}, making use of the isomorphism theorem.  Here we give a related argument, making use of the fixed-boundary isomorphism theorem (Theorem~\ref{thm:isomorphism-fixed-boundary}).

Consider a state $\sigma$ satisfying local purifiability and local extendibility, and a frustration-free Hamiltonian $H=\sum_Z(I-Q_Z)$ with $r_H<r_1$ that has $\sigma$ as a ground state and satisfies $(A,\ell)$-LTQO for every $r_1$-ball $A$, for a fixed $\ell>0$. An excitation, localized to a small ball $X$, refers to a state $\sigma^{(X)}$ that has zero energy outside $X$: $Q_Z\sigma^{(X)}=\sigma^{(X)}$ for every term whose support $Z$ is disjoint from $X$.  We say that the excitation is mobile (in any direction) if there exist local operations that can move the energy to any other small ball $Y$, by acting only along a path from $X$ to $Y$, which may be chosen arbitrarily.  

\begin{Proposition}
 \label{prop:mobility}
    Let $X=\bar{\Ba}_r(x_0)$ and $Y=\bar{\Ba}_r(x_1)$ be balls of radius $r\ge r_2$. For any state $\sigma^{(X)}$ with zero energy outside $X$, there exists a quantum channel $M_{B\to B}$ for some ball $B\supset X\cup Y$ such that $\id_{\RR^d\setminus B}\otimes M_{B\to B}[\sigma^{(X)}]$ has zero energy outside $Y$.
\end{Proposition}

The region $B$ is interpreted as a neighborhood of a path between $X$ and $Y$; see Fig.~\ref{fig:general-growth}(b). The channel $M$ is like a string operator that moves energy from $X$ to $Y$.\begin{proof}
Let $\bar{x}=(x_0+x_1)/2$ and $B=\Ba_R(\bar{x})$ for a sufficiently large $R$, say $R>4r+|x_1-x_0|/2$. Let $X^+:=X^{+(\ell+2r_1)}$. Since $\sigma^{(X)}$ has zero energy outside $X$, LTQO gives $\sigma^{(X)}|_A=\sigma_A$ for every $r_1$-ball $A$ at distance at least $\ell$ from $X$; hence $\sigma^{(X)}|_{B\setminus X^+}\in\IC(B\setminus X^+)$, with $\sigma^{(X)}$ itself as the $r_1$-consistent extension. The annuli $B\setminus\bar{\Ba}_{s(t)}(x_0+t(x_1-x_0))$, with $s(t)$ decreasing from $r+\ell+2r_1$ to $r$, are tame for all $t\in[0,1]$ while keeping $\partial B$ fixed. By Theorem~\ref{thm:isomorphism-fixed-boundary} and Remark~\ref{rmk:iso-steps}, with the complement of $B$ as the ancilla and the state on the outer collar of $B$ held fixed, there is a channel $M$ acting inside $B$ and trivially near $\partial B$ that maps $\sigma^{(X)}|_{B\setminus X^+}$ to an element of $\IC(B\setminus Y)$. (On $Y$ it may prepare any state.) Hence $\id_{\RR^d\setminus B}\otimes M[\sigma^{(X)}]$ has zero energy outside $Y$. Inside $B\setminus Y$, this is ensured by $r_1$-consistency, since $r_H<r_1$ places every term inside an $r_1$-ball. Outside, this is true because $M$ acts trivially there.
\end{proof}

\section{Circuit-invariant definition of topological entanglement entropy}
\label{sec:tee_invariant}

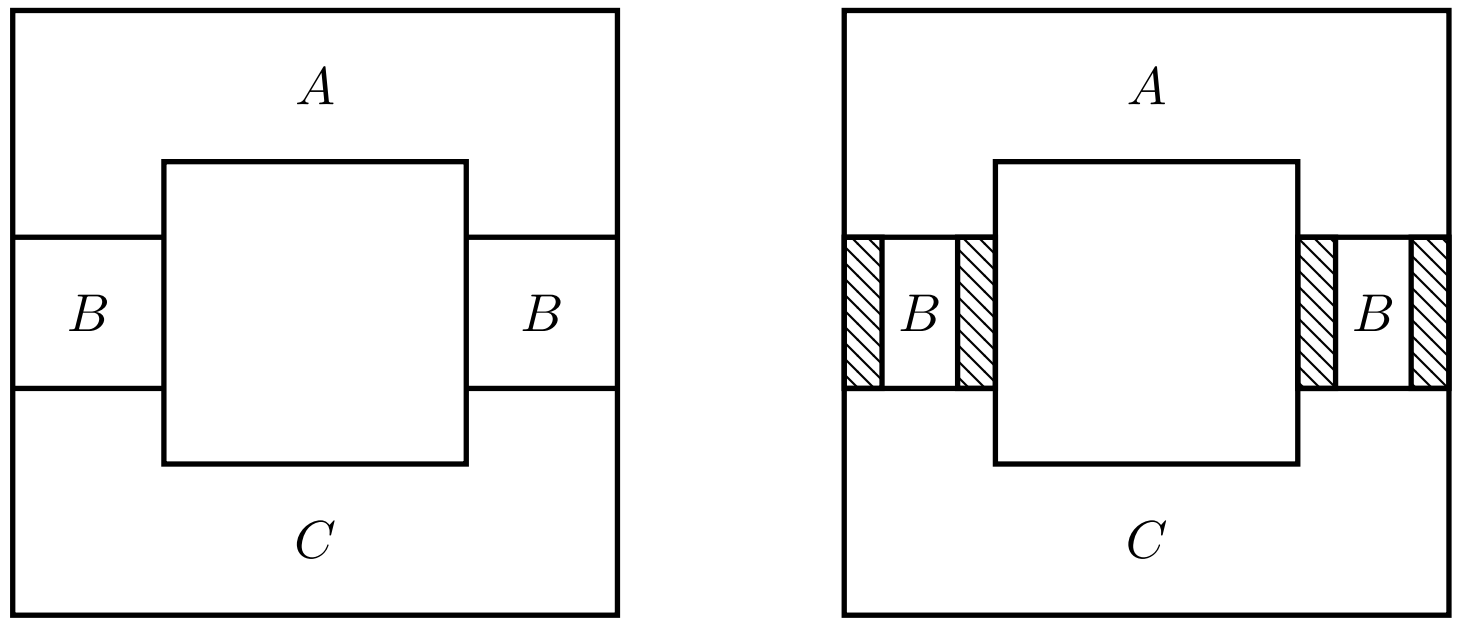
\begin{figure}[t]
    \centering
    \begin{tikzpicture}[line width=1pt]
        \draw[] (-2, -2) -- ++ (4,0) -- ++ (0, 4) -- ++ (-4, 0) -- cycle;
        \draw[] (-1, -1) -- ++ (2,0) -- ++ (0, 2) -- ++ (-2, 0) -- cycle;
        \draw[] (-1, -0.5) -- ++ (-1, 0);
        \draw[] (-1, 0.5) -- ++ (-1, 0);
        \draw[] (1, -0.5) -- ++ (1, 0);
        \draw[] (1, 0.5) -- ++ (1, 0);
        \node[] () at (0, 1.5) {$A$};
        \node[] () at (0, -1.5) {$C$};
        \node[] () at (-1.5, 0) {$B$};
        \node[] () at (1.5, 0) {$B$};
        \node[] () at (0, -2.5) {(a)};
        \begin{scope}[xshift=5.5cm]
        
        \draw[] (-2, -2) -- ++ (4,0) -- ++ (0, 4) -- ++ (-4, 0) -- cycle;
        \draw[] (-1, -1) -- ++ (2,0) -- ++ (0, 2) -- ++ (-2, 0) -- cycle;
        \draw[] (-1, -0.5) -- ++ (-1, 0);
        \draw[] (-1, 0.5) -- ++ (-1, 0);
        \draw[] (1, -0.5) -- ++ (1, 0);
        \draw[] (1, 0.5) -- ++ (1, 0);

        \draw[pattern=north west lines] (-1.25, -0.5) -- ++ (0, 1) -- ++ (0.25, 0) -- ++ (0, -1)  -- cycle;
        \draw[pattern=north west lines] (-2, -0.5) -- ++ (0, 1) -- ++ (0.25, 0) -- ++ (0, -1)  -- cycle;
        \begin{scope}[xscale=-1]
        \draw[pattern=north west lines] (-1.25, -0.5) -- ++ (0, 1) -- ++ (0.25, 0) -- ++ (0, -1)  -- cycle;
        \draw[pattern=north west lines] (-2, -0.5) -- ++ (0, 1) -- ++ (0.25, 0) -- ++ (0, -1)  -- cycle;    
        \end{scope}
        
        \node[] () at (0, 1.5) {$A$};
        \node[] () at (0, -1.5) {$C$};
        \node[] () at (-1.5, 0) {$B$};
        \node[] () at (1.5, 0) {$B$};
        \node[] () at (0, -2.5) {(b)};
            
        \end{scope}
    \end{tikzpicture}
    \caption{\textbf{(a)} Subsystems used in the definition of topological entanglement entropy. \textbf{(b)} Subsystems used in a new definition of topological entanglement entropy. The dashed region is $E$.}
    \label{fig:tee}
\end{figure}

For gapped ground states, the entanglement entropy of a ball is expected to obey an area law 
\begin{align}  \label{eq:area-law}
    S(A) \approx \alpha |\partial A| - \gamma
\end{align} 
for large regions, where $\alpha$ is some constant independent of the region~\cite{Kitaev2006,Levin2006}.  The constant term $\gamma$ is known as the topological entanglement entropy, though its definition through this formula is only heuristic. One can extract $\gamma$ by considering a judicious linear combination of entanglement entropies. For instance, 
\begin{equation} \label{eq:CMI_TEE}
    I(A:C|B)_{\rho} \approx 2\gamma 
\end{equation}
for a set of subsystems shown in Fig.~\ref{fig:tee}(a), whenever the area law of Eq.~\eqref{eq:area-law} holds. Thus Eq.~\eqref{eq:CMI_TEE} is often taken as the definition of topological entanglement entropy.  However, this topological entanglement entropy is not a true invariant of a phase, since a constant-depth circuit can change its value~\cite{Zou2016}.  This phenomenon is known as ``spurious'' topological entanglement entropy \cite{Zou2016,Williamson2019,Kato2020,kim2023universal}.

Using the notion of recoverability introduced in this paper,  we describe a new definition of topological entanglement entropy which is invariant under constant-depth circuits. This definition is formulated in terms of the \emph{fidelity of recovery}, defined as 
\begin{equation}
    F_{Y\to Z} (\rho_{XZ}\|\sigma_{XY}) := \max_{\Phi: \St_Y\to \St_Z} F(\rho_{XZ}, \id_X\otimes \Phi(\sigma_{XY})).
\end{equation}
Here $F(\cdot, \cdot)$ is the fidelity and the maximization is taken over CPTP  maps~\cite{Seshadreesan2015}. The domain and the image of the map is denoted using Roman letters for brevity.

We propose the following definition of topological entanglement entropy:
\begin{equation}
\boxed{
    \gamma := -\frac{1}{2} \log F_{BE\to BC} (\rho_{ABC} \| \rho_{ABE}),\label{eq:tee_def}
    }
\end{equation}
where the subsystems are shown in Fig.~\ref{fig:tee}(b). ($E$ is the union of dashed regions.) The main result of this section is that $\gamma$ defined this way enjoys a certain stability property. If $\gamma$ is invariant under topology-preserving deformations, it retains the same value for $\rho':=U\rho U^{\dagger}$ for any constant-depth circuit $U$, provided that the chosen subsystems are sufficiently thick compared to the depth of $U$. In order to make this statement precise, one would need to consider tame deformations (Definition~\ref{def:macro-region}). We leave such a treatment for future work.

Heuristically, the above circuit-invariance implies that Eq.~\eqref{eq:tee_def} becomes universal in the limit that the thicknesses of the subsystems become large compared to the correlation length.  To be more precise, when considering states with nonzero correlation length, we should consider adiabatic evolution connecting two states in the same phase. Here is a brief sketch of an argument that $\gamma$ in~\eqref{eq:tee_def} is approximately invariant under adiabatic evolution of the state, for large regions.  Note $\gamma$ is a function of the local state, and under adiabatic evolution, the new local state on a region of size $L$ can be well-approximated (e.g., in trace distance) by a circuit of depth $\text{poly}(\log(L))$ acting on the original state \cite{hastings2005quasiadiabatic}.  Then $\gamma$ is invariant under this circuit. The argument concludes by noting that the fidelity of recovery is continuous with respect to changes in the underlying state, without dimensional factors.  

A nontrivial hypothesis underlying this result is that there is a state for which $\gamma$ is invariant under topology-preserving deformations. We will show this in two spatial dimensions, assuming the entanglement bootstrap axioms~\cite{shi2020fusion}, including the structure of the information convex sets, the string operators, and the fusion rules. (It would be desirable to show this hypothesis also follows from the weaker extendibility property.) Therefore, our definition of topological entanglement entropy is an invariant of a phase if there is a representative wavefunction in the phase that satisfies the entanglement bootstrap axioms.

An earlier circuit-invariant definition for the topological entanglement entropy was introduced in Ref.~\cite{kato2020entropic}.  That definition requires a constant-range, commuting parent Hamiltonian, which implies zero correlation length, and it involves certain algebras defined using those Hamiltonian terms.   Therefore it is not immediately clear how to extend the definition to states with nonzero correlation length.  In contrast, Eq.~\eqref{eq:tee_def} may be calculated for such a state. 

Meanwhile, in Ref.~\cite{kim2023universal}, where it was shown that the spurious contribution to the TEE of \eqref{eq:CMI_TEE} is always positive, they gave another circuit-invariant formulation of TEE.  However, that definition involves an unwieldy minimization over circuits.  As a naive algorithm, it would involve a brute force search with a runtime doubly exponential in the region size or correlation length.  In contrast, the fidelity of recovery in Eq.~\eqref{eq:tee_def} is amenable to semidefinite program, hence with runtime only exponential in the region size.

\subsection{Fidelity of recovery}
\label{sec:for}

In this section, we review various properties of fidelity of recovery~\cite{Seshadreesan2015,Berta2016}. We will consider the following quantity:
\begin{equation}
    F_{BE\to BC}(\rho_{ABC} \| \rho_{ABE}) = \max_{\Phi:\St_{BE}\to \St_{BC}} F(\rho_{ABC}, \id_A\otimes \Phi(\rho_{ABE})),
\end{equation}
which we shall denote as $F(A:C|B;E)_{\rho}$ from now on. Here are the relevant properties.

\textbf{Unitary invariance}: For any unitary $U= U_A\otimes U_B\otimes U_C \otimes U_E$,
\begin{equation}
F(A:C|B;E)_{\rho} = F(A:C|B;E)_{U \rho U^{\dagger}}.
    \label{eq:unitary_invariance}
\end{equation}

\textbf{Monotonicity}: The following inequalities hold:
\begin{align}
    F(A:C|B;E)_{\rho} \leq F(A':C'|B';E')_{\rho} & \qquad\qquad (A'\subseteq A, B'C'\subseteq BC\text{, and } B'E'\supseteq BE),
    \label{eq:monotonicity1} \\
    F(A:C|B;E)_{\rho} \leq F(A':C'|B';E')_{\rho} & \qquad\qquad (C'\subseteq C, A'B'\subseteq AB\text{, and } B'E'\supseteq BE). \label{eq:monotonicity2}
\end{align}

These properties follow straightforwardly from the definition of the fidelity of the recovery. If $E=E'=\varnothing$, Eq.~\eqref{eq:unitary_invariance}, ~\eqref{eq:monotonicity1}, and~\eqref{eq:monotonicity2} are similar to the properties of the conditional mutual information $I(A:C|B)_{\rho} = S(\rho_{AB}) + S(\rho_{BC}) - S(\rho_B) - S(\rho_{ABC})$:
\begin{equation}
    \begin{aligned}
        I(A:C|B)_{\rho} &= I(A:C|B)_{U\rho U^{\dagger}} \qquad & (U = U_A\otimes U_B \otimes U_C), \\
        I(A':C'|B')_{\rho} &\leq I(A:C|B)_{\rho} \qquad & (A'\subseteq A, B'C'\subseteq BC, \text{ and } B'\supseteq B),\\
        I(A':C'|B')_{\rho} &\leq I(A:C|B)_{\rho} \qquad & (C'\subseteq C, A'B'\subseteq AB, \text{ and } B'\supseteq B).
    \end{aligned}
\end{equation}

\subsection{Stability under circuits}
\label{sec:stability}

Here we consider a set of subsystems topologically equivalent to what is shown in Fig.~\ref{fig:tee}(b). Deformations of a subsystem (say $X$) shall be denoted as $X_1, X_2, \widetilde{X},$ etc. Dashed regions in the diagrams represent $E$ or its deformations. 

Let $\rho$ be a state such that $F(A:C|B;E)_{\rho}$ attains the same value (say $F_0$) for any $ABCE$ topologically equivalent to what is shown in Fig.~\ref{fig:tee}(b). We prove that $F(A:C|B;E)_{U\rho U^{\dagger}}=F_0$, provided that (i) $U$ is a geometrically local depth-$d$ circuit and (ii) the thicknesses of the subsystems are at least $cd$ for some constant $c>1$.

The main idea is to use the properties of the fidelity of recovery to derive upper and lower bounds for $F(A:C|B;E)_{U\rho U^{\dagger}}$. These bounds will turn out to be of the form of $F(\widetilde{A}:\widetilde{C}|\widetilde{B};\widetilde{E})_{\rho}$, where $\widetilde{A}\widetilde{B}\widetilde{C}\widetilde{E}$ are topologically equivalent to $ABCE$ (Fig.~\ref{fig:tee}(b)). Then, by the hypothesis that $F(A:C|B;E)_{\rho} = F_0$, the claim follows.

\begin{figure}[t]
    \centering
    \begin{tikzpicture}[line width=1pt, xscale=0.8, scale=0.9]
    \draw[line width=4pt, draw=tikzblue] (-2.5, -2) -- ++ (5, 0) -- ++ (0, 4) -- ++ (-5, 0) -- cycle;
    \draw[line width=4pt, draw=tikzblue] (-0.5, -1) -- ++ (1, 0) -- ++ (0, 2) -- ++ (-1, 0) -- cycle;
    \draw[line width=4pt, draw=tikzblue] (-2.5, -0.5) -- ++ (2,0) -- ++ (0, 1) -- ++ (-2, 0) -- cycle;
    \draw[line width=4pt, draw=tikzblue] (0.5, -0.5) -- ++ (2,0) -- ++ (0, 1) -- ++ (-2, 0) -- cycle;
    \draw[line width=4pt, draw=tikzblue] (1, -0.5) -- ++ (0, 1);
    \draw[line width=4pt, draw=tikzblue] (2, -0.5) -- ++ (0, 1);
    \draw[line width=4pt, draw=tikzblue] (-1, -0.5) -- ++ (0, 1);
    \draw[line width=4pt, draw=tikzblue] (-2, -0.5) -- ++ (0, 1);
    
    \draw[] (-2.5, -2) -- ++ (5, 0) -- ++ (0, 4) -- ++ (-5, 0) -- cycle;
    \draw[] (-0.5, -1) -- ++ (1, 0) -- ++ (0, 2) -- ++ (-1, 0) -- cycle;
    \draw[] (-2.5, -0.5) -- ++ (2,0) -- ++ (0, 1) -- ++ (-2, 0) -- cycle;
    \draw[] (0.5, -0.5) -- ++ (2,0) -- ++ (0, 1) -- ++ (-2, 0) -- cycle;
    \draw[] (1, -0.5) -- ++ (0, 1);
    \draw[] (2, -0.5) -- ++ (0, 1);
    \draw[] (-1, -0.5) -- ++ (0, 1);
    \draw[] (-2, -0.5) -- ++ (0, 1);

    \draw[pattern=north west lines] (-2.5, -0.5) -- ++ (0.5, 0) -- ++ (0, 1) -- ++ (-0.5, 0) -- cycle;
    \draw[pattern=north west lines] (-1, -0.5) -- ++ (0.5, 0) -- ++ (0, 1) -- ++ (-0.5, 0) -- cycle;
    \begin{scope}[xscale=-1]
    \draw[pattern=north west lines] (-2.5, -0.5) -- ++ (0.5, 0) -- ++ (0, 1) -- ++ (-0.5, 0) -- cycle;
    \draw[pattern=north west lines] (-1, -0.5) -- ++ (0.5, 0) -- ++ (0, 1) -- ++ (-0.5, 0) -- cycle;
    \end{scope}

    \node[] () at (0, 1.5) {$A$};
    \node[] () at (-1.5, 0) {$B$};
    \node[] () at (1.5, 0) {$B$};
    \node[] () at (0, -1.5) {$C$};
    \node[] () at (0, -2.5) {(a)};

    \begin{scope}[xshift=6cm]
    \draw[line width=4pt, draw=tikzblue] (-2.375, -0.5) -- ++ (0.625,0);
    \draw[line width=4pt, draw=tikzblue] (-0.625, -0.5) -- ++ (-0.625,0);
    \draw[line width=4pt, draw=tikzblue] (-2.375, 0.5) -- ++ (1.75,0);

    \draw[line width=4pt, draw=tikzblue] (2.375, -0.5) -- ++ (-0.625,0);
    \draw[line width=4pt, draw=tikzblue] (0.625, -0.5) -- ++ (0.625,0);
    \draw[line width=4pt, draw=tikzblue] (2.375, 0.5) -- ++ (-1.75,0);

    \draw[] (-2.25, 0.5) -- (-2.25, 1.75) -- (2.25, 1.75) -- (2.25, 0.5);
    \draw[] (-2.25, -0.5) -- (-2.25, -1.75) -- (2.25, -1.75) -- (2.25, -0.5);
    \draw[] (-0.75, 0.5) -- (-0.75, 1.25) -- (0.75, 1.25) -- (0.75, 0.5);
    \draw[] (-0.75, -0.5) -- (-0.75, -1.25) -- (0.75, -1.25) -- (0.75, -0.5);

    \draw[] (-2.25, 0.5) -- ++ (1.5,0);
    \draw[] (-2.25, -0.5) -- ++ (0.375, 0);
    \draw[] (-1.125, -0.5) -- ++ (0.375, 0);
    \draw[] (-2.25, -0.75) -- ++ (1.5,0);

    \draw[] (2.25, 0.5) -- ++ (-1.5,0);
    \draw[] (2.25, -0.5) -- ++ (-0.375, 0);
    \draw[] (1.125, -0.5) -- ++ (-0.375, 0);
    \draw[] (2.25, -0.75) -- ++ (-1.5,0);

    \draw[] (1.125, -0.5) -- ++ (0, 1);
    \draw[] (1.875, -0.5) -- ++ (0, 1);
    \draw[] (-1.125, -0.5) -- ++ (0, 1);
    \draw[] (-1.875, -0.5) -- ++ (0, 1);

    \draw[pattern=north west lines] (-2.25, 0.75) -- ++ (-0.375, 0) -- ++ (0, -1.5) -- ++ (0.375, 0) -- ++ (0, 0.25) -- ++ (0.375, 0)--++ (0, 1) -- ++ (-0.375, 0) -- cycle;
    \draw[pattern=north west lines] (-0.75, 0.75) -- ++ (0.375, 0) -- ++ (0, -1.5) -- ++ (-0.375, 0) -- ++ (0, 0.25) -- ++ (-0.375, 0) -- ++ (0, 1) -- ++ (0.375, 0) -- cycle;

    \draw[pattern=north west lines] (2.25, 0.75) -- ++ (0.375, 0) -- ++ (0, -1.5) -- ++ (-0.375, 0) -- ++ (0, 0.25) -- ++ (-0.375, 0)--++ (0, 1) -- ++ (0.375, 0) -- cycle;
    \draw[pattern=north west lines] (0.75, 0.75) -- ++ (-0.375, 0) -- ++ (0, -1.5) -- ++ (0.375, 0) -- ++ (0, 0.25) -- ++ (0.375, 0) -- ++ (0, 1) -- ++ (-0.375, 0) -- cycle;

    \node[] () at (0, 1.5) {$A_1$};
    \node[] () at (-1.5, 0) {$B_1$};
    \node[] () at (1.5, 0) {$B_1$};
    \node[] () at (0, -1.5) {$C_1$};
    \node[] () at (0, -2.5) {(b)};
    \end{scope}

    \begin{scope}[xshift=12cm]
    \draw[line width=4pt, draw=tikzblue] (-2.375, -0.5) -- ++ (0.625,0);
    \draw[line width=4pt, draw=tikzblue] (-0.625, -0.5) -- ++ (-0.625,0);
    \draw[line width=4pt, draw=tikzblue] (-2.375, 0.5) -- ++ (0.625,0);
    \draw[line width=4pt, draw=tikzblue] (-0.625, 0.5) -- ++ (-0.625,0);

    \draw[line width=4pt, draw=tikzblue] (2.375, -0.5) -- ++ (-0.625,0);
    \draw[line width=4pt, draw=tikzblue] (0.625, -0.5) -- ++ (0.625,0);
    \draw[line width=4pt, draw=tikzblue] (2.375, 0.5) -- ++ (-0.625,0);
    \draw[line width=4pt, draw=tikzblue] (0.625, 0.5) -- ++ (0.625,0);

    \draw[] (-2.25, 0.5) -- (-2.25, 1.75) -- (2.25, 1.75) -- (2.25, 0.5);
    \draw[] (-2.25, -0.5) -- (-2.25, -1.75) -- (2.25, -1.75) -- (2.25, -0.5);
    \draw[] (-0.75, 0.5) -- (-0.75, 1.25) -- (0.75, 1.25) -- (0.75, 0.5);
    \draw[] (-0.75, -0.5) -- (-0.75, -1.25) -- (0.75, -1.25) -- (0.75, -0.5);

    \draw[] (-2.25, 0.75) -- ++ (1.5,0);
    \draw[] (-2.25, 0.5) -- ++ (0.375, 0);
    \draw[] (-1.125, 0.5) -- ++ (0.375, 0);
    \draw[] (-2.25, -0.5) -- ++ (0.375, 0);
    \draw[] (-1.125, -0.5) -- ++ (0.375, 0);
    \draw[] (-2.25, -0.75) -- ++ (1.5,0);

    \draw[] (2.25, 0.75) -- ++ (-1.5,0);
    \draw[] (2.25, 0.5) -- ++ (-0.375, 0);
    \draw[] (1.125, 0.5) -- ++ (-0.375, 0);
    \draw[] (2.25, -0.5) -- ++ (-0.375, 0);
    \draw[] (1.125, -0.5) -- ++ (-0.375, 0);
    \draw[] (2.25, -0.75) -- ++ (-1.5,0);

    \draw[] (1.125, -0.5) -- ++ (0, 1);
    \draw[] (1.875, -0.5) -- ++ (0, 1);
    \draw[] (-1.125, -0.5) -- ++ (0, 1);
    \draw[] (-1.875, -0.5) -- ++ (0, 1);
    
    \draw[pattern=north west lines] (-2.25, 0.75) -- ++ (-0.375, 0) -- ++ (0, -1.5) -- ++ (0.375, 0) -- ++ (0, 0.25) -- ++ (0.375, 0)--++ (0, 1) -- ++ (-0.375, 0) -- cycle;
    \draw[pattern=north west lines] (-0.75, 0.75) -- ++ (0.375, 0) -- ++ (0, -1.5) -- ++ (-0.375, 0) -- ++ (0, 0.25) -- ++ (-0.375, 0) -- ++ (0, 1) -- ++ (0.375, 0) -- cycle;

    \draw[pattern=north west lines] (2.25, 0.75) -- ++ (0.375, 0) -- ++ (0, -1.5) -- ++ (-0.375, 0) -- ++ (0, 0.25) -- ++ (-0.375, 0)--++ (0, 1) -- ++ (0.375, 0) -- cycle;
    \draw[pattern=north west lines] (0.75, 0.75) -- ++ (-0.375, 0) -- ++ (0, -1.5) -- ++ (0.375, 0) -- ++ (0, 0.25) -- ++ (0.375, 0) -- ++ (0, 1) -- ++ (-0.375, 0) -- cycle;

    \node[] () at (0, 1.5) {$A_2$};
    \node[] () at (-1.5, 0) {$B_2$};
    \node[] () at (1.5, 0) {$B_2$};
    \node[] () at (0, -1.5) {$C_2$};
    \node[] () at (0, -2.5) {(c)};
    \end{scope}

    \begin{scope}[xshift=18cm]

    \draw[] (-2.25, 0.625) -- (-2.25, 1.75) -- (2.25, 1.75) -- (2.25, 0.625);
    \draw[] (-2.25, -0.625) -- (-2.25, -1.75) -- (2.25, -1.75) -- (2.25, -0.625);
    \draw[] (-0.75, 0.625) -- (-0.75, 1.25) -- (0.75, 1.25) -- (0.75, 0.625);
    \draw[] (-0.75, -0.625) -- (-0.75, -1.25) -- (0.75, -1.25) -- (0.75, -0.625);

    \draw[] (-2.25, 0.75) -- ++ (1.5,0);
    \draw[] (-2.25, -0.75) -- ++ (1.5,0);

    \draw[] (2.25, 0.75) -- ++ (-1.5,0);
    \draw[] (2.25, -0.75) -- ++ (-1.5,0);

    \draw[] (1.125, -0.375) -- ++ (0, 0.75);
    \draw[] (1.875, -0.375) -- ++ (0, 0.75);
    \draw[] (-1.125, -0.375) -- ++ (0, 0.75);
    \draw[] (-1.875, -0.375) -- ++ (0, 0.75);
    
    \draw[pattern=north west lines] (-2.25, 0.75) -- ++ (-0.375, 0) -- ++ (0, -1.5) -- ++ (0.375, 0) -- ++ (0, 0.125) -- ++ (0.5, 0)--++ (0, 0.25) -- ++ (-0.125, 0) -- ++ (0, 0.75) -- ++ (0.125, 0) -- ++ (0, 0.25) -- ++  (-0.5, 0) -- cycle;
    \draw[pattern=north west lines] (-0.75, 0.75) -- ++ (0.375, 0) -- ++ (0, -1.5) -- ++ (-0.375, 0) -- ++ (0, 0.125) -- ++ (-0.5, 0) -- ++ (0, 0.25) -- ++ (0.125, 0) -- ++ (0, 0.75) -- ++ (-0.125, 0) -- ++ (0, 0.25) -- ++ (0.5, 0) -- cycle;

    \begin{scope}[xscale=-1]
    \draw[pattern=north west lines] (-2.25, 0.75) -- ++ (-0.375, 0) -- ++ (0, -1.5) -- ++ (0.375, 0) -- ++ (0, 0.125) -- ++ (0.5, 0)--++ (0, 0.25) -- ++ (-0.125, 0) -- ++ (0, 0.75) -- ++ (0.125, 0) -- ++ (0, 0.25) -- ++  (-0.5, 0) -- cycle;
    \draw[pattern=north west lines] (-0.75, 0.75) -- ++ (0.375, 0) -- ++ (0, -1.5) -- ++ (-0.375, 0) -- ++ (0, 0.125) -- ++ (-0.5, 0) -- ++ (0, 0.25) -- ++ (0.125, 0) -- ++ (0, 0.75) -- ++ (-0.125, 0) -- ++ (0, 0.25) -- ++ (0.5, 0) -- cycle;
    \end{scope}

    \node[] () at (0, 1.5) {$A_3$};
    \node[] () at (-1.5, 0) {$B_3$};
    \node[] () at (1.5, 0) {$B_3$};
    \node[] () at (0, -1.5) {$C_3$};
    \node[] () at (0, -2.5) {(d)};
    \end{scope}
        
    \end{tikzpicture}
    \caption{A sequence of subsystems used in the proof of the stability under circuit. These subsystems satisfy the following relations. (i) $A_1\subset A, B_1C_1\subset BC, B_1E_1\supset BE$, (ii) $C_2= C_1, A_2B_2 = A_1B_1, B_2E_2\supseteq B_1E_1$, and (iii) $C_3= C_2, A_3B_3 \subseteq A_2B_2, B_3E_3= B_2E_2$.}
    \label{fig:tee_newproof}
\end{figure}
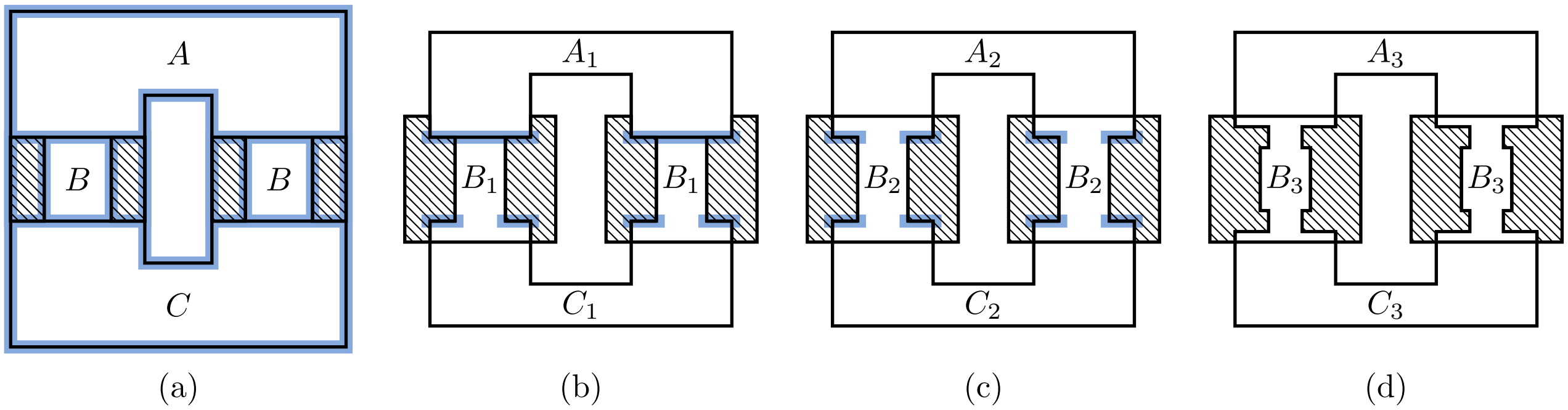

The bounds are derived as follows. Note Eq.~\eqref{eq:unitary_invariance} implies that gates in the interior of each region can be removed (Fig.~\ref{fig:tee_newproof}(a)). Let the remaining unitary to be $U_0$. Eq.~\eqref{eq:monotonicity1} implies 
\begin{equation}
    F(A:C|B;E)_{U_0\rho U^{\dagger}_0} \leq F(A_1:C_1|B_1;E_1)_{U_0 \rho U_0^{\dagger}}
\end{equation}
for the subsystems in Fig.~\ref{fig:tee_newproof}(b). Then gates in the interior of $A_1, B_1, C_1,$ and $E_1$ can be removed, resulting in a unitary $U_1$, shown in blue in Fig.~\ref{fig:tee_newproof}(b). We can similarly apply Eq.~\eqref{eq:monotonicity2} to obtain
\begin{equation}
    F(A_k:C_k|B_k;E_k)_{U_k\rho U^{\dagger}_k} \leq F(A_{k+1}:C_{k+1}|B_{k+1};E_{k+1})_{U_{k+1} \rho U_{k+1}^{\dagger}},
\end{equation}
for $k=1, 2$, where $U_{k+1}$ is obtained from $U_k$ by removing gates in the interior of $A_{k+1}, B_{k+1}, C_{k+1},$ and $E_{k+1}$ (Fig.~\ref{fig:tee_newproof}(c,d)). Since $U_3$ is the identity, we obtain 
\begin{equation}
    F(A:C|B;E)_{U\rho U^{\dagger}} \leq F(A_3:C_3|B_3;E_3)_{\rho}.
\end{equation}

Thus we get
\begin{equation}
    F(A:C|B;E)_{U\rho U^{\dagger}}\leq F(\widetilde{A}:\widetilde{C}|\widetilde{B};\widetilde{E})_{\rho}
\end{equation}
for some $\widetilde{A}\widetilde{B}\widetilde{C}\widetilde{E}$ topologically equivalent to $ABCE$. Similarly, viewing $\rho$ as $\rho= U^{\dagger}(U\rho U^{\dagger})U$, we can also obtain a lower bound:
\begin{equation}
   F(\widetilde{A}':\widetilde{C}'|\widetilde{B}';\widetilde{E}')_{\rho} \leq F(A:C|B;E)_{U\rho U^{\dagger}}
\end{equation}
for some $\widetilde{A}'\widetilde{B}'\widetilde{C}'\widetilde{E}'$ topologically equivalent to $ABCE$. As such, under the hypothesis that $F(\widetilde{A}:\widetilde{C}|\widetilde{B};\widetilde{E})_{\rho}= F(\widetilde{A}':\widetilde{C}'|\widetilde{B}';\widetilde{E}')_{\rho}=F_0$, we conclude $F(A:C|B;E)_{U\rho U^{\dagger}}=F_0$.

\subsection{Fidelity of recovery in entanglement bootstrap}
\label{sec:deformation_invariance}

Let $\rho$ be a state that satisfies the entanglement bootstrap axioms~\cite{shi2020fusion}. \textbf{A1} implies
\begin{equation}
    F(A:C|B; E) = F(A:C|B; \varnothing).
\end{equation}
In particular, we can obtain a lower bound to $F(A:C|B; \varnothing)$ by choosing a specific channel, namely the Petz map. This yields a maximum-entropy state of the information convex set on an annulus, yielding the following lower bound
\begin{equation}
\begin{aligned}
    F(A:C|B;\varnothing) &\geq  F\left(\sum_a \frac{d_a^2}{\mathcal{D}^2}\rho_a, \rho_1\right) \\
    &=\frac{1}{\mathcal{D}^2},
\end{aligned}
\label{eq:lower_bound}
\end{equation}
where $\{ \rho_a\}$ is a set of extreme points, $N_{ab}^c$ is the fusion multiplicity, $d_a$ is the quantum dimension, and $\mathcal{D} = \sqrt{\sum_a d_a^2}$ is the total quantum dimension. 

\begin{figure}[ht]
    \centering
    \begin{tikzpicture}[line width=1pt, scale=0.9]
    \draw[] (-2, 0.5) -- ++ (0, 1.5) -- ++ (4, 0) -- ++ (0, -1.5) -- ++ (-1, 0) -- ++ (0, 0.5) -- ++ (-2, 0) -- ++ (0, -0.5) -- cycle;
    \draw[] (-2, 0.5) -- ++ (1,0) -- ++ (0, -1) -- ++ (-1, 0) -- cycle;
    \draw[] (2, 0.5) -- ++ (-1,0) -- ++ (0, -1) -- ++ (1, 0) -- cycle;
    \node[] () at (-1.5, 0) {$B$};
    \node[] () at (1.5, 0) {$B$};
    \node[] () at (0, 1.5) {$A$};
    \node[] () at (0, -1) {(a)};
    
    \begin{scope}[xshift=4.5cm]
    \draw[] (-2, 0.5) -- ++ (0, 1.5) -- ++ (4, 0) -- ++ (0, -1.5) -- ++ (-1, 0) -- ++ (0, 0.5) -- ++ (-2, 0) -- ++ (0, -0.5) -- cycle;
    \node[] () at (0, 1.5) {$A$};
    \draw[] (-0.5,2) arc (180:0:0.5cm) -- cycle;
    \draw[] (-0.5,1) arc (180:360:0.5cm) -- cycle;
    \node[] () at (0, 2.25) {$\delta A$};
    \node[] () at (0, 0.75) {$\delta A$};
    \draw[] (-2, 0.5) -- ++ (1,0) -- ++ (0, -1) -- ++ (-1, 0) -- cycle;
    \draw[] (2, 0.5) -- ++ (-1,0) -- ++ (0, -1) -- ++ (1, 0) -- cycle;
    \node[] () at (-1.5, 0) {$B$};
    \node[] () at (1.5, 0) {$B$};
    \node[] () at (0, -1) {(b)};
    \end{scope}

    \begin{scope}[xshift=9cm]
    \draw[] (-2, 0.5) -- ++ (0, 1.5) -- ++ (4, 0) -- ++ (0, -1.5) -- ++ (-1, 0) -- ++ (0, 0.5) -- ++ (-2, 0) -- ++ (0, -0.5) -- cycle;
    \draw[] (-0.5,2) arc (180:0:0.5cm) -- cycle;
    \draw[] (-0.5,1) arc (180:360:0.5cm) -- cycle;
    \draw[red] plot[smooth,tension=.5]
  coordinates{(0,2.25)  (0.125, 1.75) (0, 1.5) (-0.125, 1.25) (0, 0.75)};
    \filldraw[red] (0, 2.25)  circle (2pt);
    \filldraw[red] (0, 0.75)  circle (2pt);
    \node[left] () at (-0.375, 0.75)  {$b$};
    \node[left] () at (-0.375, 2.25)  {$\overline{b}$};
    \draw[] (-2, 0.5) -- ++ (1,0) -- ++ (0, -1) -- ++ (-1, 0) -- cycle;
    \draw[] (2, 0.5) -- ++ (-1,0) -- ++ (0, -1) -- ++ (1, 0) -- cycle;
    \node[] () at (-1.5, 0) {$B$};
    \node[] () at (1.5, 0) {$B$};
    \node[] () at (0, -1) {(c)};
    \end{scope}

    \begin{scope}[xshift=13.5cm]
    \draw[] (-2, 0.5) -- ++ (0, 1.5) -- ++ (4, 0) -- ++ (0, -1.5) -- ++ (-1, 0) -- ++ (0, 0.5) -- ++ (-2, 0) -- ++ (0, -0.5) -- cycle;
    \node[] () at (0, 1.5) {$A$};
    \draw[] (-2, 0.5) -- ++ (1,0) -- ++ (0, -1) -- ++ (-1, 0) -- cycle;
    \draw[] (2, 0.5) -- ++ (-1,0) -- ++ (0, -1) -- ++ (1, 0) -- cycle;
    \node[] () at (-1.5, 0) {$B$};
    \node[] () at (1.5, 0) {$B$};
    \node[] () at (0, -1) {(d)};    
    \end{scope}
    \end{tikzpicture}
    \caption{\textbf{(a)} A disk-like subsystem. \textbf{(b)} One can extend the subsystem by two balls $\delta A$, by applying a local recovery map. \textbf{(c)} A string operator is applied, with the endpoints located in the extended regions. \textbf{(d)} The extended region is traced out.}
    \label{fig:twirling}
\end{figure}
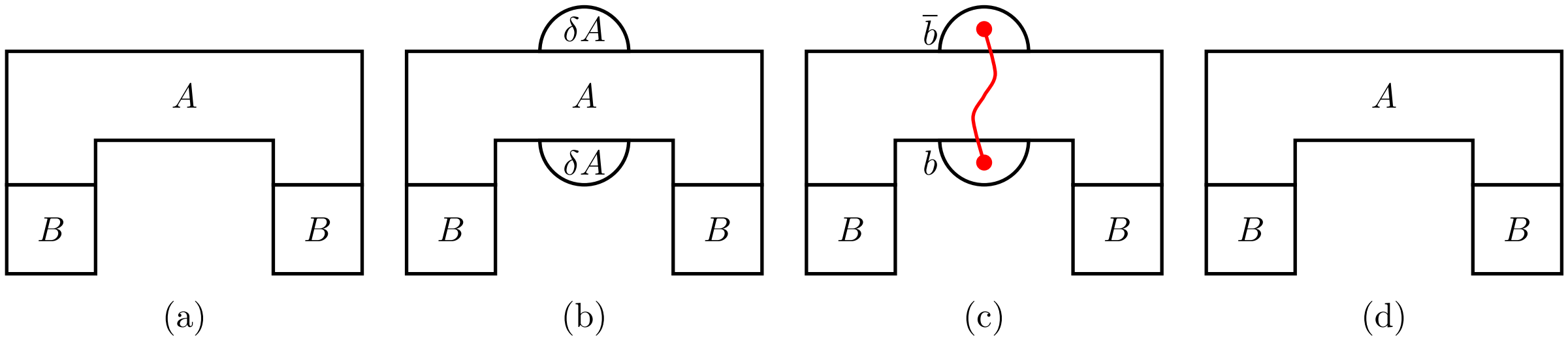

Now we aim to prove a matching upper bound. To that end, we consider the following \emph{twirling operation} $\mathcal{T}$ (Fig.~\ref{fig:twirling}). It consists of (i) extending $A$ by two balls (using \textbf{A1}), (ii) applying a string operator that creates anyon $b$ and $\overline{b}$ in the interior and the exterior region of the annulus, and (iii) tracing out the balls. Each string operator is applied with probability $\frac{d_b^2}{\mathcal{D}^2}$. After applying this map to $\rho_a$, we obtain
\begin{equation}
    \sum_{b, c} \frac{d_b^2}{\mathcal{D}^2} \frac{d_c N_{ab}^c}{d_a d_b} \rho_c = \sum_c \frac{d_c^2}{\mathcal{D}^2} \rho_c,
    \label{eq:twirl_action}
\end{equation}
independent of $a$. Thus the twirling operation maps any element in the information convex set to the maximum-entropy state. On the other hand, $\mathcal{T}(\rho_{AB}) = \rho_{AB}$ because the string operator is deformable.

Now we can upper bound $F(A:C|B;\varnothing)$ as follows:
\begin{equation}
\begin{aligned}
    F(A:C|B;\varnothing) &= \max_{\Phi: \St_B\to \St_{BC}} F(\rho_{ABC}, \id_A\otimes \Phi(\rho_{AB}))  \\ 
    &\leq \max_{\Phi:\St_B\to \St_{BC}} F(\mathcal{M}(\rho_{ABC}), \mathcal{M}(\id_A\otimes \Phi(\rho_{AB})) ),
\end{aligned}
\end{equation}
where $\mathcal{M}:=  \mathcal{M}_2 \circ \mathcal{M}_1$ is a composition of CPTP maps. Below are the descriptions of each map:
\begin{itemize}
    \item $\mathcal{M}_1$: One measures a parent Hamiltonian localized to a region $ABC\cap (BC)^+$. The measurement outcome is stored in a separate register. 
    \item $\mathcal{M}_2$: One traces out the boundary regions of $BC$ non-adjacent to $A$.
\end{itemize}
Under the entanglement bootstrap axioms, there is a local commuting parent Hamiltonian~\cite{kim2024strict} (whose ground state energy we set to zero). Therefore, after applying $\mathcal{M}$ to $\id_A\otimes \Phi(\rho_{AB})$, for one possible post-measurement state, one obtains an element of the information convex set on a thinner annulus (plus an extra register that stores the measurement outcome). Let $X$ be this thinner annulus. Under $\mathcal{M}$, $\rho_{ABC}$ becomes $\rho_{X,1}\otimes |0{\rangle}{\langle} 0|$, where $\rho_{X,a}$ denotes the sector-$a$ extreme point of the information convex set of the thinner annulus, so that $\rho_{X,1}$ is its vacuum sector.  Also, because $\Phi$ and $\mathcal{M}$ commute with $\mathcal{T}$, 
\begin{equation}
\begin{aligned}
    \mathcal{M}(\id_A\otimes \Phi(\rho_{AB})) &= \mathcal{M}(\id_A\otimes \Phi( \mathcal{T}(\rho_{AB}))) \\
    &= \mathcal{T}(\mathcal{M}(\id_A\otimes \Phi(\rho_{AB}))),
\end{aligned}
\end{equation}
which must be of the following form~\eqref{eq:twirl_action}:
\begin{equation}
    \mathcal{T}(\mathcal{M}(\id_A\otimes \Phi(\rho_{AB}))) = q \left(\sum_a \frac{d_a^2}{\mathcal{D}^2} \rho_{X,a} \otimes |0{\rangle}{\langle}0| \right) + (1-q)\widetilde{\rho},
\end{equation}
where $\widetilde{\rho} (I\otimes |0\rangle)=0$ for some $0\leq q\leq 1$. Thus we have
\begin{equation}
    F(A:C|B;\varnothing) \leq q \frac{1}{\mathcal{D}^2}.
\end{equation}
Due to the lower bound~\eqref{eq:lower_bound}, the maximum value of $q$ must be $1$, and we obtain $F(A:C|B;\varnothing) = \frac{1}{\mathcal{D}^2}$. Thus
\begin{equation}
    F(A:C|B;E) = \frac{1}{\mathcal{D}^2}.
\end{equation}

Recall the conventional entropy-based topological entanglement entropy $\gamma$ is given by $I(A:C|BE)_{\rho}\approx 2\gamma$ (Fig.~\ref{fig:tee}(b)). From Ref.~\cite{fawzi2015quantum}, it follows that
\begin{equation}
    -\log F(A:C|BE; \varnothing) \leq I(A:C|BE)_{\rho}.
\end{equation}
Moreover, since $F(A:C|BE; \varnothing) \leq F(A:C|B;E)$, we get
\begin{equation}
    -\log F(A:C|B; E) \leq I(A:C|BE)_{\rho}.
\end{equation}
Therefore, we also obtain a universal lower bound on the conventional definition of topological entanglement entropy~\cite{kim2023universal}.

\vspace{\baselineskip}
\noindent\textbf{AI Statement.} The original main text, statements, and the proofs were developed by the authors. The AI models (Claude and ChatGPT) were used to revise the manuscript and correct some statements and proofs. In particular, the models found an error in the original version of Lemma~\ref{lem:retraction_step} and proposed a simpler lemma; this manuscript uses that lemma. Both models were used to formalize selected results in Lean~\cite{extendibility_lean}; the constants used in Section~\ref{sec:rev_extendibility_merging} were adjusted to assist the formalization. The authors take responsibility for all content.

\section*{Acknowledgments}
We thank Milo Moses for collaboration on future related work and extensive comments. DR and AK acknowledge support by the Simons Foundation under grant 376205. AK also acknowledges funding provided by the Institute of Quantum Information and Matter, an NSF Physics Frontiers Center. IK acknowledges support from NSF under award number PHY-2337931 and Sloan Research Fellowship. This research was supported in part by grant NSF PHY-2309135 to the Kavli Institute for Theoretical Physics (KITP) and also performed in part at Aspen Center for Physics, which is supported by National Science Foundation grant PHY-2210452. 

\bibliographystyle{unsrtlinks}
{\hypersetup{urlcolor=black}
\bibliography{references}}

\end{document}